# Spectroscopic Studies of Iron-based Superconductors, Multi-ferroic Oxides and Double-perovskite: Phonons, Electronic and Spin Excitations

A thesis

Submitted for the Degree of

**Doctor of Philosophy**

in the Faculty of Science

by

**Pradeep Kumar**

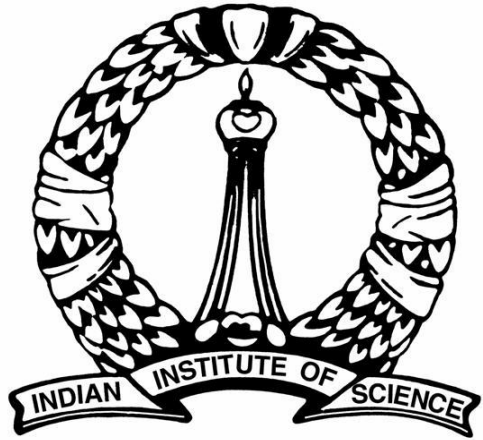

DEPARTMENT OF PHYSICS

INDIAN INSTITUTE OF SCIENCE

BANGALORE - 560012

INDIA

JANUARY 2014

i

**Dedicated to my Parents**

# Acknowledgments

I have spent exciting and enjoyable six and half years at IISc, largely due to the wonderful people that I have had the good fortune to associate with. I take this opportunity to sincerely thank them all for their help and support.

Firstly, I would like to thank my supervisors *Padam Shree* Prof. Ajay Sood and Dr. D. Victor S. Muthu for giving me the opportunity to work with them. I am very grateful to Prof. Ajay Sood for his advice, encouragement, and support over these years. His enthusiasm for new ideas is infectious, and has greatly helped me to broaden my horizons in physics. I have enjoyed learning from him various academic and non academic lessons, and consider myself very fortunate to be his student. I remember clearly, when I was doing my first experiment on iron-based superconductor and could not get the signal for some days even at low temperatures, I was becoming discouraged. It was Prof. Ajay Sood who came and told me not to worry and that I need to try a bit more to get the signal. That was very motivating for me. When I was doing my low temperature experiments, which usually run for 50-60 hours each, he used to come to lab at dawn during his morning walks, and ask about the experiment. This used to act like morning coffee for me and would refresh as well as energise me. I am very grateful to Dr. D. Victor S. Muthu for teaching me the technique of Raman spectroscopy and every possible minute detail of the instrument, optical alignment and his easy availability for all the troubleshooting in Dilor room.

Secondly, I thank all my collaborators. I would like to thank *Bharat Ratna* Prof. C.N.R. Rao and his students (JNCASR) for $TbMnO_3$ and $AlFeO_3$ samples. I am very grateful to Prof. Umesh Waghmare (JNCASR) and his student Dr. Anil Kumar and Sharmila for very fruitful collaboration. I thank Prof. Ashok Ganguli (IIT Delhi) and his student Dr. Jai Prakash; Prof. Surjit Singh, now at IISER Pune, and Prof. B. Buchner (IFW Dresden); Prof. A. Iyo and Dr. P.M. Shirage (Tsukuba, Japan) for superconducting samples and Prof. A. Sundaresan and his students (JNCASR) for double-perovskite samples.

I thank the office staff of the department, Mr. Srivatsa, Ms. Rakma, Ms. Meena, Ms. Bhargavi, Ms. Rekha and Mr. Mariyappa for their timely help. I thank Mr. Sherif in the department workshop for his help in fabrication of connector for the magnetic field system. I thank the staff, especially Raju sir, of Cryogenic facility of the Institute for the timely supply of liquid nitrogen and helium for my experiments and for their help during troubleshooting of cryogenic related problems.

I thank all my labmates past (Ajay Negi, Surajit Saha, Manas Khan, Kamaraju, Anindya (now Prof. in our dept.), Karthikeyan, Prem, Biswanath, Sunil, Sayantan, Sudakshina, Vijay Ravi, Harini, Harsh and Debaleena) and present (Vikram Rathee, Vasu (who never lies!!), Nitin, Satya, Sudheesh, Gupta, Anup, Shreyas, Achintya, Pradeep Bera, Swetha, Gyan, Shrabani, Ajoy, Dileep, Suresh, Rajesh, Dipti, Rema and Anitha Mam) for providing me with a lively and enjoyable environment during my stay. I am very grateful to Surajit for teaching me the basics of Raman technique and low temperature experiments; I enjoyed every bit of discussion I had ever with him. Special thanks to Rathee (the all weather friend) for being with me during high and low times of my six and half years of stay, discussing anything under the sun and providing all the help. He is really a true friend and has proved that a friend in need is friend indeed.

I thank my IISc batch-mates – Anupam, Alok, Kaushik, Anji, Geetanjali, and Mohan for their help and support. I would like to sincerely thank all the hockey club members, especially Vikram, Harish, Baba, Himanshu and Bharath, for allowing me to dump my frustration in hockey ground and will seriously miss the gossiping before and after the game. My special thanks to my friends and well wishers Sunil, Sat, Parveen, Satdev, Manoj, Vijaypal, Parvish, Praveen[2], Bhupender sir and Virender sir who have made my stay at IISc a pleasant and an unforgettable experience.

My parents have been a great source of inspiration for me throughout my life and are responsible for inculcating in me the ethics for hard work and I cannot thank them enough for that; my siblings, especially my brother, and other family members and relatives have given tremendous support and encouragement to me. I thank my wife for her love, patience and unwavering support. She has made my life richer in countless ways. This work would not have been possible without their absolute trust in me.

# List of Publications

1. **Pradeep Kumar**, A. Bera, D.V.S Muthu, P.M. Shirage, A. Iyo and A.K. Sood, "Superconducting fluctuations and anomalous phonon renormalization much above superconducting transition temperature in $Ca_4Al_2O_{5.7}Fe_2As_2$". **Appl. Phys. Lett.** 100, 222602 (2012). (Chapter 3)

2. **Pradeep Kumar**, A. Bera, D.V.S. Muthu, S.N. Shirodkar, R. Saha, A. Shireen, A. Sundaresan, U.V. Waghmare, A.K. Sood and C.N.R. Rao, "Coupled phonons, magnetic excitations and ferroelectricity in $AlFeO_3$: Raman and First-principles Studies". **Phys. Rev. B** 85, 134449 (2012). (Chapter 4)

3. **Pradeep Kumar**, A. Bera, D.V.S. Muthu, A. Kumar, U.V. Waghmare, L. Harnagea, C. Hess, S. Wurmehl, S. Singh, B. Buchner and A.K. Sood, "Raman evidence for superconducting gap and spin-phonon coupling in superconductor $Ca(Fe_{0.95}Co_{0.05})_2As_2$". **J. Phys. : Condens. Matter** 23, 255403 (2011). (Chapter 3)

4. **Pradeep Kumar**, A. Kumar, S. Saha, D.V.S. Muthu, J. Prakash, U.V. Waghmare, A.K. Ganguli and A.K. Sood, "Temperature-dependent Raman study of $CeFeAsO_{0.9}F_{0.1}$ superconductor: crystal field excitations, phonons and their coupling". **J. Phys. : Condens. Matter** 22, 255402 (2010). (Chapter 3)

• Also selected for the "*Lab Talk Section*" in J. Phys. : Condense. Matter titled *"Phonons, quasiparticle excitations and their coupling in iron-pnictides"*.

5. **Pradeep Kumar**, A. Kumar, S. Saha, D.V.S. Muthu, J. Prakash, S. Patnaik, U.V. Waghmare, A.K. Ganguli and A.K. Sood, "Anomalous Raman scattering from phonons and electrons of superconducting $FeSe_{0.82}$". **Solid State Commun.** (**Fast Track**) 150, 557 (2010). (Chapter 3)

6. **Pradeep Kumar**, S. Saha, D.V.S. Muthu, J.R. Sahu, A.K. Sood and C.N.R. Rao, "Raman evidence for orbiton-mediated multiphonon scattering in multiferroic $TbMnO_3$". **J. Phys. : Condens. Matter** 22, 115403 (2010). (Chapter 4)

7. **Pradeep Kumar**, S. Saha, C.R. Serrao, A.K. Sood and C.N.R. Rao, "Temperature-Dependent infrared reflectivity studies of multiferroic $TbMnO_3$: Evidence for spin-phonon coupling". **Pramana J. Phys.** 74, 281 (2010). (Chapter 4)

8. **Pradeep Kumar**, D.V.S. Muthu, J. Prakash, A.K. Ganguli and A.K. Sood, "Raman evidence for Superconducting gap in superconductor $Ce_{0.6}Y_{0.4}FeAsO_{0.8}F_{0.2}$". (**arXiv: 1310.4920,** DAE SSPS - Conference Proceding-2013). (Chapter 3)

9. A. Kumar, **Pradeep Kumar**, U.V. Waghmare and A.K. Sood, "First-principles analysis of electron correlation, spin ordering and phonons in the normal state of $FeSe_{1-x}$". **J. Phys. : Condens. Matter** 22, 385701 (2010).

10. G.K. Pradhan, A. Bera, **Pradeep Kumar**, D.V.S. Muthu and A.K. Sood, "Raman signature of pressure induced electronic topological and structural transition in $Bi_2Te_3$". **Solid State Commun.** 152, 284 (2012).

11. **Pradeep Kumar**, S. Ghara, B. Rajeshwaran, D.V.S. Muthu, A. Sundaresan and A.K. Sood, "Spin-phonon coupling and spin-glass state in double perovskite $La_2NiMnO_6$". (**arXiv: 1312.7058, Solid State Commun**. - 2014). (Chapter 4)

*Manuscripts submitted/under preparation:*

12. **Pradeep Kumar**, D. V. S. Muthu, L. Harnagea, C. Hess, S. Wurmehl, S. Singh, B. Buchner and A. K. Sood, "Orbital-ordering and electron-phonon coupling through spin-channels in iron pnictide $Ca(Fe_{1-x}Co_x)_2As_2$". (**Submitted**-2014). (Chapter 3)

13. A. Bera, **Pradeep Kumar**, D.V.S. Muthu and A.K. Sood, "Raman study of phonon anomalies in single crystal $SrTiO_3$". (To be submitted).

# Thesis Synopsis

Raman spectroscopy is a very powerful probe to study the nature of quasi-particle excitations in condensed matter physics. The work presented in this thesis is focused on two different families of novel materials, namely the iron-based superconductors (FeBS), multiferroic oxides and double perovskite. Although the properties of these two systems are quite different, some comparison can still be drawn between them. For instance, in both of these systems magnetism plays a crucial role and intricate coupling between phononic, magnetic and orbital degrees of freedom is crucial to understand their underlying physics responsible for their various exotic physical properties. Understanding the microscopic origin of quasi-particle excitations, such as phonons, magnons, orbitons, plasmons etc., and coupling between them in these complex materials, has been an intense field of research because it is believed that these excitations hold the key for explaining their rich physics. The systems studied in the thesis include (A) FeBS - (i) $FeSe_{0.82}$ (ii) $Ce_{1-z}Y_zFeAsO_{1-x}F_x$ (z = 0, 0.4; x = 0.1, 0.2) (iii) $Ca_4Al_2O_{5.7}Fe_2As_2$ (iv) $Ca(Fe_{1-x}Co_x)_2As_2$ (x = 0.03, 0.05). (B) Multiferroic oxides - (i) $AlFeO_3$ (ii) $TbMnO_3$ and double perovskite (iii) $La_2NiMnO_6$.

In **Chapter 1** we briefly introduce the systems studied in this thesis, i.e. iron-based superconductors, multiferroic oxides and double perovskite.

**Chapter 2** gives a brief introduction to the Raman scattering process. In this chapter we also describe the effect of temperature on Raman spectra since all of our studies have been done as a function of temperature. Here we also describe various experimental techniques used to carry out the research work presented in this thesis.

In **Chapter 3** we discuss our extensive temperature-dependent Raman studies on different families of iron-based superconductors. This chapter is divided into six parts.

In **Part 3.1** we present interesting anomalies in the temperature-dependent Raman studies of superconducting FeSe$_{0.82}$ measured from 3 K to 300 K, covering the superconducting ($T_c$ ~ 12 K) as well structural transition temperature ($T_s$ ~ 100 K), in the spectral range from 60 to 1800 cm$^{-1}$ and determine their origin using first-principles density functional calculations. A phonon mode, associated with the Se vibrations, near 100 cm$^{-1}$ exhibits a sharp increase by ~ 5% in frequency below a transition temperature $T_s$ attributed to the strong spin-phonon coupling and onset of short-range antiferromagnetic order. In addition, we observed two high frequency modes at 1350 cm$^{-1}$ and 1600 cm$^{-1}$, attributed to electronic Raman scattering from ($x^2$-$y^2$) to $xz/yz$ 3$d$-orbitals of Fe, suggesting the electronic nematicity of Fe 3$d$-orbitals.

**Part 3.2** describes our detailed Raman study of CeFeAsO$_{0.9}$F$_{0.1}$ superconductor in the spectral range of 60 to 1800 cm$^{-1}$ and interprets it using estimates of phonon frequencies obtained from first-principles density functional calculations. We find evidence of a strong coupling between the phonons and crystal field excitations; in particular Ce$^{3+}$ crystal field excitation at 432 cm$^{-1}$ couples strongly with the E$_g$ oxygen vibration at 389 cm$^{-1}$. Below $T_c$ (~ 38 K) the phonon mode near 280 cm$^{-1}$ (E$_g$, Fe) shows anomalous softening, signaling its coupling with the superconducting gap. The estimated ratio of the superconducting gap to $T_c$ is ~ 10 suggests CeFeAsO$_{0.9}$F$_{0.1}$ as a strong coupling superconductor. In addition, two high frequency modes observed at 1342 cm$^{-1}$ and 1600 cm$^{-1}$ are attributed to electronic Raman scattering from ($x^2$-$y^2$) to $xz/yz$ 3$d$-orbitals of Fe, as described in case of FeSe$_{0.82}$ (Part 3.1).

The optimal doping of 'Y' at the place of 'Ce' in CeFeAsO$_{1-x}$F$_x$ leads to a significant increase, by ~ 25%, in superconducting transition temperature. Increase in $T_c$ via 'Y' doping may be linked with the change in FeAs$_4$ tetrahedral structure, where ideal tetrahedral angle leads to maximum $T_c$ in these systems. The significant increase in $T_c$ motivated us to study the role of phonon dynamics and effect of superconducting transition on elementary excitations. We found strong signature of the superconductivity induced phonon renormalization for the $A_{1g}$ phonon mode near 150 cm$^{-1}$ associated with the Ce/Y vibrations as reflected in the anomalous red-shift and decrease in the linewidth below $T_c$. Invoking the coupling of this mode with the superconducting gap, the superconducting gap ($2\varDelta$) is estimated to be ~ 20 meV i.e. the ratio *$2\varDelta/K_BT_c$ is* ~ 5, suggesting Ce$_{0.6}$Y$_{0.4}$FeAsO$_{0.8}$F$_{0.2}$ also belongs to the class of strong coupling superconductors. This work is presented in **Part 3.3**.

In **Part 3.4** we present our Raman studies on Ca$_4$Al$_2$O$_{5.7}$Fe$_2$As$_2$ superconductor ($T_c$ = 28.3 K) in the temperature range of 5 K to 300 K. Our studies reveal that the Raman mode observed at ~ 230 cm$^{-1}$ shows a sharp jump in frequency by ~ 2 % and linewidth increases by ~ 175 % at $T_o$ ~ 60 K. Below $T_o$, anomalous softening of the mode frequency and a large decrease by ~ 10 cm$^{-1}$ in the linewidth are observed. These precursor effects at $T_0$ (~ $2T_c$) are attributed to significant superconducting fluctuations, possibly enhanced due to reduced dimensionality arising from weak coupling between the well separated *Fe-As* layers (~ 15 Å) in the unit cell. A large blue-shift of the mode frequency between 300 K to 60 K (~ 7 %) indicates strong spin-phonon coupling in this superconductor.

In **Part 3.5** we present our Raman and first-principle density functional studies on the single crystal of electron-doped Ca(Fe$_{0.95}$Co$_{0.05}$)$_2$As$_2$ superconductor, also termed as "122" system, having $T_c$ ~ 23 K. This system undergoes structural, from high temperature tetragonal to low temperature orthorhombic phase, as well as magnetic transition at $T_{sm}$ ~ 140 K. Our studies reveal strong evidence for superconductivity-induced phonon renormalization; in particular the phonon mode observed near 260 cm$^{-1}$ ($E_g$, Fe and As) shows anomalous hardening below $T_c$, signaling its coupling with the superconducting gap. All the observed Raman active phonon modes show anomalous temperature dependence between room temperature and $T_c$ i.e. phonon frequency decreases with lowering temperature attributed to the strong spin-phonon coupling. Using first-principles calculations, we show that the low temperature phase ($T_c < T < T_{sm}$) exhibits short-ranged stripe anti-ferromagnetic ordering, and estimate the spin-phonon couplings that are responsible for these phonon anomalies.

**Part 3.6** describes our detailed temperature dependent Raman study on another "122" system Ca(Fe$_{0.95}$Co$_{0.03}$)$_2$As$_2$ in a wide spectral range of 120-5200 cm$^{-1}$, covering the tetragonal to orthorhombic structural transition coinciding with magnetic transition temperature at $T_{sm}$ ~ 160 K. We note that Ca(Fe$_{0.97}$Co$_{0.03}$)$_2$As$_2$ does not show superconducting transition, however there is a drop in resistivity below ~ 20 K. The mode frequencies of two first-order Raman modes B$_{1g}$ and E$_g$, both involving displacement of Fe atoms, show sharp increase below $T_{sm}$. Concomitantly, the linewidths of all the first-order Raman modes show anomalous broadening below $T_{sm}$, attributed to strong spin-phonon coupling. In addition, we observed four high frequency modes between 400-1200 cm$^{-1}$ attributed to the electronic Raman scattering involving the crystal field levels of 3$d$-orbitals of Fe$^{2+}$. The splitting between $xz$ and $yz$ 3$d$-orbital levels is found to be ~ 25 meV which increases as temperature decreases below $T_{sm}$, suggesting the electronic nematicity of Fe 3$d$-orbitals similar to that of our earlier studies on FeSe$_{0.82}$ and CeFeAsO$_{1-x}$F$_x$. A broad Raman band observed at ~ 3200 cm$^{-1}$ shows anomalous temperature dependence and is assigned to coupled spin-orbital excitations.

In **Chapter 4** we describe our detailed temperature-dependent Raman studies on multiferroic oxides and double perovskite. This chapter is divided into four parts.

In **Part 4.1** we present the nature of coupled phonons and magnetic excitations in AlFeO$_3$ using temperature dependent Raman studies from 5 K to 315 K covering a spectral range from 100-2200 cm$^{-1}$ and complementary first-principles density functional calculations. A strong spin-phonon coupling and magnetic ordering induced phonon renormalization are evident in (a) anomalous temperature dependence of many modes with frequencies below 850 cm$^{-1}$, particularly near the magnetic transition temperature $T_c$ ~ 250 K, (b) distinct changes in band positions of high frequency Raman bands between 1100-1800 cm$^{-1}$, in particular a broad mode near 1250 cm$^{-1}$ appears only below $T_c$ attributed to the two-magnon Raman scattering. We also observe weak anomalies in the mode frequencies at ~ 100 K, due to a magnetically driven ferroelectric phase transition.

In **Part 4.2** we report our temperature-dependent Raman studies on multiferroic TbMnO$_3$ from 5 K to 300 K in the spectral range of 200 to 1525 cm$^{-1}$. The intensity ratio of the second-order phonon (S8) to its first-order counterpart (S6) is unusually high and it remains constant at all temperatures. This anomalous temperature dependence of the intensity ratio is attributed to the coupling of the second-order phonon with the orbital degrees of freedom as theoretically predicted. Four of the first-order modes show normal behavior with temperature, whereas the S5 (616 cm$^{-1}$, B$_{2g}$) mode behaves anomalously below $T_N$ (~ 46 K) attributed to the strong spin-phonon coupling.

As a follow up of the work described in part 4.2, in **Part 4.3** we present far infrared reflectivity studies on single crystal of TbMnO$_3$ from 10 K to 300 K in the spectral range of 50 cm$^{-1}$ to 700 cm$^{-1}$. We identified fifteen transverse optic (TO) and longitudinal optic (LO) modes in the imaginary part of the dielectric function $\varepsilon_2(\omega)$ and energy loss function $\text{Im}(-1/\varepsilon(\omega))$, respectively. Some of the observed phonon modes show anomalous softening below the magnetic transition temperature $T_N$ (~ 46 K), attributed to the spin-phonon coupling. The effective charge of oxygen (Z$_O$) calculated using the measured LO-TO splitting increases below $T_N$.

In **Part 4.4** we present our detailed magnetic, dielectric and Raman studies on partially disordered and biphasic double perovskite La$_2$NiMnO$_6$. Magnetic susceptibility measurements show two magnetic anomalies at $T_{C1}$ ~ 270 K and $T_{C2}$ ~ 240 K, attributed to the ferromagnetic ordering of the monoclinic and rhombohedral phases, respectively. We also observed a broad peak at a lower temperature ($T_{sg}$ ~ 70 K) indicating a spin-glass transition due to partial anti-site disorder of $Ni^{2+}$ and $Mn^{4+}$ ions. Our temperature-dependent Raman studies reveal a strong renormalization of the first as well as second-order Raman modes associated with the (Ni/Mn)$O_6$ octahedra near $T_{C1}$ implying a strong spin-phonon coupling. In addition, we observed an anomaly in the vicinity of spin-glass transition temperature in the temperature dependence of the frequency of the anti-symmetric stretching vibration of the octahedra.

In **chapter 5** we summarise our results on different systems studied in this thesis, namely, iron-based superconductors, multiferroic oxides and double perovskite. We also suggest some possible studies on these systems which may be undertaken in future to explore these systems further.

# Contents

# Chapter 1

# Introduction

This chapter gives an overview of the systems studied in this thesis. Part-A deals with iron-based superconductors and Part-B with multiferroic oxides.

## 1.1 Part-A

### 1.1.1 High Temperature Superconductivity

Superconductivity, the resistance free flow of charge carriers, is one of the most exotic phenomena in condense matter physics. The phenomena of superconductivity, though discovered nearly a century ago in 1911 by Dutch physicist H. Kammelingh Onnes [1] in Hg, still posses many pertinent questions yet to be answered, in particular the microscopic pairing mechanism in high temperature superconductors beyond the Bardeen-Cooper-Schrieffer (BCS) theory [2]. The high temperature superconductivity research began in 1986 when Bednorz and Muller found evidence for superconductivity at ~ 30 K in La-Ba-Cu-O ceramics [3]. Their remarkable discovery of superconductivity phenomena led to the discovery of many cuprate systems with transition temperature $T_c$ achieved upto ~ 135 K ($T_c$ ~ 164 K under high pressure) in Hg based cuprates.

Interestingly, the recent discovery of superconductivity in Fe-based systems by Kamihara et al. [4] in 2008 at $T_c$ ~ 26 K in $LaFeAsO_{1-x}F_x$, ended the monopoly of cuprates as the only high temperature superconductors. Naively we would expect that superconductivity and magnetism are incompatible which is reflected in the cooper pair breaking by magnetic impurity [5]. The discovery of superconductivity in iron containing materials was a welcome surprise. Since 2008, many new familes of compound have been discovered with maximum $T_c \sim 56$ K in $Gd_{0.8}Th_{0.2}FeAsO$ [6]. The maximum $T_c$ so far achieved in iron-based superconductors (FeBS) is not as high as in the case of cuprates but their discovery is fundamentally important in that the glue for the Cooper pair formation may have magnetic origin. Although iron-based superconductors share some common properties with cuprates, such as layered crystal structure and antiferromagnetic ordering in the parent compounds, many differences exist between these two classes related to their electronic structure. Though there is a general consensus about the unconventional nature of the pairing mechanism in FeBS but still many central questions remain unanswered -- role of magnetism in Cooper pair formation, nature of chemical and structural tuning, and the pairing symmetry. Their phase diagram looks very much similar to those of unconventional superconductors such as cuprates, organic and heavy Fermion superconductors [7]. The existing literature does point towards the magnetically mediated superconductivity in these systems similar to that in cuprates, but still there is no broad consensus.

## 1.1.2 Superconductivity in iron-based superconductors

The superconductivity in iron pnictides reported by Kamihara et al. [4] at 26 K in $LaFeAsO_{1-x}F_x$ was preceded by the discovery of superconductivity in LaFePO [8] at $\sim 5$ K. Following the work of Kamihara et al. $T_c$ was rapidly increased by substitution of rare earth ions and appropriate carrier doping [9-11] and soon many new families were discovered. Broadly these families can be divided into two major categories: the iron pnictides and iron chalcogenides. These can be further divided into many families based on their chemical composition and crystal symmetry. Till now six different families have been discovered namely RFeAsO (R=La,Ce,Sm and Nd) (abbreviated as 1111 family for 1:1:1:1 ratio of the four elements) [4], $AFe_2As_2$ (A = Ba,Ca and Sr; 122 family) [12], AFeAs [A = Na and Li; 111 family] [13], $FeSe_{1-x}Te_x$ (11 family) [14], $Sr_2MO_3FePn$ (M = Sc, V and Cr; Pn = P and As; 21311 or 42622 family) [15,16] and defect structure $A_{0.8}Fe_{1.6}Se_2$ (A = K, Rb, Cs and Tl; 122* family) [17].

The stoichiometric parent compounds are antiferromagnetic semimetals and superconductivity is induced by the suppression of long range antiferromagnetic ordering via chemical doping or external pressure. The basic building blocks of the crystal structure of all the FeBS are quasi two-dimensional (2D) layers consisting of Fe square lattice where the Fe is tetrahedrally coordinated with the pnictides (As/P) or chalcogenides (Se/Te). For example Fig. 1.1 shows the crystal structural of $LaFeAsO_{1-x}F_x$, where it can be seen that FeAs layer is sandwiched between LaO spacer layers. The LaO layers donate electrons to the FeAs layers, and Fe - sublattice controls the electronic, magnetic as well as superconducting properties of these systems, hence the name quasi 2D - systems. The geometry of $FeAs_4$ tetrahedra plays a crucial role in determining the electronic as well as magnetic properties of these systems. In fact the tetrahedral angle plays an important role in optimising $T_c$ with highest $T_c$ achieved near ideal tetrahedral angle [18-20].

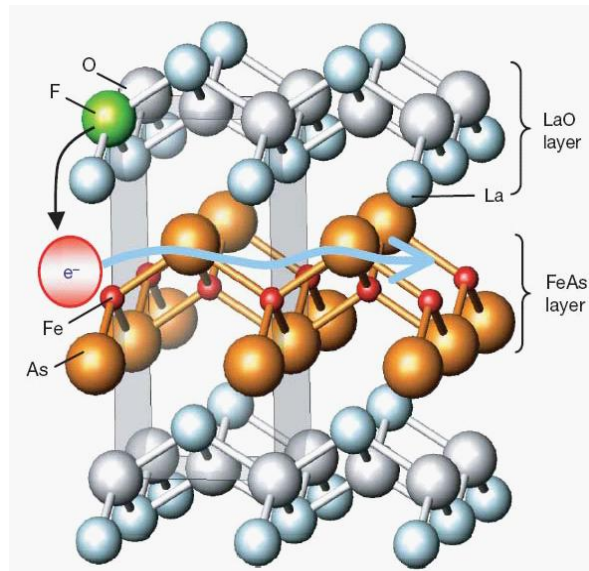

Figure 1.1: Crystal structure of $LaFeAsO_{1-x}F_x$ superconductor [4].

## 1.1.3 Crystal Structure of FeBS

Near room temperature, all the members of FeBS family crystallize in tetragonal symmetry with no magnetic order. Their crystal structures share a common two-dimensional FePn/Ch (Pn - Pnictides, Ch - Chalcogenides) layer, which is central to the electronic, magnetic as well as superconducting properties of these systems, where Fe atoms form a 2D square sublattice with Pn/Ch atoms sitting at the center of each Fe square but sticking out above and below the planes of Fe atoms as shown in Fig. 1.1, setting these materials apart from the 2D Cu-O planes in cuprates where Cu and O lie in the same plane. The Fe and O atoms lie in planes, while Pn/Ch and R (R=La,Ce,Sm and Nd) atoms are distributed on each side of these planes.

**1111 Family**:

LaFeAsO and other members of this family crystallize in the tetragonal ZrCuSiAs type structure (space group *P4/nmm*, Z = 2) with 2D layers of FePn (Pn = As and P) as shown in Fig. 1.2 (a) [21]. These FeAs layers are separated by spacer/charge donation layers along c-axis by RO (R = La, Ce, Sm, Nd and Gd) layers.

**122 Family:**

This family of FeBS also has similar tetragonal layered $ThCr_2Si_2$ type structure (space group *I4/mmm*, Z = 2). The crystal structure is shown in Fig. 1.2 (b) [22]. The FePn layers are separated by A layers (A = Ca, Ba and Sr) instead of RO layers as in the case of 1111 family. This is a well studied structure, and interestingly it is the same structure where superconductivity was first discovered in heavy fermion systems, $CeCu_2Si_2$ [23], way back in late seventies.

**111 and 11 Families:**

These families of superconductors crystallize in $Cu_2Sb$ and PbO type tetragonal structure with 2D FePn (Pn = As and P) planes, respectively. Both of these families belong to space group *P4/nmm* with two units per unit cell. The crystal structure for 11 and 111 families are shown in Figs. 1.2 (c) **[24]** and 1.2 (d) [14], respectively.

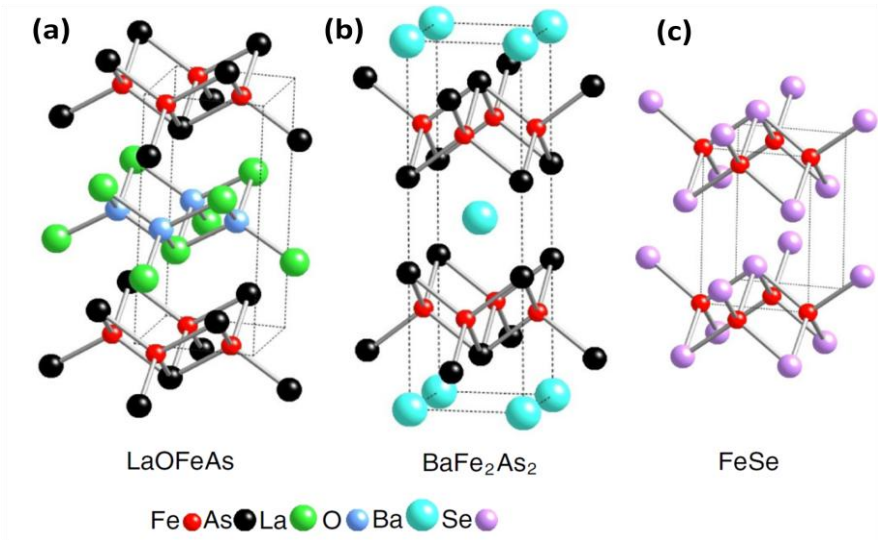

Figure 1.2 (a), (b) and (c): Crystal structure of 1111, 122 and 11 systems, respectively [21,22,24].

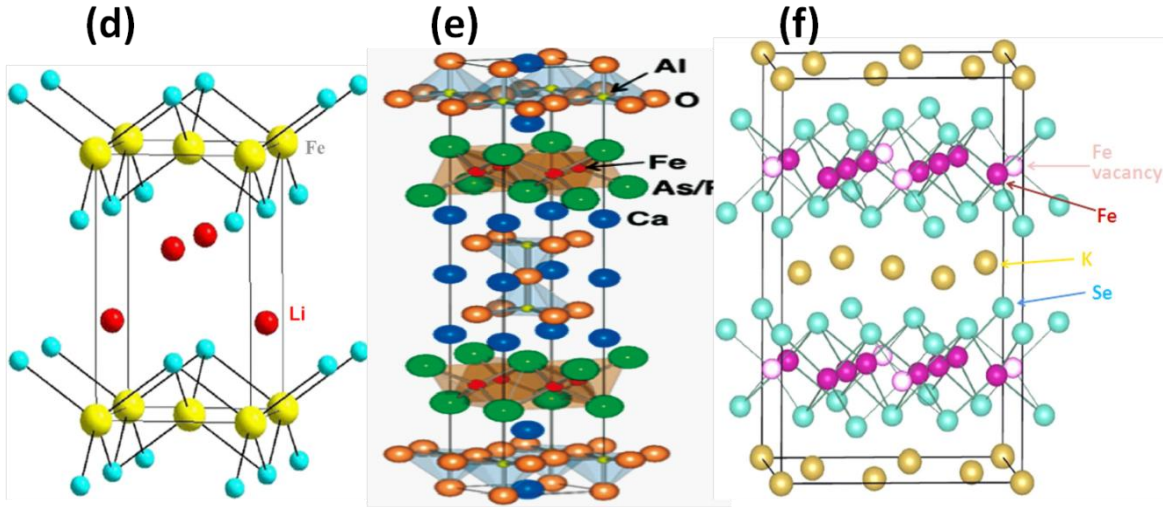

Figure 1.2 (d), (e) and (f): Crystal structure of 111, 42622 and 122* systems, respectively [14,25,26].

**21311 ( or 42622 ) Family:**

The fifth structure with FePn (Pn = As and P) planes to join the FeBS family is the so called 42622 structure. The first system which joined this family was $Sr_2ScO_3FeP$ [15], having

*I4/mmm* space group same as that of 122 systems. The crystal structure is shown in Fig. 1.2 (e) [25], which can be visualised as the layers of $SrFe_2As_2$ alternating with perovskites $Sr_3Sc_2O_6$ layers. The families of materials which have spacer layers of perovskites structure allow controlled tailoring of the distance between FePn layers, which can be used to tune the transition temperature. This tuning of distance between FePn layers has resulted in the significant increases in transition temperature from initial 17 K to 47 K [25].

**122* Family:**

The most recent addition to the family of iron based superconductors is an ordered-defect variant of the 122 $BaFe_2As_2$ structure with space group *I4/m*, also written as $A_xFe_{2-x}Se_2$ (A = K, Rb,Cs and Tl). The crystal structure is shown in Fig. 1.2 (f) [26]. The ordered iron vacancies on Fe sites (Fe2 sites are fully occupied and Fe1 sites are fully unoccupied in an ideal case) play a central role in determining various physical properties including superconducting transition temperature.

### 1.1.4  Electronic Structure and Correlation in FeBS

The electronic structure of these iron based superconductor systems have been extensively studied both theoretically and experimentally [27-35]. All the FeBS belonging to different families discovered so far have short Fe-Fe distances, ~ 2.76 Å in 11, 2.77 Å in 122, 2.84 Å in 21311 and 2.85 Å in 1111 systems [15, 17, 36-38], signalling that the Fe 3*d* electrons play a central role in band formation and their electronic properties. Due to short $Fe^{2+}$ ion separation, which is much smaller than the $Cu^{2+}$ separation (~ 3.8Å) in cuprates, the direct hopping between iron atoms leads to a metallic ground state containing equal number of electrons and holes. In fact various theoretical and experimental studies on these systems have shown that the Fe 3*d*-orbitals dominate the density of states near the Fermi surface, whereas the electronic states associated with pnictogen/chalcogen 4*p* orbital lies ~ 2eV below the Fermi level [31, 39-41]. In particular, Fe $t_{2g}$ ($d_{xy}$, $d_{xz/yz}$) orbitals contribute significantly as compared to the $e_g$ ($d_z^2$, $d_{x^2-y^2}$) orbitals. Figure 1.3 shows the typical density of states (DOS) based on density functional theory calculation for 11 [34] (top panel) and 122 systems [42] (bottom panel) showing the major contribution of Fe 3d-orbitals at the Fermi surface. In comparison, in case of cuprates, oxygen is sitting in the middle of each pair of Cu-Cu atoms, therefore in addition to the Cu 3d orbitals, O 2p orbitals also have significant contribution to the density of states at the Fermi surface.

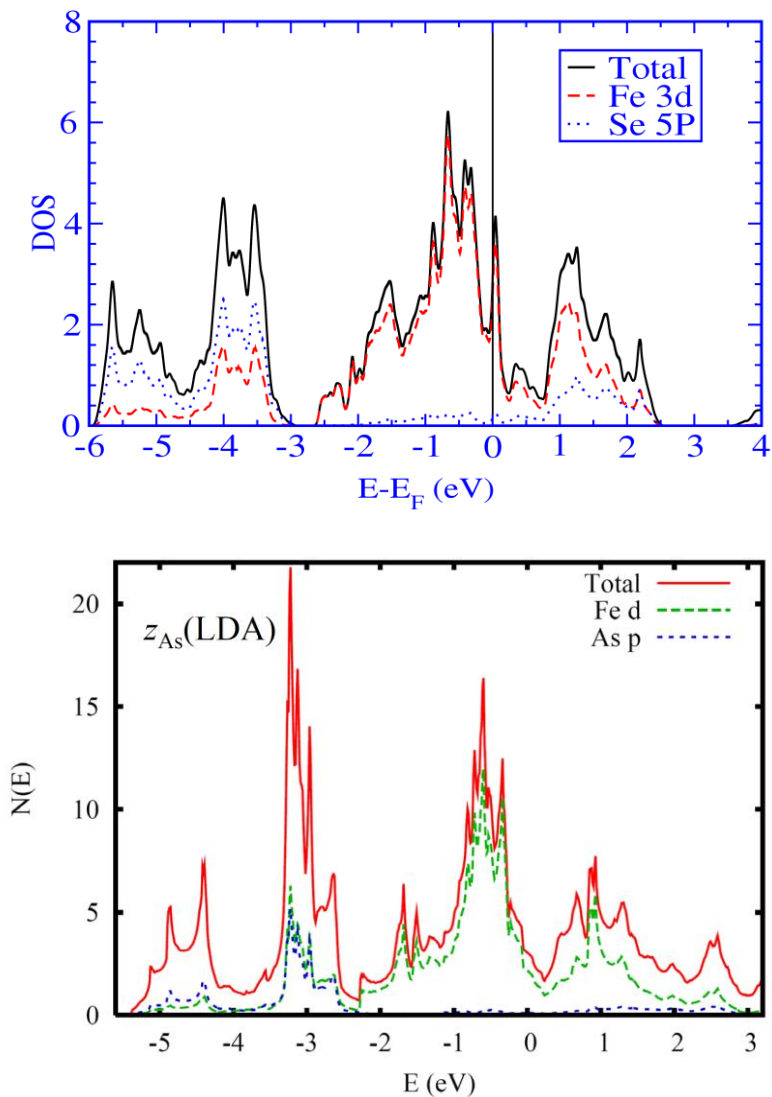

Figure 1.3: Electronic DOS for FeSe$_{1-x}$ [34] (top panel) and BaFe$_2$As$_2$ [42] (bottom panel) superconductor showing the major contribution of Fe 3d states near the Fermi surface.

In spite of different chemical compositions of the various classes of FeBS, they have fairly similar electronic structure. The common features are (i) the band structure is composed of two electron pockets near the M point and two or more hole pockets near the Γ point in the Brillouin zone. (ii) The hole pockets at the Γ point are formed by the Fe $d_{xz}$ and $d_{yz}$ orbitals, whereas the electron pockets are formed by the $d_{xz}$, $d_{yz}$ and $d_{xy}$ orbitals. Figure 1.4 shows the typical Fermi surfaces for LaFeAsO [43]. These electron and hole pockets can be well nested by the wave vector (π, π), and interestingly this wave-vector is also the ordering wave-vector of the striped antiferromagnetic phase at low temperature [44]. The numbers of Fermi surfaces and their structures are strong functions of doping (electron/hole); for electron doping, the hole cylinders at the Γ point disappear and the electron cylinders grow in size and merge; and for hole doping, the electron cylinder decreases in size as the hole cylinders grow and distort into diamond shape [30, 46], which differs significantly from the cuprates where the topology of the Fermi surface is nearly independent of doping [47]. Their electronic structure also depends on the distance of Pn/Ch from the Fe planes because of the sizeable hybridization between Fe 3$d$ and Pn/Ch 2$p$ states.

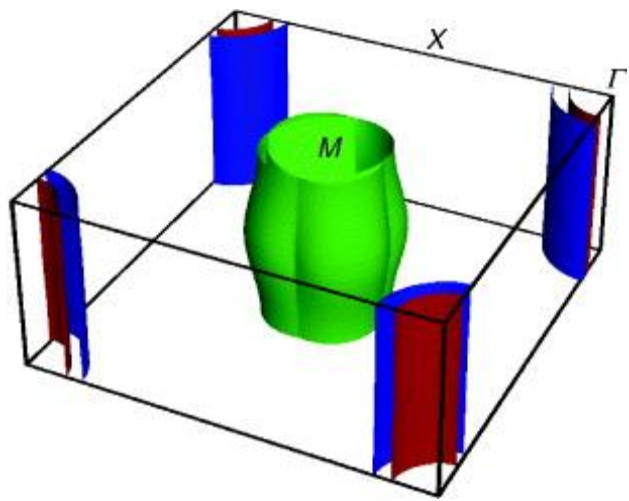

Figure 1.4: Electron and Hole Fermi surfaces for LaFeAsO superconductor [43].

Since the Fermi surface in these systems is mainly composed of $t_{2g}$ orbitals of $Fe^{2+}$, the role of Fe 3$d$-orbitals is undoubtedly crucial in providing the glue for the Cooper pair formation. The important question is related to the degenerate/non-degenerate nature of $d_{xz/yz}$ orbitals which are central to their electronic, magnetic as well as superconducting properties [28-30]. Some recent reports [103-104] have also shown the existence of electronic nematicity between Fe $d_{xz}$ and $d_{yz}$ orbitals evidencing the crucial role of orbital degrees of freedom.

Electron correlations in a material are often connected with whether the material exhibits itinerant (weak correlation) or local moment (strong correlation) magnetism. In a parent pnictide compound, with iron in $Fe^{2+}$ state, one would expect an ordered moment of $4\mu_B$ which is in sharp contrast to the low observed value in these systems. The observed magnetic moment in FeBS ranges from 0.25 $\mu_B$ for NdFeASO [48] to 0.94 $\mu_B$ in $SrFe_2As_2$ [49]. The relatively small value of the magnetic moments suggests that antiferromagnetic ground state arises from the itinerant electrons. However, alternate views based on spin fluctuations, orbital fluctuations and/or magnetic frustration have been advocated theoretically [28-29, 50-56] and arguments have been put forward that these materials should be viewed as weakly correlated [50-51], strongly correlated [28-29, 52-55] or in the intermediate regime [56].

Experiments on optical conductivity of these systems clearly show Drude peak like features, suggesting weak or moderate correlations in these systems [57-61]. A recent experimental study by Qazilbash et al. [62] on various materials have estimated the degree of electron correlations from the ratio of experimentally determined optical spectral weight (low frequency Drude contribution, $K_{exp}$) to that of theoretically predicted by band theory ($K_{band}$). From their detailed study they have placed FeBS in the intermediate regime of electron correlation. Also, a similar study by Chen et al. [63] on LaFeAsO found that electron correlations in these systems are weak as compared to cuprates and are moderately correlated. It is to be noted that the Te rich compound $Fe_{1+y}Se_{1-x}Te_x$ belonging to the 11 family of FeBS appears to be local moments antiferromagnetic with stronger electron correlation as compared to other members of FeBS [58, 64].

## 1.1.5 Antiferromagnetic Ordering and Superconductivity in FeBS

The parent compound LaFeAsO shows an anomaly near 150 K [4]. It was shown [65] that below 150 K there is no new mode or splitting of existing phonon mode appears, indicating that the structural transition associated with this anomaly is subtle. The results [65] suggested that the structural phase transition in these systems is driven by the spin density wave (SDW) instability. Also, neutron scattering experiment on the same systems [66] evidenced the existence of long range order spin density wave below 137 K and tetragonal to orthorhombic structural transition at ~ 155 K. Later a similar study on $BaFe_2As_2$ system evidenced the existence of structural as well as SDW transition [67]. Interestingly both transitions in $BaFe_2As_2$ systems occur simultaneously in contrast to that in the case of LaFeAsO, where both transitions are separated by a few Kelvin whose origin is still debated. The discovery of SDW like antiferromagnetic (AFM) transition at low temperature has led to the search of ground state for these materials.

FeBS systems show phase transitions such as tetragonal to orthorhombic structural transition, striped AFM ordering/nematic electronic phase and superconducting phase under various external perturbations such as temperature, doping, pressure. Figure 1.5 shows a generic phase diagram for FeBS as a function of tuning parameters (doping - $x$, or pressure - $p$). Parent compounds exhibit antiferromagnetic ordering and structural transition, which is suppressed with doping, and superconductivity appears at some non-zero doping and $T_c$ reaches a maximum, and with further doping superconductivity disappears, such that $T_c$ forms a dome like structure similar to that in cuprates.

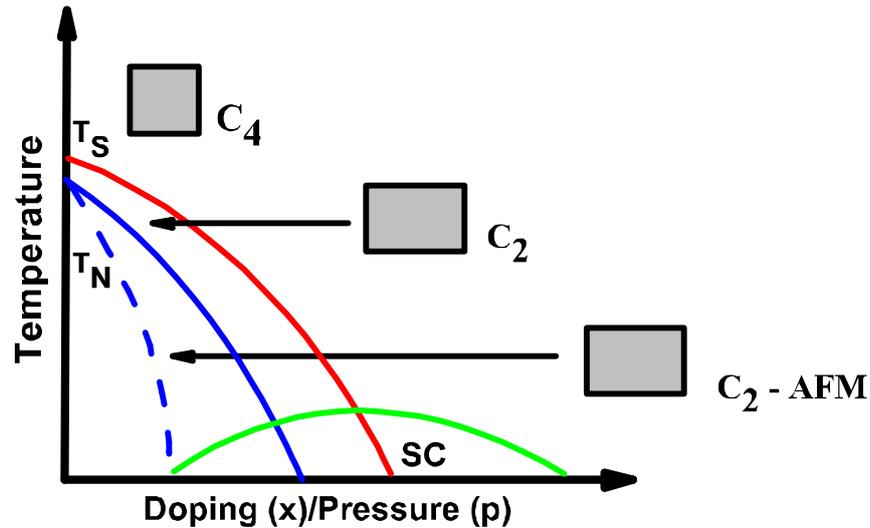

Figure 1.5: Generic phase diagram of the FeBS as a function of doping (x) and pressure (p). At room temperature these systems are four-fold symmetric (Tetragonal symmetry-$C_4$). As the temperature is lowered it undergoes structural transition into two-fold symmetry ($C_2$) (Orthorhombic), and at lower temperature AFM ordering sets in. Dotted line separates the region where AFM and superconductivity doesn't coexist.

In case of 122 systems AFM and superconductivity coexist as shown in Fig. 1.6 (c) whereas for 1111 materials it is still not clear whether these phases coexist or not (see Fig. 1.6 (a) and (b)). Below the SDW transition magnetic moments align antiferromagnetically along *a*-axis and ferromagnetically along *b*-axis resulting in the stripe AFM ground state [68]. Interestingly for these systems, AFM ordering is always preceded by or coincident (i.e. $T_s \geq T_N$) with tetragonal to orthorhombic structural transition [26, 65-67, 69-73]. However, for 122* FeBS the low temperature structure remains tetragonal, although with a lower symmetry ($I4/m$) than the high temperature tetragonal phase ($I4/mmm$) [26]. Because of the subtle change of lattice parameter (< 0.4 % [75]) in going from high temperature tetragonal phase to low temperature orthorhombic phase, it was theoretically proposed that the structural transition is not the principle order parameter and the phase transition is actually driven by an instability in the electronic systems, i.e. either the spin fluctuation [76-78] and/or orbital ordering [79-80].

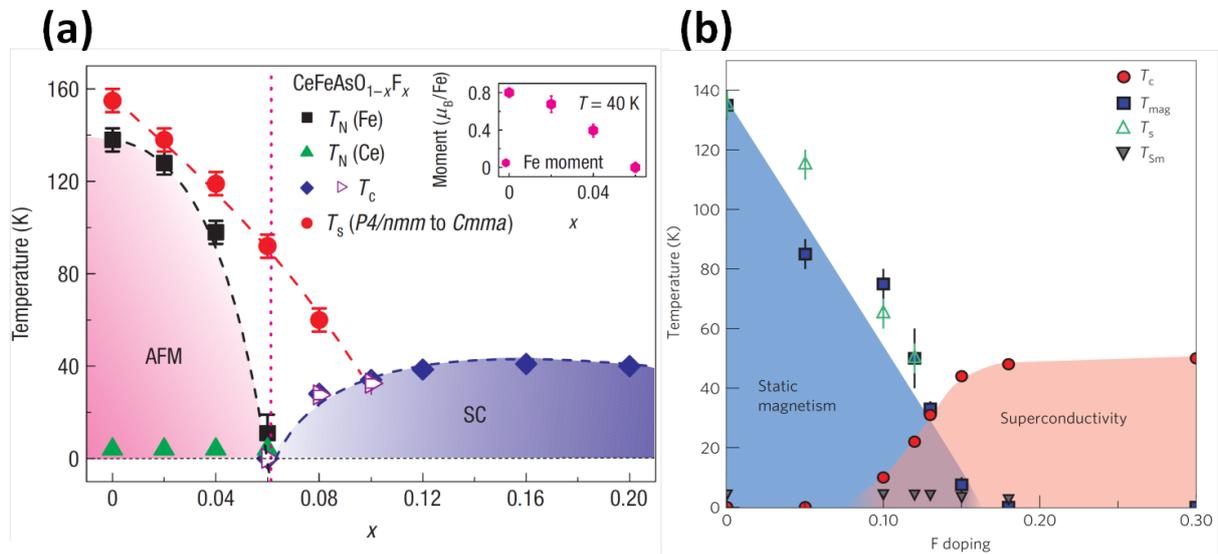

Figure 1.6: (a) Phase diagram of CeFeAsO$_{1-x}$F$_x$ [18]. (b) Phase diagram of SmFeAsO$_{1-x}$F$_x$, showing the coexistence of magnetism and superconductivity for doping x = 0.10 -0.15 [81].

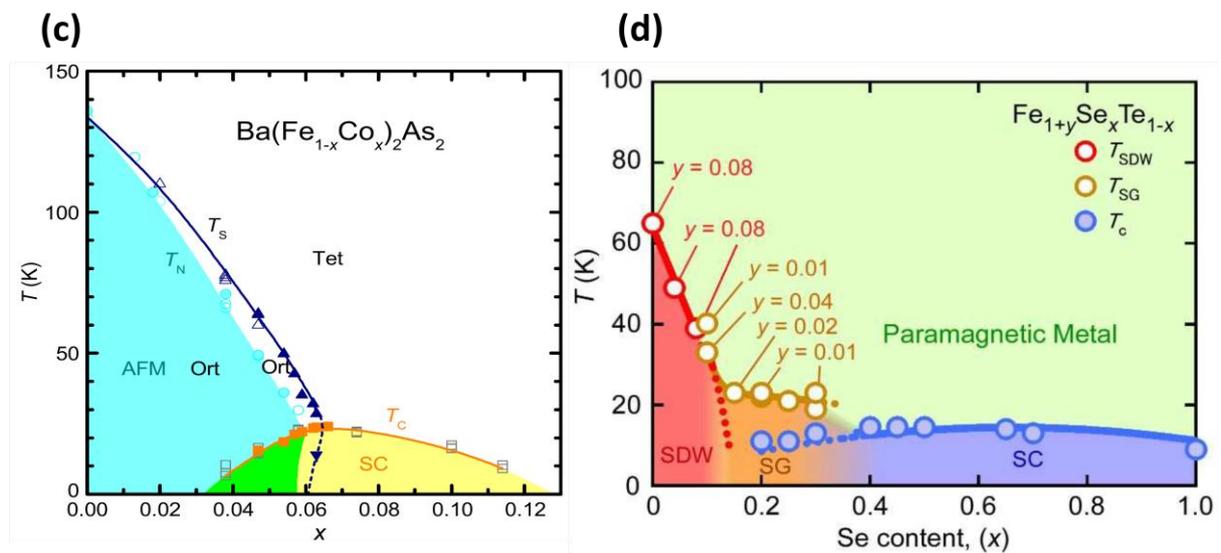

Figure 1.6: (c) Phase diagram of Ba(Fe$_{1-x}$Co$_x$)$_2$As$_2$ showing the coexistence of "often believed to be mutually exclusive phases" superconductivity and magnetism [82]. (d) Phase diagram of Fe$_{1+y}$Se$_x$Te$_{1-x}$ showing the existence of a spin glass (SG) regime between SDW and superconducting phases [84].

As these systems are chemically doped (electron/hole), both structural and SDW transitions are suppressed and superconductivity appears, giving rise to a very rich phase diagram similar to cuprates. Figure 1.6 (a, b-1111 [18, 81], c-122 [82], d-11 [83-84]) shows the rich phase diagram for FeBS. For CeFeAsO$_{1-x}$F$_x$ (see Fig. 1.6 (a)) $T_c$ becomes finite only after structural as well as magnetic transition temperature goes to zero. However, for the case of SmFeAsO$_{1-x}$F$_x$, (see Fig. 1.6 (b)) the question whether the magnetic order disappears completely before superconductivity emerges, is still not resolved [81, 85]. From the existing experimental studies it may be concluded that 1111 family of FeBS doesn't show clear evidence of coexistence of superconductivity and magnetism, with possible exception of SmFeAsO$_{1-x}$F$_x$. On the other hand, 122 members of FeBS show apparent signature of coexistence of often "mutually exclusive" superconductivity and magnetism in the underdoped region of the superconducting dome (see Fig. 1.6 (c)). Also the phase diagram for 11 system reported by Martinelli et al. [83] shows the coexistence of these two phenomena, however Katayama et al. [84] do not report the coexistent of these two order parameters for any doping range (see Fig. 1.6 (d)).

The question whether the coexistence of these two order parameters, which are often intertwined with each other, is at the microscopic level or phase separated, or whether both of these phenomena evolve from the same electrons, is still hotly debated. Existing experiments as well as theoretical studies suggest that the same (Fe-3$d$) electrons are responsible for the AFM state as well as superconductivity at low temperature, suggesting the microscopic coexistence of these two phases [82, 86-92].

## 1.1.6 Superconducting Order Parameter and Pairing Mechanism

Often one thinks of the electronic origin of the pairing mechanism in new superconductors as soon as there is some evidence that conventional phonon mediated mechanism is not strong enough to give the required transition temperature. For FeBS also, the conventional phonon mediated mechanism was ruled out [93], and an unconventional pairing mechanism mediated by AFM spin fluctuation was proposed [94]. The alternative model, which has received much attention recently, is the orbital fluctuations model, put forward by Kontani et al. [95-97]. The importance of orbital fluctuations for the pairing mechanism was appreciated due to the possibility of orbital ordering of the Fe *d*-orbitals (i.e. $d_{xz/yz}$) and it is believed that the same electrons are responsible for superconductivity [77, 99-104]. In fact orbital fluctuations can give rise to attractive interactions required to form Cooper pairs. However, it is difficult to separate orbital and spin degrees of freedom, suggesting the coupled spin-orbital fluctuations for pairing mechanism. It is believed that the pairing mechanism in FeBS is magnetically mediated similar to that in the case of cuprates; but still there is no broad consensus among the scientific community.

One of the important quantity for understanding the pairing mechanism in high temperature superconductors is the superconducting gap, whose magnitude and structure (symmetry) are intimately linked to the microscopic pairing mechanism. Therefore, understanding of the superconducting order parameters in detail should provide a way forward for the understanding of pairing mechanism in FeBS. Within the BCS theory of superconductivity, developed for conventional phonon mediated superconductors, the superconducting phase consists of Cooper pairs in which both the spin and wave-vectors of both the electrons are in opposite direction to each other giving zero net spin to Cooper pair. However, in the absence of spin-orbit coupling, the net spin of a Cooper pair can be zero (singlet state) or one (triplet state). Most of the known superconductors are spin-singlet superconductors. The wave function of the Cooper pair (for singlet state) has to be antisymmetric, which puts constraint on the orbital angular momentum (*L*) i.e. *L* can only be zero or even positive multiple of Planck constant (ℏ). If *L* is 0 or 2ℏ, in that case one has *s*-wave or *d*-wave pairing, respectively. All the conventional superconductors are spin singlet *s*-wave superconductors, and interestingly, cuprates fall in the regime of spin singlet *d*-wave superconductors [105]. On the other hand, if net spin of the Cooper pair is one (triplet state) then $L$ can only be odd positive integer multiple of ℏ. If $L$ is ℏ or 3ℏ then one can have *p*-wave or *f*-wave pairing, respectively. The known example of spin triplet *p*-wave superconductor is $Sr_2RuO_4$ [106]. All the existing superconductors have Cooper pairs with net spin zero or one; however majority of them have zero spin and fall in the regime of spin-singlet superconductors.

Existing experimental studies on FeBS have ruled out the spin triplet state in these materials and suggest spin singlet superconductivity [107-114]. Therefore, orbital momentum associated with it can only be 0, 2ℏ, 4ℏ etc., and it should correspond to *s*-wave or *d*-wave pairing superconductivity. Josephson tunnelling experiments by Zhou [115] and Zhang et al. [116] on these systems have ruled out *d*-wave and *p*-wave superconducting pairing symmetry and their results are consistent with the *s*-wave symmetry, including the $s^{\pm}$ type, which is currently favoured where the order parameter changes sign in going from hole pockets to electron pockets. Also the tunnelling measurements on 1111 systems, $NdFeASO_{1-x}F_x$, provide strong evidence in favour of $s^{\pm}$ pairing [117], which was first proposed theoretically by Mazin et al. [94]. Interestingly, the discovery of superconductivity in the 122* materials with large local moments (~ 3.3 $\mu_B$/Fe) and different Fermi surfaces, having no hole pockets but a large electron pocket at the M-point and a small electron pocket at the Γ point [17], argue against the $s^{\pm}$ pairing.

Within the BCS theory, the $2\Delta/K_bT_c$ ratio is ~ 3.53, where $K_b$ is Boltzaman constant and 2Δ is identified with the energy gap at the Fermi surface below the superconducting transition temperature. The energy of 2Δ is produced by breaking of a Cooper pair which results in the generation of two quasi-particles. In case of *s*-wave superconductor (e.g. Pb, Sn) there are no nodes in the superconducting gap around the Fermi surface. In case of FeBS which are believed to have $s^{\pm}$ pairing symmetry, uniform *s*-wave order parameters develop around the electron and hole pockets but have opposite signs. Fermiology in these materials is a strong function of doping and pressure, therefore electron and hole pockets need not have same electronic density of states at the Fermi surface. As a result of this asymmetry, two gaps associated with hole and electron pockets need not have the same magnitude. In fact these systems are grouped into multigap superconductors with $2\Delta/K_bT_c$ ratio ranging from 2.5 to 15 [118-122] in both electron and hole doped systems. However, the origin of this large spread in ratio, ranging from weak coupling to strong coupling regime, is still not understood.

## 1.1.7 Vibrational Properties and Phonon Dynamics

The 1111 (RFeAsO, R = La, Ce, Sm and Nd) family has layered structure that belongs to the tetragonal *P4/nmm* space group (point group *D4h*) containing two formula units per unit cell. Therefore, there will be 24 (8*3) normal modes (3 acoustic and 21 optical). Out of these only eight are Raman active with irreducible representations as $\Gamma_{Raman} = 2A_{1g} + 2B_{1g} + 4E_g$, and six are infrared active with irreducible representations as $\Gamma_{IR} = 3A_{2u} + 3E_u$ [123]. The phonon modes are localised in R-O and Fe-As layers with the dominant displacement of atoms classified as R: $A_{1g} + E_g$; O/F: $B_{1g} + E_g$; Fe: $B_{1g} + E_g$ and As: $A_{1g} + E_g$. The first order phonon modes are observed below 600 cm$^{-1}$ where phonon modes above 300 cm$^{-1}$ are dominated by oxygen atom. The typical frequencies for the Raman active modes are ~ 100 cm$^{-1}$ ( R, $E_g$ ), ~ 140 cm$^{-1}$ (As and Fe, $E_g$ ), ~ 160 cm$^{-1}$ ( R, $A_{1g}$ ), ~ 200 cm$^{-1}$ ( As, $A_{1g}$ ), ~ 210 cm$^{-1}$ ( Fe, $B_{1g}$ ), ~ 270 cm$^{-1}$ ( Fe, $E_g$ ), ~ 350 cm$^{-1}$ ( O, $B_{1g}$ ) and ~ 440 cm$^{-1}$ ( O, $E_g$) [122-127].

The 11 (FeSe$_{1-x}$Te$_x$) family belongs to space group *P4/nmm* having two formula units per unit cell. It has four Raman active and two infrared active modes belonging to the irreducible representations $\Gamma_{Raman} = A_{1g} + B_{1g} + 2E_g$ and $\Gamma_{IR} = A_{2u} + E_u$ [128]. The typical frequencies of Raman active phonon modes are ~ 100 cm$^{-1}$ (Se/Te, E$_g$), ~ 160 cm$^{-1}$ (Se/Te, A$_{1g}$), ~ 220 cm$^{-1}$ (Fe, E$_g$) and ~ 230 cm$^{-1}$ (Fe, B$_{1g}$) [102, 128-130].

The 122 family (AFe$_2$As$_2$, A = Ba, Ca and Sr) belongs to the space group *I4/mmm*. There are four Raman active and four infrared active modes belonging to the irreducible representations $\Gamma_{Raman} = A_{1g} + B_{1g} + 2E_g$ and $\Gamma_{IR} = 2A_{2u} + 2E_u$ [131]. Figure 1.7 shows the eigen vectors for the Raman active modes. The observed modes for these systems are found at ~ 110 cm$^{-1}$ (As and Fe, E$_g$), ~ 180 cm$^{-1}$ (As, A$_{1g}$), 200 cm$^{-1}$ (Fe, B$_{1g}$) and 260 cm$^{-1}$ (As and Fe, E$_g$) [121,131-135].

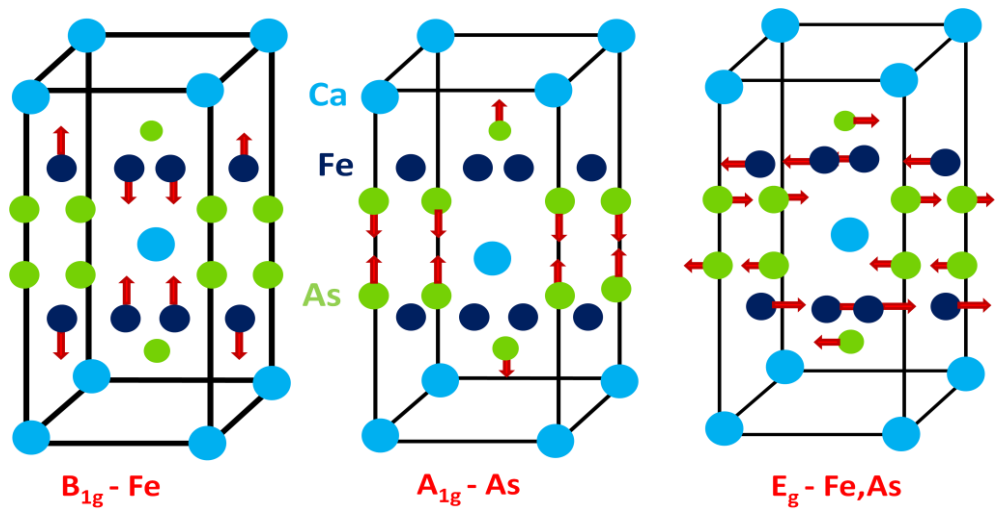

Figure 1.7: Eigen vectors for the 122 systems.

The recently discovered 42622 family (Ca$_4$Al$_2$O$_{6-x}$Fe$_2$As$_2$) also belongs to the *P4/nmm* space group, with two formula units per unit cell. There are 96 (32*3) normal modes, out of these three are acoustic and 93 are optical modes. It has sixteen Raman active and fifteen infrared active modes with irreducible representations $\Gamma_{Raman} = 6A_{1g} + 2B_{1g} + 8E_g$ and $\Gamma_{IR} = 7A_{2u} + 8E_u$.

Many recent experimental [121-141] as well as theoretical studies [34, 50-51, 78, 82-83, 94] on these systems have discussed the very important issue of spin-phonon coupling, which is ubiquitous in FeBS, and coupling of the phonon modes with quasi-particle excitations below the transition temperatures ($T_c$, $T_{SDW}$). Indeed spin degrees of freedom play a central role in FeBS due to the presence of magnetically ordered states, therefore understanding the nature of coupling between spin and lattice vibrations will help to unravel the underlying physics of these materials. In fact the onset of magnetic ordering below the magnetic transition temperature has a profound effect on the electronic structure of these materials which in-turn can affect the phonon self energy via coupling with quasi-particle excitations, and this coupling between different degrees of freedom may help in understanding the pairing mechanism.

## 1.2 Part-B

### 1.2.1 Introduction to Multiferroics

In recent years, a lot of research has focussed on multifunctional materials, which provide a way to combine several physical properties simultaneously in single phase. A key question is what is the nature of coupling between different functionalities. Multiferroics also fall in the class of multifunctional materials as they possess several so called ferroic orders such as (anti)ferroelectricity, (anti)ferromagnetism and ferroelasticity in the same phase. It has been observed that coexistence of often thought to be mutually exclusive phenomena gives rise to phenomena which can be harnessed for the better control of the functionalities of multifunctional materials. Multiferroics belongs to this class where the mutually exclusive ferroic order parameters coexist in the same phase.

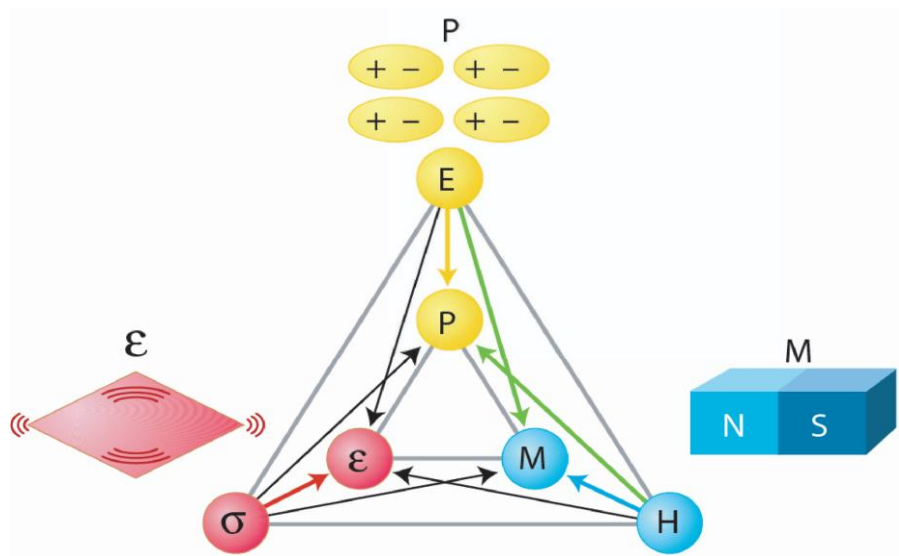

Figure 1.8: Schematic showing the possible cross coupling between different order parameters of a multiferroic material [146].

The idea of coupling between electric and magnetic degrees of freedom in the same phase was put forward for the first time by Landau and Lifshitz in late fifties [142], which triggered the research in this field and soon after it was proved experimentally [143]. However, the term 'multiferroic' was initially coined by Schmid [144] to define a material possessing two or more ferroic order parameters, namely ferroelectricity, ferromagnetism and ferroelasticity in the same phase. Such materials have been shown to exhibit cross-coupling between these different order parameters (see Fig. 1.8), making the field of multiferroics interesting for basic science and applications [145-149]. In a ferroic material, the order parameters have non-zero value below a characteristic transition temperature and it can be switched by external perturbation (such as electric field, magnetic field and stress). Ferroics can be classified mainly into three categories namely ferroelastic, ferroelectric and ferromagnetic where the corresponding order parameters are strain ($\varepsilon$), polarization (P) and magnetization (M), respectively. A ferromagnet has a spontaneous magnetisation (M) and shows hysteresis under an applied magnetic field (H). A ferroelectric has a spontaneous polarisation (P) and shows hysteresis under an applied electric field (E). Similarly, a ferroelastic material undergoes spontaneous deformation or strain ($\varepsilon$) below a transition temperature and shows hysteresis under stress ($\sigma$). When stress is applied, the strain can be switched hysteretically. These spontaneous order parameters typically occur as a result of a phase transition below the characteristic temperature.

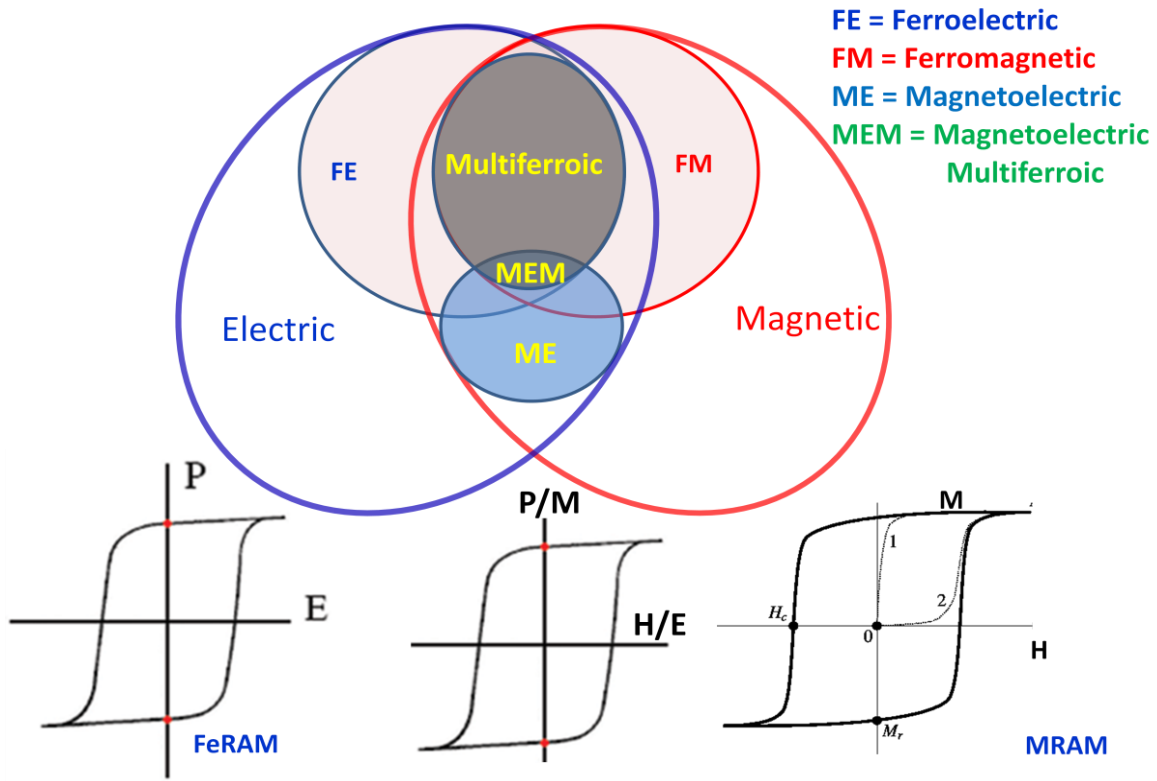

Figure 1.9: Electrically and magnetically polarizable materials are shown as elliptic region (with blue and red outline respectively). Ferroelectrics (FE) and Ferromagnets (FM) which fall under the family of electrically polarisable and magnetically polarisable, respectively, are shown as circular regions. The materials in the overlapped region (grey) between FE and FM materials are multiferroic. Magnetoelectric (ME) effect which need not arise in any of the materials that are both magnetically and electrically polarizable is shown by light blue circular region. The materials belonging to the region MEM are magnetoelectric multiferroic.

### 1.2.2  Ferroelectricity

Materials where the dipoles can be polarized with an application of external electric field fall in the class of electrically polarizable materials as shown in Fig. 1.9. Ferroelectricity is a phenomenon where a non-zero electric polarization exists below a characteristic transition temperature even in the absence of any external field and it can be switched with an external electric field showing the hysteresis with the field. The phenomenon of ferroelectricity is intimately related to symmetry of the systems, therefore only the crystal with non-centrosymmetric structure can show the polar nature. The polarisation in a material arises due to lack of inversion symmetry. For example, consider the perovskite of the form $ABO_3$, (e.g. $BaTiO_3$) [150-51] in which a central positive B-ion (a transition metal element) is surrounded by negatively charged oxygen ions (see Fig. 1.10). A slight shift in the position of the negatively charged B-site ion below the transition temperature would break the inversion symmetry and result in non-zero polarisation i.e. ferroelectric order below the transition temperature. Ferroelectricity can also arise due to other effects such as lone pair effect [152], geometric ferroelectricity [153], charge-ordered ferroelectric [154-55] and magnetic ordering below the transition temperature [149,156].

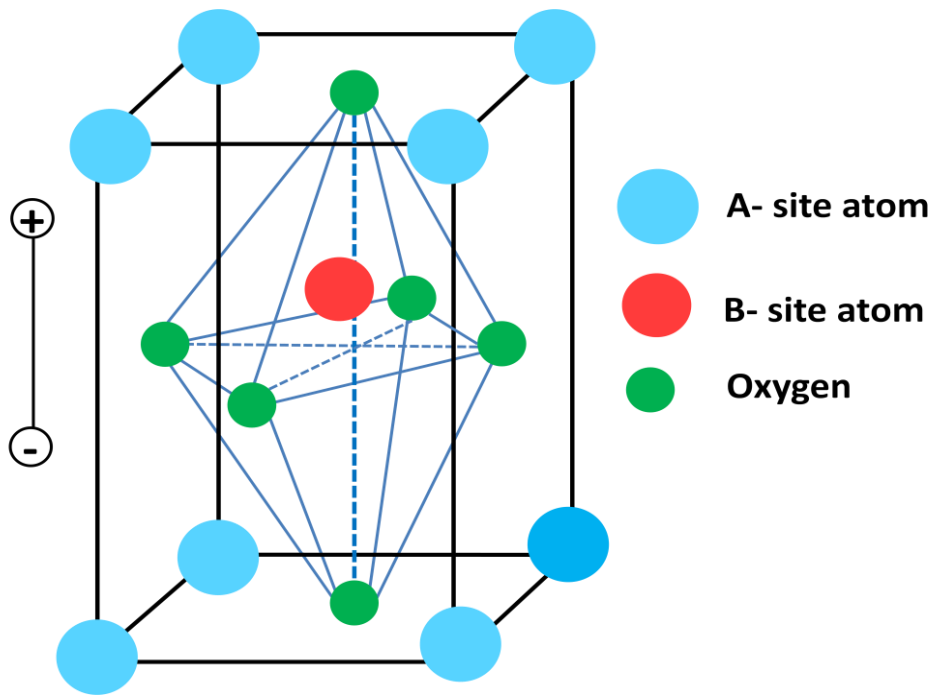

Figure 1.10: Schematic representation of the generation of dipole moment (polarisation) with the displacement of central B-site atom in perovskite crystal structure.

## 1.2.3 Magnetism

Exchange interaction between the localised magnetic moments in a material leads to magnetic order. The interaction between the neighbouring spins, which gives rise to the magnetic moments, can be described using Heisenberg Hamiltonian and is given as

$$\hat{H} = -\sum_{ij} J_{ij} S_i S_j \quad (1.1)$$

where the $J_{ij}$ is the exchange constant which describes the nature of interaction between the neighbouring spins. The interaction is ferromagnetic i.e. neighbouring spins favour parallel alignment (if $J_{ij}$ is negative); or antiferromagnetic i.e. neighbouring spins favour anti-parallel alignment (if $J_{ij}$ is positive) and results in zero net magnetic moment below the transition temperature known as Neel temperature, $T_N$. Ferromagnetism can be considered as the magnetic analogue of ferroelectricity, where the magnetic moment is non zero below the Curie temperature and is switchable with the application of external magnetic field. There exists other type of magnetic ordering known as ferrimagnetism where the spins with different magnetic moments couple antiferromagnetically resulting in finite magnetic moments. The above described three different types of magnetism are illustrated in Fig. 1.11.

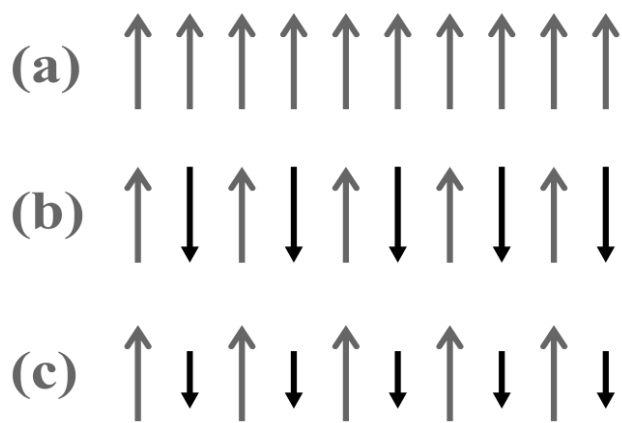

Figure 1.11: Schematic representation of the different magnetic order (a) ferromagnetism (b) antiferromagnetism and (c) ferrimagnetism.

Antiferromagnetism can exist in many forms, however, the net magnetic moment remains zero below the Neel temperature. The spin ordering in the antiferromagnetic state can be described as commensurate or incommensurate depending upon how the periodicity of spins is connected to the crystal structure. If the spin ordering is connected to the crystal structure as shown in Fig. 1.12 then it is termed as commensurate e.g. A-type, G-type antiferromagnetism etc., and if it not connected to the crystal structure then it is termed as incommensurate as shown in Fig. 1.13, e.g. sinusoidally modulated spin density wave, and the cycloidal order where the spins change their orientation along the direction of propagation.

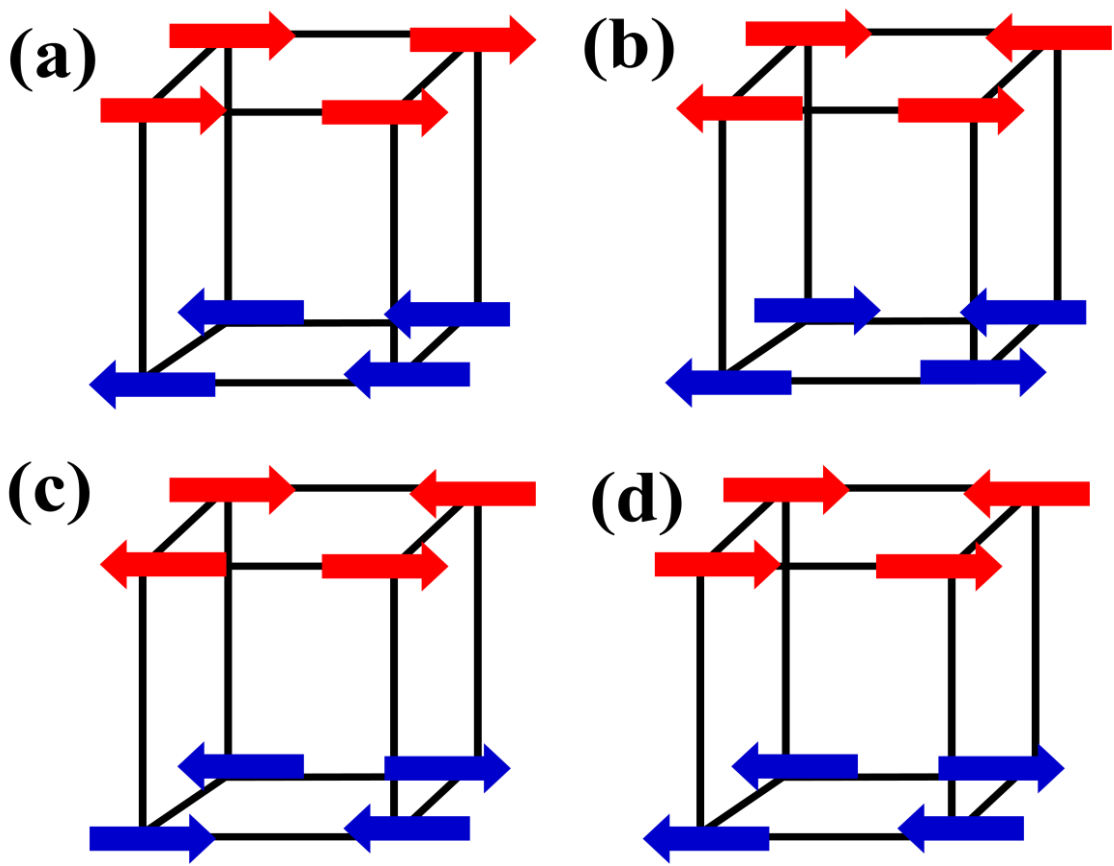

Figure 1.12: Showing the schematic representation of different types of commensurate antiferromagnetic ordering (a) A-type (b) C-type (c) G-type and (d) E-type.

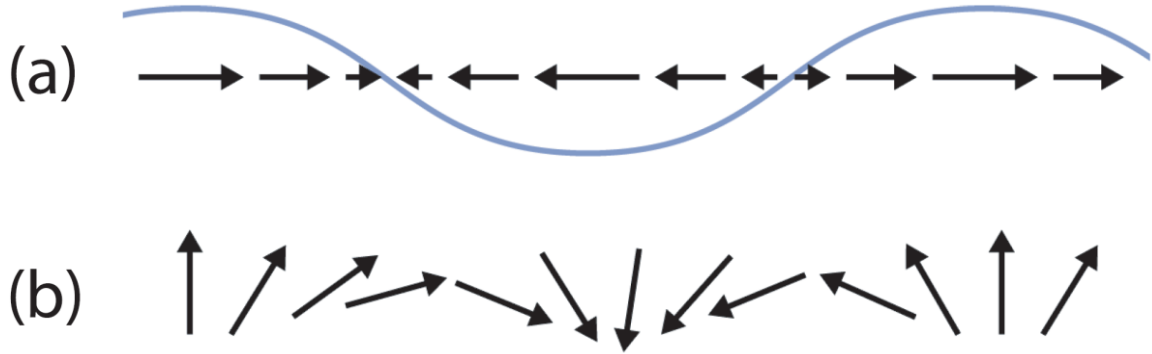

Figure 1.13: Schematic representation of the different types of incommensurate antiferromagnetic ordering (a) sinusoidal and (b) cycloidal.

### 1.2.4 Magneto-electric Effect

Magnetoelectric effect (ME) is a phenomenon where the polarisation can be induced by the application of an external magnetic field and vice-versa. ME effect is intimately linked to the cross-coupling between electric and magnetic order parameter i.e. polarisation and magnetisation, respectively, and also indirectly via strain, which plays a key role in composite multiferroics where strain alone is responsible for a non-zero value of ferroics order-parameters below the characteristic temperature. Within the Landau theory, the phenomena of ME effect is treated by expressing the free energy (F) of a homogeneous and stress free system in terms of magnetic field (H) and electric field (E) as [145]

$$F(\vec{E},\vec{H}) = (F_0 - P_i^s E_i - M_i^s H_i) - (\frac{1}{2}\varepsilon_o\varepsilon_{ij}E_iE_j + \frac{1}{2}\mu_o\mu_{ij}H_iH_j + \alpha_{ij}E_iH_j)$$
$$-\frac{1}{2}\beta_{ijk}E_iH_jH_k - \frac{1}{2}\gamma_{ijk}H_iE_jE_k - \ldots \quad (1.2)$$

where ε and μ are electric and magnetic susceptibilities and α is a second rank tensor known as the linear magneto-electric coefficient. One can differentiate the above equation w.r.t.

electric and magnetic fields to get polarisation ($P = -\partial F/\partial E$) and magnetisation ($M = -\partial F/\partial H$) respectively, (superscript $s$ in above equation denotes the spontaneous component) and can be written as,

$$P_i(\vec{E},\vec{H}) = -\frac{\partial F}{\partial E_i} = P_i^s + \varepsilon_o \varepsilon_{ij} E_j + \alpha_{ij} H_j + \frac{1}{2}\beta_{ijk} H_j H_k + \gamma_{ijk} H_i E_j + \ldots \quad (1.3)$$

$$M_i(\vec{E},\vec{H}) = -\frac{\partial F}{\partial H_i} = M_i^s + \mu_o \mu_{ij} H_j + \alpha_{ij} E_j + \beta_{ijk} E_i H_j + \frac{1}{2}\gamma_{ijk} E_i E_k + \ldots \quad (1.4)$$

It can be clearly seen from the above two equations that in the case of pure electric and magnetic systems only the first two terms are non-zero. The coefficient of third term (i.e. α) contains the information about the cross coupling between electric and magnetic order parameters, and it is responsible for linear ME effect. The higher order coefficients (i.e. β and γ) describe higher order coupling and result in non-linear ME effect.

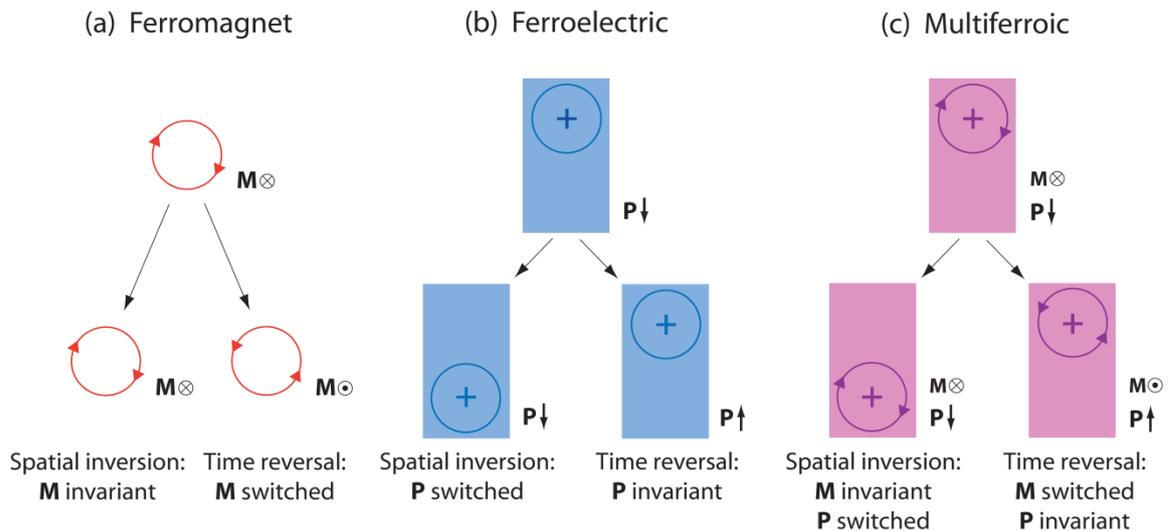

Figure 1.14: Showing the effect of spatial inversion and time reversal on (a) ferromagnet (b) ferroelectric and (c) multiferroics [157]. Multiferroics don't possess any symmetry i.e. symmetry breaking is intrinsic to the multiferroic materials.

Majority of existing materials have zero cross coupling between their electric and magnetic order parameters. This can be explained very simply on the basis of symmetry argument, for example electric polarisation (P) changes sign under spatial inversion and is invariant under time reversal, and magnetisation changes sign under time reversal and is invariant under spatial inversion (schematically shown in Fig. 1.14). The magnetoelectric susceptibility tensor is also invariant for inversion symmetry [157]. Now, considering only the cross coupling term, neglecting higher order trems, in equation 1.3 i.e. P = α H, under spatial inversion it becomes -P = α H. This is consistent with the initial state only if α is zero i.e. no cross coupling. However, α can be non-zero only for materials which break both time and spatial inversion symmetry (Fig. 1.14) i.e. materials having multiferroics property have strong ME effect. The first experimental evidence of a magenetoelectric effect was reported in antiferromagnetic $Cr_2O_3$ [143,158-159], although the strength of cross-coupling was not much and limited their potential application. However, this discovery triggered a great deal of research in the scientific community to study these systems experimentally as well as theoretically.

## 1.2.5 Multiferroics

Multiferroics are the materials where multiple ferroic order parameters can coexist (see Fig. 1.9). The coexistence of these order parameters often leads to an intricate coupling and this leads to the genesis of potential device applications based on these materials. Majority of the existing device applications are based on materials where their electric and magnetic properties are controlled only by electric and magnetic field, respectively. However, cross coupling between electric and magnetic field offers a renewed opportunity in the coming years both from fundamental physics' point of view as well as potential applications. In majority of existing ferroelectric materials, e.g. $BaTiO_3$, ferroelectricty is caused by the hybridisation between empty *d*-orbitals (*d⁰*) of Ti with the occupied *p*-orbitals of the surrounded oxygen atoms [150-51,160]. On the other hand, magnetic ordering required a partially filled *d*-orbitals (*dⁿ*) therefore these two order parameters seem incompatible with each other. Hill [161] and Khomeski [149] illustrated that this mutually exclusive property is not a theorem and therefore there certainly exists some way which can be used to combine these two order parameters in a single phase. Although multiferroic materials are known since 1960s, the reinvestigation of multiferroicity in BiFeO$_3$ and TbMnO$_3$ rejuvenated this field because ferroelectricity and magnetism not only coexist in these systems but surprisingly magnetism is responsible for ferroelectricity [162-163]. At present, the materials showing multiferroics nature are increasing rapidly, ranging from oxides to fluorides and oxyfluorides [164].

## 1.2.6 Classification of Multiferroics

After the discovery of multiferroics materials in last decade, e.g. BiFeO$_3$, TbMnO$_3$ etc., the Schmid's definition of multiferroic [144] has been evolving constantly. Broadly multiferroics can be divided in to two main classes [149]:

**Type- I:** Type- I multiferroics are those in which ferroelectric and magnetic orderings occur at different temperatures, usually the ferroelectric order parameter sets in at a much higher temperature than the magnetic transition temperature and both of these orderings have different microscopic origin. Owing to different microscopic origins there is a very weak coupling between these order parameters, e.g. BiFeO$_3$ is a type- I multiferroic with $T_{FE}$ ~ 1100 K, $T_N$ ~ 643 K and P ~ 90μc/cm$^2$ [163].

**Type- II:** This family of multiferroics is a relatively new class of materials where ferroelectricity is induced by magnetism implying a strong coupling between electric and magnetic degrees of freedom. However, the polarisation is of an order of magnitude smaller than the type-I multiferroics, which limits their practical application [162,165]. The famous example of this family is TbMnO$_3$ ($T_N$ ~ 46 K, $T_{FE}$ ~ 26 K), where the electric polarisation arises as a result of cycloidal antiferromagnetic ordering below ~ 26 K.

A challenge in the field of multiferroics is to identify the underlying mechanism and search for a lead free material that provides strong coupling, like type - II multiferroics, between magnetization and polarisation at room temperature. Multiferroics continue to reveal rich physics and their potential applications also extend far beyond conventional magnetic control of polarisation and vice-versa, which further widens the boundary that is still effectively unexplored.

## 1.3  Part-C

### 1.3.1  Systems Studied in the Thesis

Systems studied in this thesis and described in detailed in part (A) cover iron-based superconductors namely - (i) FeSe$_{0.82}$ (ii) Ce$_{1-z}$Y$_z$FeAsO$_{1-x}$F$_x$ (z = 0, 0.4; x = 0.1, 0.2) (iii) Ca$_4$Al$_2$O$_{5.7}$Fe$_2$As$_2$ (iv) Ca(Fe$_{1-x}$Co$_x$)$_2$As$_2$ (x = 0.03, 0.05), and part (B) include multiferroic oxides - (i) TbMnO$_3$ (ii) AlFeO$_3$ and double-perovskite (iii) La$_2$NiMnO$_6$ .

(1) TbMnO$_3$ : Rare earth manganites RMnO$_3$ (R = Lu, Yb, Er, Ho, Dy, Tb, Nd, Pr and La ) can form in two possible phases depending on the ionic radius of the 'R' [166]. For large ionic radii Re = La - Dy, the orthorhombic perovskite structure is stable and for smaller radii they form hexagonal structure. TbMnO$_3$ is stable in the orthorhombic structure. The orthorhombic structure of TbMnO$_3$ exhibits corner-sharing MnO$_6$ octahedra (see Fig. 1.15) [167], which undergoes Jahn-Teller and rotational distortions. Jahn-Teller distortion results from the electronically degenerated Mn$^{3+}$ $d$-orbitals and different size of ionic radii of 'R' and 'Mn' results in rotational distortions [168]. RMnO$_3$ exhibits rich phase diagram owing to their complex magnetic structure (see Fig. 1.16) [169]. TbMnO$_3$ undergoes multiple magnetic and ferroelectric phase transitions, it undergo antiferromagnetic phase transition at ~ 46 K and at lower temperature (~27 K) it develops ferroelectric order due to spiral magnetic ordering of Mn$^{3+}$ spins, and at further lower temperature (~7 K) Tb$^{3+}$ spins order antiferromagnetically [170]. Number of Raman active modes are 24 with irreducible representation as $\Gamma_{Raman} = 7A_g + 5B_{1g} + 7B_{2g} + 5B_{3g}$ and the total number of infra-red active modes are 25 and these are classified as $\Gamma_{IR} = 9B_{1u} + 7B_{2u} + 9B_{3u}$.

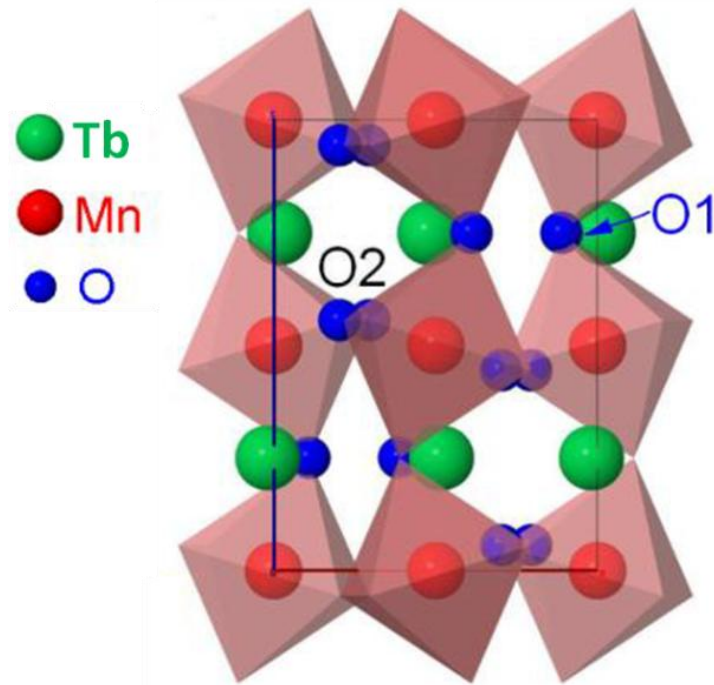

Figure 1.15: Schematic of the orthorhombic crystal structure of TbMnO$_3$. O1 and O2 refer to the apical and planar oxygen atoms, respectively [167].

(2) La$_2$NiMnO$_6$ : Crystal structure of double perovskites systems is very similar to that of perovskites crystal structures as these are extensions of perovskites, here Ni and Mn occupy the centre of corning sharing (Mn/Ni)$O_6$ octahedra in a distorted perovskite structure (see Fig. 1.17) [171]. These systems being ferromagnetic insulators with transition temperature close to room temperature offer a potential solution for practical device applications. They also provide an opportunity to control multiferroics properties since these systems are ferromagnetic and are governed by super-exchange interactions which can generate a non-zero polarization [172]. $La_2NiMnO_6$ typically exists in two phases at room temperature i.e. rhombohedral and monoclinic due to the local inhomogeneities arising from anti-site disorder between Ni/Mn sites. However with stringent synthesis condition one can control the phases [171]. For the monoclinic phase (*P2₁/n*) number of Raman active modes are 24 belonging to the irreducible presentation $\Gamma_{Raman} = 12A_g + 12B_g$, however the rhombohedral (R-3) structure have 8 modes belonging to the irreducible presentation $\Gamma_{Raman} = 4A_g + 4B_g$ [173].

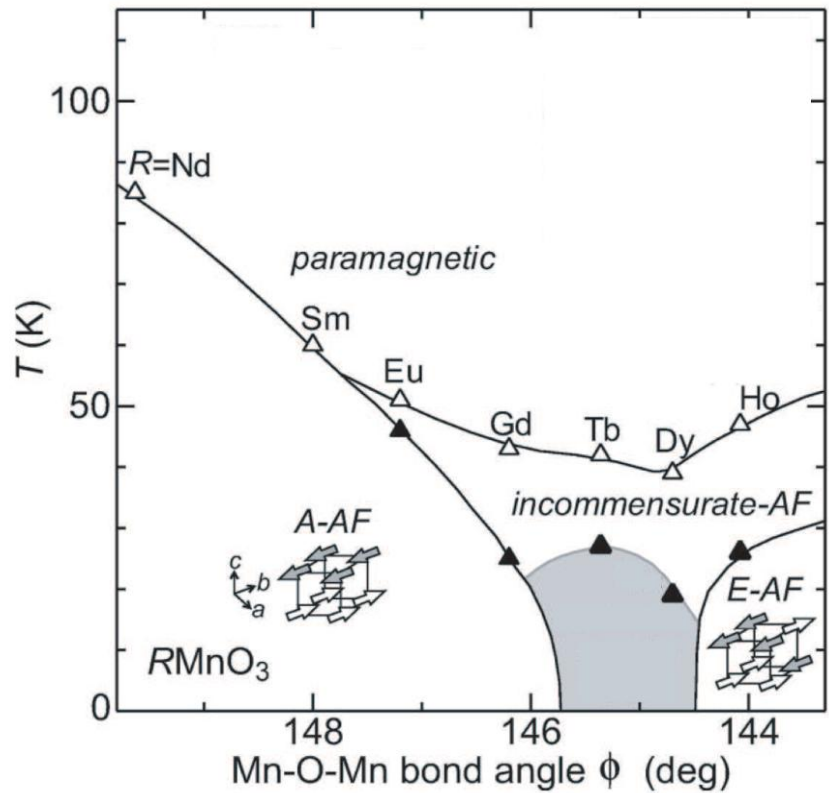

Figure 1.16: Magnetic phase diagram for orthorhombic $RMnO_3$. Grey region indicates the region where Mn spin order cycloidally [169].

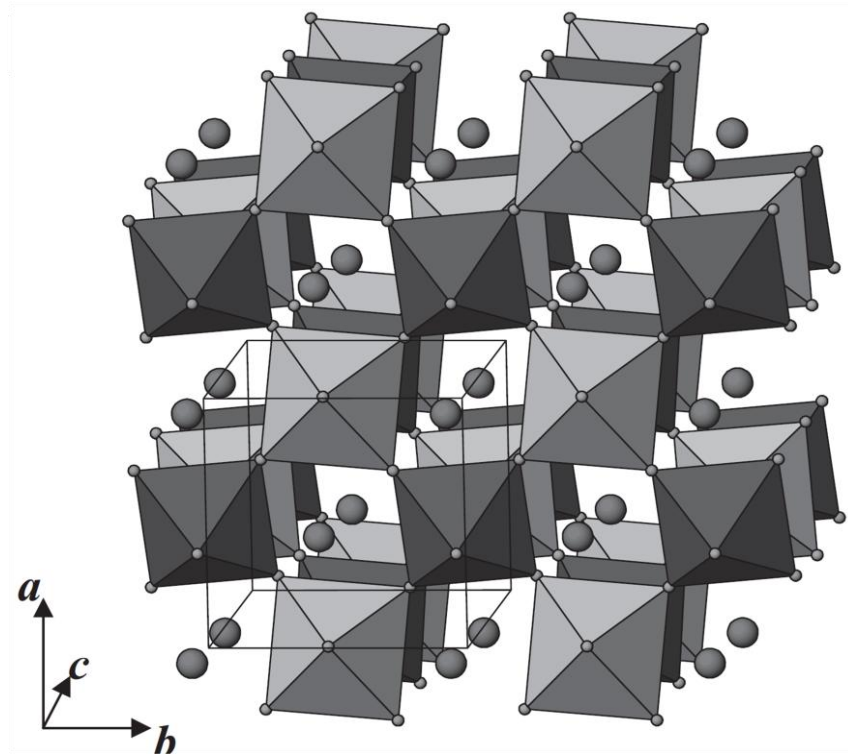

Figure 1.17: Schematic of the crystal structure of La$_2$NiMnO$_6$. Dark and light octahedra are for MnO$_6$ and NiO$_6$, respectively. Large and small spheres represent La and O, respectively [171].

(3) AlFeO$_3$ : AlFeO$_3$ is a new addition to the family of multiferroics, where the anti-site cationic disorder between Fe and Al gives rise to changes in the magnetic order and subsequently to ferroelectricity through spin-phonon coupling [156]. The transition temperature is close to the room temperature (~250 K) which can be tuned with the doping with 'Ga' at the place of Fe [156]. The room temperature structure crystallises in orthorhombic phase (*Pna2₁*) (see Fig. 1.18). Number of Raman active modes are 117 with irreducible representation $\Gamma_{Fe}$= 6$A_1$ + 6$A_2$ + 6$B_1$ + 6$B_2$, $\Gamma_{Al}$ = 6$A_1$ + 6$A_2$ + 6$B_1$ + 6$B_2$ and $\Gamma_O$ = 18$A_1$ + 18$A_2$ + 18$B_1$ + 18$B_2$, while $A_1$ + $B_1$ + $B_2$ are acoustic modes [174].

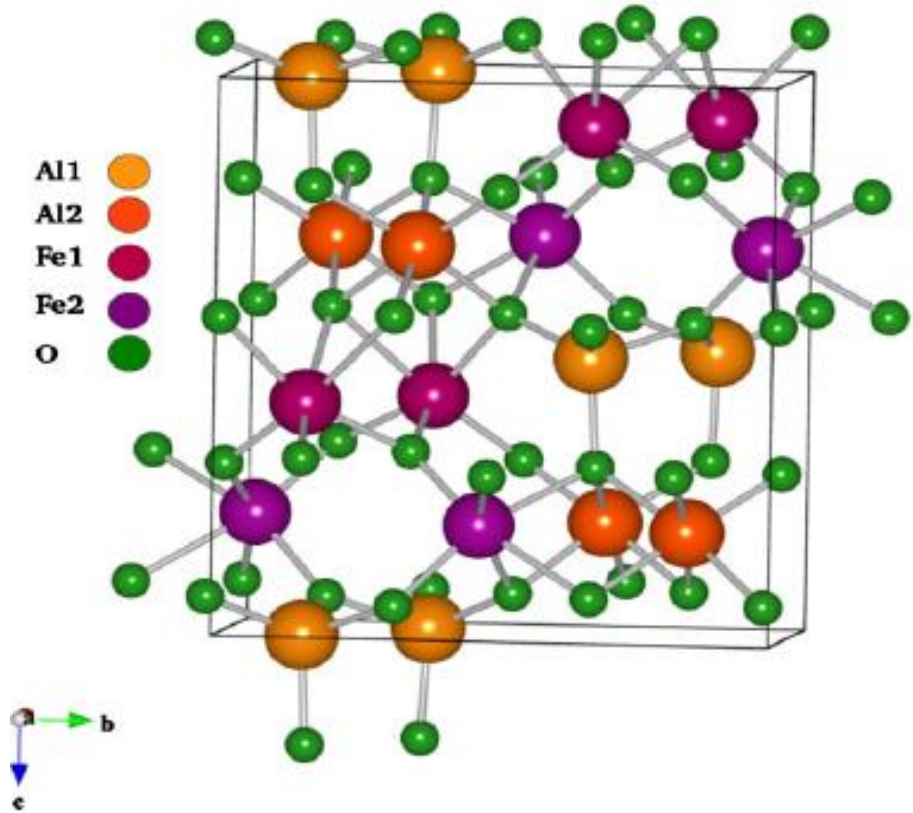

Figure 1.18: Schematic of the orthorhombic structure of AlFeO$_3$.

# Bibliography:

[1] H. K. Onnes, Commun. Phys. Lab, Univ. Leiden 120b, 122b, 124c (1911).

[2] J. Bardeen, L. N. Cooper, and J. R. Schrieffer, Phys. Rev. **106**, 162-164; ibid **108**, 1175 (1957).

[3] J. G. Bednorz and K. A. Muller, Z. Phys. B **64,** 189 (1986).

[4] Y. Kamihara, T. Watanabe, M. Hirano and H. Hosono, J. Am. Chem. Soc. **130,** 3296 (2008).

[5] A. A. Abrikosov and L. P. Gorkov, Sov. Phys. JETP **12**, 1243 (1961).

[6] C. Wang, L. Li, S. Chi, Z. Zhu, Z. Ren, Y. Li, Y. Wang, X. Lin, Y. Luo, S. Jiang, X. Xu, G. Cao and Z. Xu, Euro Phys. Lett. **83,** 67006 (2008).

[7] J. Paglione and R. L. Greene, Nature Phys. **6**, 645 (2010).

[8] Y. Kamihara, H. Hiramatsu, M. Hirano, R. Kawamura, H. Yanagi, T. Kamiya and H. Hosono, J. Am. Chem. Soc. **128**, 10012 (2006).

[9] X. H. Chen, T. Wu, G. Wu, R. H. Liu, H. Chen and D. F. Fang, Nature **453**, 761 (2008).

[10] G. F. Chen, Z. Li, D. Wu, G. Li, W. Z. Hu, J. Dong, P. Zheng, J. L. Luo and N. L. Wang, Phys. Rev. Lett **100**, 247002 (2008).

[11] Z.A. Ren, J. Yang, W. Lu, W. Yi, X.-L. Shen, Z.-C. Li, G.-C. Che, X.-L. Dong, L.-L. Sun, F. Zhou and Z.-X. Zhao, Euro Phys. Lett. **82**, 57002 (2008).

[12] M. Rotter, M. Tegel and D. Johrendt, Phys. Rev. Lett. **101**,107006 (2008).

[13] X. C. Wang, Q. Q. Liu, Y. X. Lv, W. B. Gao, L. X. Yang, R. C. Yu, F. Y. Li and C. Q. Jin, Solid State Commun. **148**, 538 (2008).

[14] F.C. Hsu, J.-Y. Luo, K.-W. Yeh, T.-K. Chen, T.-W. Huang, P.M. Wu, Y.-C. Lee, Y.-L. Huang, Y.-Y. Chu, D.-C. Yan, and M.-K. Wu, PNAS **105**, 14262 (2008).

[15] H. Ogino, Y. Matsumara, Y. Katsura, K. Ushiyama, S. Horii, K. Kishio, and J.-I. Shimoyama, Super. Sci. Technol. **22**, 075008 (2009).

[16] X. Zhu, F. Han, G. Mu, P. Cheng, B. Shen, B. Zeng and H.-H. Wen, Phys. Rev. B **79**, 220512 (2009).

[17] J. Guo, S. Jin, G. Wang, S. Wang, K. Zhu, T. Zhou, M. He and X. Chen, Phys. Rev. B **82**, 180520 (2010).

[18] J. Zhao, Q. Huang, C. de la Cruz, S. Li, J. W. Lynn, Y. Chen, M. A. Green, G. F. Chen, G. Li, Z. Li, J. L. Luo, N. L. Wang and P. Dai, Nature Mater. **7**, 953 (2008).

[19] C. H. Lee, A. Iyo, H. Eisaki, H. Kito, M. T. Fernandez-Diaz, T. Ito, K. Kihou, H. Matsuhata, M. Braden and K. Yamada, J. Phys. Soc. Jpn. **77,** 083704 (2008).

[20] C.H. Lee, H. Eisaki, H. Kito, M. T. Fernandez-diaz, R. Kumai, K. Miyazawa, K. Kihou, H. Matsuhata, M. Braden and K. Yamada, J. Phys. Soc. Jpn. **77**, 44 (Suppl. C) (2008).

[21] K. H. Chen, T. Wu, G. Wu, R. H. Liu, H. Chen and D. F. Fang, Nature **453**, 07045 (2008).

[22] I. R. Shein and A. L. Ivanovskii, Solid State Commun. **149**, 1860 (2009).

[23] F. Steglich, J. Aarts, C. D. Bredl, W. Lieke, D. Meschede, W. Franz and H. Schaefer, Phys. Rev. Lett. **43**, 1892 (1979).

[24] Z. Deng, X. C. Wang, Q. Q. Liu, S. J. Zhang, Y. X. Lv, J. L. Zhu, R. C. Yu and C. Q. Jin, Euro Phys. Lett. **87**, 37004 (2009).

[25] H. Ogino, Y. Shimizu, K. Ushiyama, N. Kawaguchi, K. Kishio and J.-I. Shimoyama, App. Phys. Expre. **3**, 063103 (2010).

[26] W. Bao, G. N. Li, H. Q. Zhen, C. G. Fu, H. J. Bao, W. Du-Ming, M. A. Green, Q. Yi-Ming, L. Jian-Lin and W. M. Mei, Chin. Phys. Lett. **30**, 027402 (2013).

[27] A. Subedi, L. Zhang, D. J. Singh and M. H. Du, Phys. Rev. B **78**, 134514 (2008).

[28] K. Haule and G. Kotliar, New. J. Phys. **11**, 025021 (2009).

[29] Q. Si and E. Abrahmas, Phys. Rev. Lett. **101**, 076401 (2008).

[30] Z. P. Yin, S. Lebegue, M. J. Han, B. P. Neal, S. Y. Savrasov and W. E. Pickett, Phys. Rev. Lett. **101**, 047001 (2008).

[31] D. J. Singh and M. H. Du, Phys. Rev. Lett. **100**, 237003 (2008).

[32] D. J. Singh, Physica C **469**, 418 (2009).

[33] S. Raghu, X. L. Qi, C.-X. Liu, D. J. Scalapino and S.-C. Zhang, Phys. Rev. B **77**, 220503 (2008).

[34] A. Kumar, P. Kumar, U. V. Waghmare and A. K. Sood, J. Phys. Cond. Matt. **22**, 385701 (2010).

[35] D. H. Lu, M. Yi, S.-K. Mo, A. S. Erickson, J. Analytis, J.-H. Chu, D. J. Singh, Z. Hussain, T. H. Geballe, I. R. Fisher and Z.-X. Shen, Nature **455**, 81 (2008).

[36] H. Ikeda, R. Arita and J. Kunes, Phys. Rev. B **81**, 054502 (2010).

[37] Y. Nakai, S. Kitagawa, K. Ishida, Y. Kamihara, M. Hirano and H. Hosono, Phys. Rev. B **79**, 212506 (2009).

[38] H. Hosono, J. Phys. Soc. Jap. **77**, Suppl. C1 (2008).

[39] W. Malaeb, T. Yoshida, T. Kataoka, A. Fujimor, M. Kubota, K. Ono, H. Usui, K. Kuroki, R. Arita, H. Aoki, Y. Kamihara, M. Hirano and H. Hosono, J. Phy. Soc. Jap. **77**, 093714 (2008).

[40] C. Cao, P. J. Hirschfeld and H. P. Cheng, Phys. Rev. B **77**, 220506 (2008).

[42] D. J. Singh, Phys. Rev. B **78**, 094511 (2008).

[43] S. Graser, T. A. Maier, P. J. Hirschfeld and D. J. Scalapino, New J. Phys. **11**, 025016 (2009).

[44] M. D. Lumsden and A. D. Christianson, J. Phys. Cond. Matt. **22** (20), 203203 (2010).

[46] S. Lebegue, Z. P. Yin and W. E. Pickett, New J. Phys. **11**, 025004 (2009).

[47] K. Kuroki, H. Usui, S. Onari, R. Arita and H. Aoki, Phys. Rev. B **79**, 224511 (2009).

[48] Y. Qui, W. Bao, Q. Huang, T. Yildirim, J.M. Simmons, M. A. Green, J. W. Lynn, Y. C. Gasparovic, J. Li, T. Wu, G. Wu and X.H. Chen, Phys. Rev. Lett. **101**, 257002 (2008).

[49] Y. Chen, J. W. Lynn, J. Li, G. Li, G. F. Chen, J. L. Luo, N. L. Wang, P. Dai, C. de la Cruz and H. A. Mook, Phys. Rev. B **78**, 064515 (2008).

[50] T. Yildirim, Phys. Rev. Lett. **101**, 057010 (2008).

[51] T. Yildirim, Physica C **469**, 425 (2009).

[52] J. Dai, Q. Si, J. X. Zhuc and E. Abrahamsd, PNAS **106**, 4118 (2009).

[53] Q. Si, E. Abrahms, J. Dai and J. X. Zhu, New J. Phys. **11**, 045001 (2009).

[54] K. Haule, J. H. Shim and G. Kotliar, Phys. Rev. Lett. **100**, 226402 (2008).

[55] L. Craco, M. S. Laad, S. Leoni and H. Rosner, Phys. Rev. B **78**, 134511 (2008).

[56] M. D. Johannes and I. I. Mazin, Phys. Rev. B **79**, 220510 (2009).

[57] W. Z. Hu, J. Dong, G. Li, Z. Li, P. Zheng, G. F. Chen, J. L. Luo, and N. L. Wang, Phys. Rev. Lett. **101**, 257005 (2008).

[58] C. C. Homes, A. Akrap, J. Wen, Z. Xu, Z. W. Lin, Q. Li and G. Gu, arXiv: 1007.1447.

[59] N. Barasic, D. Wu, M. Dressel, L. J. Li, G. H. Cao and Z. A. Xu, Phys. Rev. B **82**, 054518 (2010).

[60] D. Wu, N. Barisic, P. Kallina, A. Faridian, B. Gorshunov, N. Drichko, L. J. Li, X. Lin, G. H. Cao, Z. A. Xu, N. L. Wang and M. Dressel, Phys. Rev. B **81**, 100512 (2010).

[61] A. Lucarelli and A. Dusza, New J. Phys. **12**, 094506 (2009).

[62] M. M. Qazilbash, J. J. Hamlin, R. E. Baumbach, L. Zhang, D. J. Singh, M. B. Maple and D. N. Basov, Nature Phys. **5**, 647 (2009).

[63] Z. G. Chen, R. H. Yuan, T. Dong and N. L. Wang, Phys. Rev. B **81**, 100502 (2010).

[64] G. F. Chen, Z. G. Chen, J. Dong, W. Z. Hu, G. Li, X. D. Zhang, P. Zheng, J. L. Luo, and N. L. Wang, Phys. Rev. B **79**, 140509 (2009).

[65] J. Dong, H. J. Zhang, G. Xu, Z. Li, G. Li, W. Z. Hu, D. Wu, G. F. Chen, X. Dai, J. L. Luo, Z. Fang, N. L. Wang, Euro Phys. Lett. **83**, 27006 (2008).

[66] C. de La Cruz, Q. Huang, J. W. Lynn, J. Li, W. Ratcliff, J. L. Zarestky, H. A. Mook, G. F. Chen, J. L. Luo, N. L. Wang and P. Dai, Nature **453**, 899 (2008).

[67] Q. Huang, Y. Qui, W. Bao, M. A. Green, J. W. Lynn, Y. C. Gasparovic, T. Wu, G. Wu and X. H. Chen, Phys. Rev. Lett. **101**, 257003 (2008).

[68] F. Wang and D. H. Lee, Science **332**, 200 (2011).

[69] D. K. Pratt, W. Tian, A. Kreyssig, J. L. Zarestky, S. Nandi, N. Ni, S. L. Budko, P. C. Canfield, A. I. Goldman and R. J. McQueeney, Phys. Phys. Lett. **103**, 087001 (2009).

[70] R. Viennois, E.Giannini, D. van der Marel and R. Cerny, J. Soli. Sta. Chem. **183**, 769 (2010).

[71] G. F. Chen, W. Z. Hu, J. L. Luo and N. L. Wang, Phys. Rev. Lett. **102**, 227004 (2009).

[72] D. R. Parker, M. J. Pitcher, P. J. Baker, I. Franke, T. Lancaster, S. J. Blundell and S. J. Clarke, Chem. Commu. 2189 (2009).

[73] S. Li, C. de La Cruz, Q. Huang, G. F. Chen, T.-L. Xia, J. L. Luo, N. L. Wang and P. Dai, Phys. Rev. B **80**, 020504 (2009).

[74] T. M. Mcqueen, A. J. Williams, P. W. Stephens, J. Tao, Y. Zhu, V. Ksenofontov, F. Casper, C. Felser and R. J. Cava, Phys. Rev. Lett. **103**, 057002 (2009).

[75] R. Prozorov, M. A. Tanatar, N. Ni, A. Kreyssig, S. Nandi, S. L. Budko, A. I. Goldman and P. C. Canfield, Phys. Rev. B **80**, 174517 (2009).

[76] C. Fang, H. Yao, W. F. Tsai, J. P. Hu and S. A. Kivelson, Phys. Rev. B **77**, 224509 (2008).

[77] C. Xu, M. Muller and S. Sachdev, Phys. Rev. B **78**, 020501 (2008).

[78] T. Yildirim, Phys. Rev. Lett. **102**, 037003 (2009).

[79] C. C. Chen, B. Moritz, J. van den Brink, T. P. Devereaux and R. R. P. Singh, Phys. Rev. B **80**, 180418 (2009).

[80] W. Lv, J. Wu and P. Phillips, Phys. Rev. B **80**, 224506 (2009).

[81] A. J. Drew, C. Niedermayer, P. J. Baker, F. L. Pratt, S. J. Blundell, T. Lancaster, R. H. Liu, G. Wu, X. H. Chen, I. Watanabe, V. K. Malik, A. Dubroka, M. Rossle, K. W. Kim, C. Baines and C. Bernhard, Nature Mater. **8**, 2396 (2009).

[82] S. Nandi, M. G. Kim, A. Kreyssig, R. M. Fernandes, D. K. Pratt, A. Thaler, N. Ni, S. L. Budko, P. C. Canfield, J. Schmalian, R. J. McQueeney and A. I. Goldman, Phys. Rev. Lett. **104**, 057006 (2010).

[83] A. Martinelli, A. Palenzona, M. Tropeano, C. Ferdeghini, M. Putti, M. R. Cimberle, T. D. Nguyen, M. Affronte and C. Ritter, Phys. Rev. B **81**, 094115 (2010).

[84] N. Katayama, S. Ji, D. Louca, S.-H. Lee, M. Fujita, T. J. Sato, J. S. Wen, Z. J. Xu, G. D. Gu, G. Xu, Z. W. Lin, M. Enoki, S. Chang, K. Yamada and J. M. Tranquada, arXiv: 1003.4525 (2010).

[85] Y. Kamihara, T. Nomura, M. Hirano, J. E. Kim, K. Kato, M. Takata, Y. Kobayashi, S. Kitao, S. Higashitaniguchi, Y. Yoda, M. Seto and H. Hosono, New J. Phys. **12**, 033005 (2005).

[86] A. B. Voronstov, M. G. Varilov and A. V. Chubukov, Phys. Rev. B **79**, 060508 (2009).

[87] Y. Laplace, J. Bobroff, F. R.-Albenque, D. Colson and A. Forget, Phys. Rev. B **80**, 140501 (2009).

[88] R. M. Fernandes, D. K. Pratt, W. Tian, J. Zarestky, A. Kreyssig, S. Nandi, M. G. Kim, A. Thaler, N. Ni, P. C. Canfield, R. J. McQueeney, J. Schmalian and A. I. Goldman, Phys. Rev. B **81**, 140501 (2010).

[89] R. Prozorov, M. A. Tanatar, E. C. Blomberg, P. Prommapan, R. T. Gordon, N. Ni, S. L. Budko and P. C. Canfield, Physica C **469**, 667 (2009).

[90] D. K. Pratt, W. Tian, A. Kreyssig, J. L. Zarestky, S. Nandi, N. Ni, S. L. Budko, P. C. Canfield, A. I. Goldman and R. J. McQueeney, Phys. Rev. Lett. **103**, 087001 (2009).

[91] N. Ni, M. E. Tillman, J.-Q. Yan, A. Kracher, S. T. Hannahs, S. L. Budko and P. C. Canfield, Phys. Rev. B **78**, 214515 (2008).

[92] Z. Shermadini, A. K. Maziopa, M. Bendele, R. Khasanov, H. Luetkens, K. Conder, E. Pomjakushina, S. Weyeneth, V. Pomjakushin, O. Bossen and A. Amato, Phys. Rev. Lett. **106**, 117602 (2011).

[93] L. Boeri, O. V. Dolgov and A.A. Golubov, Phys. Rev. Lett. **101**, 026403 (2008).

[94] I. I. Mazin, D. J. Singh, M. D. Johannes and M. H. Du, Phys. Rev. Lett. **101**, 057003 (2008).

[95] H. Kontani and S. Onari, Phys. Rev. Lett. **104**, 157001 ( 2010).

[96] S. Onari, H. Kontano and M. Sato, Phys. Rev. B **81**, 060504 (2010).

[98] T. Saito, S. Onari and H. Kontani, Phys. Rev. B **82**, 144510 (2010).

[99] F. Kruger, S. Kumar, J. Zaanen and J. Van den Brink, Phys. Rev. B **79**, 054504 (2009).

[100] C. C. Lee, W. G. Yin and W. Ku, Phys. Rev. Lett. **103**, 267001 (2009).

[101] V. Barzykin and L. P. Gorkov, Phys. Rev. B **79**, 134510 (2009).

[102] P. Kumar, A. Kumar, S. Saha, J. Prakash, S. Patnaik, U. V. Waghmare, A. K. Ganguly and A. K. Sood, Solid State Commun. **150**, 557 (2010).

[103] T. M. Chuang, M. P. Allan, J. Lee,Y. Xie, N. Ni, S. L. Budko, G. S. Boebinger, P. C. Canfield and J. C. Davis, Science **327**, 181 (2010).

[104] J. H. Chu, J. G. Analytis, K. De Greve, P. L. McMahon, Z. Islam, Y. Yamamoto and I. R. Fisher, Science **329**, 824 (2010).

[105] C. C. Tsuei and J. R. Kirtely, Rev. Mod. Phys. **72**, 969 (2000).

[106] Y. Maeno, T. M. Rice and M. Sigrist, Phys. Today **54**, 42 (2001).

[107] F. Ning, K. Ahilan, K. Ahilan, T. Imai, A. S. Sefat, R. Jin, M. A. McGuire, B. C. Sales and D. Mandrus, J. Phys. Soc. Jpn. **77**, 103705 (2008).

[108] H. J. Grafe, D. Paar, G. Lang, N. J. Curro, G. Behr, J. Werner, J. H.-Borrero, C. Hess, N. Leps, R. Klingeler and B. Buechner, Phys. Rev. Lett. **101**, 047003 (2008).

[109] K. Matano, Z. A. Ren, Z. A. Ren, X. L. Dong, L. L. Sun, Z. X. Zhao and G.-Q. Zheng, Euro Phys. Lett. **83**, 57001 (2008).

[110] K. Matano, Z. Li, G. L. Sun, D. L. Sun, C. T. Lin, M. Ichioka and G.-Q. Zheng, Euro Phys. Lett. **87**, 27012 (2009).

[111] Y. Yashima, H. Nishimura, H. Mukuda, Y. Kitaoka, K. Miyazawa, P. M. Shirage, K. Kiho, H. Kito and H. Eisaki, J. Phys. Soc. Jpn. **78**, 103702 (2009).

[112] P. Jeglic, A. Potocnik, M. Klanjsek, M. Bobnar, M. Jagodic, K. Koch, H. Rosner, S. Margadonna, B. Lv, A. M. Guloy and D. Arcon, Phys. Rev. B **81**, 140511 (2010).

[113] Z. Li, Y. Ooe, X. C. Wang, Q. Q. Liu, C. Q. Jin, M. Ichioka and G.-Q. Zheng, J. Phys. Soc. Jpn. **79**, 083702 (2010).

[114] Y. Nakai, T. Iye, S. Kitagawa, K. Ishida, S. Kasahara, T. Shibauchi, Y. Matsuda and T. Terashima, Phys. Rev. B **81**, 020503 (2010).

[115] Y. R. Zhou, Y. R. Li, J. W. Zuo, R. Y. Liu, S. K. Su, G. F. Chen, J. L. Lu, N. L. Wang, Y.-P. Wang, arXiv: 0812.3295.

[116] X. Zhang, Y. S. Oh, Y. Liu, L. Yan, K. H. Kim, R. L. Greene and I. Takeuchi, Phys. Rev. Lett. **102**, 147002 (2009).

[117] C. T. Chen, C. C. Tsuei, M. B. Ketchen, Z.-A. Ren and Z. X. Zhao, Nature Phys. **6**, 260 (2010).

[118] D. C. Johnston, Adv. Phys. **59**, 803 (2010).

[119] D. V. Evtushinsky, D. S. Inosov, V. B. Zabolotnyy, M. S. Viazovska, R Khasanov, A. Amato, H.-H. Klauss, H. Luetkens, C. Niedermayer, G. L. Sun, V. Hinkov, C. T. Lin, A. Varykhalov, A. Koitzsch, M. Knupfer, B Buchner, A. A. Kordyuk, and S. V. Borisenko, New J. Phys. **11**, 055069 (2009).

[120] P. Kumar, D. V. S. Muthu, P. M. Shirage, A. Iyo and A. K. Sood, Appl. Phys. Lett. **100**, 222602 (2012).

[121] P. Kumar, D. V. S. Muthu, A. Kumar, U. V. Waghmare, L. Harnagea, C. Hess, SWurmehl, S. Singh, B. Buchner and A. K. Sood, J. Phys. Cond. Matt. **23**, 255403 (2011).

[122] P. Kumar, A. Kumar, S. Saha, D. V. S. Muthu, J. Prakash, U. V. Waghmare, A. K. Ganguli and A. K. Sood, J. Phys. Cond. Matt. **22**, 255402 (2010).

[123] V. G. Hadjiev, M. N. Iliev, K. Sasmal, Y.-Y. Sun and C. W. Chu, Phys. Rev. B **77**, 220505 (2008).

[124] S. C. Zhao, D. Hou, Y. Wu, T. L. Xia, A. M. Zhang, G. F. Chen, J. L. Luo, N. L. Wang, J. H. Wei, Z. Y. Lu and Q. M. Zhang, Super. Sci. Techn. **22**, 015017 (2009).

[125] L. Zhang, T. Fujita, F. Chen, D. L. Feng, S. Maekawa and M. W. Chen, Phys. Rev. B **79**, 052507 (2009).

[126] Y. Gallais, A. Sacuto, M. Cazayous, P. Cheng, L. Fang and H. H. Wen, Phys. Rev. B **78**, 132509 (2008).

[127] C. Marini, C. Mirri, G. Profeta, S. Lupi, D. D. Castro, R. Sopracase, P. Postorino, P. Calvani, A. Perucchi, S. Massidda, G. M. Tropeano, M. Putti, A. Martinelli, A. Palenzona and P. Dore, Euro Phys. Lett. **84**, 67013 (2008).

[128] T. L. Xia, D. Hou, S. C. Zhao, A. M. Zhang, G. F. Chen, J. L. Luo, N. L. Wang, J. H. Wei, Z.-Y. Lu and Q. M. Zhang, Phys. Rev. B **79**, 140510 (2009).

[129] V. Gnezdilov, Y. Pashkevich, P. Lemmens, A. Gusev, K. Lamonova, T. Shevtsova, I. Vitebskiy, O. Afanasiev, S. Gnatchenko, V. Tsurkan, J. Deisenhofer and A. Loidl, Phys. Rev. B **83**, 245127 (2011).

[130] K. Okazaki, S. Sugai, S. Niitaka and H. Takagi, Phys. Rev. B **83**, 035103 (2011).

[131] K. Y. Choi, D. Wulferding, P. Lemmens, N. Ni, S. L. Budko and P. C. Canfield, Phys. Rev. B **78**, 212503 (2008).

[132] A. P. Litvinchuk, V. G. Hadjiev, M. N. Iliev, B. Lv, A. M. Guloy and C. W. Chu, Phys. Rev. B **78**, 060503 (2008).

[133] M. Rahlenbeck, G. L. Sun, D. L. Sun, C. T. Lin, B. Keimer and C. Ulrich, Phys. Rev. B **80**, 064509 (2009).

[134] L. Chauviere, Y. Gallais, M. Cazayous, A. Sacuto, M. A. Measson, D. Colson and A. Forget, Phys. Rev. B **80**, 094504 (2009).

[135] S. Sugai, Y. Mizuno, K. Kiho, M. Nakajima, C. H. Lee, A. Iyo, H. Eisaki and S. Uchida, Phys. Rev. B **82**, 140504 (2010).

[136] M. Zbiri, H. Schober, M. R. Johnson, S. Rols, R. Mittal, Y. Su, M. Rotter and D. Johrendt, Phys. Rev. B **79**, 064511 (2009).

[137] T. Shimojima, K. Ishizaka, Y. Ishida, N. Katayama, K. Ohgushi, T. Kiss, M. Okawa, T. Togashi, X. Y. Wang, C. T. Chen, S. Watanabe, R. Kadota, T. Oguchi, A. Chainani and S. Shin, Phys. Rev. Lett. **104**, 057002 (2010).

[138] S. E. Sebastian, J. Gillett, N. Harrison, P. H. C. Lau, D. J. Singh, C. H. Mielke and G. G. Lonzarich, J. Phys. Cond. Matt. **20**, 422203 (2008).

[139] J. G. Analytis, R. D. McDonald, J. H. Chu, S. C. Riggs, A. F. Bangura, C. Kucharczyk, M. Johannes and I. R. Fisher, Phys. Rev. B **80**, 064507 (2009).

[140] A. Kreyssig, M. A. Green, Y. Lee, G. D. Samolyuk, P. Zajdel, J. W. Lynn, S. L. Budko, M. S. Torikachvili, N. Ni, S. Nandi, J. B. Leao, S. J. Poulton, D. N. Argyriou, B. N. Harmon, R. J. McQueeney, P. C. Canfield and A. I. Goldman, Phys. Rev. B **78**, 184517 (2008).

[141] A. I. Goldman, A. Kreyssig, K. Prokes, D. K. Pratt, D. N. Argyriou, J. W. Lynn, S. Nandi, S. A. J. Kimber, Y. Chen, Y. B. Lee, G. Samolyuk, J. B. Leao, S. J. Poulton, S. L. Budko, N. Ni, P. C. Canfield, B. N. Harmon and R. J. McQueeney, Phys. Rev. B **79**, 024513 (2009).

[142] L. D. Landau and E. M. Lifshitz, Electrodynamics of continuous medium (Fizmatgiz, Moscow 1959).

[143] D. N. Astrov, Sov. Phys. JETP **11**, 708 (1960).

[144] H. Schmid, Ferroelectrics, **162**, 317 (1994).

[145] M. Fiebig, Journal of Physics D: Applied Physics **38**, R123 (2005).

[146] N. A. Spaldin and M. Fiebig, Science **309**, 391 (2005).

[147] R. Ramesh and N. A. Spaldin, Nature Materials **6**, 21 (2007).

[148] S. W. Cheong and M. Mostovoy, Nature Materials **6**, 13 (2007).

[149] D. Khomskii, Physics **2**, 20 (2009).

[150] W. J. Merz, Phys. Rev. **76**, 1221 (1949).

[151] W. J. Merz, Phys. Rev. **91**, 513 (1953).

[152] J. Moreau and C. Michel, Journal of Phys. and Chem. of Solids **32**, 1315 (1971).

[153] B. B. V. Aken and T. T. Palstra, Nature Materials **3**, 164 (2004).

[154] T. Portengen, T. Ostreich and L. J. Sham, Phys. Rev. B **54**, 17452 (1996).

[155] N. Ikeda, H. Ohsumi, K. Ohwada, K. Ishii, T. Inami, K. Kakurai, Y. Murakami, K. Yoshii, S. Mori, Y. Horibe and H. Kito, Nature **436**, 1136 (2005).

[156] P. Kumar, D. V. S. Muthu, S. N. Shirodkar, R. Saha, A. Shireen, A. Sundaresan, U. V. Waghmare, A. K. Sood and C. N. R. Rao, Phys. Rev. B **85,** 134449 (2012).

[157] W. Eerenstein, N. D. Mathur and J. F. Scott, Nature **442**, 759 (2006).

[158] V. J. Folen, G. T. Rado and E. W. Stalder, Phys. Rev. Lett. **6**, 607 (1961).

[159] G. T. Rado and V. J. Folen, Phys. Rev. Lett. **7**, 310 (1961).

[160] R. E. Cohen, Nature **358**, 136 (1992).

[161] N. A. Hill, Journal of Phys. Chem. B **104**, 6694 (2000).

[162] T. Kimura, T. Goto, H. Shintani, K. Ishizaka, T. Arima and Y. Tokura, Nature **426**, 55 (2003).

[163] J. Wang, J. B. Neaton, H. Zheng, V. Nagarajan, S. B. Ogale, B. Liu, D. Viehland, V. Vaithyanathan, D. G. Schlom, U. V. Waghmare, N. A. Spaldin, K. M. Rabe, M. Wuttig and R. Ramesh, Science **299**, 1719 (2003).

[164] J. F. Scott and R. Blinc, J. Phys. Cond. Matt. **23**, 113202 (2011).

[165] N. Hur, S. Park, P. A. Sharma, J. S. Ahn, S. Guha and S.-W. Cheong, Nature **429**, 392 (2004).

[166] W. Prellier, M. P. Singh and P. Murugavel, J. Phys. Cond. Matt. **17** (30), R803 (2005).

[167] Jmol: http://www.jmol.org/

[168] M. N. Iliev and M. V. Abrashev, J. of Raman Spectroscopy **32** (10), 805-811 (2001).

[169] T. Goto, T. Kimura, G. Lawes, A. P. Ramirez and Y. Tokura, Phys. Rev. Lett. **92**, 257201 (2004).

[170] T. Kimura, T. Goto, H. Shintani, K. Ishizaka, T. Arima and Y. Tokura, Nature **426**, 55 (2003).

[171] N. S. Rogado, J. Li, A. W. Sleight and M. A. Subramanian, Adv. Mater.**17**, 2225 (2005).

[172] Y. Du, Z. X. Cheng, S. X. Dou, X. L. Wang, H. Y. Zhao and H. Kimura, Phys. Rev. Lett. **97**, 122502 (2010).

[173] M. N. Iliev, M. M. Gospodinov, M. P. Singh, J. Meen, K. D. Truong, P. Fournier and S. Jandl, arXiv: 0905.0202.

[174] K. Sharma, V. R. Reddy, D. Kothari, A. Gupta, A. Banerjee and V. G. Sathe, J. Phys. Cond. Matt. **22,** 146005 (2010).

# Chapter 2

# Methodology and Instrumentation

## 2.1 Raman Scattering

Raman scattering is inelastic scattering of light which originates from the change in the electronic polarizability of the system by the quasi-particle excitations. When the incident radiation (photons) interacts with the medium, it can be scattered either elastically or inelastically. If the scattered photons have the same energy as that of incident one, it is called Rayleigh scattering; on the other hand if the scattered photons energy is more/less than the incident one, it is called Anti-Stokes/Stokes Raman scattering. Figure 2.1 shows the schematic representation of the above mentioned scattering processes. Raman scattering is a very powerful technique to study the phonon dynamics and coupling of lattice vibrations with quasi-particle excitations (such as magnons, orbitons, plasmons etc.).

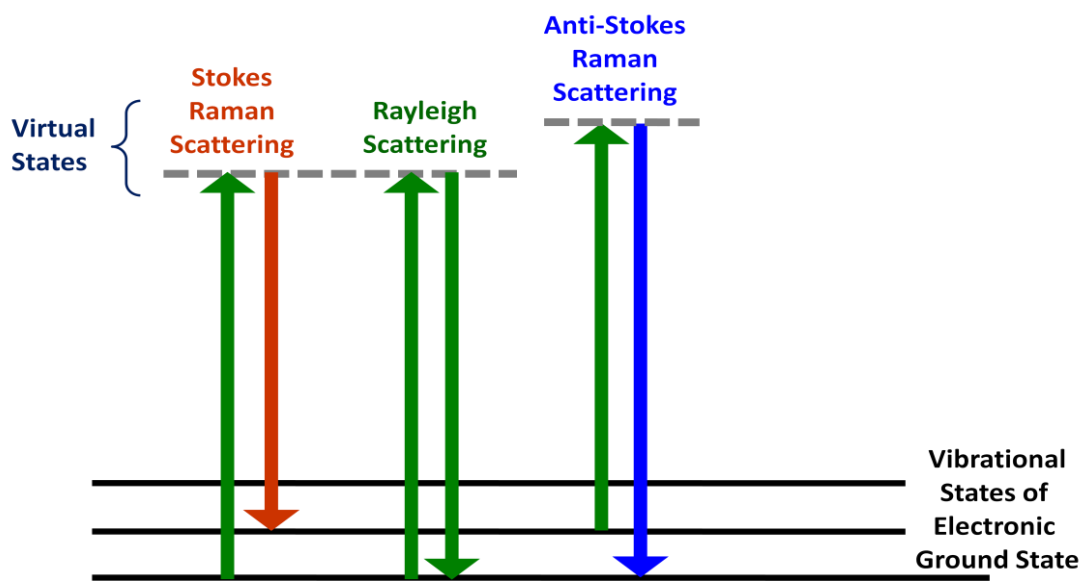

Figure 2.1: Schematic representation of the inelastic Stokes, Anti-Stokes Raman scattering process and elastic Rayleigh scattering process.

The three important steps involved in the first-order Raman scattering process are: (i) The incident photon excites an electron from its ground state, creating electron-hole pairs (ii) The excited electron or hole is scattered by the creation or annihilation of phonon (quanta of lattice vibrations); and (iii) Finally, the electron recombines with the hole emitting a scattered photon (see Fig. 2.2).

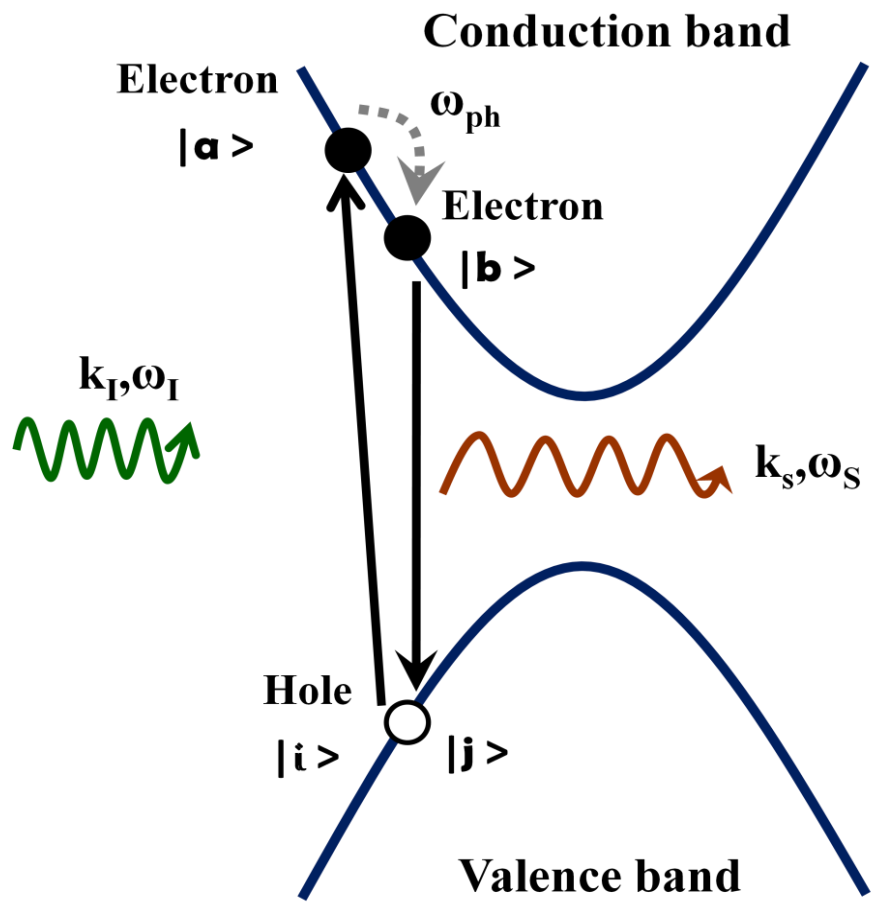

Figure 2.2: The three processes, i.e. the excitation of the electron from the valence band to conduction band by incoming photon, creating an electron-hole pair, creation of a phonon for Stokes process (represented by dotted line) and finally emission of a photon due to electron hole recombination. The initial, intermediate and final states are represented by $|i\rangle, |a\rangle, |b\rangle$ and $|f\rangle$, respectively.

From the above described quantum mechanical picture for Raman scattering, one can see that the process of first-order Raman scattering is dominated by three interactions (third order process) i.e. interaction of incident and scattered radiation with electron ($\hat{H}_{ER}$), and interaction between lattice and electron ($\hat{H}_{EL}$), as depicted pictorially in Fig. 2.3.

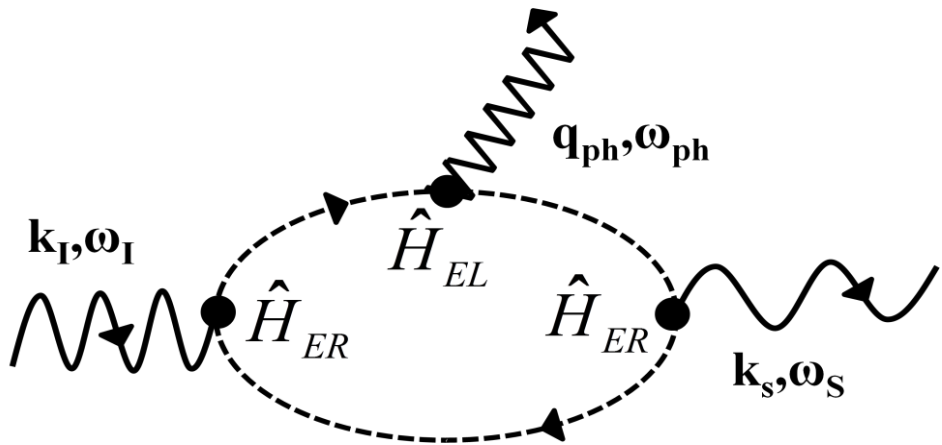

Figure 2.3: Feynman diagram of one of the possible mechanisms of the first-order Raman scattering process, it corresponds to the third-order perturbation process. The interaction vertices are: electron-radiation interaction ($\hat{H}_{ER}$) and electron-lattice interaction ($\hat{H}_{EL}$).

In the presence of the electromagnetic field, the electron momentum ($\vec{P}$) becomes ($\vec{P} - e\vec{A}$), where $\vec{A}$ is vector potential of the field which is linear in photon creation ($\hat{a}^+$) and annihilation ($\hat{a}$) operator. The resulting electron-radiation interaction Hamiltonian ($\hat{H}_{ER}$) is given as [1]

$$\hat{H}_{ER} = \hat{H}'_{ER} + \hat{H}''_{ER} \qquad (2.1)$$

$$\hat{H}_{ER} = \frac{e^2}{2m} \sum_j \vec{A}(\vec{r}_i).\vec{A}(\vec{r}_j) - \frac{e}{m} \sum_j \vec{A}(\vec{r}_i).\vec{P}_j \qquad (2.2)$$

where $m$ is the mass of the electron and $e$ is the charge. Both the terms in equation (2.1) contribute to the Raman scattering, which are referred to as $A^2$ and $\vec{A}.\vec{P}$ parts. Contribution of first term in equation (2.1) is usually negligible and it is substantial only in special cases like (i) scattering by free carriers, (ii) by collective electronic excitations like plasmon and (iii) conduction electrons in a medium. The second term makes significant contribution to most of the scattering processes.

The scattering cross section for the Stokes process within the solid angle $d\Omega$ having frequency between $\omega_s$ and $\omega_s + d\omega_s$ can be given as [1-2]

$$\frac{d^2S}{d\Omega d\omega_s} \propto \sum_{a,b} \frac{\left|\langle f|H_{ER}''|b\rangle\langle b|H_{EL}|a\rangle\langle a|H_{ER}''|i\rangle\right|^2}{(\omega_a - \omega_I)(\omega_b - \omega_s)} \delta(\omega_I - \omega - \omega_s) \qquad (2.3)$$

where $|a\rangle$ and $|b\rangle$ are intermediate states which may be real or virtual. Real intermediate states give rise to Resonance Raman scattering resulting in large increase in the Raman cross section. The initial $|i\rangle$ and final states $|f\rangle$ are given by $|n_I, n_s, n_0, g\rangle$ and $|n_I - 1, n_s + 1, n_0 + 1, g\rangle$ respectively, where $n_I, n_s, n_0$ and $g$ are the number of incident photons, scattered photons, optical phonons and electronic ground state respectively. $\omega_I, \omega_s$ and $\omega$ are the frequencies of the incident photons, scattered photons and phonons, respectively. $\hat{H}_{EL}$ is the electron-lattice interaction Hamiltonian. The delta function ($\delta$) in equation (2.3) takes care of the energy conservation in the scattering process. The conservation rules for the scattering process are: (i) wave vector conservation, $\vec{K}_I = \vec{K}_S \pm \vec{q}$, ( $\vec{K}_I, \vec{K}_s$ are the incident and scattered photon wave vector, respectively and $\vec{q}$ is the phonon wave vector) and (ii) energy conservation $\omega_I = \omega_S \pm \omega$. Here +ve and −ve signs correspond to Stokes and Anti-Stokes Raman scattering, respectively.

In principle, phonon wave vector ($\vec{q}$) can take any value within the first Brillouin zone, with maximum value $= \pi/a$ ( a = Lattice constant ), depending on the incident photon energy.

For incident radiation in the optical region $|\vec{q}| \sim (\frac{4\pi n}{\lambda_I}) Sin(\Theta/2)$ ($\lambda_I$ = wavelength of incident radiation, n is refractive index of the material and $\Theta$ is the scattering angle) is ~ $10^5$ cm$^{-1}$ and is much smaller than the Brillouin zone boundary (~ $10^8$ cm$^{-1}$), as pictorially shown in Fig. 2.4. Hence, only $\Gamma$-point phonons with nearly zero wave vector ($q \sim 0$) can participate in the first order process. However it is the crystal symmetry which decides which mode will be Raman active.

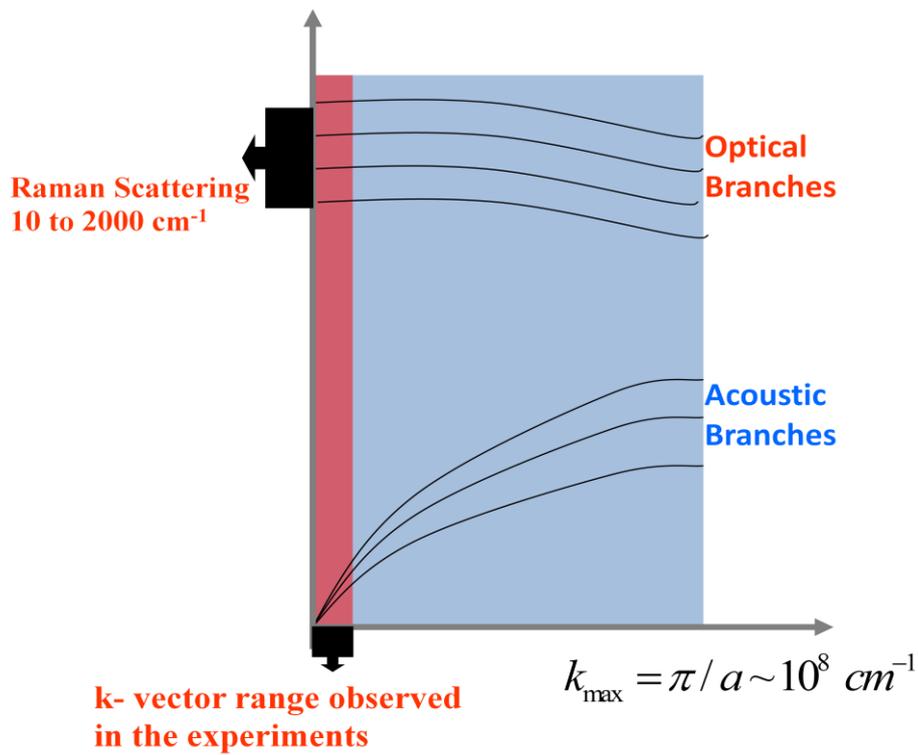

Figure 2.4: Schematic representation of the Raman scattering process shows that only the zero wave vector ($q \sim 0$) phonons can participate in the first-order scattering process.

The scattering cross-section in the Raman scattering process is related to the power spectrum of the quasi-particle excitations in the medium as [1-2]

$$\frac{d^2 S}{d\Omega d\omega_s} \propto \left| \varepsilon_0 \hat{e}_s^i \hat{e}_I^j \chi^{ij}(\omega_I, -\omega) \right|^2 . \langle X(\vec{q}).X^*(\vec{q}) \rangle_\omega \qquad (2.4)$$

where $\hat{e}_I(\hat{e}_s)$ is a unit vector parallel to the incident (scattered) radiation, $\chi^{ij}(\omega_I,-\omega)$ is the second order susceptibility which is a function of the incident frequency and frequency of the quasi-particle excitations. $\langle X(\vec{q}).X^*(\vec{q})\rangle_\omega$ is the power spectrum of the fluctuations of the relevant quantity of the scattering medium. Here $X$ can be phonon displacement coordinates, electron/charge density or fluctuation in the magnetisation for the magnetic excitations. The power spectrum described in equation (2.4) can be calculated using fluctuation-dissipation theorem which relates the power spectrum to the imaginary part of the linear response function [1]. Considering $X$ being phonon displacement coordinates, represented by damped harmonic oscillator, the power spectrum for Stokes lines is given as

$$\langle X(\vec{q}).X^*(\vec{q})\rangle_\omega = c.[n(\omega)+1].g(\omega) \tag{2.5}$$

where c is a constant, $n(\omega)$ is the Bose-Einstein factor and $g(\omega)$ is a Lorentzian function given by:

$$g(\omega) = \frac{A}{2\pi}\left[\frac{\Gamma}{4(\omega-\omega_0)^2+\Gamma^2}\right] \tag{2.6}$$

where $A$ is the area under the curve (integrated intensity) and $\Gamma$ is full width at half maxima (FWHM). The second order susceptibility ($\chi^{ij}(\omega_I,-\omega)$) has additional selection rule depending on the symmetry of the crystal lattice. The phonon modes in a crystal are indentified with irreducible representations. The irreducible representations of $\chi^{ij}(\omega_I,-\omega)$ tensor are also determined by the crystal symmetry. Group theoretical calculations tell us that a phonon mode would be Raman active if its irreducible representation is same as one of the irreducible representations of $\chi^{ij}(\omega_I,-\omega)$.

## 2.1.1 Scattering from Magnetic excitations and second-order Raman Scattering

Here we describe the process of inelastic scattering of light by magnetic quasi-particle excitations namely magnons, in ferromagnetic as well as antiferromagnetic systems, in the language of Fleury and Loudon [3]. The electric field of the incident radiation does not couple directly with the spin degrees of freedom (DOF); however it couples with the system through an indirect mechanism which arises due to mixing of spin and orbital DOF, namely $\vec{L}.\vec{S}$ coupling. The mechanism of coupling between photons and spin degrees of freedom, described by Elliot and Loudon [3-5] for AFM as well as ferromagnetic systems can be understood as follows:

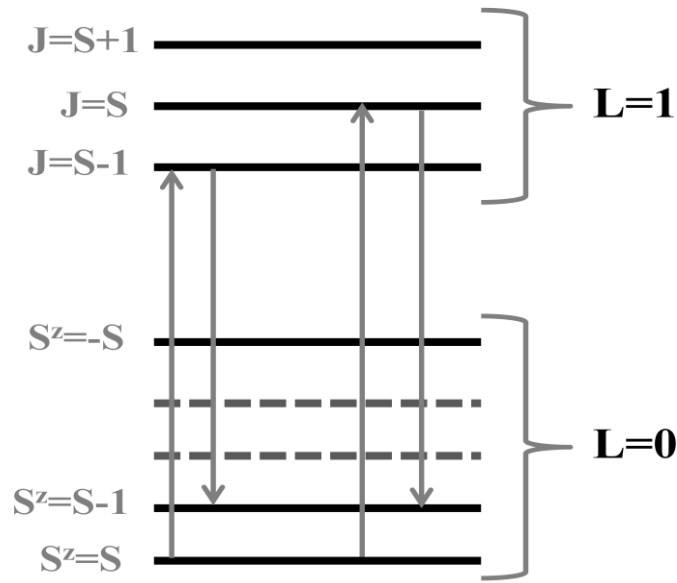

Figure 2.5: Schematic showing the one-magnon light scattering mechanism [5].

Assume a system where ground state of the magnetic ion has zero orbital angular momentum ($L=0$) and a net spin S. The ground state is split into 2S+1 different components owing to the exchange field. If there is an excited state with orbital angular momentum L=1, then this will split into three components corresponding to J = S+1, S and S-1 by the spin-orbit coupling. The Stokes scattering process by magnetic excitations involves the following process: transition from $S^z = S$ ground state to the $S^z = S-1$ spin-orbit excited state. Now the excited state decays to $S^z = S-1$ ground state emitting a magnon (schematically shown in Fig. 2.5), giving rise to a non-zero contribution to the Raman scattering cross-section.

In second order Raman scattering, two phonons (or in general two quasi-particle) with wave vector within the entire Brillouin zone can participate, in comparison to the first-order scattering process where phonons only near the centre of the Brillouin zone are involved. This is because any two phonons with wave vector $\vec{q}_1$ and $\vec{q}_2$ from the entire Brillouin zone can satisfy the conservation laws of energy and momentum ($\vec{q}_1 + \vec{q}_2 \sim 0$). The scattering cross section for the second-order process depends on two-phonon density of states and is given as [1,3]

$$\frac{d^2 S}{d\Omega d\omega_s} \propto \sum_{i,j} \sum_{\vec{q}} \delta(\omega - \omega_{i,\vec{q}} - \omega_{j,-\vec{q}}) \qquad (2.7)$$

where i, j runs over all the branches in the first Brillouin zone and $\vec{q}$ runs over the Brillouin zone.

## 2.1.2 Instrumentation

Raman spectrometer consists of (i) a monochromatic source of light, e.g. laser (ii) optics to focus the incident laser beam onto the sample and to couple the scattered light to the monochromator (iii) monochromator(s) to disperse the scattered light (iv) light detector, to count the numbers of scattered photons, which can be a single channel detector like a photomultiplier tube or a multichannel detector like charge coupled device (CCD).

The experiments reported in this thesis have been done using a DILOR XY triple grating spectrometer coupled with a liquid nitrogen cooled CCD detector. The monochromatic source of light used is an argon ion Laser (Coherent Innova model 300), which can be tuned to different wavelengths, e.g. 488 nm, 514.5 nm, 496 nm.

### 2.1.2.1 DILOR XY Spectrometer

DILOR XY spectrometer consists of three monochromators, in which the first two monochromators (called fore-monochromator) are coupled either in additive or subtractive mode (acting as band pass filter), while the third one (called spectrograph) disperses the scattered light onto the detector [6]. In additive mode there is a significant gain in the resolution at the expense of bandpass scanned and throughput in comparison to the subtractive mode. In the subtractive mode with multichannel detection, the resolution at 514.5 nm is 0.8 cm$^{-1}$ with a bandpass of ~ 350 cm$^{-1}$, on the other hand in additive mode it is ~ 0.3 cm$^{-1}$ with a bandpass of ~ 130 cm$^{-1}$. The spectrometer with triple monochromators, focal length of 500 mm and focal speed f/6, is equipped with three holographic gratings (1800 line/mm) blazed for 500 nm [6]. The spectral range is from 11200 to 30500 cm$^{-1}$. Figure 2.6 shows the optical layout of DILOR XY triple grating spectrometer. The plasma lines from the laser are eliminated using a prism based monochromator and an interference filter.

We have performed our experiments in the subtractive mode of the spectrometer using CCD as the detector. The CCD which is coupled to DILOR XY spectrometer is a 2D array of Indium Gallium Arsenide detector manufactured by Spectrum One. The CCD of dimensions ~ 25 mm × 8 mm with pixels 1024 × 256 pixels, is cooled with liquid nitrogen and controlled by the CCD-300 controller.

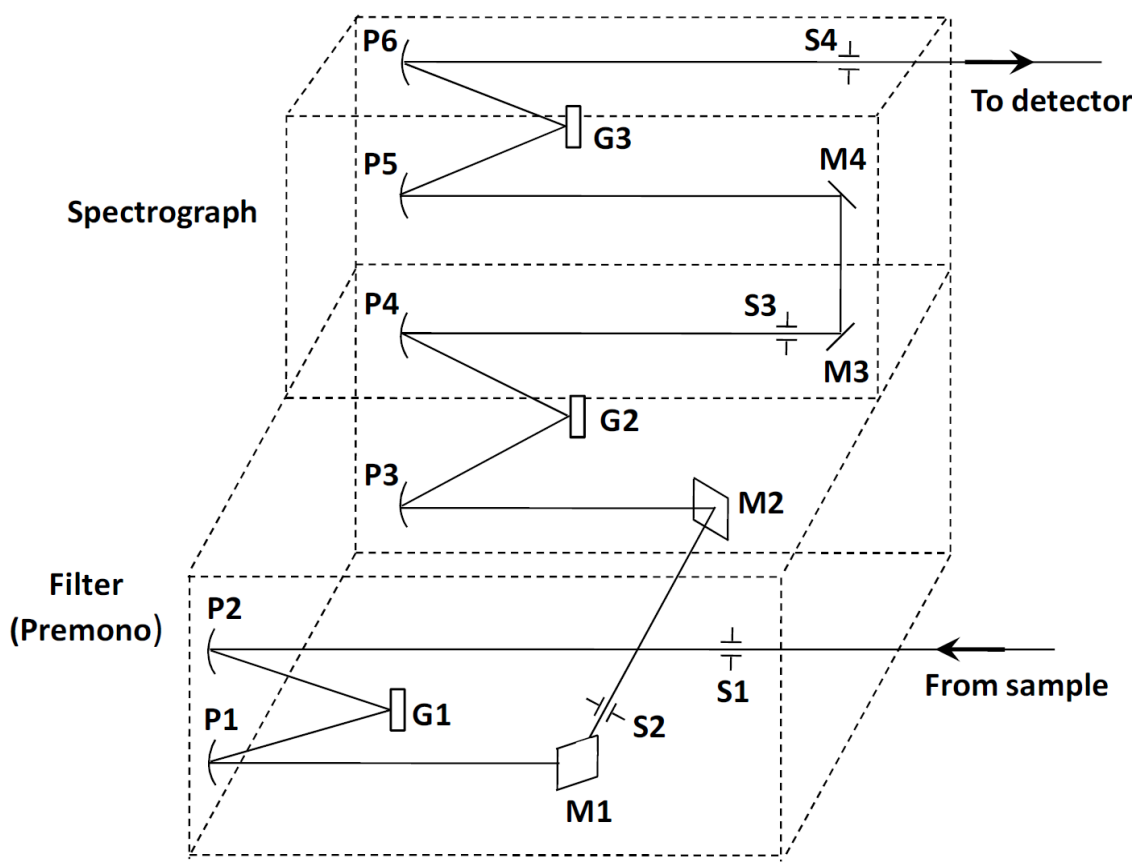

Figure 2.6: Optical layout of the DILOR XY Raman spectrometer. Slits, Lens, parabolic mirrors, plane mirrors and gratings are denoted by S, L, P, M and G respectively.

The spectrometer can be used in macro as well as micro configuration. In macro configuration, the laser beam is focussed onto the sample using a Minolta camera lens of focal length 50 mm and speed f/1.8 giving a Laser spot size of ~ 40 μm. The same camera lens is used to collect the scattered light and fed to the spectrometer. On the other hand, for micro configuration, an objective of high numerical aperture and magnification (100x/NA=0.9, 50x/NA=0.45) is used to illuminate the sample with a spot size of ~ 1μm or less, depending on the wavelength used, and collect the scattered light in the back scattering geometry. For majority of our experiments, we have used micro Raman configuration.

## 2.2 Infrared Spectroscopy

We have also performed temperature-dependent infrared transmission measurements on multiferroic TbMnO$_3$. In a similar formalism as that discussed for Raman scattering process above, in terms of power spectrum of quasi-particle excitations, the scattering cross-section for infrared activity can be written as

$$\frac{d^2 S}{d\Omega d\omega_s} \propto \left\langle \vec{\mu}_T(\vec{q}) \cdot \vec{\mu}_T^*(\vec{q}) \right\rangle_\omega \tag{2.8}$$

where $\vec{\mu}_T$ is the sum of the permanent and induced dipole moments. The power spectrum resulting from permanent dipole moments leads to infrared activity in the far infrared region while induced dipole moments give rise to the infrared activity of the lattice vibrations. The contribution due to induced dipole moments to the scattering cross-section is given as

$$\frac{d^2 S}{d\Omega d\omega_s} \propto \left\langle \left(\frac{\partial \vec{\mu}}{\partial Q}\right)_0 Q \left(\frac{\partial \vec{\mu}}{\partial Q}\right)_0 Q^* \right\rangle_\omega \tag{2.9}$$

Hence, a vibration will be infrared active only if the derivative of the induced dipole moments is non-zero. If a molecule is centrosymmetric, i.e. having centre of symmetry, then Raman modes are infrared inactive and vice-versa.

### 2.2.1 Instrumentation

The spectrometer used in our measurements on TbMnO$_3$ systems is a Fourier Transform Infra Red (FTIR) spectrometer manufactured by Bruker Optics Inc. (model IFS 66v/s). The spectrometer consists of a light source, a Michelson interferometer, a beam-splitter and a detector. The available spectrum range is from 20 cm$^{-1}$ to 50,000 cm$^{-1}$. Depending on the range of frequency a suitable combination of light source, beam-splitter and detector are selected. We have performed our experiments within the spectral range of 50 to 700 cm$^{-1}$, where we have used GLOBAR as light source, Mylar (Ge) 6μm film as beam-splitter and DTGS as the detector. Using the OPUS software, we have used the spectrometer at a resolution of 2 cm$^{-1}$. The low temperature measurements in the range of 4 K to 300 K were performed using a continuous flow Helium Oxford cryostat controlled by Oxford intelligent temperature controller.

## 2.3 Low Temperature Technique

### 2.3.1 Instrumentation

Most of our studies have been carried out as a function of temperature and have been done using a commercial cryostat and temperature controller. The cryostats used are closed cycle helium refrigerator (CCR) systems manufactured by CTI-cryogenics (model 8200) and a continuous flow helium cryostat (cools down to 4 K) manufactured by Oxford instruments (model ITC-502), equipped with optical windows. CCR works on the principle of adiabatic expansion of gases and cools down to ~ 12 K, and consists of an air compressor and a two stage cold head where the sample holder with the sample can be mounted. The temperature in both the cryostats is controlled by a temperature controller manufactured by CRYO-CON (model-34, for CCR) and Oxford instruments (model ITC-502).

### 2.3.2 Effects of Temperature

The change of temperature as an external perturbation in a crystal changes its unit cell volume (also known as quasi-harmonic effect) and also phonon population (called intrinsic anharmonic effect). The changes in the unit cell volume and phonon population are reflected in the mode frequencies and linewidths. These changes can be understood in terms of correction to the phonon self-energy, where the real part of the self-energy corresponds to the change in phonon frequency and its imaginary part to the phonon linewidth. The linewidth of a phonon mode in a perfect crystal, i.e. free of disorder/defects or free carriers, is determined by anharmonic interactions. The anharmonic interactions in a crystal can be taken care of by adding higher order terms to the harmonic potential ($U_0 = \frac{1}{2}Kx^2$), i.e. $U = \frac{1}{2}Kx^2 + ax^3 + bx^4 + .....$, where first term is pure harmonic potential and second and third terms are referred to as cubic and quartic correction, and $x$ is the displacement of the atom from its equilibrium position. In most of the cases it is sufficient to consider only the third order term (i.e. $ax^3$) of the lattice potential, which corresponds to the cubic anharmonicity of the phonons mode. Assuming only the cubic anharmonicity, where each phonon of frequency (wavevector) $\omega_0(K_0)$ decays into two phonons of energy $\omega_1$ and $\omega_2$ such that $\omega_0 = \omega_1 + \omega_2$ and $\vec{K}_0 = \vec{K}_1 + \vec{K}_2$. Considering only the simplest decay channel i.e. each phonon decays into two phonons of equal energy ($\omega_0/2 = \omega_1 = \omega_2$) [7], the phonon linewidth and frequency can be given as

$$\Gamma_{cubic} = \Gamma_0 + D[1+2n(\omega_0/2)] \quad \text{and,} \quad (2.10)$$

$$\omega_{cubic} = \omega_0 + C[1+2n(\omega_0/2)] \quad (2.11)$$

where $\omega_0$ and $\Gamma_0$ are the frequency and linewidth at absolute zero and $n(\omega)$ is the Bose-Einstein factor. $\Gamma_0$ arises from lattice disorder/defects. Here C and D are assumed to be constant depending on the three phonon interaction vertex. The coefficient C is usually negative, i.e. phonon frequency decreases with increasing temperature, and D is positive, i.e. phonon linewidth increases with increasing temperature. In the high temperature limits (i.e. $k_B T \gg \hbar\omega$ and $\theta_D > T$, $\theta_D$ is the Debye temperature) linewidth and frequency become linear functions of temperature i.e. $\Gamma_{cubic} = \Gamma_0 + D'T$ and $\omega_{cubic} = \omega_0 + C'T$. However for $T > \theta_D$, there is an additional contribution arising from the quartic term in the harmonic potential and as a result linewidth and frequency become quadric functions of temperature $\Gamma_{cubic} = \Gamma_0 + D'T + D''T^2$ and $\omega_{cubic} = \omega_0 + C'T + C''T^2$. The change in phonon frequency due to quasiharmonic effect can be expressed in terms of mode Grüneisen parameter $\gamma$ as [8]

$$\omega_i(T) - \omega_i(0) = -\gamma_i \frac{\Delta V}{V} \qquad (2.12)$$

where $\gamma_i$ is Grüneisen parameter for the i$^{th}$ mode and $\frac{\Delta V}{V}$ is the fractional change in volume. Grüneisen parameter for the i$^{th}$ mode is defined as $\gamma_i = \frac{B}{\omega_i} \frac{d\omega_i}{dP}$, where $B$ is the bulk modulus and $\frac{d\omega_i}{dP}$ is change of frequency with pressure.

We will present results in the thesis to show anomalous temperature dependence of the phonon parameters, thereby bringing out contributions to the phonon self-energy other than anharmonic interactions. In particular, spin-phonon coupling will be involved to understand the anomalous temperature dependence of the phonon frequencies and linewidth.

# Bibliography:

[1] W. Hayes and R. Loudon, Scattering of Light by Crystals, John Wiley and Sons (1979).

[2] R. Loudon, Adv. Phys. **13**, 423 (1964).

[3] P.A. Fleury and R. Loudon, Phys. Rev. **166**, 514 (1968).

[4] R. J. Elliot and R. Loudon, Phys. Lett. **3**, 189 (1963).

[5] M. G. Cottam and D. J. Lockwood, Light Scattering in Magnetic Solids, John Wiley and Sons (1986).

[6] Dilor XY users manual 1991.

[7] P. G. Klemens, Phys. Rev. **148**, 845 (1966).

[8] M. Born and K. Huang, Dynamical theory of crystal lattice. Oxford University Press, Oxford, (1954).

# Chapter 3

## 3.1 Part-A

## Anomalous Raman Scattering from Phonons and Electrons of Superconducting $FeSe_{0.82}$

### 3.1.1 Introduction

The discovery of superconductivity in rare earth iron pnictides $RFeAsO_{1-x}F_x$ (R = La, Sm, Ce, Nd, Pr, and Gd [1-3]) was soon expanded to include the related alkali doped $A_xM_{1-x}Fe_2As_2$ (A = K and Na; M = Ca, Sr and Ba) [4-5] and iron-chalcogenides $Fe_{1+\delta}Se_{1-x}Te_x$ [6-7]. Iron pnictides and iron chalcogenides share the common feature of tetragonal structure with FeAs or FeSe layers, wherein Fe is tetrahedrally coordinated with As or Se neighbours. The superconducting transition temperature ($T_c$) for iron chalcogenides has increased from initial 8 K [6] to 14 K [7] with suitable amount of Te substitution, and to 27 K [8] under high pressure. An interesting observation in the $Fe_{1+\delta}Se_{1-x}Te_x$ system is the occurrence of a structural phase transition from tetragonal to orthorhombic at $T_s$ ~ 100 K [6,9], accompanied by an anomaly in magnetic susceptibility [7]. Wang et al. [10] have shown that only the thin films of tetragonal $FeSe_{1-x}$, which show such a low temperature structural transition are superconducting, thus suggesting a crucial link between structural transition and superconductivity. Earlier reports of superconductivity in tetragonal $FeSe_{1-x}$ compound termed PbO-tetragonal phase as α-phase [6]. However, some other reports have termed the PbO-tetragonal phase as β-FeSe as well [11-12], causing some confusion in the nomenclature. A recent report of the phase diagram [13] (where the tetragonal phase is designated as β-phase), claims that only compounds with the stoichiometry in the narrow range of $Fe_{1.01}Se$ to $Fe_{1.03}Se$ are superconducting. Understanding the primary origin of these co-occurring structural transition and magnetic anomaly at $T_s$ is essential in uncovering important couplings in the normal phase that are relevant to superconductivity in FeSe$_{1-x}$ and related materials.

Here, we present Raman scattering from tetragonal FeSe$_{0.82}$ with onset $T_c$ of ~ 12 K. There are two motivating factors behind this work: first, to identify the phonons relevant to the observed structural and magnetic phase transition, and use first-principles calculations to determine the possible coupling between these phonons and accompanying changes in magnetic structure via spin-phonon coupling, if any. This may throw light on the possible role of electron-phonon coupling through spin channel in the mechanism of superconductivity [14-15]. Second, to determine the changes in electronic states and excitations near Fermi level through the transition(s) at $T_s$, which may set the stage for superconductivity at lower temperatures. As emphasized earlier [16], Fe 3$d$-orbitals contribute significantly to the electronic states near the Fermi level and hence they are expected to play a crucial role in the mechanism of superconductivity. In particular, it is not clear experimentally if $d_{xz}$ and $d_{yz}$ orbitals are split or not [17-20].

From the temperature dependence of Raman scattering from phonons and electrons (in crystal-field split 3$d$-orbitals of Fe) in tetragonal-FeSe$_{0.82}$, we present two significant results: (i) The lowest frequency phonon, associated with Se vibration in the $ab$ plane, shows anomalously large blue shift of ~ 5% in frequency below $T_s$; (ii) The $d_{xz}$ and $d_{yz}$ orbitals of Fe are non-degenerate with a splitting of ~ 30 meV. Our study on this superconducting system gave first evidence of the possible splitting of $d_{xz}$ and $d_{yz}$ energy levels that is being discussed in recent theoretical studies [17-20] in the general class of iron based superconductors. In ref. [19], related to LaFeAsO, the splitting of 3$d$ orbitals of Fe results from combined effects of tetrahedral crystal field, spin-orbit coupling, strong hybridization between Fe 3$d$ and As 4$p$ orbitals and lattice compression along z-axis. Also, it has been suggested [20] that, since $d_{xz}$ and $d_{yz}$ orbitals are roughly half filled, spontaneous symmetry breaking can lead to the lifting of the degeneracy of $d_{xz}$ and $d_{yz}$ orbitals. Using first-principles Density Functional Theory (DFT) based calculations our collaborators (Prof. Umesh Waghmare and his group) show that our experimenatl observations can be readily understood within a single picture: there is a transition to a phase with short-range stripe [21] antiferromagnetic order below $T_s$, with orthorhombic structural distortion as a secondary order parameter.

We note that there are a few temperature dependent Raman studies on $NdFeAsO_{1-x}F_x$ [22, 23], $Sr_{1-x}K_xFe_2As_2$ (x = 0 and 0.4) [24], $CaFe_2As_2$ [25], $R_{1-x}K_xFe_2As_2$ (R = Ba, Sr) [26-27] and $Ba(Fe_{1-x}Co_x)_2As_2$ [28]. In case of $NdFeAsO_{1-x}F_x$ [22-23] and $Sr_{1-x}K_xFe_2As_2$ [24] none of the observed phonon modes show any anomaly as a function of temperature. However, in another study of $R_{1-x}K_xFe_2As_2$ (R= Ba and Sr) [26] the linewidths of the phonon modes involving Fe and As near 185 $cm^{-1}$ ($A_{1g}$) and 210 $cm^{-1}$ ($B_{1g}$) show a significant decrease below the spin density wave transition temperature $T_s$ ~ 150 K, arising from the opening of the spin density wave gap. Also, the phonon frequency of the 185 $cm^{-1}$ mode shows a discontinuous change at $T_s$, signaling the first order structural transition accompanying the spin-density-wave transition at $T_s$. Similar results for the $B_{1g}$ mode (near 210 $cm^{-1}$) are seen for $Sr_{0.85}K_{0.15}Fe_2As_2$ and $Ba_{0.72}K_{0.28}Fe_2As_2$ ($T_s$ ~ 140 K) [27]. In parent compound $CaFe_2As_2$, the $B_{1g}$ phonon frequency (210 $cm^{-1}$) shows a discontinuous decrease at $T_s$ ~ 173 K and $A_{1g}$ phonon (near 190 $cm^{-1}$) intensity is zero above $T_s$, attributed to the first order structural phase transition and a drastic change of charge distribution within the FeAs plane [25]. The $E_g$ (Fe, As) phonon (~ 135 $cm^{-1}$) in $Ba(Fe_{1-x}Co_x)_2As_2$ (x < 0.06) splits into two modes near the structural transition temperature ($T_s$ ~ 100 to 130 K) linked to strong spin-phonon coupling [28].

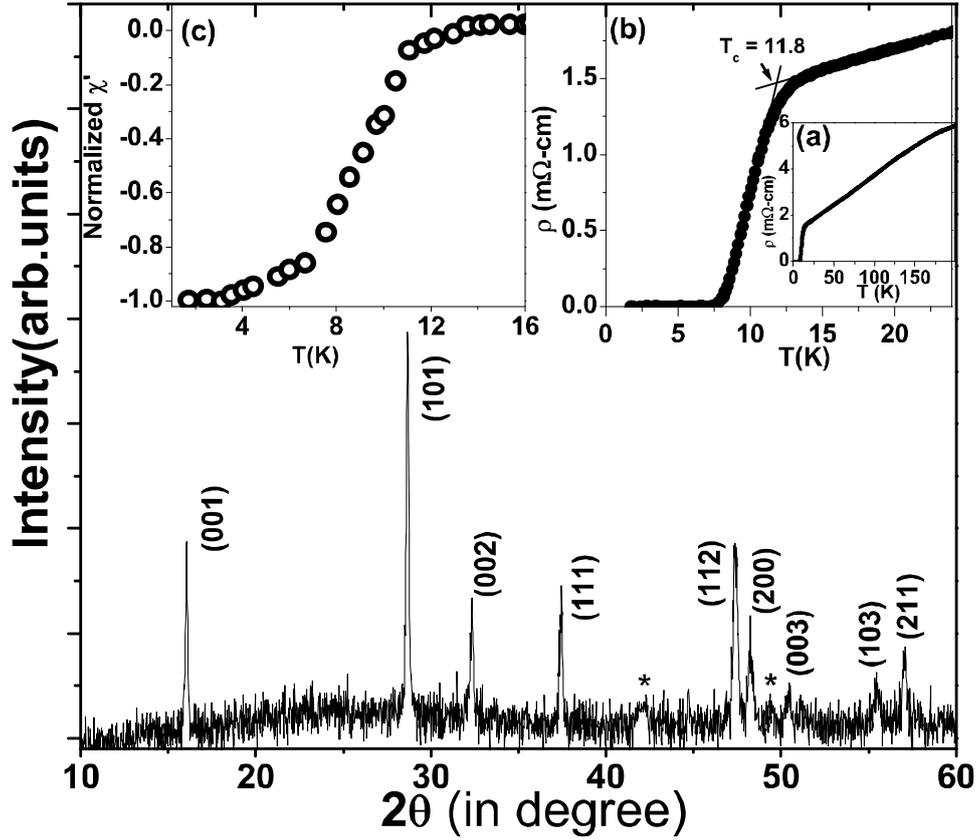

Figure 3.1: x-ray diffraction pattern of FeSe$_{0.82}$ sintered at 720 $^0$C. Secondary hexagonal-FeSe phase is marked by asterisk (*). Inset (a) Temperature dependence of resistivity (ρ) up to 200 K, (b) Temperature dependence of resistivity showing the criterion used to determine $T_C$, (c) Magnetic susceptibility as a function of temperature.

### 3.1.2 Experimental Details

Polycrystalline samples with nominal composition of FeSe$_{0.82}$ were synthesized in IIT, Delhi (by Prof. A. K. Ganguly and his group) by sealed tube method (690 ºC/24h and 720 ºC/24h) using the elements as starting materials. The powder *x*-ray diffraction pattern of the compound shows the presence of the tetragonal FeSe phase (space group *P4/nmm*) with a very small amount of hexagonal FeSe (NiAs type) phase (< 5%) (diffraction lines marked by * in Fig. 3.1). The onset of superconducting transition occurs at 11.8 K (see inset (b) of Fig. 3.1). The resistivity decreases with temperature and the onset of superconducting transition occurs at 11.8 K with residual resistivity value (RRR = $\rho_{300}/\rho_{13}$) of 4.9. The criterion used for the determination of $T_c$ is shown in the inset (b) of Fig. 3.1. The inductive part of the rf magnetic susceptibility attesting the onset of bulk diamagnetic state with transition temperature near 11.8 K is shown in the inset (c) of Fig. 3.1. Unpolarised micro-Raman measurements were performed on sintered pellets of $FeSe_{0.82}$ in backscattering geometry, using 514.5 nm line of an Ar-ion Laser (Coherent Innova 300), from 3 K to 300 K using a continuous flow liquid helium cryostat with a temperature accuracy of ± 0.1 K. The scattered light was analysed using a Raman spectrometer (DILOR XY) coupled to liquid nitrogen cooled CCD detector.

### 3.1.3 Results and Discussion
### 3.1.3.1 Raman Scattering

There are four Raman active phonon modes belonging to the irreducible representations $A_{1g}$ + $B_{1g}$ + $2E_g$ [14]. Figure 3.2 shows Raman spectrum recorded at 3 K where the spectral range is divided into two parts: low frequency range (80 to 350 $cm^{-1}$) showing 5 Raman bands, labeled as S1 to S5 and high frequency range (800 to 1800 $cm^{-1}$) displaying weak Raman bands, S6 and S7. Following the earlier Raman studies of tetragonal-FeTe and $Fe_{1.03}Se_{0.3}Te_{0.7}$ [14] and our calculations (to be discussed below), the four Raman active modes are S1: 106 $cm^{-1}$, symmetry $E_g$, mainly Se vibrations; S2: 160 $cm^{-1}$, $A_{1g}$, Se vibrations; S3: 224 $cm^{-1}$, $B_{1g}$, Fe vibrations; and S4: 234 $cm^{-1}$, $E_g$, Fe vibrations. Mode S5 is weak and its intensity depends on the spot on the sample. We assign the mode S5 (254 $cm^{-1}$) to the hexagonal phase of FeSe [29]. Raman spectra were fitted with a sum of Lorentzian functions. The individual modes are shown by thin lines and resultant fit by thick line. It can be seen in Fig. 3.3 that while the frequencies of modes S2, S3 and S4 show normal temperature dependence, i.e.

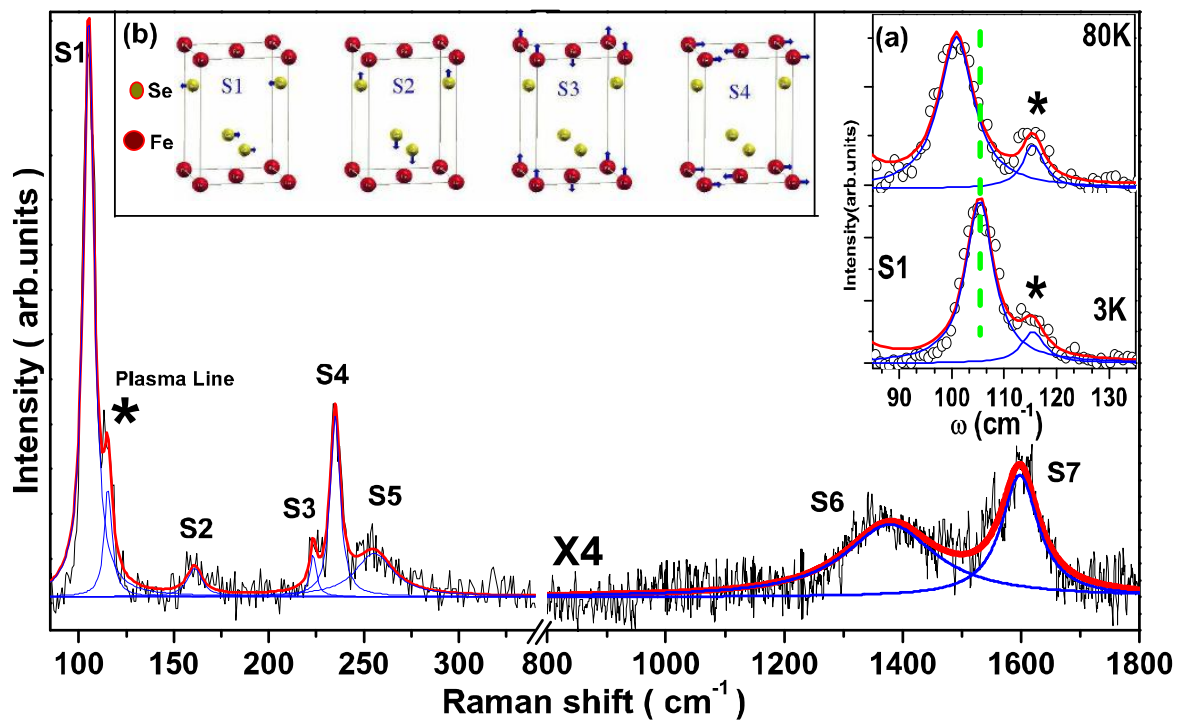

Figure 3.2: Raman spectrum of $FeSe_{0.82}$ at 3 K. Thick solid line (red) shows the total fit and thin solid lines (blue) show the individual fit. Inset (a) shows the S1 mode at two temperatures. Inset (b) shows the eigen vectors of the phonon modes.

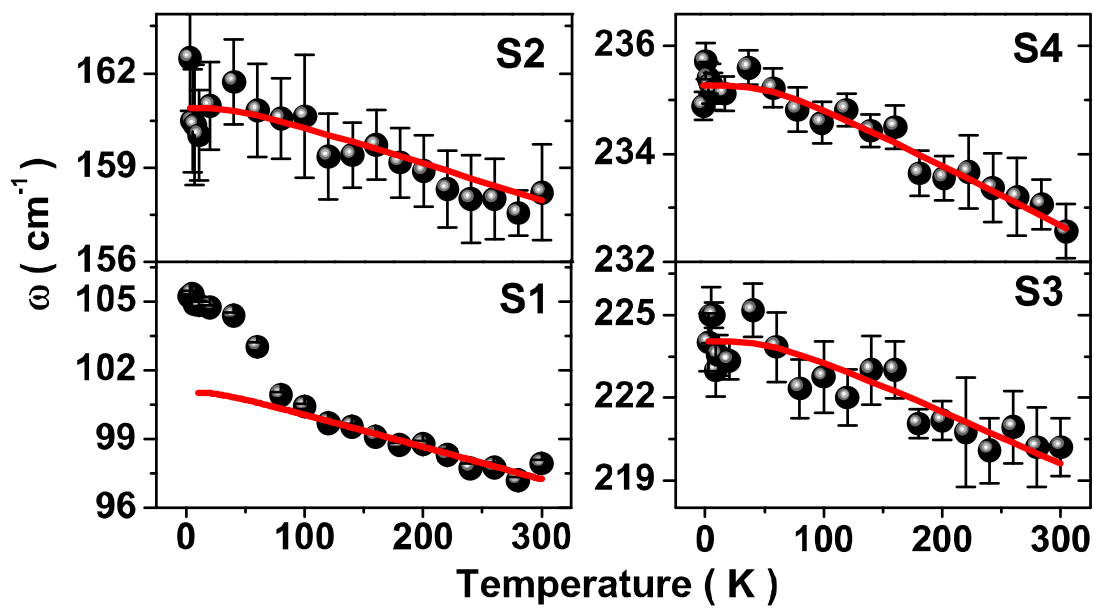

Figure 3.3: Temperature dependence of the modes S1, S2, S3 and S4. Solid lines are fitted lines as described in the text.

phonon frequency increases with decreasing temperature, mode S1 shows a sharp change near $T_s \sim 100$ K, where a tetragonal to orthorhombic phase transition is expected [6,7,9]. The frequency $\omega(T)$ of the S1 mode increases by a large amount, by ~ 5% below $T_s$. The solid lines in Fig. 3.3 are fitted with an equation (2.11) described in Chapter 2. The fitting parameters $\omega_o$ and C are given in Table-3.1. We will come back to discuss the anomalous change in frequency of the S1 mode after presenting our density functional calculations.

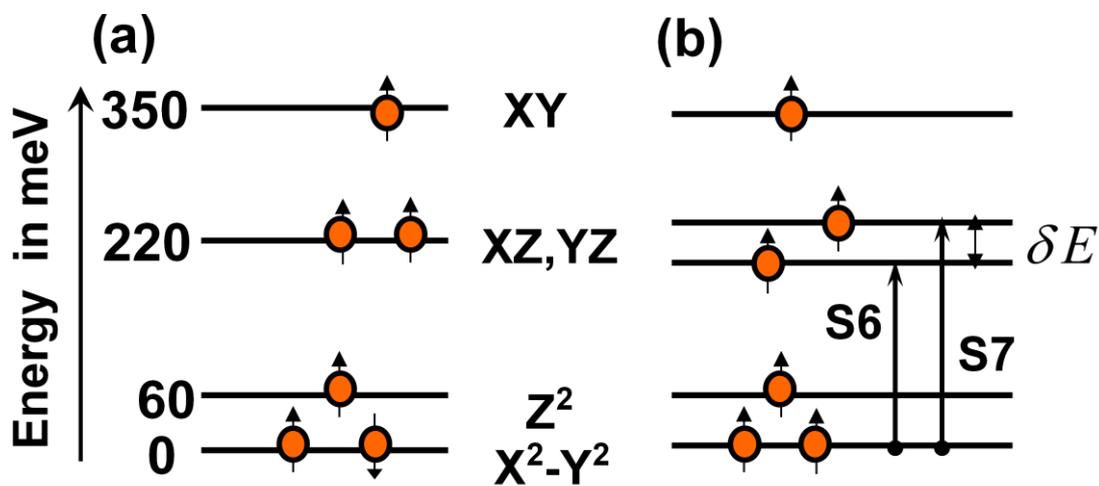

Figure 3.4: Crystal-field split energy level diagram for the Fe 3*d*-orbitals (see ref. 17 for energy values). (a) Degeneracy of *xz* and *yz* orbitals is not lifted. (b) Degeneracy of *xz* and *yz* orbitals is lifted (Energy values are not to scale).

In addition to the expected Raman active phonon modes, Fig. 3.2 shows two additional modes S6 (1350 cm$^{-1}$) and S7 (1600 cm$^{-1}$). These modes are very weak at all temperatures and hence their quantitative temperature dependence was not studied. In earlier Raman studies [30] of CeFeAsO$_{1-x}$F$_x$ at room temperature, weak Raman modes have been noticed at 846 and 1300 cm$^{-1}$. It was suggested [30] that the high frequency modes may be related to the electronic Raman scattering involving the 3*d*-orbitals of Fe. We attempt to understand S6 and S7 modes in terms of electronic Raman scattering involving Fe 3*d*-orbitals. In the case of x =

0.0, Fe atoms are tetrahedrally coordinated with Se atoms and the crystal field (CF) leads to the splitting of 3d-orbitals of Fe into four energy levels $x^2$-$y^2$, $z^2$, $xy$ and $xz/yz$ ($xz$ and $yz$ orbitals remain degenerate). Figure 3.4(a) shows the energy diagram as given in ref. [17-18]. The mode at 1350 cm$^{-1}$ is very close to the energy difference between $x^2$-$y^2$ and $xz/yz$. Our clear observation of two modes (S6 and S7) with an energy difference of δE ~ 240 cm$^{-1}$ (30 meV) suggests that the 3d-orbitals $xz$ and $yz$ are split with this energy difference as theoretically argued in ref. [19-20]. This is schematically shown in Fig. 3.4(b).

### 3.1.3.2 First-Principles Calculations

Our first-principles calculations are based on DFT as implemented in PWSCF [31] package. We use ultrasoft pseudopotentials [32] to describe the interaction between the ionic cores and the valence electrons, and a plane wave basis with energy cutoffs of 40 Ry for wave functions and 480 Ry for charge density. To model Se vacancies, we use $\sqrt{2} \times \sqrt{2} \times 1$ and 2×2×1 super cells of FeSe$_{1-x}$ for x = 0.00 and 0.0125 in FeSe$_{1-x}$, respectively. Structural optimization is carried through minimization of energy using Hellman-Feynman forces and the Broyden-Flecher-Goldfarb-Shanno based method. Frequencies of the zone centre (q = 0,0,0) phonons are determined using a frozen phonon method for the relaxed structure with experimental lattice constants. We have also carried out calculations to include effects of correlations, likely to be important in these iron pnictides [33], with an LDA+U correction, and find that phonon frequencies do change sizably with U [34], confirming the conclusion of ref. [33] in the context of phonons. However, the picture (relevant to Raman measurement here) developed here based on LDA calculations does not change qualitatively.

For the non-magnetic (NM) state of FeSe, our optimized internal structure agrees within 0.5 % with the experimental one, and estimates of phonon frequencies are slightly lower for x = 0.125 as compared to those of undoped FeSe (see Table-3.1). While the discrepancy of the calculated and measured values of phonon frequencies is larger than the typical errors of DFT calculations, it is comparable (or slightly better) to that in a recent DFT calculation of phonons in tetragonal-FeTe [14]. This could originate from various factors such as change in magnetic ordering, electronic correlations, disorder in Se vacancies and possibly strong spin-phonon couplings in the FeSe$_{1-x}$.

Table-3.1: Experimentally observed, calculated phonons frequencies and fitted parameters in cm$^{-1}$.

| Assignment | Experi--mental ω at 3K | Fitted-parameters ω$_0$ | C | Calculated Frequency x=0 | x=0.125 |
|---|---|---|---|---|---|
| S1 E$_g$ (Se) | 106 | $^a$101.5 ± 0.3 | -1.1±0.1 | 147 | 135 |
| S2 A$_{1g}$ (Se) | 160 | 161.8 ± 0.3 | -1.7±0.2 | 226 | 222 |
| S3 B$_{1g}$ (Fe) | 224 | 225.6 ± 0.4 | -3.2±0.4 | 251 | 239 |
| S4 E$_g$ (Fe) | 234 | 236.3 ± 0.1 | -2.1±0.1 | 315 | 304 |
| S5 | 254 | | | | |
| S6 CF (Fe) | 1350 | | | | |
| S7 CF (Fe) | 1600 | | | | |

$^a$Fitted Parameters to data from 100 K to 300 K.

Interestingly, our calculations on FeSe with initial state of ferromagnetic (FM) or antiferromagnetic (opposite spins at nearest neighbors, AFM1) ordering of spins at Fe sites converged to a NM state upon achieving self-consistency. On the other hand, antiferromagnetically ordered stripe [21] phase (AFM2) of FeSe is stable and lower in energy than the NM phase by 50 meV/cell. In the stripe AFM2 phase, S$_i$.S$_j$ vanishes for the nearest neighbors (NN) and is negative for the second nearest neighbor (SNN) spins. The stability of AFM2 and instability of FM and AFM1 states can be understood if the exchange coupling between SNN spins is antiferromagnetic and that between NN spins is ferromagnetic of roughly the same magnitude: FM and AFM1 states are frustrated. The origin of this can be understood through analysis of super-exchange interaction between Fe sites, mediated via p-states of Se. Opposite signs of the exchange coupling for NN and SNN spins arise from the fact that Fe-Se-Fe angles relevant to their super-exchange interactions are closer to 90 and 180 degrees, respectively. Our DFT calculations for x ≠ 0 reveal that the AFM1 state is locally stable as the magnetic frustration is partially relieved by Se vacancies. From the energy difference between AFM1 and AFM2 states (for x ≠ 0), we estimate the exchange coupling to be about 10 meV. Due to frustrated magnetic interactions, we expect there to be many states close in energy to the stripe AFM2 state. To this end, we carried out Monte Carlo simulations with a 2-D Ising model upto second nearest neighbor interactions ($-J_1 = J_2 = 10$ meV), and found that the system develops only short-range order of the stripe kind [34] below about 100 K and that statistical averages are rather demanding numerically due to frustrated J-couplings.

Interestingly, our calculation shows that the tetragonal symmetry of FeSe is broken with this stripe-AFM order, as manifested in the stresses $\sigma_{xx} \neq \sigma_{yy}$ for the AFM-stripe phase, whereas $\sigma_{xx} = \sigma_{yy}$ for the NM phase. A weakly orthorhombic symmetry of the stress on the unit cell induced by the AFM stripe order is expected to result in orthorhombic strain, $b\text{-}a/a \neq 0$, in the crystal as a secondary order parameter in a transition to AFM-stripe phase, similar to that in improper ferroelectrics (we note that there are no unstable phonons in the nonmagnetic tetragonal phase ruling out a primarily structural transition).

We now attempt to understand the large observed change of ~ 5% in the frequency of the S1 ($E_g$) mode below $T_s$. To this end, we first determined the change in frequency of phonons as a function of orthorhombic strain within DFT. It has been shown experimentally that maximum value of *(b-a)/a* is ~ 0.5% at the lowest temperature of 5 K [9]. Our calculations show that phonon frequencies hardly change with an orthorhombic strain of ~ 0.5% in the non-magnetic state, indicating a weak strain-phonon coupling and ruling out its role in *T*-dependence of the S1 mode frequency. Thus, it has to be a strong spin-phonon coupling that is responsible for the anomalous behavior of the mode S1 below $T_s$. To this effect, our DFT calculations show that the mode S1 frequency in FeSe$_{0.875}$ indeed hardens by ~ 7 % (to 126 cm$^{-1}$) in the AFM stripe phase as compared to that (118 cm$^{-1}$) in the AFM1 state. This hardening is a theoretical (DFT) measure of spin-phonon coupling: $\omega = \omega_0 + \lambda S_i S_j$ [35], $\lambda$ being negative. Using $<S_i S_j>$ for SNN spins obtained from Monte Carlo simulations, which is negative and sharply increases in magnitude below 100 K [34], we confirm that the hardening of S1 mode below $T_s$ arises from the strong spin-phonon coupling and emergence of short-range stripe AFM2 order. As the ordering is short-range, we do not predict the splitting of S1 mode, as otherwise expected from a long-range AFM2 ordering. The coupling of E$_g$ and B$_{1g}$ modes with spin degrees of freedom originates from the mode induced changes in Fe-Se-Fe bond-angle and consequent variation of the superexchange interactions. We note that a strong spin-phonon coupling has been inferred in iron pnictides (CeFeAs$_{1-x}$P$_x$O) through phenomenological analysis of the dependence of the measured magnetic moment of Fe on Fe-As layer separation and $T_c$ [15]. Our calculations show that the higher energy E$_g$ mode (S4 mode at 315 cm$^{-1}$ involving Fe displacements) also couples with spins with a comparable strength to the S1 mode. However, the perturbative analysis shows that its correction to frequency of S4 mode is much weaker (since the mode is of higher frequency, and correction depends inversely on the frequency of the mode).

### 3.1.4 Conclusion

To summarize, we have presented Raman measurements of FeSe$_{0.82}$, showing all the four Raman active modes and electronic Raman scattering involving 3$d$-orbitals of Fe. A picture consistent with our measured Raman spectra and DFT calculations is as follows: as the temperature is lowered, there is a transition to a phase with short range AFM2 stripe order at $T_s$ ~ 100 K, (as reflected in the anomaly observed in magnetic susceptibility [7]), long-range order being suppressed due to frustrated magnetic interactions. This change in spin ordering is accompanied by a weak orthorhombic distortion as a secondary order parameter. By symmetry, the coupling between this AFM2 order and $E_g$ (S1 and S4) modes is non-zero, which leads to anomalous change in frequency of the lower of the two $E_g$ modes (S1). Thus, a strong spin-phonon coupling (estimated from the phonon frequencies of AFM1 and AFM2 stripe phases) is responsible for the observed hardening of S1 mode in the normal state of FeSe. This may be relevant to the recent proposal of strong electron-phonon coupling through the spin-channel being discussed as a mechanism of superconductivity [15]. In addition, the high frequency modes at 1350 cm$^{-1}$ and 1600 cm$^{-1}$ are attributed to electronic Raman scattering involving $x^2$-$y^2$ to ($xz$, $yz$) $3d$-orbitals of Fe. We conclude that spin polarized $xz$ and $yz$ orbitals of Fe, $E_g$ modes with in-plane displacement of Se atoms, orthorhombic strain and their couplings are relevant to the observed temperature dependent Raman spectra presented here.

## 3.2 Part-B

## Crystal Field Excitations, Phonons and their Coupling in Superconductor $CeFeAsO_{0.9}F_{0.1}$

### 3.2.1 Introduction

Discovery of superconductivity in $LaFeAsO_{1-x}F_x$ was very soon extended to other member of '1111' family [1-3]. A few room temperature and temperature dependent Raman studies on $RFeAsO_{1-x}F_x$ (R = Ce, La, Nd and Sm) [22-23,30,36-37] systems were reported in literature before our work, however no phonon anomalies were reported as a function of temperature in these systems [22-23]. This motivated us to do a detailed temperature dependent Raman study on $CeFeAsO_{0.9}F_{0.1}$ having $T_c \sim 38$ K [38] because: (i) Raman study of phonons in these materials can provide information on the superconducting state through coupling of phonons to Raman active electronic excitations, and (ii) electronic excitations between the crystal field split $f$-orbitals of $Ce^{3+}$ and $3d$-orbitals of Fe can be probed. This is important because Fe $3d$-orbitals contribute significantly to the electronic states at the Fermi level and hence they are expected to play a crucial role in the mechanism of superconductivity. In particular, it is not clear experimentally if $d_{xz}$ and $d_{yz}$ orbitals are split or not [17-20, 39] in the general class of iron based superconductors as also described in detail in Part A-3.1.

### 3.2.2 Experimental Details

Polycrystalline samples of $CeFeAsO_{0.9}F_{0.1}$ with a superconducting transition temperature of 38 K and without any impurity phase were prepared and characterized as described in ref. 38. Unpolarised micro-Raman measurements were performed similar to those discussed in Part A -3.1.

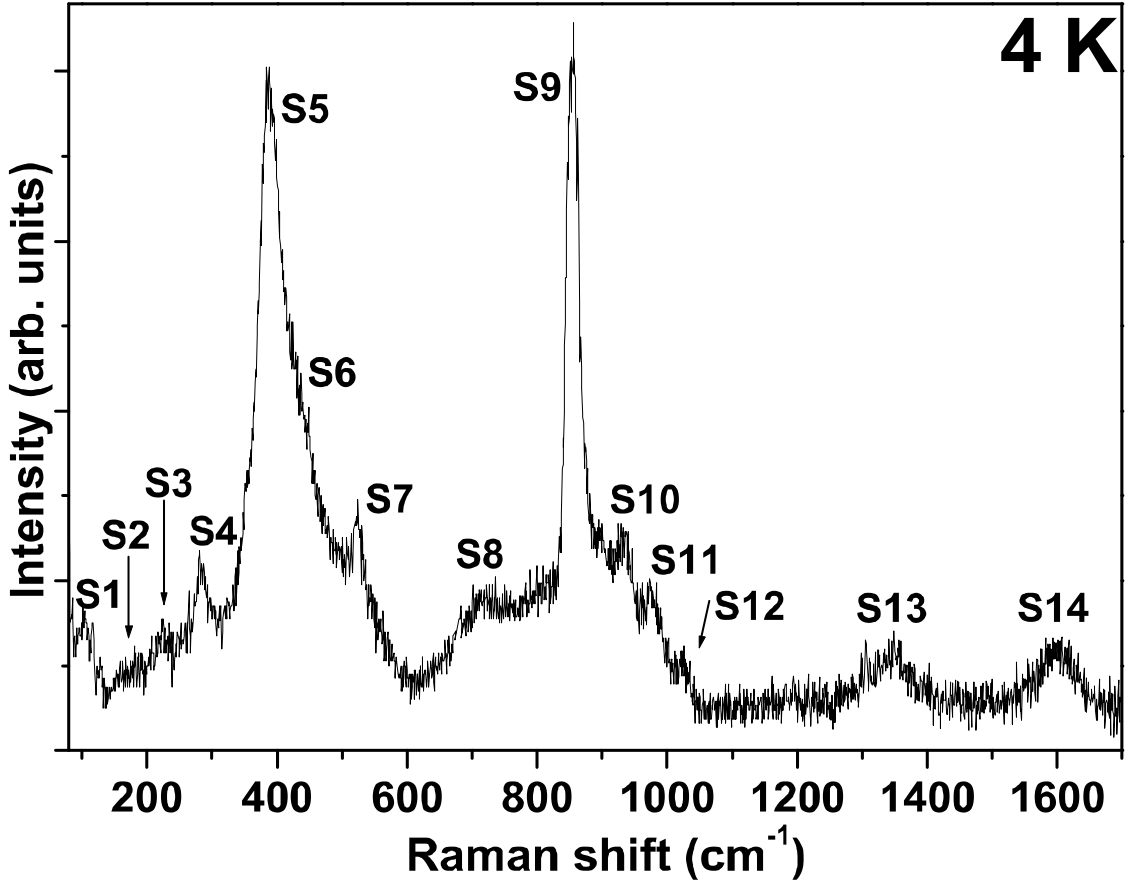

Figure 3.5: Raman spectra of CeFeAsO$_{0.9}$F$_{0.1}$ at 4 K.

## 3.2.3 Results and Discussion

### 3.2.3.1 Raman Scattering from Phonons

CeFeAsO has a layered structure belonging to the tetragonal *P4/nmm* space group containing two CeFeAsO units per unit cell. There are eight Raman active phonon modes belonging to the irreducible representation $2A_{1g} + 2B_{1g} + 4E_g$, with the dominant displacement of atoms classified as La: $A_{1g} + E_g$; O/F: $B_{1g} + E_g$; Fe: $B_{1g} + E_g$; and As: $A_{1g} + E_g$ [36]. Figure 3.5 shows Raman spectrum at 4 K, revealing 14 modes labeled as S1 to S14 in the spectral range of 80-1700 cm$^{-1}$. Raman spectra at a few temperatures are shown in frequency range 200 to 600 cm$^{-1}$ in Fig. 3.6 (a) and from 600 to 1700 cm$^{-1}$ in Fig. 3.6 (b). Spectra are fitted to a sum of Lorentzian functions. The individual modes are shown by thin lines and the resultant fit by thick line. It can be seen from Fig. 3.6 (a) that the mode near 400 cm$^{-1}$ is highly asymmetric and can be fitted well to a sum of two Lorentzian functions, representing modes S5 and S6. The frequencies measured at 4 K are tabulated in Table-3.2.

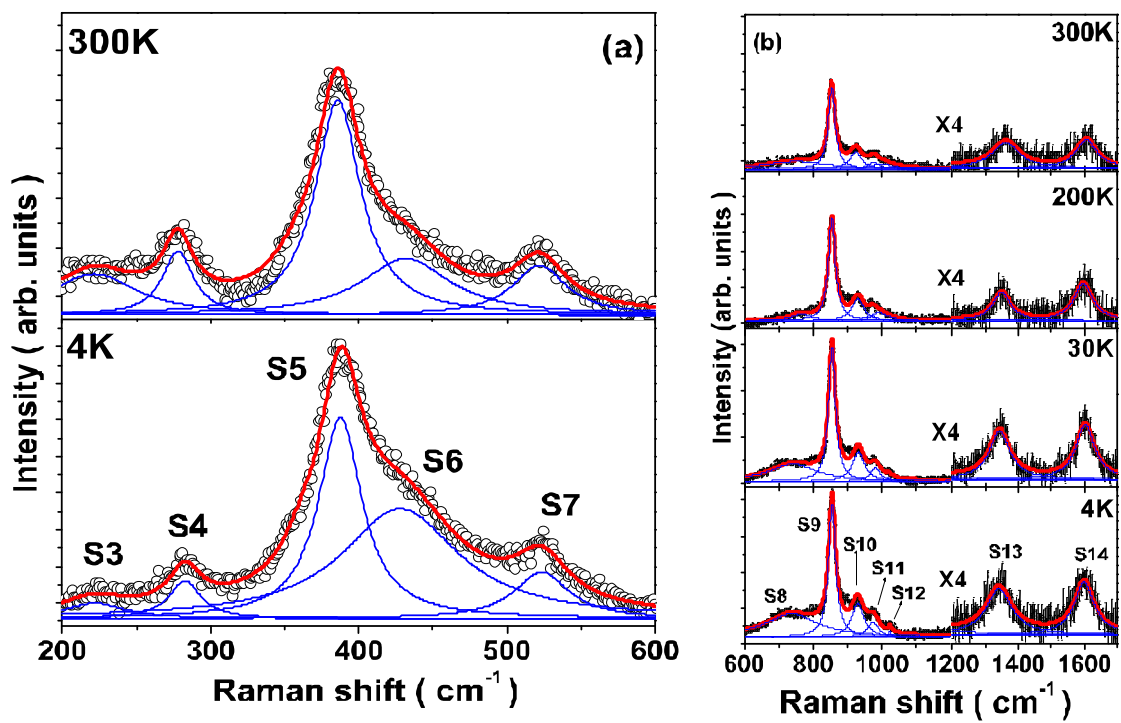

Figure 3.6: (a) Raman spectra from 200 to 600 cm$^{-1}$. (b) Temperature evolution of the high frequency (600 to 1700 cm$^{-1}$) modes. Solid (thin) lines are fit of individual modes and solid (thick) line shows the total fit to the experimental data.

A noticeable feature of our data is the observation of modes S8 to S14 above 600 cm$^{-1}$. The available experiments [22-28,30,36-37,39] and theoretical calculation [40] of phonons in the iron-based superconductors show that the first-order Raman phonons are observed only below 550 cm$^{-1}$. Before we discuss the assignment of modes S1 to S5 and mode S7 as phonon modes, we review the assignments of the Raman modes observed so far in "1111" systems [22-23,30,36-37] to help us in associating the observed modes given in Table-3.2. Out of the eight Raman active modes, three distinct modes observed at room temperature in CeFeAsO$_{0.84}$F$_{0.16}$ are 220 cm$^{-1}$ (B$_{1g}$, Fe), 280 cm$^{-1}$ (E$_g$, Fe) and 450 cm$^{-1}$ (E$_g$, O) [30]. In addition, a weak mode could be seen at 396 cm$^{-1}$ as a shoulder of the 450 cm$^{-1}$ mode [30]. In NdFeAsO$_{0.9}$F$_{0.1}$, five Raman active modes have been observed and assigned as 252.5 cm$^{-1}$ (mixed mode of Fe and As in *c*-direction), 262.5 cm$^{-1}$ (mixed mode of Fe and As in *a* or *b* direction), 337.8 cm$^{-1}$ (oxygen mode), 368 cm$^{-1}$ (Nd-O or Nd-F vibration) and 487 cm$^{-1}$ (Nd-O stretching mode) [23]. In LaFeAsO system, six observed modes have been assigned as 96 cm$^{-1}$ (E$_g$, La), 137 cm$^{-1}$ (E$_g$, As and Fe), 161 cm$^{-1}$ (A$_{1g}$, La), 214 cm$^{-1}$ (B$_{1g}$, Fe), 278 cm$^{-1}$ (E$_g$, Fe) and 423 cm$^{-1}$ (E$_g$, O) [30]. In SmFeAsO four Raman modes have been identified as 170 cm$^{-1}$ (A$_{1g}$, Sm), 201 cm$^{-1}$ (A$_{1g}$, As), 208 cm$^{-1}$ (B$_{1g}$, Fe) and 345 cm$^{-1}$ (B$_{1g}$, O) [36]. The calculated value of (E$_g$,O) phonon frequency for SmFeAsO is 503 cm$^{-1}$ [37]. Keeping these data in view and our density functional calculations (discussed below), we assign the modes S1 to S5 to be the first order Raman active modes as S1 (102 cm$^{-1}$, E$_g$, Ce ), S2 (162 cm$^{-1}$, A$_{1g}$, Ce), S3 (223 cm$^{-1}$, B$_{1g}$ ,Fe), S4 (281 cm$^{-1}$, E$_g$, Fe ) and S5 (389 cm$^{-1}$, E$_g$ , O) (see Table-3.2).

Table-3.2: List of the experimentally observed frequencies at 4 K in CeFeAsO$_{0.9}$F$_{0.1}$ and calculated phonons mode frequencies in CeFeAsO$_{1-x}$F$_x$.

| Mode Assignment | Experimental ω (cm$^{-1}$) | Calculated ω (cm$^{-1}$) | | | |
|---|---|---|---|---|---|
| | | Experimental Lattice constant | | Theoretical Lattice constant | |
| | | x = 0.0 | x = 0.25 | x = 0.0 | x = 0.25 |
| S1  E$_g$ (Ce) | 102 | 101 | 90 | 127 | 112 |
| S2  A$_{1g}$ (Ce) | 162 | 171 | 162 | 200 | 189 |
| S3  B$_{1g}$ (Fe) | 223 | 201 | 202 | 233 | 241 |
| S4  E$_g$ (Fe) | 281 | 287 | 285 | 323 | 322 |
| S5  E$_g$ (O) | 389 | 394 | 417 | 467 | 486 |
| S6  CF of Ce$^{3+}$ levels | 432 | | | | |
| S7  Two-Phonon | 524 | | | | |
| S8  DOS | 720 | | | | |
| S9  (S5 + S6) | 855 | | | | |
| S10 Multiphonon | 932 | | | | |
| S11 Multiphonon | 975 | | | | |
| S12 Multiphonon | 1026 | | | | |
| S13 CF of Fe-d levels | 1342 | | | | |
| S14 CF of Fe-d levels | 1600 | | | | |

## 3.2.3.2 First-Principles Calculations

Our first-principles calculations are based on density functional theory as implemented in the PWSCF [31] package. We use optimized norm-conserving pseudopotential for Ce [41-42] constructed with $Ce^{3+}$ as a reference state and ultrasoft pseudopotentials [32] for other elements (O, Fe, As) to describe the interaction between ionic cores and valence electrons, and exchange correlation energy functional with a local density approximation. We use plane wave basis with a kinetic energy cutoff of 60 Ry in representation of wavefunctions and a cutoff of 360 Ry for representing the charge density. Convergence with respect to basis set and grid used in the Brillouin zone sampling has been carefully checked. We sampled integration over the Brillouin zone (of single unit cell) with 12x12x6 Monkhorst Pack Mesh [43]. For Fluorine-substituted compound, we used a $\sqrt{2}x\sqrt{2}x1$ supercell. Structural optimizations of CeFeAsO and $CeFeAsO_{0.75}F_{0.25}$ are carried through minimization of the total energy using Hellman-Feynman forces and the Broyden-Flecher-Goldfarb-Shanno based method. Zone center (q = 0,0,0) phonon spectra are determined using a frozen phonon method (with atomic displacements of 0.04 Å) for the relaxed structure obtained at experimental lattice constants. In addition, we used DFT-linear response [31] to obtain phonon spectra at wave-vectors on a 2x2x2 mesh and Fourier interpolation of the interatomic force constants to determine phonons at wave-vectors on a finer mesh (8x8x8), and obtain one and two-phonon density of states.

We have calculated phonons at gamma point (q = 0,0,0) of $CeFeAsO_{1-x}F_x$ for x = 0.00 and x = 0.25 using first-principles density functional theory described above. Calculated phonon frequencies for both experimental and theoretical lattice constants are listed in Table-3.2. The lattice constant used are as follows, Experimental: $a$ = 3.996 Å, $c$ = 8.648 Å and Theoretical: $a$ = 3.808 Å, $c$ = 8.242 Å. Our first-principles theoretical estimates of the optimized lattice constants are 4-5% smaller than the experimental values (as found typically for pnictides, and consistent with the earlier reports [44]). To identify phonon modes observed in the Raman spectra, we calculated phonons (using both experimental and theoretical lattice constants) at gamma point (q = (0,0,0)) of CeFeAsO$_{1-x}$F$_x$ for x = 0.00 and x = 0.25 using frozen phonon method. Calculated phonon frequencies at the experimental lattice constants are in fairly good agreement with the experimental values (see Table-3.2) and corresponding eigen modes are shown in Fig. 3.7.

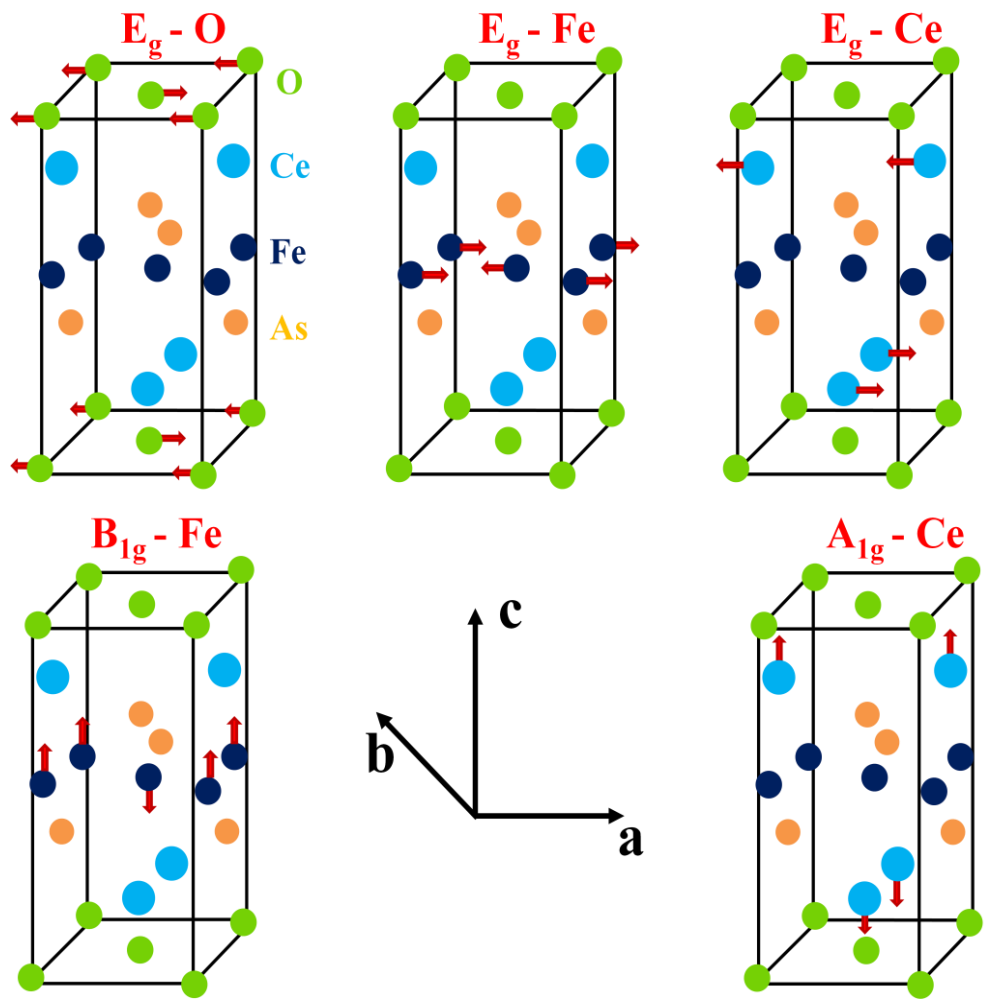

Figure 3.7: Eigen modes corresponding to different Raman modes in Table-3.2.

Generally, we find that modes are localized in Ce-O and Fe-As layers. We find that the Raman-active E$_g$ modes with Ce and O character change notably with F-concentration *x*, while $E_g$ (Fe) mode is quite insensitive to *x*. A similar trend is found for the $B_{1g}$ modes. This is expected as Ce is coordinated with oxygen while Fe is coordinated with As. Also, all the phonon frequencies increase by 10-25 % with reduction in the lattice constant by about 5 % (from the experimental value to the LDA value). This gives us some idea about the errors involved in the theoretical estimates of phonon frequencies. In order to explore the possibility of second order Raman scattering, we also determine (using experimental lattice constants) one phonon density of states ($D(\omega) = \sum_i \delta(\omega - \omega_i)$) and two-phonon density of states (DOS) ($D(\omega) = \sum_{i,j} \delta(\omega - \omega_i + \omega_j)$) by calculating phonons at 2×2×2 grid of *q*-points using linear response method. We find peaks at ~560 and 700 cm$^{-1}$ that correlate with S7 and S8 modes observed in Raman spectra. Based on our estimates, the modes above 800 cm$^{-1}$, are likely to arise from the electronic Raman scattering, such as ones associated with crystal field splitting of Fe 3*d*-levels.

### 3.2.3.3  Raman Scattering from Crystal Field Split Excitations of $Ce^{3+}$

We now discuss why we assign S6 mode (~ 435 cm$^{-1}$) as electronic Raman scattering involving the crystal field split excitations of $Ce^{3+}$ *f*-levels (J = 5/2). Inelastic neutron scattering from $CeFeAsO_{1-x}F_x$ (x = 0, 0.16) shows that the $Ce^{3+}$ crystal field levels are three magnetic doublets at 0 ($E_0$), 18.7 meV ($E_1$) and 58.4 meV ($E_2$) [45], thereby allowing three possible transitions: $E_0$ to $E_1$ (~ 150 cm$^{-1}$), $E_0$ to $E_2$ (~ 470 cm$^{-1}$) and $E_1$ to $E_2$ (~ 320 cm$^{-1}$) [45]. At low temperatures only the first two transitions can be seen. Raman cross section for the crystal field excitation (CFE) is often weak and these have been observed mixed with nearby Raman-active phonons via electron-phonon interaction [46]. The frequency of the mode S6 is very close to the transition from ground state ($E_0$) to the second excited state ($E_2$). Moreover, the integrated intensity of the S6 mode decreases with increasing temperature (shown in Fig. 3.8 (a)), which is typical for Raman scattering associated with crystal field transitions (because the ground state population decreases with increasing temperature). We, therefore, assign S6 to electronic Raman scattering from the crystal filed split levels of $Ce^{3+}$. The transition from $E_0$ to $E_1$ is not clearly seen probably because of its overlap with Raman phonon S2.

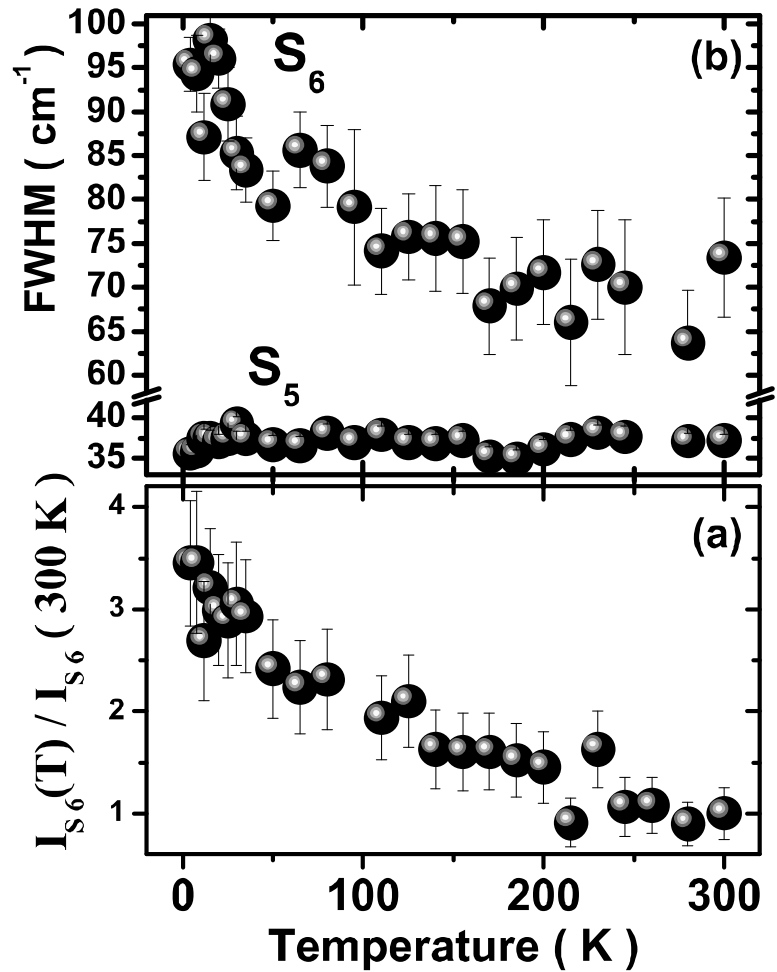

Figure 3.8: (a) Normalized integrated intensity of mode S6. (b) Temperature dependence of the Linewidth of modes S5, S6.

A coupling between CFE and phonon mode is expected only if they have comparable energies and are of the same symmetry [46]. Since $Ce^{3+}$ ion is located between planes of oxygen and arsenic, phonon modes involving oxygen and arsenic will be expected to couple strongly to the CFE via their modulation of the crystal field. The electric field due to surrounding ions split the $Ce^{3+}$ ion $^2F_{5/2}$ multiplets into one $M^{6-}$ and two $M^{7-}$ (Koster notation [47]) Kramers doublets, denoted by $E_0$ ($M^{6-}$), $E_1$ ($M^{7-}$) and $E_2$ ($M^{7-}$). The symmetry of the CF excitation within this multiplet is given by the direct product of the irreducible representation of the doublets: $E_0$ to $E_1$ or $E_2$: $M^{6-} \otimes M^{7-} = B_{1g} + B_{2g} + E_g$ and $E_1$ to $E_2$: $M^{7-} \otimes M^{7-} = A_{1g} + A_{2g} + E_g$. Therefore, the CF excitation from the ground state ($E_0$) to second excited state ($E_2$) is group theoretically allowed to couple with the mode S5 ($E_g$).

Figure 3.8 (b) shows the temperature dependence of full width at half maxima (FWHM) of modes S5 and S6. It can be seen that FWHM of S5 mode is independent of temperature whereas that of S6 is anomalous, in the sense that the linewidth increases on decreasing temperature. The frequency of mode S6 remains nearly independent of temperature (see Fig. 3.9). Similar anomalies have been reported in cuprates for the crystal field modes [48]. These observations suggest that $Ce^{3+}$ CF mode (S6) is coupled strongly with the phonon mode at 389 cm$^{-1}$ (S5) associated with the oxygen vibrations.

### 3.2.3.4 Temperature Dependence of the Phonon Frequencies

Mode S1, S2 and S3 are weak and their temperature dependence is difficult to quantify; we, therefore, focus on the temperature dependence of peak positions of S4 to S7 and S9, as shown in Fig. 3.9. The solid line for mode S5 in Fig. 3.9 is fitted to an equation (2.11) described in detail in Chapter 2. Fitted parameters for the mode S5 are $\omega_o = 391.14$ cm$^{-1}$, C = $-5.12 \pm 0.44$ cm$^{-1}$.

Phonons which couple strongly to electronic states near Fermi surface can be easily influenced by changes in the neighborhood of the Fermi surface. In superconductors, the opening of the superconducting gap redistributes the electronic states in the neighborhood of the Fermi surface and hence will change phonon self-energy seen in cuprate superconductors [49]. Qualitatively, phonons below the gap show a softening whereas phonons with frequency above the gap value can harden. The latter can turn into a small softening if impurity scattering is taken into account [50]. The anomalous softening of the S4 mode below $T_c$ (shown in Fig. 3.9) indicates a coupling of this phonon to the electronic system. As superconducting gap pushes the phonon towards lower energy below $T_c$, we suggest that superconducting gap in this system should lie close to this phonon towards high energy side i.e. $2\Delta \geq 281$ cm$^{-1}$, similar to a softening of the Raman active phonon mode below $T_c$ in cuprate superconductors [49]. This gives an estimate of the ratio of gap ($2\Delta$) to the transition temperature $2\Delta/K_B T_c \sim 10$. This ratio suggests that this system belongs to the class of strong-coupling superconductors. We note that the present value of $2\Delta/K_B T_c$ is close to the value found for (Nd, Sm, La)FeAsO$_{1-x}$F$_x$ ($2\Delta/K_BT_c$ ~ 8) [51-53] and also (SrBa)$_{1-x}$(K, Na)$_x$Fe$_2$As$_2$ ($2\Delta/K_BT_c$ ~ 8) [54]. There are reports of multiple gaps in "1111" and "122" systems with values of $2\Delta/K_BT_c$ ranging from 3 to 9 [53, 55-56].

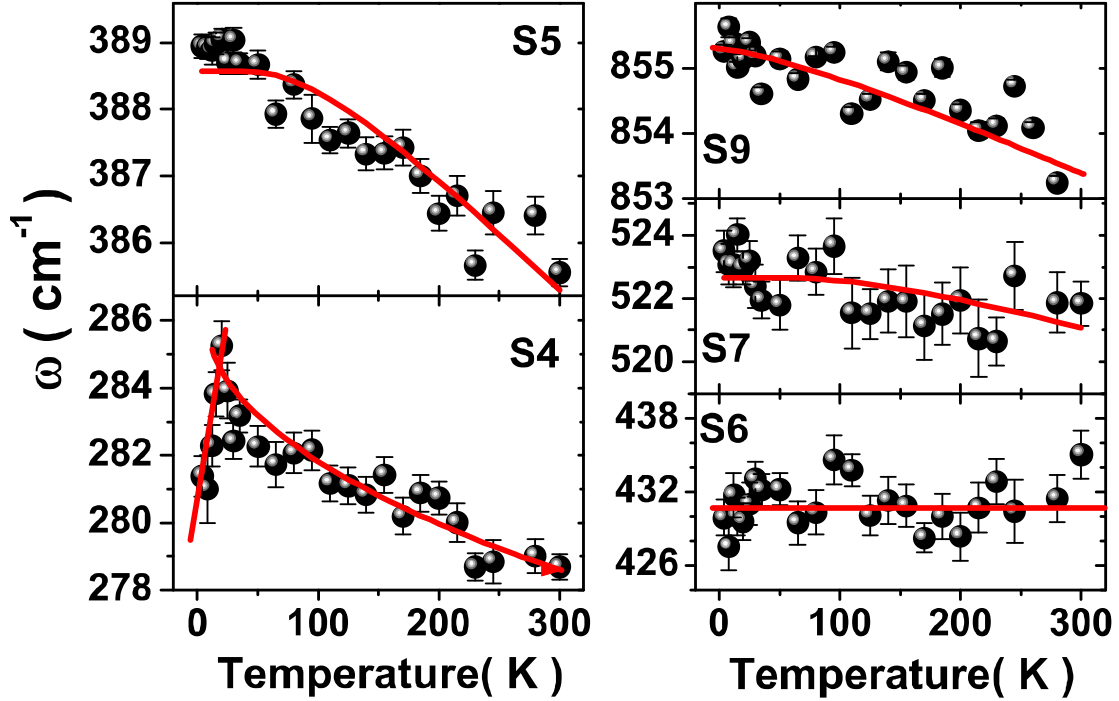

Figure 3.9: Temperature dependence of the modes S4, S5, S6, S7, and S9. Solid lines for S4, S6, S7, and S9 are drawn as guide to the eye. The solid line for S5 is the fitted curve as described in text.

### 3.2.3.5 Origin of the High Frequency Modes S7-S14

In addition to the expected first-order Raman bands, we observe modes S7 to S14. A weak mode was observed at the frequency close to that of mode S7 in the room temperature Raman study of CeFeAsO$_{1-x}$F$_x$ [30], attributed to multiphonon Raman scattering. We make similar assignment for the mode S7. A broad mode around 720 cm$^{-1}$ (S8) is similar to a broad band in Raman spectra of superconductor MgBr$_2$ [57], which has been attributed to a maximum in density of states (DOS) arising from the disorder induced contributions from phonons away from the zone centre.

The coupled CFE with the longitudinal optical phonons have been observed in case of UO$_2$ [58]. Following this, the mode S9 (855 cm$^{-1}$) is assigned as a second-order Raman scattering from a combination of modes S5 and S6. The weak modes S10, S11 and S12 can be multiphonon Raman scattering. In addition, we observe two high frequency modes S13 (~1342 cm$^{-1}$) and S14 (~1600 cm$^{-1}$) which are attributed to electronic Raman scattering from ($x^2$-$y^2$) to $xz/yz$ 3$d$-orbitals of Fe in line with our observation in FeSe$_{0.82}$ described in detail in Part A-3.1.

### 3.2.4 Conclusion

In conclusion, we have presented Raman measurements of CeFeAsO$_{0.9}$F$_{0.1}$ as a function of temperature. We suggest that the softening of the Raman active phonon mode (S4) below $T_c$ is due to an opening of a superconducting gap, yielding $2\Delta/K_BT_c$ ~ 10. A Raman mode at ~ 432 cm$^{-1}$ is attributed to electronic Raman scattering involving Ce$^{3+}$ crystal field levels, coupled strongly to $E_g$ oxygen phonons. A strong mode is observed at 855 cm$^{-1}$ which has been attributed to a combination mode of the above crystal field excitation and the $E_g$ oxygen vibration. High frequency modes observed at ~ 1342 cm$^{-1}$ and ~ 1600 cm$^{-1}$ are attributed to electronic Raman scattering involving $x^2$-$y^2$ to ($xz$, $yz$) 3$d$-orbitals of Fe.

# 3.3 Part-C

# Raman Evidence for Coupling of Superconducting quasi-particles with a Phonon and Crystal Field Excitation in $Ce_{0.6}Y_{0.4}FeAsO_{0.8}F_{0.2}$

## 3.3.1 Introduction

The optimal doping of 'Y' in $CeFeAsO_{0.8}F_{0.2}$ leads to a significant increase in the transition temperature, which in the present case is ~ 48.6 K as compared to $T_c$ ~ 42.7 K in case of system without 'Y'-doping [59]. We note that $T_c$ of CeFeAs(O/F) increases on increasing chemical pressure by substituting smaller $Y^{3+}$ ion in place of $Ce^{3+}$, which leads to shrinking of both *a* and *c* lattice parameters. On the other hand $T_c$ is suppressed on increasing external pressure and is lowered to 1.1 K at 265 Kbar [60]. The increase of $T_c$ in 'Y'-doped system may be linked with the change in $FeAs_4$ tetrahedral structure, with doping the tetrahedral angle between As-Fe-As may approach its ideal value i.e. close to 109 degrees and as a result of this maximum $T_c$ is achieved as also observed in other iron-based superconductors (FeBS). The effect of 'Y' doping which leads to maximum $T_c$ in CeFeAs(O/F) system motivated us to study the phonon dynamics and we note that the effect of doping is also reflected in the observed phonon frequencies of the Raman active modes discussed later, where some of the phonon modes are red-shifted and some are blue-shifted as compared to the earlier Raman studies on $CeFeAsO_{0.9}F_{0.1}$ superconductor [61]. Another motivation was to study the potential effect of superconducting transition on the phonon dynamics and understanding their role in the pairing mechanism. Here, we present a study of superconductor $Ce_{0.6}Fe_{0.4}AsO_{0.8}F_{0.2}$, focusing mainly on the temperature dependence of the lowest frequency phonon mode and coupling of crystal field excitation with the superconducting quasi-particles below $T_c$.

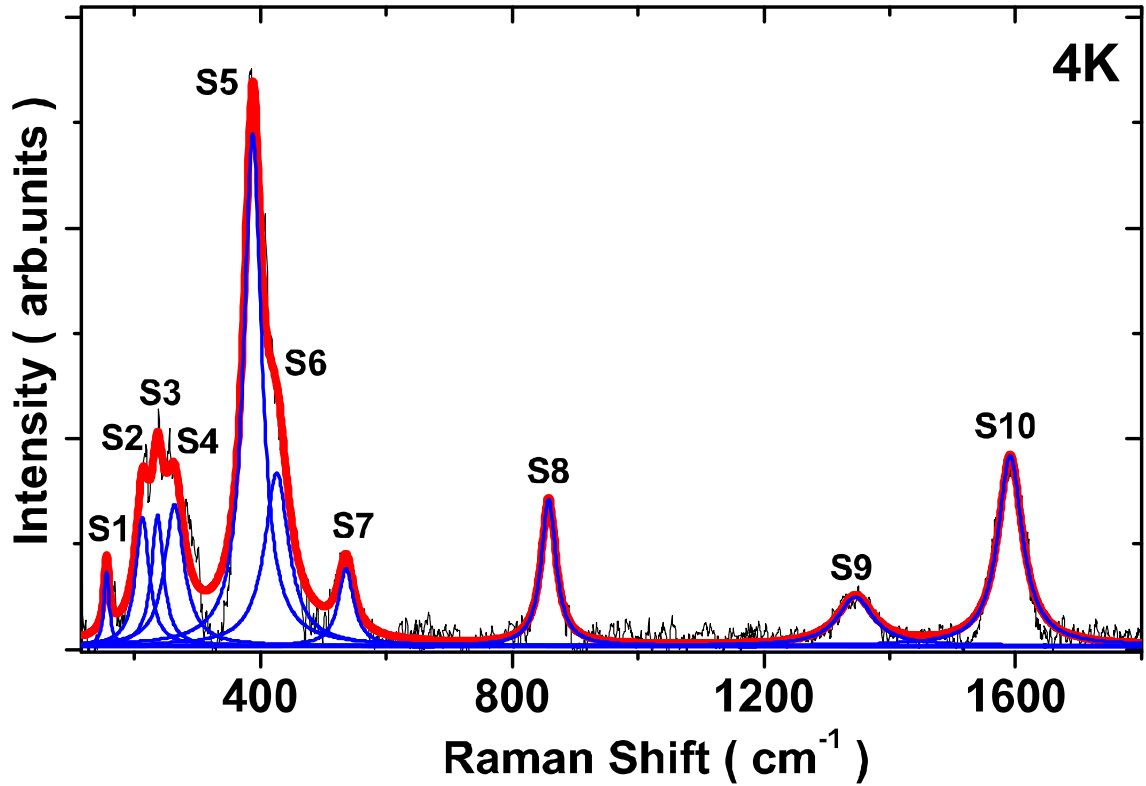

Figure 3.10: Raman spectrum at 4 K. Thin solid lines are Lorentzian fit to individual modes and the solid thick line shows the total fit to the experimental data. The raw data at 4 K is denoised using FFT filtering process.

### 3.3.2 Results and Discussion

### 3.3.2.1 Raman Scattering from Phonons

Figure 3.10 shows Raman spectrum at 4 K, revealing 10 modes labeled as S1 to S10 in the spectral range 100 - 1800 cm$^{-1}$. Spectra are fitted to a sum of Lorentzian functions. The individual modes are shown by thin lines and the resultant fit by thick line. The frequencies measured at 4 K are tabulated in Table-3.3. The existing experiments and theoretical calculation [22-24,27,30,34,36-37,39,61-62] of phonons show that first-order Raman phonons are observed only below 550 cm$^{-1}$. However, we clearly observed three modes (S8,

S9 and S10) above 600 cm$^{-1}$. Following the earlier Raman studies on "1111" systems [23,30,36-37,61], we assign the modes S1 to S5 to be the first-order Raman active modes as S1 (~ 146 cm$^{-1}$, A$_{1g}$, Ce/Y), S2 (~ 220 cm$^{-1}$, B$_{1g}$, Fe), S3 (~ 255 cm$^{-1}$, E$_g$, Fe), S4 (~ 290 cm$^{-1}$, B$_{1g}$, O) and S5 (~ 394 cm$^{-1}$, E$_g$, O) (see Table-3.3). The mode S6 is assigned to the crystal field excitation of Ce$^{3+}$ (to be explained later), S7 and S8 as second-order Raman modes and S9, S10 as crystal field excitations of Fe$^{2+}$ 3$d$-levels. We note that mode S1 (A$_{1g}$ - Ce,Y), S2 (B$_{1g}$ - Fe) and S3 (E$_g$ - Fe) are red-shifted and mode S5 (E$_g$ - O) is blue-shifted with maximum shift (Δω) ~ 25 cm$^{-1}$ as compared to those in without 'Y'-doped sample [61], thus suggesting the significant effect of chemical pressure on phonon dynamics.

Table-3.3: List of the experimentally observed frequencies at 4 K in Ce$_{0.6}$Y$_{0.4}$FeAsO$_{0.8}$F$_{0.2}$.

| Mode Assignment | Experimental ω (cm$^{-1}$) |
|---|---|
| S1  A$_{1g}$ (Ce/Y) | 146 |
| S2  B$_{1g}$ (Fe) | 220 |
| S3  E$_g$ (Fe) | 255 |
| S4  B$_{1g}$ (O) | 290 |
| S5  E$_g$ (O) | 394 |
| S6  CF of Ce$^{3+}$ levels | 430 |
| S7  Two-Phonon | 534 |
| S8  (S5 + S6) | 860 |
| S9  CF of Fe-d levels | 1354 |
| S10 CF of Fe-d levels | 1592 |

### 3.3.2.2 Temperature Dependence of the Phonon Frequencies

The anomalous change in the frequency and linewidth of the phonons across the superconducting transition temperature can be attributed to change in the self energies, $\Delta\Sigma = \Delta\omega + i\Delta\Gamma$, induced by the superconducting transition [50]. The real and imaginary parts of the self energy renormalize frequency and linewidth, respectively. Within the framework of strong coupling Eliasberg theory, Zeyer et al. [50] showed that a change in the phonon self energy across transition temperature is linked with the interaction of the phonons with the superconducting quasi-particle excitations. Qualitatively, phonons below the superconducting gap ($2\Delta$) become sharp and show frequency softening whereas the phonons above the gap may broaden and show anomalous hardening. The renormalization of a phonon mode is strongest when the phonon energy is close to the gap. Figure 3.11 (a) shows the frequency and linewidth of mode S1 as a function of temperature. Figure 3.11 (b) shows a typical spectrum in the range of mode S1 to reflect the anomalous behavior. The anomalous sharpening and frequency softening of the mode below $T_c$ indicates a strong coupling of this phonon of $A_{1g}$ symmetry involving the vibration of Ce/Y with the electronic states. A microscopic understanding of this coupling can help in understanding the symmetry of the superconducting gap. Equaling the phonon energy with the gap energy at zero temperature gives an estimate of $2\Delta/K_B T_c \sim 5$, suggesting this system to be a strong coupling superconductor. We note that in "1111" systems experimental evidence of single and multiple gaps has already been reported [53,61,63], where ratio varies from 3 to 10. The origin of this large variation is still not understood.

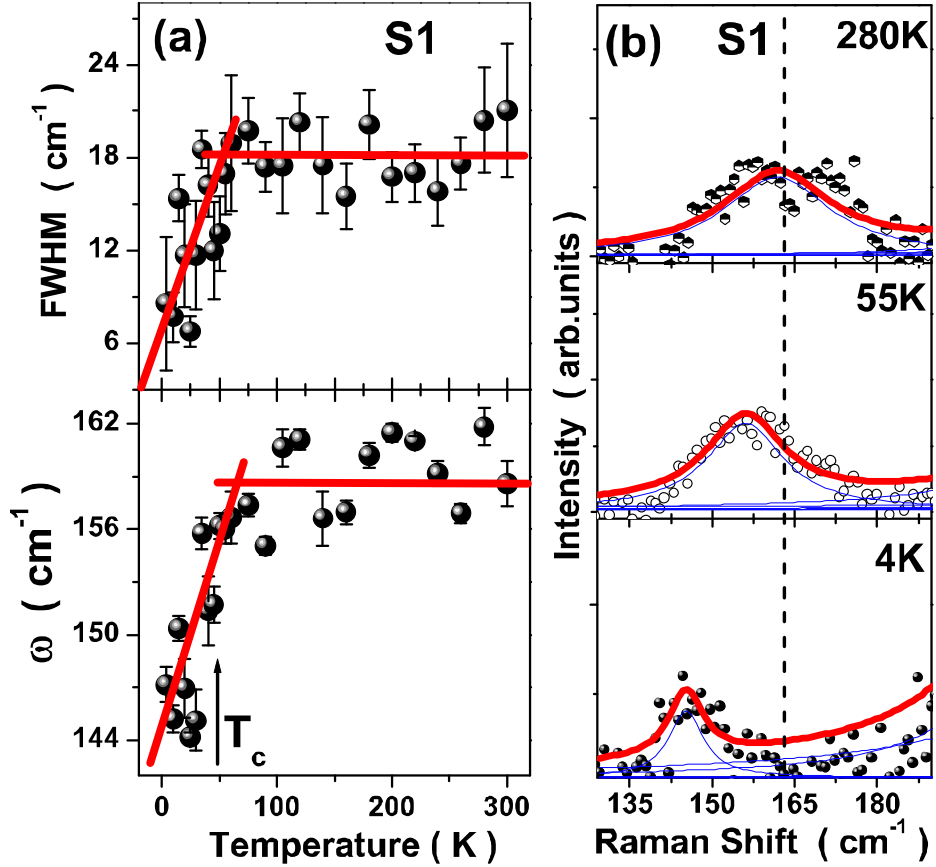

Figure 3.11: (a) Temperature-dependence of phonon mode S1. (b) Temperature evolution of mode S1 at a few temperatures.

In comparison to mode S1, Fig. 3.12 (a) shows that temperature dependence of the modes S5, S7 and S8 is normal as expected based on anharmonic and quasi-harmonic effects. The solid line for the first-order mode S5 is fitted to an equation (2.11) described in Chapter 2. Fitted parameters for the mode S5 are $\omega_o = 396.9$ cm$^{-1}$, C = -5.4 ± 1.4 cm$^{-1}$.

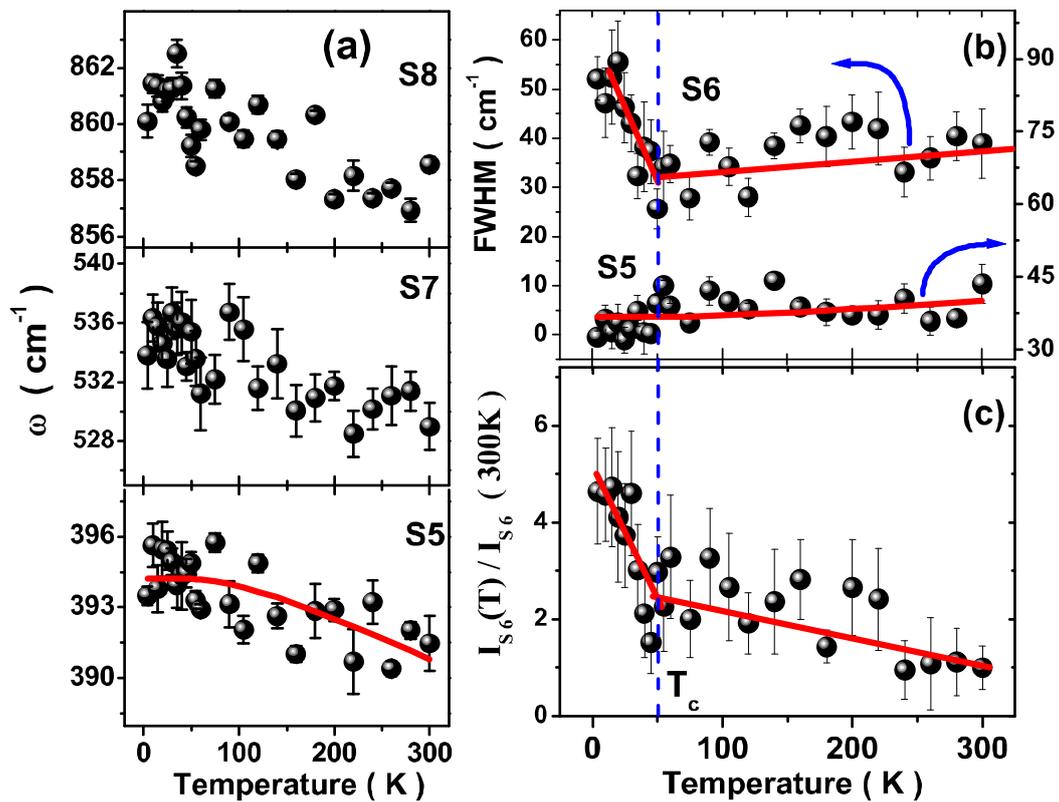

Figure 3.12: (a) Temperature dependence of the modes S5, S7 and S8. The solid line for mode S5 is the fitted curve as described in text. (b) Temperature dependence of the linewidth of modes S5 and S6. Solid lines for mode S6 are linear fit in two temperature ranges and for mode S5 is fitted curve as described in text. (c) Normalized integrated intensity of mode S6. Solid lines are linear fit in two temperature ranges.

### 3.3.2.3 Raman Scattering from Crystal Field Split Excitations of $Ce^{3+}$ and Origin of the High Frequency Modes S8-S10

In earlier temperature-dependent Raman study of superconductor $CeFeAsO_{0.9}F_{0.1}$ [61] a mode near 430 cm$^{-1}$ was observed and assigned to the electronic Raman scattering involving the crystal field excitations (CFE) of $Ce^{3+}$ *f*-levels (J = 5/2). Following earlier studies we have assigned mode S6 to the CFE of $Ce^{3+}$. Figure 3.12 (b) shows the temperature dependence of the full width at half maxima (FWHM) of the phonon mode S5 and the CF excitation S6. It can be seen that the linewidth of mode S5 is as expected from anharmonic effects, fitted by a solid line using $\Gamma(T) = \Gamma_o + D[1+2n(\omega(0)/2)]$, where D is a constant and $\Gamma_o$ is intrinsic linewidth at zero Kelvin. In comparison, the linewidth of the mode S6 is anomalous below $T_c$, reflecting a strong coupling of the CF excitation of $Ce^{3+}$ with the superconducting quasi-particles. This is also corroborated by the increases in the normalized intensity of the mode S6 below $T_c$ (see Fig. 3.12 (c)). The mode S8 (860 cm$^{-1}$) is assigned as a second-order Raman scattering from a combination of modes S5 and S6, as reported earlier in case of $UO_2$ and $CeFeAsO_{0.9}F_{0.1}$ [58,61].

We also observed two high energy excitations at 1354 cm$^{-1}$ (S9) and 1592 cm$^{-1}$ (S10). Our Raman studies on $CeFeAsO_{0.9}F_{0.1}$ (see Part B-3.2) and $FeSe_{0.82}$ (see Part A-3.1) superconductors have also shown similar modes and were attributed to electronic Raman scattering between crystal field split 3*d*-orbitals of $Fe^{2+}$ as ($x^2$-$y^2$) level to *xz* and *yz* levels [39,61]. We make this assignment in the present case as well, giving us a value of splitting of *d*-orbitals *xz* and *yz* as ~ 29 meV. We note that the recent experimental as well as theoretical studies on FeBS suggest that orbital and spin degrees of freedom are believed to conspire to give the much elusive glue for the Cooper pair formation [64-66]. In particular, Fe t$_{2g}$ ($d_{xy}$, $d_{xz/yz}$) orbitals play a central role in controlling their electronic as well as magnetic properties.

Further, reports on FeBS have shown the existence of electronic nematicity between Fe $d_{xz}$ and $d_{yz}$ orbitals [39,61,66] evidencing the crucial role of orbital degrees of freedom.

### 3.3.3 Conclusion

In conclusion, the anomalous behavior of the Raman active phonon mode (S1) below $T_c$ is attributed to the strong coupling between the phonon mode and superconducting quasi-particle excitations due to the opening of superconducting gap below $T_c$. The Raman mode near 430 cm$^{-1}$ associated with the CFE of Ce$^{3+}$ is also shown to be coupled strongly with the superconducting quasi-particle excitations. Our results obtained here suggest that the strong interplay between phonon and electronic degree of freedom is important to understand the pairing mechanism in iron-based superconductors.

## 3.4 Part-D

# Superconducting Fluctuations and Anomalous Phonon Renormalization in $Ca_4Al_2O_{5.7}Fe_2As_2$

### 3.4.1 Introduction

Iron-based superconductors (FeBS) discovered recently are the systems having identical FePn (Pn = As, P) layers but with various block layers [1,4,67-69]. The families of materials which have block layers of the perovskites structure allow a controlled tailoring of the distance between the neighboring FePn layers, such as in $Ba_4Sc_3O_{7.5}Fe_2As_2$ [70], $Ca_{n+m}(M,Ti)_nFe_2As_2$ (M = Sc, Mg and Al; n = 2,3,4,5 and m = 1, 2) [71-72], $Ca_3Al_2O_{5-x}Fe_2Pn_2$ (Pn = As and P) [73] and $Ca_4Al_2O_{6-x}Fe_2Pn_2$ (Pn = As and P) [74], with transition temperature up to 47 K. Of these, $Ca_3Al_2O_{5-x}Fe_2Pn_2$ and $Ca_4Al_2O_{6-x}Fe_2Pn_2$ have the smallest *a*-lattice parameters and the largest iron interlayer distance ($d_{Fe}$). Infact, emergence of superconductivity in these compounds has been ascribed to the smallness of the tetragonal *a*-axis lattice constant [73-74]. It has been further shown that increasing $d_{Fe}$ is correlated with the enhanced $T_c$, possibly due to a reduced coupling between different Fe-As layers [75]. Since Fe-As planes are believed to be the key players in the occurrence of superconductivity in FeBS, a large $d_{Fe}$ would cause the dimensionality to decrease from three *(3D)* to two *(2D)* dimensions. As a consequence, the superconducting fluctuations above $T_c$ should be enhanced - an issue which has been earlier addressed in high temperature cuprate superconductors, but not in iron-based superconductors. Here, we demonstrate the onset of superconducting fluctuations at ~ 60 K, i.e. at ~ *2T_c,* using Raman phonon renormalization finger prints.

In superconductors, the opening of a superconducting gap renormalizes the electronic states near the Fermi surface, which in turn can change the phonon self-energy. Correspondingly,

Raman and infrared spectra do reveal anomalies in vibrational modes below $T_c$ attributed to an opening of a superconducting gap. These superconductivity induced self-energy effects have been used to estimate the magnitude and symmetry of the superconducting order parameter [56,61,76-78]. On the other hand, precursor effects of superconducting fluctuations above $T_c$ on phonons can provide information on a relatively local scale in contrast to transport studies which probe average bulk properties. We here show that phonon anomalies seen at a temperature of ~ $2T_c$ imply the existence of superconducting clusters or droplets much above the critical temperature $T_c$, consistent with the Tisza-London model of a superconductor [79]. The precursor effects above $T_c$ arise from strong fluctuations of the phase of the complex superconducting order parameter thus suggesting the existence of phase-incoherent Cooper pairs above $T_c$. Within the model of phase-incoherent Cooper pairs originating from the classical fluctuations of the phase of the superconducting order parameter, pairing of the superconducting quasi-particles is expected to start much above $T_c$; however, global phase coherence is achieved only below $T_c$. The superconducting transition in classic superconductors is much sharper because the coherence length is much larger than the interatomic distance. However, if coherence length is comparable to atomic dimensions, then it will lead to much more prominent fluctuation effects. Iron-based superconductors have small coherence length similar to cuprate superconductors [80], thus setting the stage for superconducting fluctuations.

As mentioned earlier, a few temperature dependent Raman studies have been reported for "1111" [22,61], "122" [27-28,76] and "11" [39,62] systems to study the effect of superconducting transition on the phonons below $T_c$. However, no precursor effects have been seen in these studies. Here, we report Raman study of $Ca_4Al_2O_{5.7}Fe_2As_2$ with $T_c$ = 28.3 K [74] in the temperature range 5 K to 300 K covering the spectral range of 120 to 800 cm$^{-1}$. We present two significant results: (i) the Raman mode at ~ 230 cm$^{-1}$ shows a jump in frequency by ~ 5 cm$^{-1}$ (~ 2 %) and the corresponding linewidth increases from ~ 16 cm$^{-1}$ to ~ 44 cm$^{-1}$ (~ 175 %) at temperature $T_o$ ~ 60 K. The mode frequency shows anomalous softening by ~ 5 cm$^{-1}$ below $T_o$, accompanied by a large decrease of linewidth by ~ 10 cm$^{-1}$ from 60 K to 5 K. These results on phonon renormalization with an onset at ~ $2T_c$ are attributed to superconducting fluctuation induced effects on the phonon self-energy. (ii) A large increase of the phonon frequency (~ 7 %) between 300 K and 60 K suggests a strong spin-phonon coupling in $Ca_4Al_2O_{5.7}Fe_2As_2$, similar to other iron-based superconductors [28,39,76].

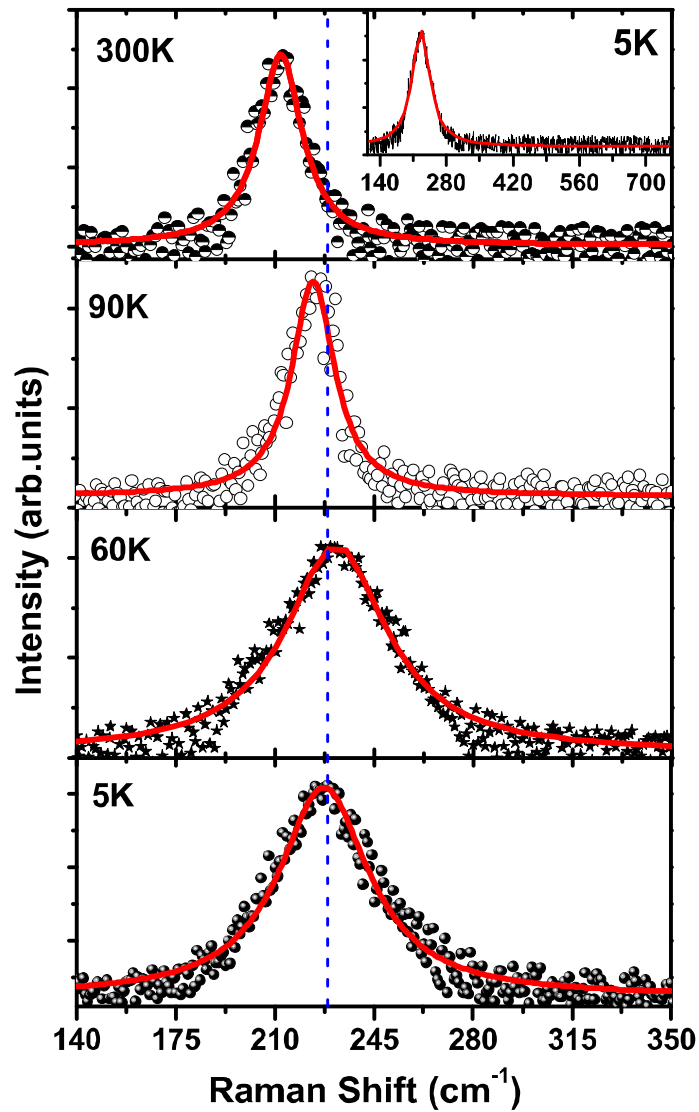

Figure 3.13: Raman spectra of $Ca_4Al_2O_{5.7}Fe_2As_2$ at a few temperatures. Solid lines are the fit to the Lorentzian function. Inset shows the spectrum at 5 K in a large spectral window.

### 3.4.2 Experimental Details and Results

Polycrystalline samples with nominal composition of $Ca_4Al_2O_{5.7}Fe_2As_2$, and with a superconducting transition temperature of $T_c$ = 28.3 K were prepared by our Japanese collaborators using high pressure synthesis. The samples were characterized as per details described in ref. 74. Unpolarised micro-Raman measurements were performed similar to our earlier Raman studies on iron-based superconductors. $Ca_4Al_2O_{5.7}Fe_2As_2$ has a layered structure belonging to the tetragonal *P4/nmm* space group. The inset in Fig. 3.13 exhibits the Raman spectrum at 5 K revealing only one mode. Raman spectra shown in Fig. 3.13 at a few temperatures are fitted with a Lorentzian function indicated by solid lines. Figure 3.14 shows the peak frequency ω (bottom panel) and full width at half maximum (FWHM) (top panel) as a function of temperature. The following observations can be made: (i) frequency of the mode shows anomalous hardening (~ 7 %) between 300 K and $T_o$ (~ 60 K), (ii) the frequency jumps abruptly by ~ 5 $cm^{-1}$ at $T_o$, (iii) the mode frequency displays a softening below $T_o$, (iv) the linewidth decreases slightly (~ 2.5 $cm^{-1}$) between 300 K to 60 K, as expected due to anharmonic interactions, and shows a very large anomalous increase (~ 175 %) at $T_o$, and (v) below $T_o$, the linewidth decreases significantly from ~ 46 $cm^{-1}$ to 35 $cm^{-1}$ in a narrow temperature range from 60 K to 5 K.

### 3.4.3 Discussion

The anomalous hardening of the mode from room temperature to 60 K (~ 7 %) is attributed to strong spin-phonon coupling, similar to other studies for similar iron-based superconductors. The coupling between phonons and the spin degrees of freedom can arise either due to modulation of exchange integral by phonon amplitude and/or by involving change in the Fermi surface by spin-waves, provided the phonon couples to that part of the Fermi surface [35,81]. We note that earlier [28,39,76] Raman studies on other iron-based superconductors have also indicated strong spin-phonon coupling, supported by theoretical calculations [34].

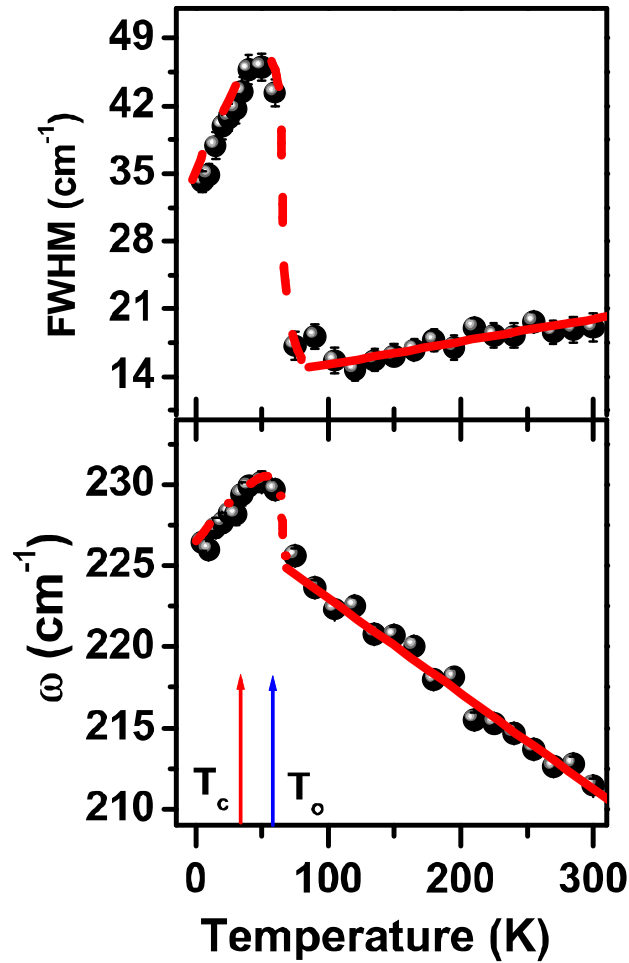

Figure 3.14: Temperature dependence of the phonon mode. Solid lines from 300 K to $T_o$ are linear fits and dash lines are guide to the eye.

Below the superconducting transition there is a redistribution of electronic states near the Fermi surface due to opening of superconducting gap. A phonon, if it couples to that part of the Fermi surface, gets renormalized reflected in its self-energy parameters i.e. phonon frequency and linewidth [50]. Qualitatively, phonons with frequency below the superconducting gap ($2\Delta$) can soften whereas the phonons with frequency above the gap can harden. The renormalization of a given phonon mode is strongest when the phonon energy is close to the gap. However, Fig. 3.14 clearly shows that the onset of renormalization of the Raman phonon is at a temperature $T_o$ much higher than $T_c$. We note that phonon anomalies have been reported for cuprate superconductors at temperature as high as ~ $2T_c$ attributed to the superconducting order-parameter fluctuations [82]. The fact that precursor effects seen in this work have not been observed in other iron-based superconductors is possibly related to the reduced dimensionality of the Cooper pair wavefunction due to large distance $d_{Fe}$ between the two Fe-As layers in the unit cell. The large anomalous increase of the frequency at $T_o$ and significant broadening of the observed mode suggests that the phonon frequency is very close to the superconducting gap (i.e. $2\Delta \sim 230$ cm$^{-1}$) in superconducting clusters existing above $T_c$. Near $T_o$, opening of superconducting gap provides additional decay channel due to the breaking of the Cooper pairs leading to an increase of the phonon linewidth. On further decreasing the temperature the phonon is inside the gap, which results in sharpening of the linewidth as decay channels are being removed [77-78]. The anomalous softening below $T_o$ and a large jump in the broadening of the mode at $T_o$ are evidence of a strong coupling between the Raman phonon and the superconducting quasi-particle excitations.

The importance of incoherent phase fluctuations may be understood by using experimentally determined quantities to evaluate the characteristic temperature $T_\theta$ which is linked to the stiffness of the phase of the superconducting order parameter. For $T_c \sim T_\theta$, the incoherent phase fluctuations are prominent even for $T > T_c$ but no global phase coherence till $T_c$ to give bulk superconductivity. The characteristic energy scale for phase fluctuations of the superconducting order parameter is the zero temperature phase stiffness given as $V = (\hbar c)^2 a / (16\pi e^2 \lambda_0^2)$, where '$a$' is the length scale, defined as the average spacing between superconducting layers, and $\lambda_0$ is the penetration depth. The characteristic temperature ($T_\theta$) is related to the phase stiffness as $T_\theta = AV (A \sim 1)$ [83]. Taking $\lambda_0 \sim 200$ nm [80,84], we estimate that $T_c \sim T_\theta$, similar to that in cuprate superconductors [83]. This suggests that incoherent phase fluctuations with the non-zero value of the superconducting gap are possible above $T_c$.

### 3.4.4 Conclusion

In conclusion, we have provided evidence for strong coupling of the observed phonon mode with the superconducting order parameter fluctuations setting in at almost ~ *2T$_c$*. Strong precursor effects due to existence of dynamic superconducting clusters at temperatures as high as *2T$_c$* are attributed to the reduced overlap of the electronic wave function of the Fe-As layers in this compound due to a large distance between the pnictides layers. It will be interesting to look for these strong precursor effects in careful transport measurements and Andreev reflection in scanning tunneling microscopy.

# 3.5 Part-E

# Evidence for Superconducting Gap and Spin-Phonon Coupling in Ca(Fe$_{0.95}$Co$_{0.05}$)$_2$As$_2$

## 3.5.1 Introduction

The "122" iron-based superconductors (FeBS) exhibit a rich phase diagram, described in detail in Chapter 1, as they undergo multiple phase transitions, e.g. tetragonal to orthorhombic structural transition simultaneously accompanied by magnetic transition and at lower temperature show superconductivity. These systems show clear evidence of coexistence of superconductivity and magnetism unlike the "1111" systems. The magnetic transition in FeBS is generally precede by the structural transition, however the separation between these two transitions depends on the quality of the sample, and it is noted that separation decreases with the improved quality of the sample [85]. Coexistence of these multiple phase transitions in this system motivated us to study the phonon dynamics and effect of magnetic as well as superconducting transition on the phonons. Here we present our detailed temperature dependent Raman studies on electron-doped single crystal of Ca(Fe$_{0.95}$Co$_{0.05}$)$_2$As$_2$ with $T_c$ ~ 23 K [86-87] in the temperature range of 4 K to 300 K, covering the tetragonal to orthorhombic structural transition as well as magnetic transition at $T_{sm}$ ~ 140 K, in the spectral range of 120 to 350 cm$^{-1}$. From our earlier study on FeSe$_{0.82}$ systems, described in Part A-3.1, we establish that magnetic transition in these systems is the primary order parameter and therefore it is likely to effect the phonon dynamics in an intense way. In this system, we found that magnetic ordering strongly affects the observed phonon modes attributed to the strong spin-phonon coupling. One of the observed phonon modes undergoes renormalization below the superconducting transition temperature suggesting coupling with the superconducting quasi-particle excitations.

## 3.5.2 Experimental Details

Single crystals of Ca(Fe$_{0.95}$Co$_{0.05}$)$_2$As$_2$ with nominal composition were prepared and characterized as described in ref. 86-87. Unpolarised micro-Raman measurements were performed in a similar way as described in earlier parts of Chapter 3.

## 3.5.3 Results and Discussion

### 3.5.3.1 Raman Scattering from Phonons

CaFe$_2$As$_2$ has a layered structure belonging to the tetragonal *I4/mmm* space group. There are four Raman active modes belonging to the irreducible representation A$_{1g}$ (As) + B$_{1g}$ (Fe) +2E$_g$ (As and Fe) [25]. Figure 3.15 shows the Raman spectrum at 4 K, revealing 3 modes labeled as S1 (205 cm$^{-1}$), S2 (215 cm$^{-1}$) and S3 (267 cm$^{-1}$). Spectra are fitted to a sum of Lorentzian functions. The individual modes are shown by thin lines and the resultant fit by thick line. Before we discuss assignment of modes S1 to S3 as phonon modes, we review the assignment of the Raman modes calculated and experimentally observed so far in "122" systems [24-28, 88]. In Sr$_{1-x}$K$_x$Fe$_2$As$_2$ (x = 0, 0.4), the four Raman active modes have been observed and assigned as 114 cm$^{-1}$ (E$_g$ : As and Fe), 182 cm$^{-1}$ (A$_{1g}$ : As), 204 cm$^{-1}$ (B$_{1g}$ : Fe) and 264 cm$^{-1}$ (E$_g$ : As and Fe) [24]. The two modes observed in CaFe$_2$As$_2$ [25] are 189 cm$^{-1}$ (A$_{1g}$ : As) and 211 cm$^{-1}$ (B$_{1g}$ : Fe). In Ba(Fe$_{1-x}$Co$_x$)$_2$As$_2$ system, three observed modes have been assigned as 124 cm$^{-1}$ (E$_g$ : As and Fe), 209 cm$^{-1}$ (B$_{1g}$ : Fe) and 264 cm$^{-1}$ (E$_g$ : As and Fe) [28]. In R$_{1-x}$K$_x$Fe$_2$As$_2$ (R = Ba and Sr), four Raman active modes have been identified as 117 cm$^{-1}$ (E$_g$ : As and Fe), 189 cm$^{-1}$ (A$_{1g}$ : As), 206 cm$^{-1}$ (B$_{1g}$ : Fe) and 268 cm$^{-1}$ (E$_g$ : As and Fe) [26]. However in another Raman study of R$_{1-x}$K$_x$Fe$_2$As$_2$ (R = Ba and Sr) only one mode was observed and assigned as 210 cm$^{-1}$ (B$_{1g}$ : Fe) [27]. In a theoretical calculation by Hou et al. [88] for SrFe$_2$As$_2$ the four calculated Raman modes in non-magnetic state are 138.9 cm$^{-1}$ (E$_g$ : As and Fe), 207.6 cm$^{-1}$ (A$_{1g}$ : As), 219.5 cm$^{-1}$ (B$_{1g}$ : Fe) and 301.2 cm$^{-1}$ (E$_g$ : As and Fe). The phonon frequencies for BaFe$_2$As$_2$ are close to the values for SrFe$_2$As$_2$. Keeping these reports in view and our density functional calculations (see Table-3.4, 3.5 and Fig. 3.16), we assign the modes S1 to S3 as S1: 205 cm$^{-1}$ (A$_{1g}$ : As), S2 : 215 cm$^{-1}$ (B$_{1g}$ : Fe) and S3: 267 cm$^{-1}$ (E$_g$ : Fe and As). We note from Fig. 3.15 that the linewidth of the mode S3 is larger than that of S1 and S2 modes. It is likely that the large linewidth of S3 (with eigen vectors in *a-b* plane) may arise from the disorder caused by the slight rotation of *a-b* plane in these layered crystals.

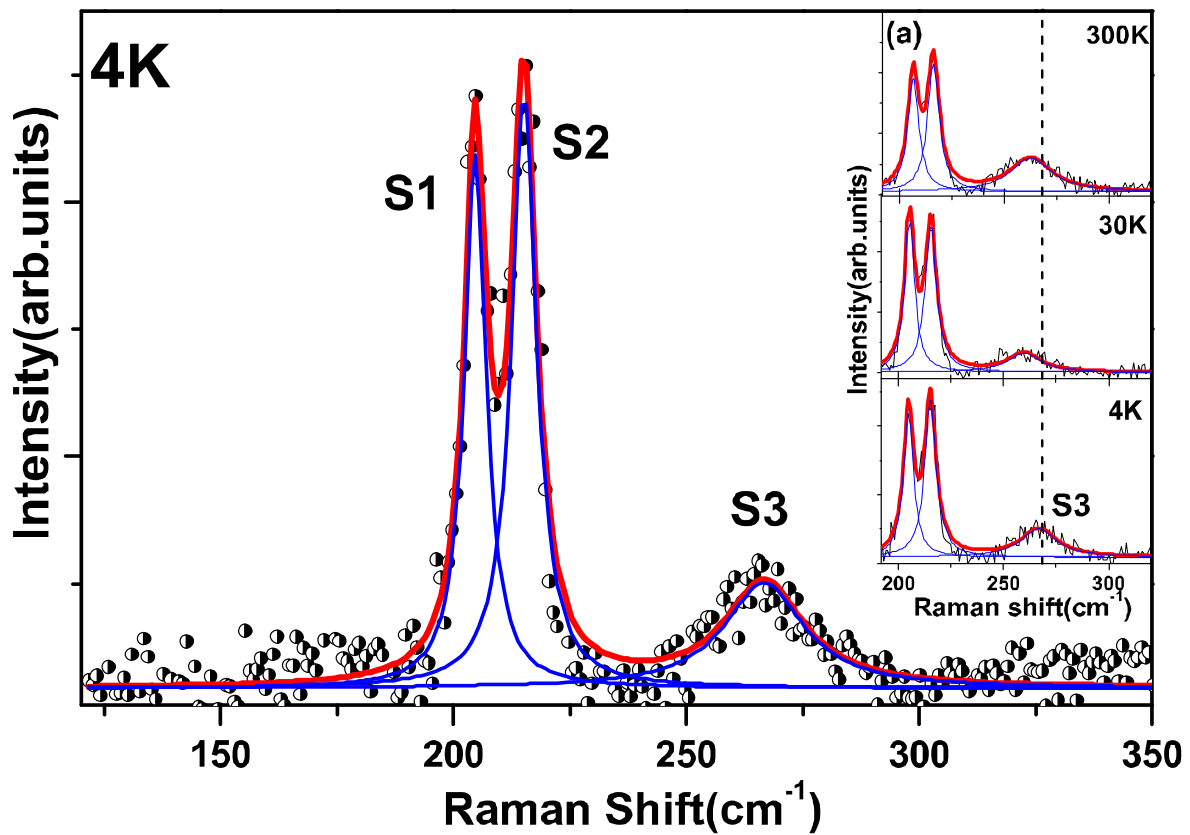

Figure 3.15: Raman spectra of Ca(Fe$_{0.95}$Co$_{0.05}$)$_2$As$_2$ at 4 K. Solid (thin) lines are fit of individual modes and solid (thick) line shows the total fit to the experimental data (circle). Inset (a) shows the mode S3 at few temperatures.

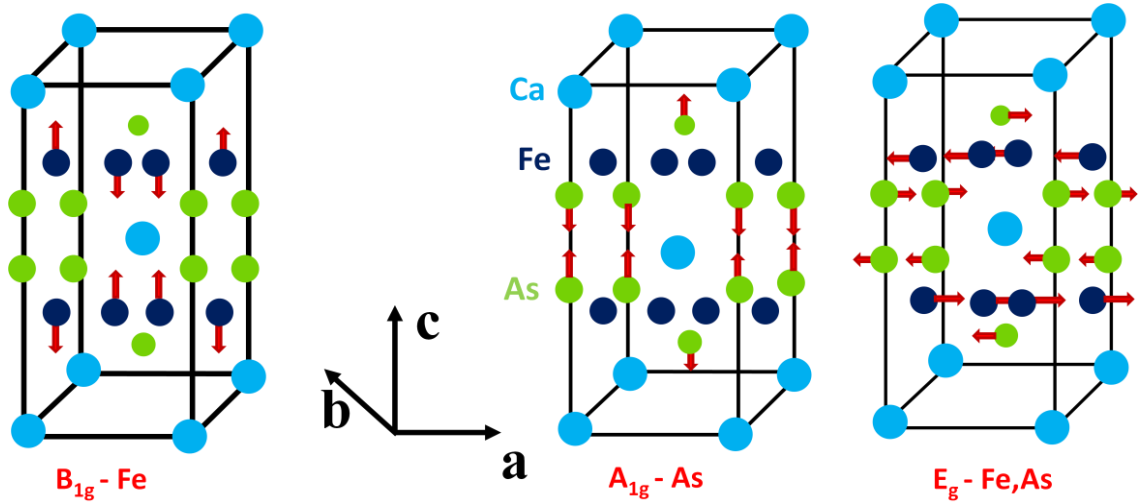

Figure 3.16: Eigen modes corresponding to Raman modes in Table -3.5.

### 3.5.3.2 Temperature Dependence of the Phonon Frequencies

Figure 3.17 shows the peak frequency and linewidths of the three phonon modes as a function of temperature. The solid lines are linear fits to the data in a given temperature window. The following observations can be made: (i) The temperature dependence of the frequency of mode S1 shows a discontinuous change at $T_{sm}$. The frequency of the mode has anomalous temperature dependence below $T_{sm}$ (i.e. the frequency decreases with lowering of the temperature). (ii) The temperature dependence of the mode S2 is anomalous in the entire temperature range of 4 K to 300 K. (iii) The frequency of the mode S3 also shows anomalous temperature dependence between $T_c$ and 300 K. Below $T_c$, the mode hardens on decreasing the temperature. (iv) The linewidths of mode S1 and S2 remain nearly constant with temperature, but on the other hand, the linewidth of mode S3 shows non-monotonic dependence on temperature below $T_{sm}$.

In superconductors the opening of the superconducting gap ($2\Delta$) below $T_c$ redistributes the electronic states in the neighborhood of the Fermi surface and hence can change the phonon self-energy as seen in cuprates and other high temperature superconductors [45,61,89-92].

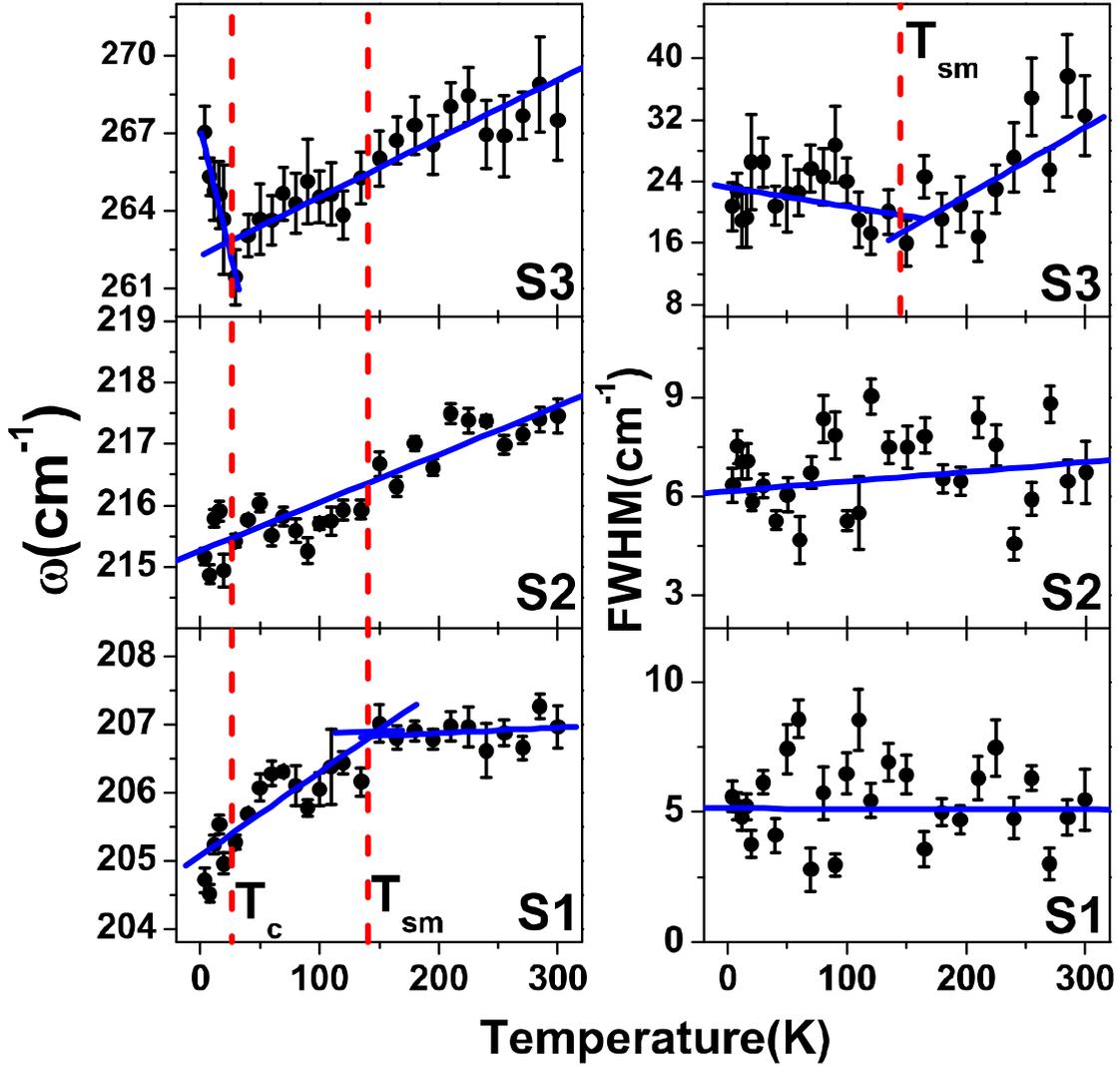

Figure 3.17: Temperature dependence of the modes S1, S2 and S3. Solid lines are linear fit in a given temperature window.

According to Zeyer et al. [50], a change in phonon self-energy below $T_c$ is linked with the interaction of the phonons with the superconducting quasi-particles. Qualitatively, based on mode repulsion coupled excitation model, phonons above $2\Delta$ show hardening below $T_c$ whereas the phonons with frequency below the gap value can soften. The anomalous hardening of mode S3 below $T_c$ (shown in Fig. 3.17) indicates a coupling of this phonon of $E_g$ symmetry involving the vibrations of Fe and As atoms to the electronic system. A microscopic understanding of this coupling may help in understanding the symmetry of the superconducting gap. Taking the phonon frequency of S3 mode as an upper limit of 2Δ, an estimate of $2\Delta/k_BT_c$ is ~ 15, pointing to strong-coupling nature of superconductor. We note that in "122" systems experimental evidences of single and multiple gaps have already been reported from infrared spectroscopy [56], angle resolved spectroscopy [55, 93], nuclear magnetic resonance [94-95] and muon spin rotation [96]. The reported gap values show a large variation from $2\Delta/k_BT_c$ ~ 1.6 to 10 in both electron-doped [55,56,96] and hole-doped systems [93,94-95]. The origin of this large spread is still not understood.

Table-3.4: List of the experimentally observed frequencies at 4 K in Ca(Fe$_{0.95}$Co$_{0.05}$)$_2$As$_2$ and calculated frequencies for Ca(Fe$_{1-x}$Co$_x$)$_2$As$_2$ using LDA+U( = 4eV for Fe and Co), obtained with Quantum Espresso based calculations. Δω represents the difference between frequency in NM and AFM2 phase.

| Mode Assignment | Experimental ω (cm$^{-1}$) | Calculated ω (cm$^{-1}$) | | | | | |
|---|---|---|---|---|---|---|---|
| | | x = 0.0 | | | x = 0.25 | | |
| | | NM | AFM2 | Δω/ω$_{nm}$ (%) | NM | AFM2 | Δω/ω$_{nm}$ (%) |
| S1  A$_{1g}$ (As) | 205 | 211 | 171 | 18.9 | 215 | 160 | 25.5 |
| S2  B$_{1g}$ (Fe) | 215 | 227 | 208 | 8 | 239 | 216 | 9.6 |
| S3  E$_g$ (As and Fe) | 267 | 320 | 279 | 12.8 | 306 | 270 | 11.7 |

Table-3.5: Zone center phonons (frequencies given in cm$^{-1}$) for NM, FM, AFM1 and AFM2 ordering, giving an estimate of the spin-phonon coupling, obtained with VASP-based calculations.

| Mode | NM | FM | AFM1 | AFM2 |
|---|---|---|---|---|
| A$_{1g}$ (S1) | 204 | 180 | 150 | 197 |
| B$_{1g}$ (S2) | 207 | 185 | 203 | 201 |
| E$_g$ (S3) | 301 | 223 | 237 | 248 |
| E$_g$ | 156 | 109 | 115 | 89 |

Table-3.6: Energies of internally relaxed structures FM, AFM1 and AFM2 (stripe) magnetic ordering and corresponding stresses on unit cell.

| Magnetic order | Energy (2*eV/fmu) | $\sigma_{xx}$ (kB) | $\sigma_{zz}$ (kB) | $\sigma_{xy}$ (kB) |
|---|---|---|---|---|
| NM | -49.501 | -76 | -131 | 0 |
| FM | -54.051 | -35 | -21 | 0 |
| AFM1 | -53.722 | 9 | -17 | 0 |
| AFM2 (Stripe) | -54.487 | -12 | 9 | 20 |

The anomalous softening of the modes with decreasing temperature for all the three modes is attributed to strong spin-phonon coupling, in line with other studies for similar iron-based superconductors. A recent study on the isotope effect in iron-pnictide "122" superconductors [97] suggests that electron-phonon interaction does play a role in the superconducting pairing mechanism via strong spin-phonon coupling. The coupling between phonons and spin degrees of freedom can arise either due to modulation of exchange integral by phonon amplitude [35,81] and/or by involving change in the Fermi surface by spin-waves provided phonon couples to that part of the Fermi surface [91]. Earlier Raman [26,28,39] and neutron scattering studies [98-99] in iron-based superconductors have also indicated strong spin-phonon coupling and have been supported by earlier theoretical calculations [15,34,88]. In order to further elucidate the importance of strong spin-phonon coupling in the present case of Ca(Fe$_{0.95}$Co$_{0.05}$)$_2$As$_2$ also, we performed detailed DFT calculations.

### 3.5.3.3 Theoretical Calculations

Our first-principles calculations are based on density functional theory as implemented in the PWSCF [31] package. We use optimized norm-conserving pseudopotential [100] for Ca, As and ultrasoft pseudopotentials [32] for Co and Fe to describe the interaction between ionic cores and valence electrons, and a local density approximation (LDA) of the exchange energy functional. We use plane wave basis with a kinetic energy cutoff of 40 Ry in representation of wavefunctions and a cutoff of 240 Ry in representation of the charge density. We sampled integration over the Brillouin zone (of single unit cell) with 12x12x8 Monkhorst Pack Mesh [43]. Structural optimizations of $CaFe_2As_2$ and $Ca(Fe_{1-x}Co_x)_2As_2$ are done with $\sqrt{2}\times\sqrt{2}\times 1$ supercells by minimizing the total energy using Hellman-Feynman forces and the Broyden-Flecher-Goldfarb-Shanno based method. Zone center (q = 0,0,0) phonon spectra are determined using a frozen phonon method (with atomic displacements of ± 0.04 Å) for the relaxed structure obtained at experimental values of the lattice constants. Self-consistent solution with different magnetic ordering, particularly ferromagnetic (FM) and G-antiferromagnetic types (AFM1), was rather demanding and was achieved with a mixing of charge density based on local density dependent Thomas-Fermi screening and a mixing factor of 0.1 (density from the new iteration with weight of 0.1). To facilitate comparison between theory and experiment within the accuracy of calculational framework and estimate the errors associated with use of pseudopotentials, we repeated all calculations for *x=0* with projector augmented wave potentials as implemented in a plane-wave package VASP [101-102].

To find out which magnetic ordering is relevant in the ground state at low temperatures, we carried out self-consistent total energy calculations for non-magnetic (NM) as well as different magnetic orderings i.e. FM, AFM1 and AFM2. Within local spin density approximation (LSDA), our calculations initialized with FM, AFM1 and AFM2 type antiferromagnetic orderings relax to a nonmagnetic structure at the self-consistency. However, local stability of these magnetically ordered states could be achieved through use of an onsite correlation (Hubbard) parameter U (= 4eV for Fe atoms) in the LSDA+U formalism. Calculated total energies of different magnetic orderings within LSDA+U description show that the stripe antiferromagnetic ordering has lowest energy (see Table-3.6) and it should be prominent at low temperatures. We note a weak shear stress $\sigma_{xy}$ appearing in the stripe phase (see Table-3.6), which should give rise to a small orthorhombic strain as the secondary order parameter below $T_{sm}$.

To estimate the strength of spin-phonon coupling, if any, we determined phonon frequencies at wave-vector q = (0,0,0) using frozen phonon method for both the NM and stripe antiferromagnetic ordering (AFM2). Effects of spin-ordering on phonons are reflected in the frequencies of these two phases (see Table-3.4). We find that the frequencies change by a large amount from NM phase to AFM2 phase (changes are of the order of 20-40 cm$^{-1}$), indicating the presence of a strong spin-phonon coupling in pure as well as doped CaFe$_2$As$_2$ systems. Calculated Raman active phonon modes are in reasonable agreement with the observed values (see Table-3.4). We find that frequencies of the three Raman active modes estimated for a state with AFM2 ordering are smaller than those estimated for the NM ordering at both the compositions x = 0 and x = 0.25 studied here (see Table-3.4). To estimate the spin-phonon couplings relevant to observed Raman spectra, we determined zone centre phonons for relaxed structures (kept at the experimental lattice constant) with AFM1, AFM2, FM and NM ordering (see Table-3.5). It is evident from the frequencies that the spin-phonon coupling for all the three modes is strong. To understand the temperature dependence of phonon frequencies arising from the spin-phonon coupling, ($\lambda^{(1)}u^2*S_iS_j$, u being phonon coordinate and $S_i$ the spin on $i^{th}$ Fe) we construct a simple Ising spin-Hamiltonian [34] H = $J_1 \sum_{NN} S_i.S_j + J_2 \sum_{2^{nd}NN} S_i.S_j$, where J$_1$ and J$_2$ are the nearest and next nearest neighbor exchange interaction parameters. We note that though Heisenberg-type model should be used to capture the spin dynamics of complex systems such as iron-based superconductors, it has been shown theoretically [21] that ground states of these systems have collinear stripe type anti-ferromagnetic spin ordering and for such systems Heisenberg and Ising models should give similar results. We use Monte Carlo (MC) simulations to obtain *T*-dependent spin-ordering.

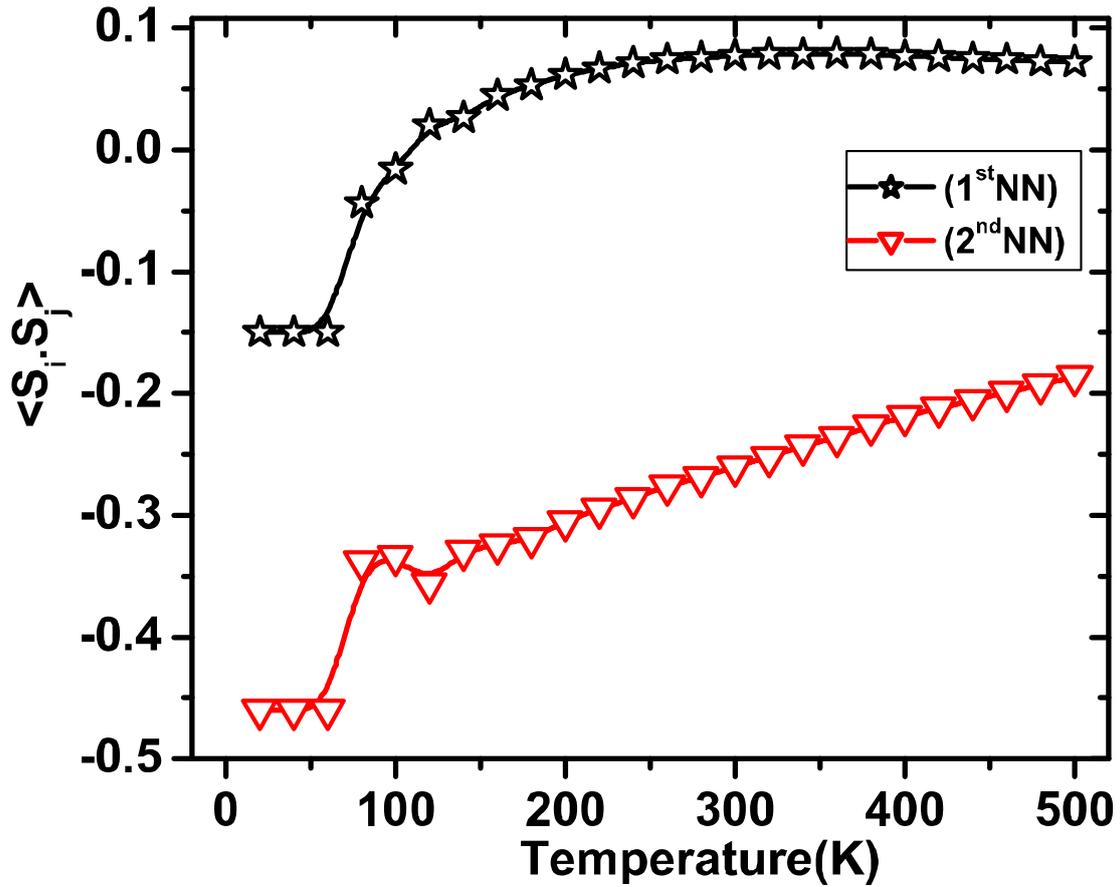

Figure 3.18: Temperature dependence of the spin correlation function ($<S_i.S_j>$) for CaFe$_2$As$_2$ obtained from Monte Carlo simulations. Solid lines are guide to the eye.

$J_1$ and $J_2$ are estimated from the energies of states with different magnetic ordering (see Table-3.6) of CaFe$_2$As$_2$: $J_1$ = -9 meV and $J_2$ = 19 meV. The temperature dependence of the nearest and next nearest neighbor spin-spin correlations, obtained from MC simulations (see Fig. 3.18), bears a monotonous decrease with reducing temperature below room temperature. Given that the modes S1, S2 and S3 couple similarly to spin (see Table-3.5) with the highest value of frequency in the NM state and the lowest one in stripe phase (AFM2), all the three Raman active modes should soften as the temperature is lowered; the *T*-dependence of S2 mode is expected to be weaker because its coupling with spin is much smaller. This is consistent with our data in Fig. 3.17 where changes in the frequency below $T_{sm}$ are minimum for the S2 mode. These theoretical predictions are consistent with the observed softening of Raman modes at low temperature in our experiment. The precise magnetic ordering in the low temperature phase is mixed as reflected in low temperature values of the spin-spin correlation (see Fig. 3.18), which should have been 0 and -1 for the first and second neighbor spin-correlations in the stripe phase. We suggest that the low temperature phase of $CaFe_2As_2$ is partially stripe anti-ferromagnetic and it is due to frustration coming from opposite signs of first and second neighbor exchange interactions between Fe. With decrease in temperature, we believe that the system undergoes a magnetic transition from paramagnetic to short-ranged AFM2 at ~ 140 K [87] and softening of Raman active modes at the low temperatures is due to the spin-phonon coupling which remains strong as a function of *x*. This is understandable because such a coupling arises from the changes in the Fe-As-Fe bond angles (subsequently the super-exchange interactions) associated with atomic displacements (see Fig. 3.16) in S1 and S2 modes.

### 3.5.4 Conclusion

In conclusion, we have shown that all the three observed modes show anomalous temperature-dependence due to strong spin-phonon coupling in $Ca(Fe_{0.95}Co_{0.05})_2As_2$. Density functional calculations of phonons in different magnetic phases show strong spin-phonon coupling in parent and doped "122" system which is responsible for the observed softening of the phonon frequency with decreasing temperature. The anomalous hardening of one of the Raman active modes below the superconducting transition temperature is attributed to the coupling of the mode with the superconducting quasi-particle excitations, yielding an upper limit of $2\Delta/k_BT_c$ ~ 15. Results obtained here suggest that the interplay between phonons and spin degrees of freedom are crucial to unravel the underlying physics responsible for the pairing mechanism in iron-based superconductors.

# 3.6 Part-F

# Orbital-ordering, Phonon Anomalies and Electronic Raman Scattering in Ca(Fe$_{0.97}$Co$_{0.03}$)$_2$As$_2$

## 3.6.1 Introduction

In Part E-3.5 we studied Ca(Fe$_{0.95}$Co$_{0.05}$)$_2$As$_2$ superconductor focusing only on the first-order phonon modes. Here we will present our detailed Raman studies on Ca(Fe$_{0.97}$Co$_{0.03}$)$_2$As$_2$ systems in a wide spectral range of 120-5200 cm$^{-1}$ from 5 K to 300 K, covering the tetragonal to orthorhombic structural transition as well as magnetic transition at $T_{sm}$ ~ 160 K. We note that Ca(Fe$_{0.97}$Co$_{0.03}$)$_2$As$_2$ does not shows superconductivity, however there is a drop in resistivity below ~ 20 K [86-87]. In this system we focused on high energy excitations along with the first-order phonon modes and found a signature of intricate coupling between electronic, magnetic and orbital degrees of freedom. Recent experimental as well theoretical studies on iron-based superconductors (FeBS) have also argued that coupled orbital and spin ordering can play an important role for the superconducting pairing mechanism [64-65,103-107]. There is a growing body of evidence that suggest that intricate interplay between spin (magnon) and orbitals (orbiton) degrees of freedom leads to novel electronic phases in FeBS.

Strongly electron correlated systems such as manganites, ruthenates and high temperature superconductors have been shown to support orbital excitations (orbitons), which are invoked to explain their interesting behavior [33,108-111]. The partially filled $d$-electron subshell ($d_{yz/xz}$) of Fe$^{2+}$ in FeBS can have orbital excitations [64,103-107] due to breaking of rotational symmetry of the $d_{xz}$ and $d_{yz}$ orbitals. Since the Fermi surface (FS) in FeBS is mainly composed of different components of the t$_{2g}$ orbitals of Fe$^{2+}$, namely $d_{xy}$, $d_{xz}$ and $d_{yz}$, these orbitals are very important in the pairing mechanism. The $d_{xz}$ and $d_{yz}$ bands have strong in-plane anisotropy and play an important role in the electronic ordering observed in Ca(Fe$_{1-x}$Co$_x$)$_2$As$_2$ [111], similar to ruthenates [112] and this nematic ordering has been interpreted as orbital ordering between $d_{xz}$ and $d_{yz}$ orbitals [110-111]. Therefore, the orbital excitations may be crucial for the coupling between the electrons which allow them to flow without resistance, suggesting an unconventional origin of superconductivity mediated by orbital and spin fluctuations [64-65,103-107,113].

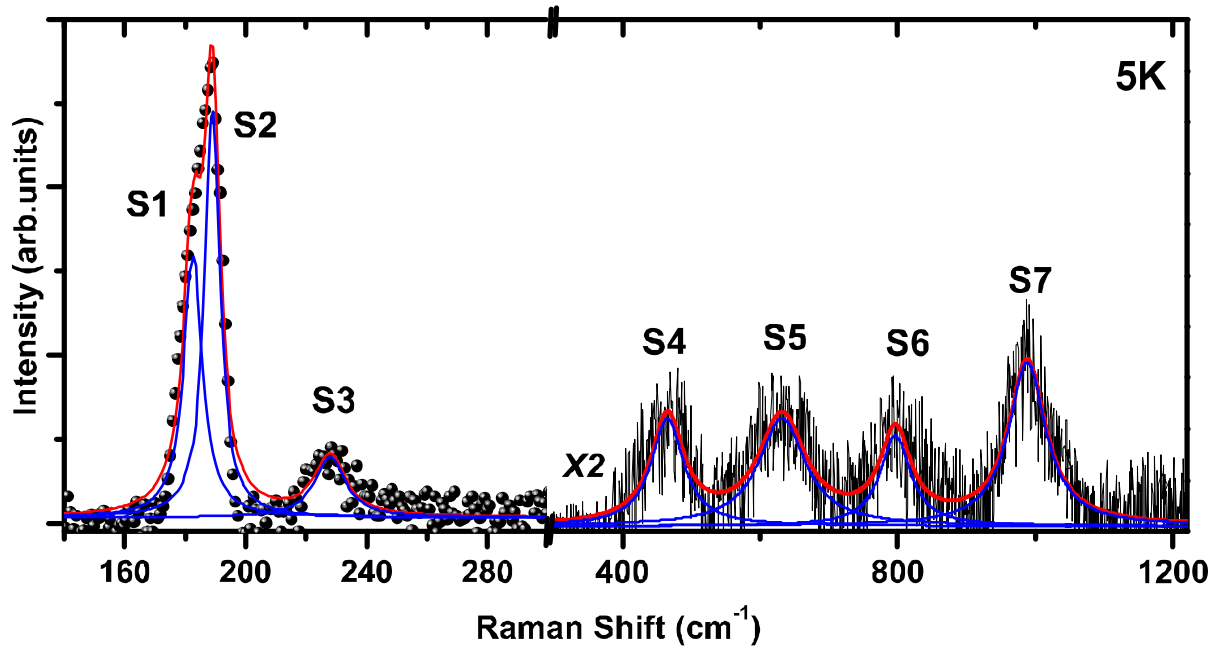

Figure 3.19: Raman spectra of Ca(Fe$_{0.97}$Co$_{0.03}$)$_2$As$_2$ at 5 K. Solid (thin) lines are fit of individual modes and solid (thick) line shows the total fit to the experimental data.

### 3.6.2 Results and Discussion

### 3.6.2.1 Raman Scattering from Phonons

Single crystals of Ca(Fe$_{0.97}$Co$_{0.03}$)$_2$As$_2$ were prepared and characterized as described in ref. 86-87. Figure 3.19 shows Raman spectrum at 5 K, revealing 7 modes labeled as S1 to S7 in the spectral range of 140-1200 cm$^{-1}$. Spectra are fitted to a sum of Lorentzian functions. The individual modes are shown by thin lines and the resultant fit by thick line. Following the previous Raman studies on "122" system, we assign the mode S1 to S3 as S1 : 181 cm$^{-1}$ ($A_{1g}$ : As) ; S2 : 189 cm$^{-1}$ ($B_{1g}$ : Fe) and S3 : ($E_g$ : Fe and As). Figure 3.20 shows the temperature evolution of the first-order modes S1, S2 and S3. Figure 3.21 shows the temperature dependence of the mode frequencies (panel-a) and their full width at half maxima (FWHM) (panel-b). The solid lines are drawn as guide to the eyes. The following observations can be made: (i) the frequencies of the modes S2 and S3 show abrupt hardening below $T_{sm}$, whereas mode S1 shows normal temperature dependence from 5 K to 300 K as expected due to anharmonic interactions. The dotted line for the mode S1 is a fit to an equation (2.11) described in Chapter 2. Fitting parameters for the mode S1 are $\omega_0$ = 181.5 ± 0.2, C = -0.66 ± 0.1 cm$^{-1}$. (ii) The FWHM decreases as temperature is decreased from 300 K to $T_{sm}$ due to reduced anharmonicity. Interestingly, below $T_{sm}$, the linewidths of all the three phonon modes show anomalous broadening.

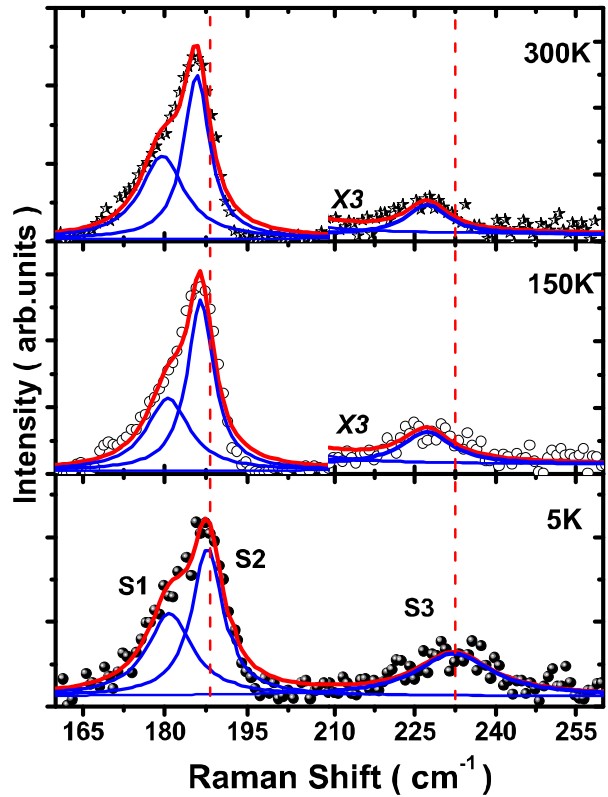

Figure 3.20: Temperature evolution of the first-order phonon modes S1 ($A_{1g}$-As), S2 ($B_{1g}$-Fe) and S3 ($E_g$-Fe and As).

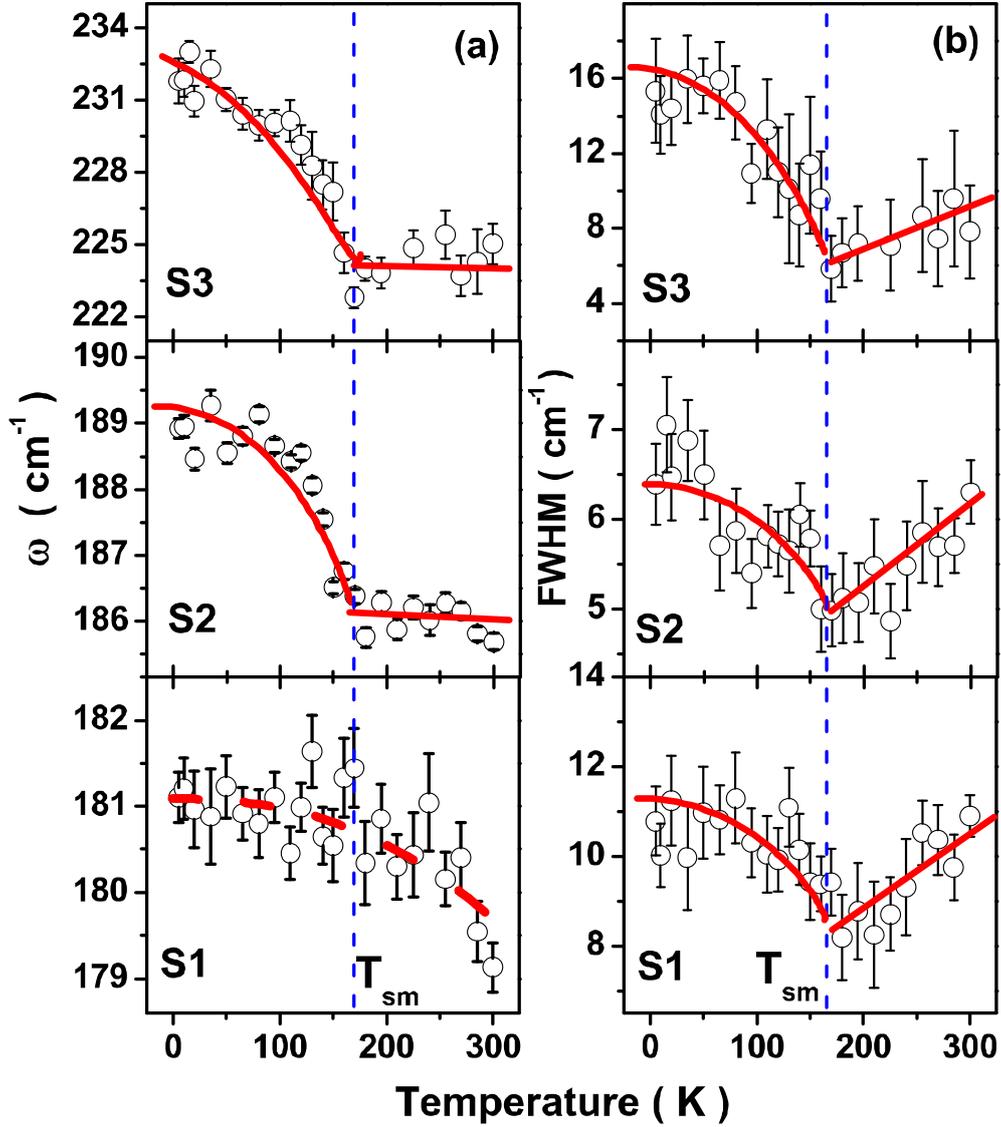

Figure 3.21: Temperature dependence of the modes S1 ($A_{1g}$-As), S2 ($B_{1g}$-Fe) and S3 ($E_g$-Fe and As) frequencies (panel-a) and linewidths (panel-b). Solid lines are guide to the eyes. Dotted line for mode S1 is fitted curve as described in the text.

The anomalous hardening of modes S2 and S3 below $T_{sm}$, both involving the displacement of magnetic ion $Fe^{2+}$, is attributed to strong spin-phonon coupling [25-28,34,39,76,88]. The coupling between the phonons and the spin degrees of freedom can arise either due to modulation of the exchange integral by the phonon amplitude [35,81] and/or by involving change in the Fermi surface by spin waves, provided the phonon couples to that part of the Fermi surface [91]. Microscopically, renormalization of the phonon frequency due to spin ordering below the magnetic transition temperature can be correlated [114] with sublattice-magnetization, M(T) as $\Delta\omega \approx [\frac{1}{8\mu_B m\omega}\frac{\partial^2 J}{\partial u^2}]M^2(T)$. Here $\frac{\partial^2 J}{\partial u^2}$ is the second derivative of the spin exchange integral '$J$' with respect to the phonon amplitude '$u$'. To our knowledge, there is no report of temperature-dependent sub-lattice magnetization measurement on this system and therefore we could not compare the renormalization of the phonon frequencies with the sub-lattice magnetization.

Since phonon linewidth decreases with decreasing temperature due to reduced anharmonicity, the increase in the phonon linewidths below $T_{sm}$ can not be explained without taking into account the intricate coupling between the phonons and spin degrees of freedom which become prominent below the magnetic transition temperature $T_{sm}$. The sharp increase in the phonon frequencies below $T_{sm}$ clearly suggest the strong coupling between these two degrees of freedom; the spin-phonon coupling will also affect the phonon decay rate below the magnetic transition temperature where the optical phonon may decay into another phonon and magnon/or two magnons, similar to the case of manganites [115].

### 3.6.2.2 Orbital-Ordering and Electronic Raman Scattering

As noted earlier, in FeBS iron 3$d$-orbitals predominantly contribute to the electronic states near the Fermi surface and hence they are expected to play a role in the superconducting pairing mechanism in these systems. In particular, whether the $d_{xz/yz}$ orbitals degeneracy is lifted or not is hotly debated experimentally as well as theoretically [17-20,56,66,116-118]. We note that a recent report [116] on detwinned Ba(Fe$_{1-x}$Co$_x$)$_2$As$_2$ shows the splitting of Fe $d_{xz/yz}$ orbital much above the $T_{sm}$. Other reports on BaFe$_2$(As$_{1-x}$P$_x$)$_2$ [56,66,117] and NaFeSe [118] show the existence of electronic nematicity well above the structural and magnetic transition temperature using magnetic torque measurements, x-ray absorption spectroscopy and angle resolved photoemission spectroscopy.

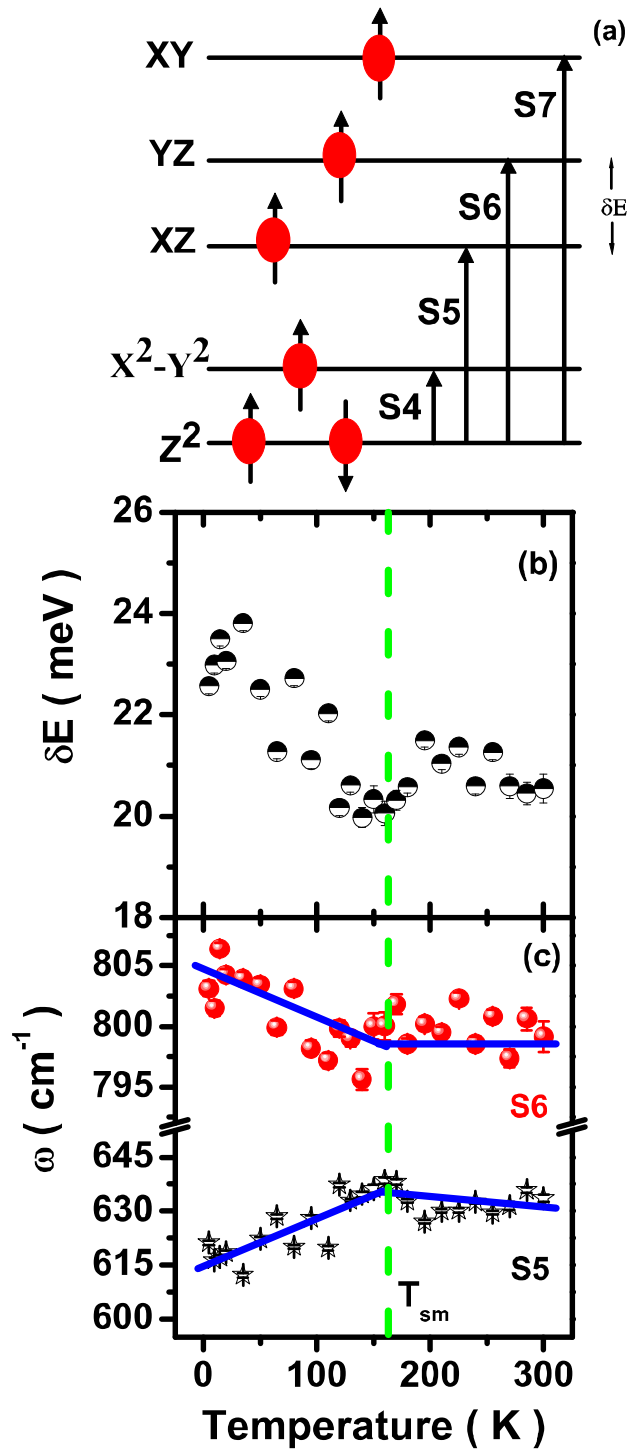

Figure 3.22: (a) Schematic diagram for the crystal field split energy level of Fe 3d-orbitals showing the electronic transitions. (b), (c) Temperature dependence of the energy difference (δE) between mode S6 and S5 and their frequency, respectively. Solid lines are guide to the eye.

Coming to our present study, we observe four weak Raman modes S4 to S7 in addition to the expected first-order Raman modes. The frequencies of the mode S4 is ~ 460 cm$^{-1}$, S5 ~ 620 cm$^{-1}$, S6 ~ 800 cm$^{-1}$ and S7 ~ 1000 cm$^{-1}$. In earlier Raman studies [30,119] on FeBS, similar weak Raman modes were observed which were attributed to the electronic Raman scattering involving 3$d$-orbitals of Fe. We follow the same assignment of the four modes S4 to S7 in terms of the electronic Raman scattering involving Fe 3$d$-orbitals.

Figure 3.22 (a) shows a schematic diagram of Fe$^{2+}$ 3$d$-levels. The calculated level splitting are close to the frequency position of the high frequency modes[17-20]. Accordingly, we assign modes S4, S5, S6 and S7 as the transition from the orbital state $z^2$ to orbital states ($x^2-y^2$), $xz$, $yz$ and $xy$, respectively. Our observation of modes S5 and S6 suggests that the $xz$, $yz$ orbitals are non-degenerate with a split of ~ 25 meV, which is close to the theoretical [19-20] and experimental value [116-118] determined using angle resolved photoemission spectroscopy. In Fig. 3.22 (c) and (b) we have plotted the frequencies of modes S5 and S6 and their energy difference ($\delta E$), respectively. It is clearly seen that their splitting ($\delta E$) increases with decreasing temperature, similar to that reported for "122" and "111" systems using angle resolved photoemission spectroscopy [116-118].

Now, we will discuss the origin of high frequency broad Raman band with center near ~ 3200 cm$^{-1}$ as shown in Fig. 3.23 (c). Inset of Fig. 3.23 (c) shows the broad band at room temperature at two different wavelengths i.e. 488 nm and 633 nm showing that the band is not related to photoluminescence, but is a Raman excitation. We note that similar broad Raman bands in earlier Raman studies on Fe$_{1+y}$Se$_x$Te$_{1-x}$ and BaFe$_{2-x}$Co$_x$As$_2$ systems have been attributed to the pure two-magnon excitations [120-121]. We suggest that this assignment needs a relook. The broad band in ref. [120-121] was observed even at a temperature much higher than the magnetic ordering temperature (as high as 5$T_{sm}$). Further, in FeBS, even though long range magnetic ordering is not expected in the superconducting phase, high frequency Raman band was observed. The frequency of the high frequency band (~ 2200 cm$^{-1}$) is similar in "11" and "122" systems even though they have different magnetic structures ("122" has a single stripe AFM structure and "11" has double stripe AFM structure). Based on experiments, Zhao et al. [122] have deduced the exchange parameters for Ca122 system as $SJ_{1a}$ = 49.9 meV, $SJ_{1b}$ = -5.7 meV and $SJ_2$ = 18.9 meV. The two-magnon energy for "122" systems can be estimated as $4S(J_{1a} - J_{1b} + 2J_2) - J_{1a}$ = ~ 324 meV (~ 2600 cm$^{-1}$), which differs significantly from our observed value for the broad Raman band (~ 3200 cm$^{-1}$) by ~ 600 cm$^{-1}$. We have fitted the broad Raman band with a Lorentzian function to extract the peak position and integrated intensity. Figure 3.23(a) shows the peak position of the broad band as a function of temperature which clearly displays significant temperature dependence i.e. ~ 100 cm$^{-1}$ shift in the temperature range of 4 K to 300 K. We note that a similar blue-shift has also been reported for the Raman band associated with the orbital excitation in LaMnO$_3$ [123]. Figure 3.23(b) shows the normalized integrated intensity of the broad band showing that $I_{300K}/I_{4K}$ ~ 0.8. It is known that pure magnetic excitations have negligible intensity above the magnetic transition temperature, as seen in systems such as AlFeO$_3$ [124]. Keeping in mind that the frequency of the mode increases by a significant amount as temperature is lowered (like the orbital excitations in case of LaMnO$_3$) and non-zero large intensity above $T_{sm}$ ($I_{300K}/I_{4K}$ ~ 0.8), the broad Raman band may not be assigned to a pure two-magnon excitation. This suggestion comes from the fact that the electronic nematic order breaks the $C_4$ rotational symmetry [125] which couples spin and orbital degrees of freedom, as observed in recent experimental [65] as well as theoretical studies [64,103-107]. More theoretical work is required to understand quantitatively our experimental data.

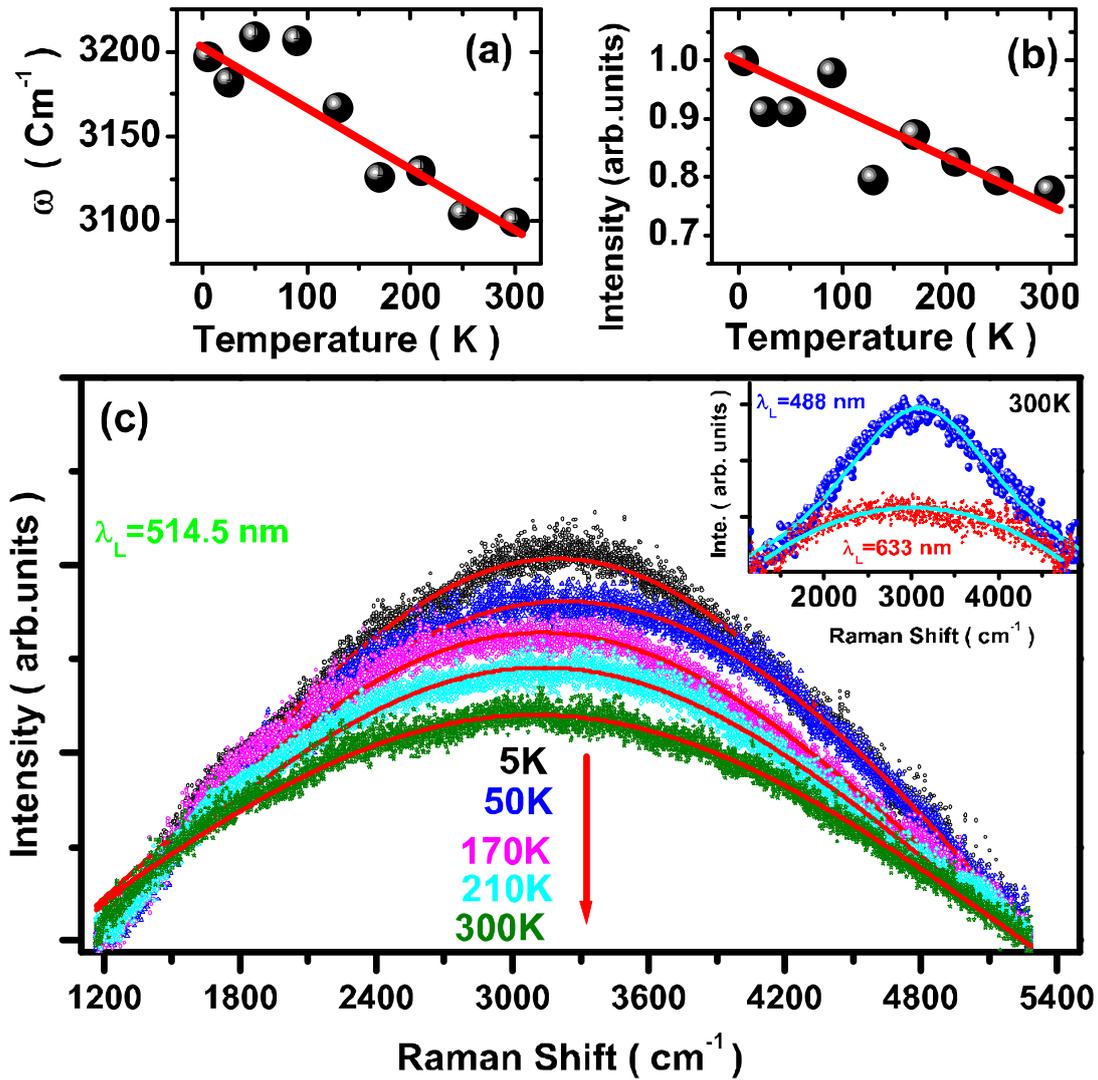

Figure 3.23: (a, b) Temperature dependence of the peak frequency and integrated intensity, respectively. Solid lines are straight line fits. (c) Temperature-dependent Raman spectra in a wide spectral range 1200-5200 cm$^{-1}$. Inset shows the broad Raman band at room temperature at two different wavelength i.e. 488 nm and 633 nm. Solid lines are Lorentzian fits.

### 3.6.3 Conclusion

In conclusion, our temperature dependent Raman study on iron-pnictide Ca(Fe$_{0.97}$Co$_{0.03}$)$_2$As$_2$ provides clear evidence of the strong spin-phonon coupling as reflected in the anomalous temperature of the first-order phonon modes. The four weak modes observed in the range 400-1200 cm$^{-1}$ are attributed to the electronic Raman scattering involving 3$d$-orbitals of Fe$^{2+}$ and suggest the non-degenerate nature of the $d_{xz/yz}$ orbitals. In addition, the high frequency Raman band is ascribed to the coupled excitations associated with spin and orbital degrees of freedom. Our results suggest that the intricate interplay between spin, charge, lattice and orbital degrees of freedom may be crucial for the pairing mechanism in iron-pnictides. We hope that our experiments will motivate further theoretical work in this direction.

# Bibliography:

[1] Y. Kamihara, T. Watanabe, M. Hirano and H. Hosono, J. Am. Chem. Soc. **130,** 3296 (2008).

[2] X. H. Chen, T. Wu, G. Wu, R. H. Liu, H. Chen and D. F. Fang, Nature **453**, 761 (2008).

[3] G. F. Chen, Z. Li, D. Wu, G. Li, W. Z. Hu, J. Dong, P. Zheng, J. L. Luo and N. L. Wang, Phys. Rev. Lett **100**, 247002 (2008).

[4] M. Rotter, M. Tegel and D. Johrendt, Phys. Rev. Lett. **101**,107006 (2008).

[5] N. Ni, S. L. Budko, A. Kreyssig, S. Nandi, G. E. Rustan, A. I. Goldman, S. Gupta, J. D. Corbett, A. Kracher and P. C. Canfield, Phys. Rev. B **78** , 014507 (2008).

[6] F. C. Hsu, J.-Y. Luo, K.-W. Yeh, T.-K. Chen, T.-W. Huang, P. M. Wu, Y.-C. Lee, Y.-L. Huang, Y.-Y. Chu, D. C. Yan and M. K. Wu, PNAS **105**, 14262 (2008).

[7] M. H. Fang, H. M. Pham, B. Qian, T. J. Liu, E. K. Vehstedt, Y. Liu, L. Spinu and Z. Q. Mao, Phys. Rev. B **78**, 224503 (2008).

[8] Y. Mizuguchi, F. Tomioka, S. Tsuda, T. Yamaguchi and Y. Takano, Appl. Phys. Lett. **93**, 152505 (2008).

[9] S. Margadonna, Y. Takabayashi, M. T. McDonald, K. Kasperkiewicz, Y. Mizuguchi, Y. Takano, A. N. Fitch, E. Suard and K. Prassides, Chem. Commun. **43,** 5607-5609 (2008).

[10] M. J. Wang, J. Y. Luo, T. W. Huang, H. H. Chang, T. K. Chen, F. C. Hsu, C. T. Wu, P. M. Wu, A. M. Chang and M. K. Wu, Phys. Rev. Lett. **103**, 117002 (2009).

[11] F. Gronvold, Acta Chemica Scandinavica **22**, 1219 (1968).

[12] B. K. Jain, A. K. Singh and K. Chandra, J. Phys. F-Metal Phys. **8**, 2625 (1978).

[13] T. M. McQueen, Q. Huang, V. Ksenofontov, C. Felser, Q. Xu, H. Zandbergen, Y. S. Hor, J. Allred, A. J. Williams, D. Qu, J. Checkelsky, N. P. Ong and R. J. Cava, Phys. Rev. B **79**, 014522 (2009).

[14] T. L. Xia, D. Hou, S. C. Zhao, A. M. Zhang, G. F. Chen, J. L. Luo, N. L. Wang, J. H. Wei1, Z.-Y. Lu and Q. M. Zhang, Phys. Rev. B **79**, 140510 (2009).

[15] T. Egami, B. V. Fine, D. J. Singh, D. Parshall, C. de la Cruz and P. Dai, arXiv : 0908.4361; 0907.2734.

[16] A. Subedi, L. Zhang, D. J. Singh and M. H. Du, Phys. Rev. B **78**, 134514 (2008).

[17] K. Haule and G. Kotliar, New J. Phys. **11**, 025021 (2009).

[18] Q. Si and E. Abrahams, Phys. Rev. Lett. **101**, 076401 (2008).

[19] J. Wu, P. Phillips and A. H. C. Neto, Phys. Rev. Lett. **101**, 126401 (2008).

[20] Z. P. Yin, S. Lebegue, M. J. Han, B. P. Neal, S. Y. Savrasov and W. E. Pickett, Phys. Rev. Lett. **101**, 047001 (2008).

[21] T. Yildirim, Phys. Rev. Lett. **101**, 057010 (2008).

[22] Y. Gallais, A. Sacuto, M. Cazayous, P. Cheng, L. Fang and H. H. Wen, Phys. Rev. B 78, 132509 (2008).

[23] L. Zhang, T. Fujita, F. Chen, D. L. Feng, S. Maekawa and M. W. Chen, Phys. Rev. B **79**, 052507 (2008).

[24] A. P. Litvinchuk, V. G. Hadjiev, M. N. Iliev, B. Lv, A. M. Guloy and C. W. Chu, Phys. Rev. B **78**, 060503 (2008).

[25] K. Y. Choi, D. Wulferding, P. Lemmens, N. Ni, S. L. Budko and P. C. Canfield, Phys. Rev. B **78**, 212503 (2008).

[26] M. Rahlenbeck, G. L. Sun, D. L. Sun, C. T. Lin, B. Keimer and C. Ulrich, Phys. Rev. B **80**, 064509 (2009).

[27] K. Y. Choi, P. Lemmens, I. Eremin, G. Zwicknagl, H. Berger, G. L. Sun, D. L. Sun and C. T. Lin, J. Phys. Cond. Matt. **22**, 115802 (2010).

[28] L. Chauviere, Y. Gallais, M. Cazayous, A. Sacuto, M. A. Measson, D. Colson and A. Forget, Phys. Rev. B **80**, 094504 (2009).

[29] C.E.M. Campos, V. Drago, J. C. De Lima, T. A. Grandi, K. D. Machado and M. R. Silva, J. Magn. Magn. Mater. **270**, 89 (2004).

[30] S.C. Zhao, D. Hou, Y. Wu, T. L. Xia, A. M. Zhang, G. F. Chen, J. L. Luo, N. L. Wang, J. H. Wei, Z. Y. Lu and Q. M. Zhang, Supercond. Sci. Technol. **22**, 015017 (2009).

[31] S. Baroni, http:// www.pwscf.org (2001).

[32] D. Vanderbilt, Phys. Rev. B **41**, 7892 (1990).

[33] M. M. Qazilbash, J. J. Hamlin, R. E. Baumbach, L. Zhang, D. J. Singh, M. B. Maple and D. N. Basov, Nature Phys. **5**, 647 (2009).

[34] A. Kumar, P. Kumar, U. V. Waghmare and A. K. Sood, J. Phys. Cond. Matt. **22**, 385701 (2010).

[35] D. J. Lockwood and M. G. Cottam, J. Appl. Phys. **64**, 5876 (1988).

[36] V. G. Hadjiev, M. N. Iliev, K. Sasmal, Y.-Y. Sun and C. W. Chu, Phys. Rev. B **77**, 220505 (2008).

[37] C. Marini, C. Mirri, G. Profeta, S. Lupi, D. D. Castro, R. Sopracase, P. Postorino, P. Calvani, A. Perucchi, S. Massidda, G. M. Tropeano, M. Putti, A. Martinelli, A. Palenzona and P. Dore, Euro Phys. Lett. **84**, 67013 (2008).

[38] J. Prakash, S. J. Singh, J. Ahmed, S. Patnaik and A.K. Ganguli, Physica C **469**, 82 (2009).

[39] P. Kumar, A. Kumar, S. Saha, J. Prakash, S. Patnaik, U. V. Waghmare, A. K. Ganguly and A. K. Sood, Solid State Commun. **150**, 557 (2010).

[40] A. D. Christianson, M. D. Lumsden, O. Delaire, M. B. Stone, D. L. Abernathy, M. A. McGuire, A. S. Sefat, R. Jin, B. C. Sales, D. Mandrus, E. D. Mun, P. C. Canfield, J. Y. Y. Lin, M. Lucas, M. Kresch, J. B. Keith, B. Fultz, E. A. Goremychkin and R. J. McQueeney, Phys. Rev. Lett. **101**, 157004 (2008).

[41] http://opium.sourceforge.net/

[42] A. M. Rappe, K. M. Rabe, E. Kaxiras and J. D. Joannopoulos, Phys. Rev. B **41**, 1227 (1990).

[43] H. J. Monkhorst and J. D. Pack, Phys. Rev. B **13**, 5188 (1977); **16**, 1748 (1976).

[44] I. I. Mazin, M. D. Johannes, L. Boeri, K. Koepernik and D. J. Singh, Phys. Rev. B **78**, 085104 (2008).

[45] S. Chi, D. T. Adroja, T. Guidi, R. Bewley, S. Li, J. Zhao, J. W. Lynn, C. M. Brown, Y. Qiu, G. F. Chen, J. L. Lou, N. L. Wang and P. Dai, Phys. Rev. Lett. **101**, 217002 (2008).

[46] Light Scattering in solids VII, Crystal - Field and Magnetic Excitation, Edited by M. Cardona and G.Guntherodt (Springer 2000).

[47] G. F. Koster, in Sol. State Phys., edited by F. Seitz and F. Turnball (Acad. Press vol.5).

[48] E.T. Heyen, R. Wegerer, E. Schonherr and M. Cardona, Phys. Rev. B **44**, 10195 (1991).

[49] B. Friedl, C. Thomsen and M. Cardona, Phys. Rev. Lett. **65**, 915 (1990).

[50] R. Zeyher and G. Zwicknagi, Z. Phys. B **78**, 175 (1990).

[51] A. Dubroka, K.W. Kim, M. Rossle, V. K. Malik, A. J. Drew, R. H. Liu, G. Wu, X. H. Chen and C. Bernhard, Phys. Rev. Lett. **101**, 097011 (2008).

[52] T. Kondo, A. F. S. Syro, O. Copie, C. Liu, M. E. Tillman, E. D. Mun, J. Schmalian, S. L. Budko, M. A. Tanatar, P. C. Canfield and A. Kaminski, Phys. Rev. Lett. **101**, 147003 (2008).

[53] S. Kawasaki, K. Shimada, G. F. Chen, J. L. Luo, N. L. Wang, G. Q. Zheng, Phys. Rev. B **78**, 220506 (2008).

[54] L. Wray, D. Qian, D. Hsieh, Y. Xia, L. Li, J. G. Checkelsky, A. Pasupathy, K. K. Gomes, C. V. Parker, A. V. Fedorov, G. F. Chen, J. L. Luo, A. Yazdani, N. P. Ong, N. L. Wang and M. Z. Hasan, Phys. Rev. B **78**, 184508 (2008).

[55] K. Terashima, Y. Sekib, J. H. Bowen, K. Nakayam, T. Kawahara, T. Sato, P. Richard, Y.-M. Xu, L. J. Li, G. H. Cao, Z.-A. Xu, H. Ding and T. Takahashi, PNAS **106**, 7330 (2009).

[56] K. W. Kim, M. Rossle, A. Dubroka, V. K. Malik, T. Wolf and C. Bernhard, Phys. Rev. B **81**, 214508 (2010).

[57] X. K. Chen, M. J. Konstantinovic, J. C. Irwin, D. D. Lawrie and J. P. Franck, Phys. Rev. Lett. **87**, 157002 (2001).

[58] T. Livneh, J. Phys. Cond. Matt. **20**, 085202 (2008).

[59] J. Prakash, S. J. Singh, A. Banerjee, S. Patnaik and A. K. Ganguli, App. Phys. Lett. **95**, 262507 (2009).

[60] D. A. Zocco, J. J. Hamlin, R. E. Baumbach, M. B. Maple, M. A. McGuire, A. Sefat, B. Sales, R. Jin, D. Mandrus, J. Jeffries, T. S. Weir and B. Vohra, Physica C **468**, 2229 (2008).

[61] P. Kumar, A. Kumar, S. Saha, D. V. S. Muthu, J. Prakash, U. V. Waghmare, A. K. Ganguli and A. K. Sood, J. Phys. Cond. Matt. **22**, 255402 (2010).

[62] K. Okazaki, S. Sugai, S. Niitaka and H. Takagi, Phys. Rev.B **83**, 035103 (2011).

[63] K. Ichimura, J. Ishioka, T. Kurosawa1, K. Inagaki, M. Oda, S. Tanda, H. Takahashi, H. Okada, Y. Kamihara, M. Hirano and H. Hosono, J. Phys. Soc. Jpn. **77**, 151 (2008).

[64] C. C. Chen, B. Moritz, J. Van den Brink, T. P. Devereaux and R. R. P. Singh, Phys. Rev. B **80**, 180418 (2009).

[65] S. H. Lee, G. Xu, W. Ku, J. S. Wen, C. C. Lee, N. Katayama, Z. J. Xu, S. Ji, Z. W. Lin, G. D. Gu, H.-B. Yang, P. D. Johnson, Z.-H. Pan, T. Valla, M. Fujita, T. J. Sato, S. Chang, K. Yamada and J. M. Tranquada, Phys. Rev. B **81**, 220502 (2010).

[66] S. Kasahara, H. J. Shi, K. Hashimoto, S. Tonegawa, Y. Mizukami, T. Shibauchi, K. Sugimoto, T. Fukuda, T. Terashima, A. H. Nevidomskyy and Y. Matsuda, Nature **486**, 382 (2012).

[67] X. C. Wang, Q. Q. Liu, Y. X. Lv, W. B. Gao, L. X. Yang, R. C. Yu, F. Y. Li and C. Q. Jin, Solid State Commun. **148**, 538 (2008).

[68] X. Zhu, F. Han, G. Mu, P. Cheng, B. Shen, B. Zeng and H. H. Wen, Phys. Rev. B **79**, 220512 (2009).

[69] J. H. Tapp, Z. Tang, B. Lv, K. Sasmal, B. Lorenz, P. C. W. Chu and A. M. Guloy, Phys. Rev. B **78**, 060505 (2008).

[70] N. Kawaguchi, H. Ogino, Y. Shimizu, K. Kishio and J. I. Shimoyama, Appl. Phys. Express **3**, 063102 (2010).

[71] H. Ogino, Y. Shimizu, N. Kawaguchi, K. Kishio, J.-I. Shimoyama, T. Tohei and Y. Ikuhara, Supercond. Sci. Technol. **24**, 085020 (2011).

[72] H. Ogino, Y. Shimizu, K. Ushiyama, N. Kawaguchi, K. Kishio and J.-I. Shimoyama, Appl. Phys. Express **3**, 063103 (2010).

[73] P. M. Shirage, K. Kihou, C.-H Lee, H. Kito, H. Eisaki and A. Iyo, J. Am. Chem. Soc. **133**, 9630 (2011).

[74] P. M. Shirage, K. Kihou, C.-H Lee, H. Kito, H. Eisaki and A. Iyo, Appl. Phys. Lett. **97**, 172506 (2010).

[75] J. Zhao, Q. Huang, C. de la Cruz, S. Li, J. W. Lynn, Y. Chen, M. A. Green, G. F. Chen, G. Li, Z. Li, J. L. Luo, N. L. Wang and P. Dai, Nature Mater. **7**, 953 (2008).

[76] P. Kumar, D.V.S. Muthu, A. Kumar, U.V. Waghmare, L. Harnagea, C. Hess, S. Wurmehl, S. Singh, B. Buchner and A. K. Sood, J. Phys. Cond. Matt. **23**, 255403 (2011).

[77] B. Friedl, C. Thomson, H. U. Habermeier and M. Cardona, Solid State Commun. **78**, 291 (1991).

[78] C. Thomson, B. Friedl, M. Cieplak and M. Cardona, Solid Stat. Comm. **78**, 727 (1991).

[79] F. London, Superfluids, Dover, New-York.

[80] M. Putti, I. Pallecchi, E. Bellingeri, M. R. Cimberle, M. Tropeano, C. Ferdeghini, A. Palenzona, C. Tarantini, A. Yamamoto, J. Jiang, J. Jaroszynski, F. Kametani, D. Abraimov, A. Polyanskii, J. D. Weiss, E. E. Hellstrom, A. Gurevich, D. C. Larbalestier, R. Jin, B. C. Sales, A. S. Sefat, M. A. McGuire, D. Mandrus, P. Cheng, Y. Jia, H. H. Wen, S. Lee and C. B. Eom, Supercon. Sci. Technol. **23**, 034003 (2010).

[81] P. Kumar, S. Saha, D. V. S. Muthu, J. R. Sahu, A. K. Sood and C. N. R. Rao, J. Phys. Cond. Matt. **22**, 115403 (2010).

[82] H. S. Obhi and E. K. H. Salje, J. Phys. Cond. Matt. **4**, 195 (1992).

[83] V.J. Emery and S. A. Kivelson, Nature **374**, 434 (1995).

[84] D. C. Johnston, Adv. Phys. **59**, 803 (2010).

[85] G. R. Stewart, Rev. Mod. Phys. **83**, 1589 (2011).

[86] L. Harnagea, S. Singh, G. Friemel, N. Leps, D. Bombor, M. A. Hafiez, A. U. B. Wolter, C. Hess, R. Klingeler, G. Behr, S. Wurmehl and B. Buchner, Phys. Rev. B **83**, 094523 (2011).

[87] R. Klingeler, N. Leps, I. Hellmann, A. Popa, U. Stockert, C. Hess, V. Kataev, H.-J. Grafe, F. Hammerath, G. Lang, S. Wurmehl, G. Behr, L. Harnagea, S. Singh and B. Buchner, Phys. Rev. B **81**, 024506 (2010).

[88] D. Hou, Q. M. Zhang, Z. Y. Lu and J. H. Wei, arXiv: 0901.1525.

[89] V. G. Hadjiev, C. Thomson, A. Erb, G. M. Vogt, M. R. Koblischka and M. Cardona, Solid State Commun. **80**, 643 (1991).

[90] C. Thomson, B. Friedl, M. Cieplak and M. Cardona, Solid State Commun. **78**, 727 (1991).

[91] A. P. Litvinchuk, C. Thomsen and M. Cardona, Solid State Commun. **83**, 343 (1992).

[92] P. Kumar, D. V. S Muthu, P. M. Shirage, A. Iyo and A. K. Sood, App. Phys. Lett. **100**, 222602 (2012).

[93] H. Ding, P. Richard, K. Nakayama, T. Sugawara, T. Arakane, Y. Sekiba, A. Takayama, S. Souma, T. Sato, T. Takahashi, Z. Wang, X. Dai, Z. Fang, G. F. Chen, J. L. Luo and N. L. Wang, Euro Phys. Lett. **83**, 47001 (2008).

[94] M. Yashima, H. Nishimura, H. Mukuda, Y. Kitaoka, K. Miyazawa, P. M. Shirage, K. Kihou, H. Kito, H. Eisaki and A. Iyo, J. Phys. Soc. Jpn. **78**, 103702 (2009).

[95] K. Matano, Z. Li, G. L. Sun, D. L. Sun, C. T. Lin, M. Ichioka and G.-Q. Zheng, Euro. Phys. Lett. **87**, 27012 (2009).

[96] T. J. Williams, A. A. Aczel, E. B. Saitovitch, S. L. Budko, P. C. Canfield, J. P. Carlo, T. Goko, J. Munevar, N. Ni, Y. J. Uemura, W. Yu and G. M. Luke, Phys. Rev. B **80**, 094501 (2009).

[97] R. H. Liu, T. Wu, G. Wu, H. Chen, X. F. Wang, Y. L. Xie, J. J. Ying, Y. J. Yan, Q. J. Li, B. C. Shi, W. S. Chu, Z. Y. Wu and X. H. Chen, Nature **459**, 64 (2009).

[98] R. Mittal, S. Rols, M. Zbiri, Y. Su, H. Schober, S. L. Chaplot, M. Johnson, M. Tegel, T. Chatterji, S. Matsuishi, H. Hosono, D. Johrendt and T. Brueckel, Phys. Rev. B **79**, 144516 (2009).

[99] M. Zbiri, R. Mittal, S. Rols, Y. Su, Y. Xiao, H. Schober, S. L. Chaplot, M. R. Johnson, T. Chatterji, Y. Inoue, S. Matsuishi, H. Hosono and T. Bruecke, J. Phys. Cond. Matt. **22**, 315701 (2010).

[100] G. B. Bachelet, D. R. Hamann and M. Schluter, Phys. Rev. B **26**, 4199 (1982).

[101] G. Kresse and J. Hafner, Phys. Rev. B **47**, 558 (1993).

[102] G. Kresse and J. Furthmller, Phys. Rev. B **54**, 11169 (1996).

[103] F. Kruger, S. Kumar, J. Zaanen and J. van den Brink, Phys. Rev. B **79**, 054504 (2009).

[104] C. C. Lee, W. G. Yin and W. Ku, Phys. Rev. Lett. **103**, 267001 (2009).

[105] W. Lv, J. Wu and P. Phillips, Phys. Rev. B **80**, 224506 (2009).

[106] A. M. Turner, F. Wang and A. Vishwanath, Phys. Rev. B **80**, 224504 (2009)

[107] F. Cricchio, O. Granes and L. Nordstrom, Phys. Rev. B **81**, 140403 (2010).

[108] M. Imada, A. Fujimori and Y. Tokura, Rev. Mod. Phys. **70**, 1039 (1998).

[109] Y. Tokura and N. Nagaosa, Science **288**, 462 (2000).

[110] W. C. Lee and C. Wu, Phys. Rev. Lett. **103**, 176101 (2009).

[111] T. M. Chuang, M. P. Allan, J. Lee, Y. Xie, N. Ni, S. L. Budko, G. S. Boebinger, P. C. Canfield and J. C. Davis, Science **327**, 181 (2010).

[112] S. A. Grigera, R. S. Perry, A. J. Schofield, M. Chiao, S. R. Julian, G. G. Lonzarich, S. I. Ikeda, Y. Maeno, A. J. Millis and A. P. Mackenzie, Science **294**, 329 (2001).

[113] I. I. Mazin, D. J. Singh, M. D. Johannes and M. H. Du, Phys. Rev. Lett. **101**, 057003 (2008).

[114] E. Grando, A. Garcia, J. A. Sanjurjo, C. Rettori, I. Torriani, F. Prado, R. D. Sanchez, A. Caneiro and S. B. Oseroff, Phys. Rev. B **60**, 11879 (1999).

[115] R. Gupta, T. V. Pai, A. K. Sood, T. V. Ramakrishnan and C. N. R. Rao, Euro Phys. Lett. **58**, 778 (2002).

[116] Y. Ming, D. Lu, J.-H. Chu, J. G. Analytis, A. P. Sorini, A. F. Kemper, B. Moritza, S. K. Mo, R. G. Moore, M. Hashimoto, W.-S. Lee, Z. Hussain, T. P. Devereaux, I. R. Fisher and Z.-X. Shen, PNAS **108**, 6878 (2011).

[117] T. Shimojima, T. Sonobe, W. Malaeb, K. Shinada, A. Chainani, S. Shin, T. Yoshida, S. Ideta, A. Fujimori, H. Kumigashira, K. Ono, Y. Nakashima, H. Anzai, M. Arita, A. Ino, H. Namatame, M. Taniguchi, M. Nakajima, S. Uchida, Y. Tomioka, T. Ito, K. Kihou, C. H. Lee, A. Iyo, H. Eisaki, K. Ohgushi, S. Kasahara, T. Terashima, H. Ikeda, T. Shibauchi, Y. Matsuda and K. Ishizaka, arXiv :1305.3875.

[118] M. Yi, D. H. Lu, R. G. Moore, K. Kihou, C.-H. Lee, A. Iyo, H. Eisaki, T. Yoshida, A. Fujimori and Z. X. Shen, New J. Phys. **14**, 073019 (2012).

[119] Z. Qin, C. Zhang, S. O. Malley, K. Lo, T. Zhou and S. W. Cheong, Solid State Commun. **150**, 768 (2010).

[120] S. Sugai, Y. Mizuno, K. Kiho, M. Nakajima, C. H. Lee, A. Iyo, H. Eisaki and S. Uchida, Phys. Rev. B **82**, 140504 (2010).

[121] S. Sugai, Y. Mizuno, R. Watanabe, T. Kawaguchi, K. Takenaka, H. Ikuta, Y. Takayanagi, N. Hayamizu and Y. Sone, arxiv : 1010.6151.

[122] J. Zhao, D. T. Adroja, D.-X. Yao, R. Bewley, S. Li, X. F. Wang, G. Wu, X. H. Chen, J. Hu and P. Dai, Nature Phys. **5**, 555 (2009).

[123] E. Saitoh, S. Okamoto, K. T. Takahashi, K. Tobe, K. Yamamoto, T. Kimura, S. Ishihara, S. Maekawa and Y. Tokura, Nature **410**, 180 (2001).

[124] P. Kumar, D. V. S. Muthu, S. N. Shirodkar, R. Saha, A. Shireen, A. Sundaresan, U. V. Waghmare, A. K. Sood and C. N. R. Rao, Phys. Rev. B **85**, 134449 (2012).

[125] W. C. Lee, W. Lv and H. Z. Arham, Int. J. Mod. Phys. B **27**, 1330014 (2013).

# Chapter 4

## 4.1 Part-A

## Coupled Phonons, Magnetic Excitations and Ferroelectricity in AlFeO$_3$

### 4.1.1 Introduction

Materials that exhibit co-occurrence of both magnetic and ferroelectric order parameters have generated enormous interest in recent years because of fundamental issues related to the coupling between spin, orbital, charge and lattice degrees of freedom as well as for their potential applications [1-5]. For applications, it is desirable to have materials with magneto-electric properties around room temperature, which is often not realized in many magneto-electric materials in which magnetic ordering is the primary driving force. In this context, AlFeO$_3$ exhibiting ferrimagnetism and possible magneto-electric coupling is very promising with a paramagnetic to ferrimagnetic transition temperature $T_c$ ~ 250 K [6-7]. Another attractive feature is its environment friendly nature as compared to other lead based multiferroics.

In AlFeO$_3$, cations occupy four distinct crystallographic sites: cations Fe$_1$, Fe$_2$ and Al$_2$ are octahedrally coordinated by oxygen whereas Al$_1$ is tetrahedrally coordinated. Structural analysis of AlFeO$_3$ [7] shows significant distortion of the FeO$_6$ octahedra, while oxygen tetrahedron around Al$_1$ is quite regular. The cause for the local deformation of lattice has been attributed to the difference between octahedral radii of Fe$^{3+}$ and Al$^{3+}$ ions and the disorder in the occupation of octahedral cation sites. Vibrational properties, which bear signatures of structure and magnetic order, are central to magneto-electric behavior of many multiferroics.

In particular, Raman spectroscopy has proved to be a powerful probe to investigate magnetic-ordering induced phonon renormalization where the observed phonon anomalies below the magnetic transition temperature have been associated with the strong spin-phonon coupling [8-11].

We note that on a related system GaFeO$_3$ first-order Raman modes are reported [12-13] and the observed modes show anomalous temperature dependence near $T_c$ (~ 210 K) attributed to the spin-phonon interactions. Here, we present a detailed temperature-dependent Raman study of AlFeO$_3$ ($T_c$ ~ 250 K) with a goal to understand the phonon renormalization due to spin-phonon coupling in magnetically ordered state below $T_c$. We have also looked for phonon signatures of a ferroelectric transition at ~ 100 K arising from magnetic interactions [14]. Our study covers first-order as well as high frequency second-order and two-magnon Raman scattering. Taking inputs from first-principles density functional theory based calculations of phonons in different magnetically ordered AlFeO$_3$, our temperature dependent Raman study reveals strong phonon renormalization below $T_c$, and that its origin is in the strong spin-phonon coupling and a coupling between two-phonon modes and magnetic excitation.

### 4.1.2 Results and Discussion

### 4.1.2.1 Raman Scattering from Phonons

Polycrystalline samples of AlFeO$_3$ were prepared and characterized as described in reference [15]. At room temperature AlFeO$_3$ crystallizes in the orthorhombic phase (*Pna2$_1$*) containing eight formula units i.e. 40 atoms in a unit cell, resulting in 120 normal modes, namely $\Gamma_{Fe}$= 6A$_1$ + 6A$_2$ + 6B$_1$ + 6B$_2$, $\Gamma_{Al}$ = 6A$_1$ + 6A$_2$ + 6B$_1$ + 6B$_2$ and $\Gamma_O$ = 18A$_1$ + 18A$_2$ + 18B$_1$ + 18B$_2$ [12]. Since the inversion symmetry is lacking, Raman modes are also infrared active. There are 117 Raman modes, while A$_1$ + B$_1$ + B$_2$ are acoustic modes. Figure 4.1 shows the Raman spectrum at 5 K, revealing 18 modes labeled as S1 to S18 in the spectral range of 100-1950 cm$^{-1}$. Spectra are fitted with a sum of Lorentzian functions; the individual modes are shown by thin lines and the resultant fit by a thick line. Our first-principles density functional calculations (discussed later) suggest that the first-order Raman phonons occur below ~ 810 cm$^{-1}$. Table-4.1 lists the experimental (at 5 K) and the calculated frequencies for disordered Anti-Ferromagnetic (AFM) state close to the experimental values. Since the intensity of the mode S15 is zero above $T_c$, it is attributed to the two-magnon Raman scattering (to be discussed later). The modes S16 to S18 are assigned to second-order Raman scattering coupled with magnetic degrees of freedom.

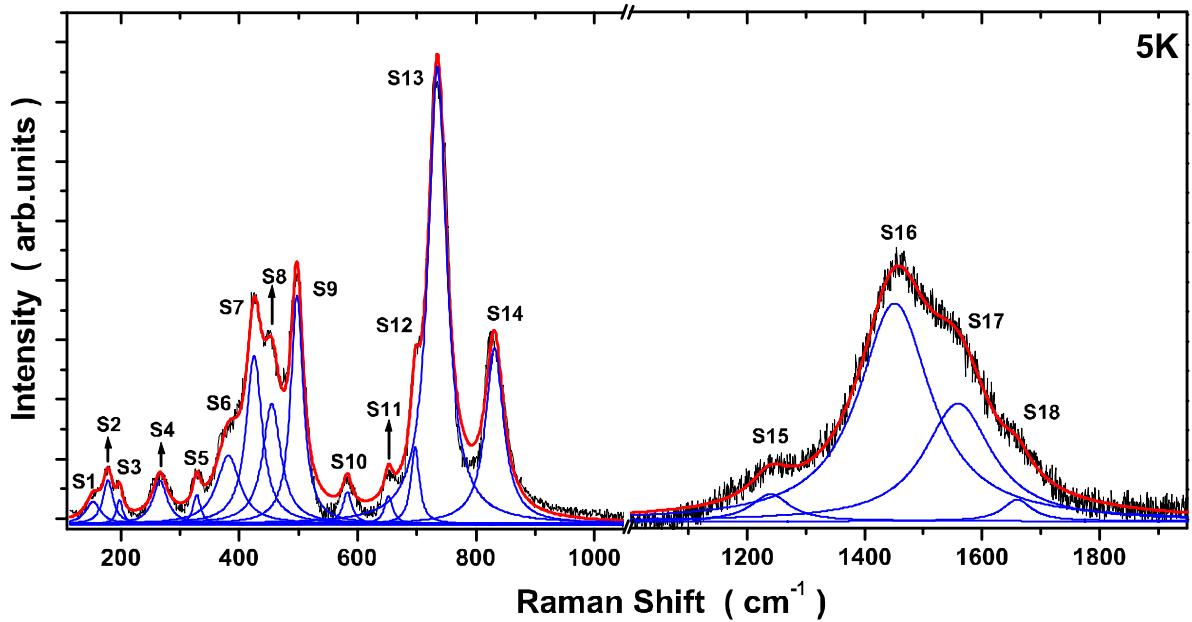

Figure 4.1: Raman spectra of AlFeO$_3$ at 5 K. Solid (thin) lines are fit of individual modes and solid (thick) line shows the total fit to the experimental data.

## 4.1.2.2 Temperature Dependence of the First-Order Phonons

Figure 4.2 shows the mode frequencies of some of the prominent first-order phonon modes S4, S7 to S10, S13 and S14 as a function of temperature. The following observations can be made: (i) The frequencies of S4, S7, S8, S9, S10 and S13 modes show a sharp change at $T_c$.

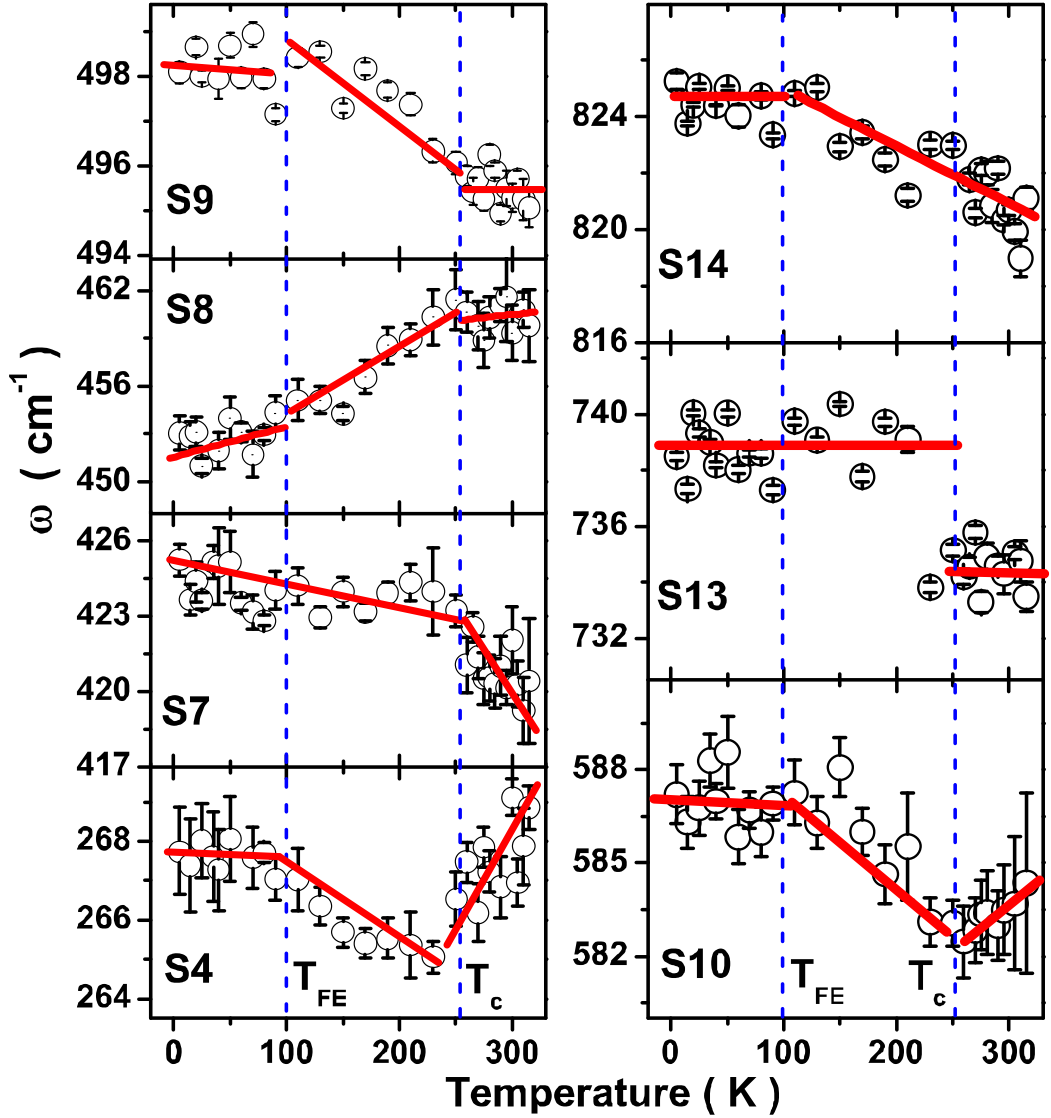

Figure 4.2: Temperature dependence of the first-order phonon modes S4, S7, S8, S9, S10, S13 and S14. Solid lines are the linear fits in different temperature region as described in the text.

The temperature derivative of modes S4 and S10 frequencies ($\partial\omega/\partial T$) change sign at $T_c$. The frequency of mode S13 shows a jump by ~ 4 cm$^{-1}$ near $T_c$. (ii) The slope of $\omega$ with respect to temperature for the S4, S8, S9, S10 and S14 modes shows changes near 100 K. We attribute these changes to a ferroelectric transition in this system at ~ 100 K, as the pyroelectric experiments showed [14] that a polar phase exists below ~ 100 K and the reversal of polarization data with changing direction of electric field during pyroelectric current measurement demonstrates that the material is indeed a ferroelectric. The solid lines in panels are linear fits in three regions i.e. 315 K to 250 K, 250 K to 100 K and 100 K to 5 K. (iii) The temperature dependence of the mode S8 is anomalous below $T_c$ i.e frequency decreases on lowering the temperature.

The anomalies in the temperature dependence of the phonon modes S4, S7, S8, S9 and S10 near $T_c$ are similar to those in $RMnO_3$ (R = Pr, Nd, Sm, Tb, Dy and La), $GaFeO_3$ and $BiFeO_3$ [8,10-13,16-19]. The sharp jump in the frequency of the mode S13 at $T_c$ can arise from subtle local structural change. Following manganites [8,10-13,16-19] and our theoretical calculations, the sharp changes in mode frequencies of S4, S7, S8, S9 and S10 are attributed to strong spin-phonon coupling in the magnetic phase below $T_c$.

### 4.1.2.3 High Frequency Modes: Second-Order Phonon and Magnon Scattering

Two-phonon Raman bands are related to two-phonon density of states having contribution from all the branches in the first Brillouin zone. For simplicity, we have fitted the high energy Raman band (1100 - 1800 cm$^{-1}$) with a sum of four Lorentzain (S15 to S18) where peak positions represent maxima in the two-phonon density of states. As the second-order Raman scattering involves the phonons over the entire Brillouin zone, the frequencies of the observed second-order phonons are not necessarily double of the first-order phonons at the Γ (q = 0,0,0) point. Accordingly, mode S17 can be assigned as a combination of S13 and S14 and mode S16 as overtone of mode S13 and mode S18 as overtone of mode S14.

Figure 4.3(a and b) shows the high frequency modes at few typical temperatures. It can be seen that the mode S15 is absent in the spectrum recorded at 265 K and above. Taking S13 mode as an internal marker, Fig. 4.3 (c) shows the intensity of S15 mode with respect to that of S13 mode. Figure 4.3 (d) shows the integrated intensity of the S15 mode with respect to its

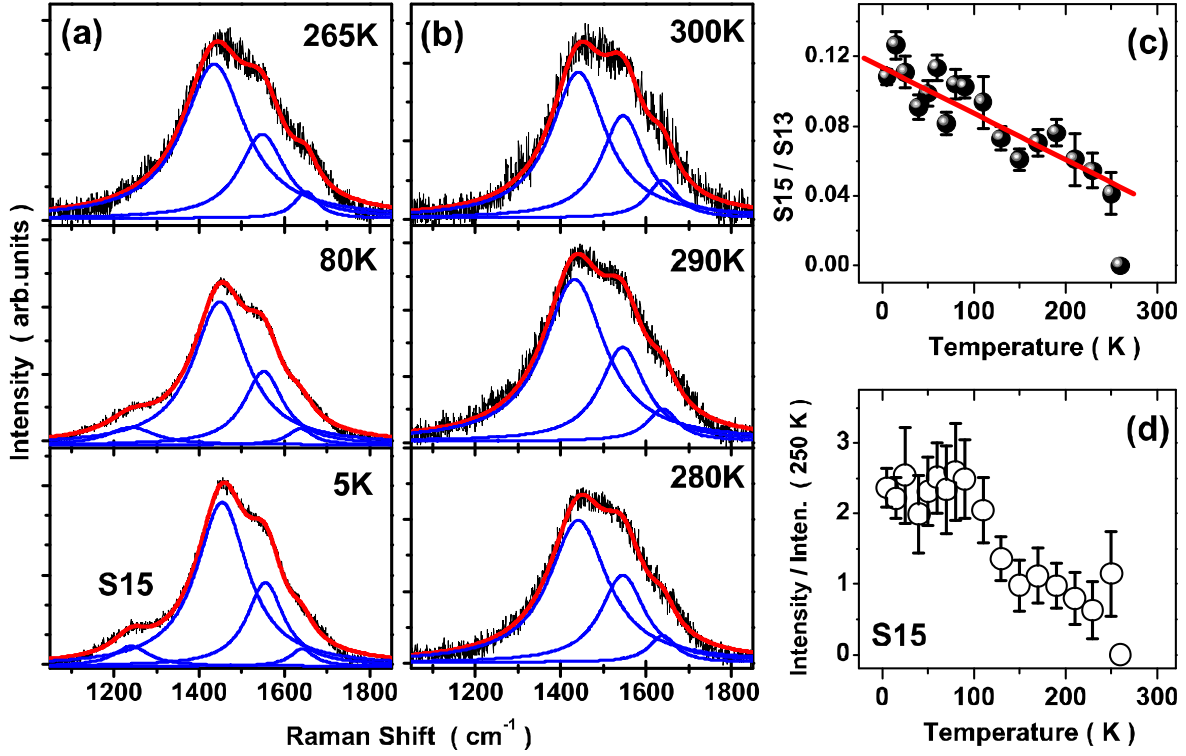

Figure 4.3: (a and b) Temperature evolution of mode S15 at few typical temperatures. (c) Intensity ratio of S15 mode w.r.t to prominent first-order S13 mode. Solid line is the linear fit. (d) Temperature-dependence of the intensity of mode S15 w.r.t to its intensity at 250 K.

its intensity at 250 K. The intensity of the mode S15 is zero above $T_c$ and its intensity builds up as we lower the temperature. The vanishing of the S15 mode above 250 K suggests that it can be associated with two-magnon Raman scattering. From the energy of the two-magnon band, an estimate of the nearest neighbour exchange coupling parameter $J_o$ can be made. If spins deviations are created on the adjacent sites, the two-magnon energy is given by $J_o(2Sz-1)$ where S is the spin on the magnetic site ($Fe^{3+}$ here with $S = 5/2$) and $z$ ($z = 6$) is the number of nearest neighbour to that site [20]. Using $\omega = 1240$ cm$^{-1}$ (at 5 K), the estimated value of the exchange parameter $J_o$ is ~ 5.3 meV. We note that this value is very close to our first-principles calculation of $J_o$ ~ 6 meV (discussed later). As temperature is lowered below $T_c$, the S15 mode frequency decreases significantly (~ 5 %) (see Fig. 4.4 (e)). The frequencies of modes S16, S17 and S18 show a change in $\partial\omega/\partial T$ near the transition temperature (see Figs. 4.4 (b), 4.4 (c) and 4.4 (d)) attributed to the possible coupling between two-phonon and magnetic excitations similar to that in other magnetic systems [8, 11, 16-23].

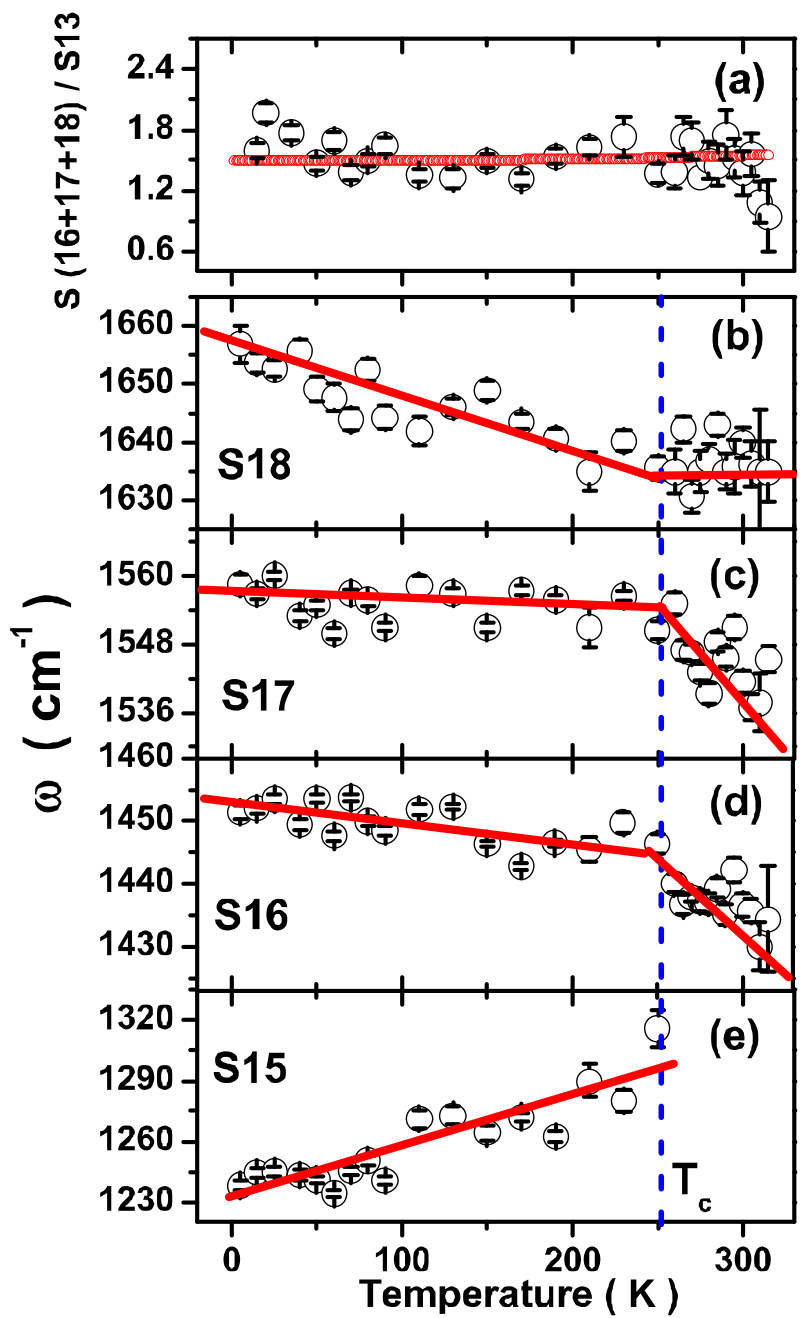

Figure 4.4: (a) Intensity ratio of the high frequency band w.r.t to the prominent first-order mode. Solid line is the fitted curve as described in the text. (b,c,d and e) Temperature dependence of the mode S15, S16, S17 and S18. Solid lines are linear fits below and above $T_c$.

To ascertain the second-order nature of high frequency bands S16, S17 and S18, we plot in Fig. 4.4 (a) the sum of their intensities with respect to the intensity of the dominant first-order S13 mode. The second-order Raman intensity for a combination mode of frequency $\omega_1 + \omega_2$ is $[n(\omega_1)+1][n(\omega_2)+1]$, where $n(\omega)$ is the Bose-Einstein mean occupation number. The ratio of the second-order band with respect to the first-order mode of frequency $\omega_1$ will be $c[n(\omega_2)+1]$, where c is the ratio depending on the matrix elements in second-order and first-order Raman scattering. The solid line in Fig. 4.4(a) is $1.5*[n(\omega=740cm^{-1})+1]$ showing that the broad band (decomposed into S16, S17 and S18 modes) is due to second-order Raman scattering. We now develop theoretical understanding of our results.

Table-4.1: List of the experimentally observed frequencies at 5 K and calculated frequencies in AlFeO$_3$ for disordered AFM (Fe$_2$-Al$_2$ anti-site disorder) state.

| Mode Assignment | Experimental ω (cm$^{-1}$) | Calculated ω (cm$^{-1}$) |
|---|---|---|
| S1 | 156 | 154 |
| S2 | 178 | 179 |
| S3 | 198 | 197 |
| S4 | 268 | 270 |
| S5 | 328 | 331 |
| S6 | 380 | 379 |
| S7 | 425 | 425 |
| S8 | 453 | 453 |
| S9 | 498 | 499 |
| S10 | 587 | 581 |
| S11 | 650 | 654 |
| S12 | 698 | 691 |
| S13 | 738 | 733 |
| S14 | 826 | 807 |
| S15 (Two-magnon) | 1240 | |
| S16 (Overtone) | 1450 | |
| S17 (Second-order) | 1560 | |
| S18 (Overtone) | 1660 | |

## 4.1.2.4 First-Principles Calculations

Our first-principles calculations are based on density functional theory (DFT) with spin-density dependent exchange correlation energy functional of a generalized gradient approximated (GGA) (PerdewWang 91 (PW 91)) form [24] as implemented in the Vienna ab-initio Simulation Package (VASP) [25-26]. The projector augmented wave (PAW) method [27] was used to capture interaction between ionic cores and valence electrons. An energy cutoff of 400 eV was used for the plane wave basis and integrations over the Brillouin zone of the orthorhombic crystal were sampled with a regular 4x2x2 mesh of k-points. Dynamical matrix and phonons at $\Gamma$-point ( q = 0,0,0 ) have been obtained with a frozen-phonon method with atomic displacements of $\pm$ 0.04 Å. Numerical errors in our calculations break the symmetry of the dynamical matrix weakly and introduce an error of about $\pm$ 12 cm$^{-1}$ in the phonon frequencies.

It is known [7] that AlFeO$_3$ has a site occupancy disorder between Fe and Al sites, with most common occurrence of anti-site disorder being between Fe$_2$ and Al$_2$ sites [7]. This disorder is taken into account by exchanging the site positions of an Fe atom at Fe$_2$ site with an Al atom at Al$_2$ site. We have also considered the anti-site disorder between Fe$_1$ and Al$_2$ sites. From the energetics, we find that the AFM state is the most stable for system with either type of anti-site disorder between Fe and Al. The AFM state with Fe$_1$-Al$_2$ anti-site disorder is higher in energy as compared to the AFM state with Fe$_2$-Al$_2$ anti-site disorder by 5.7 meV/atom, confirming the higher occurrence of Fe$_2$-Al$_2$ anti-site disorder. To facilitate a meaningful comparison with experimental Raman spectra, we simulate the structure with experimental lattice constants and relaxing internally the atomic positions using conjugate gradients algorithm.

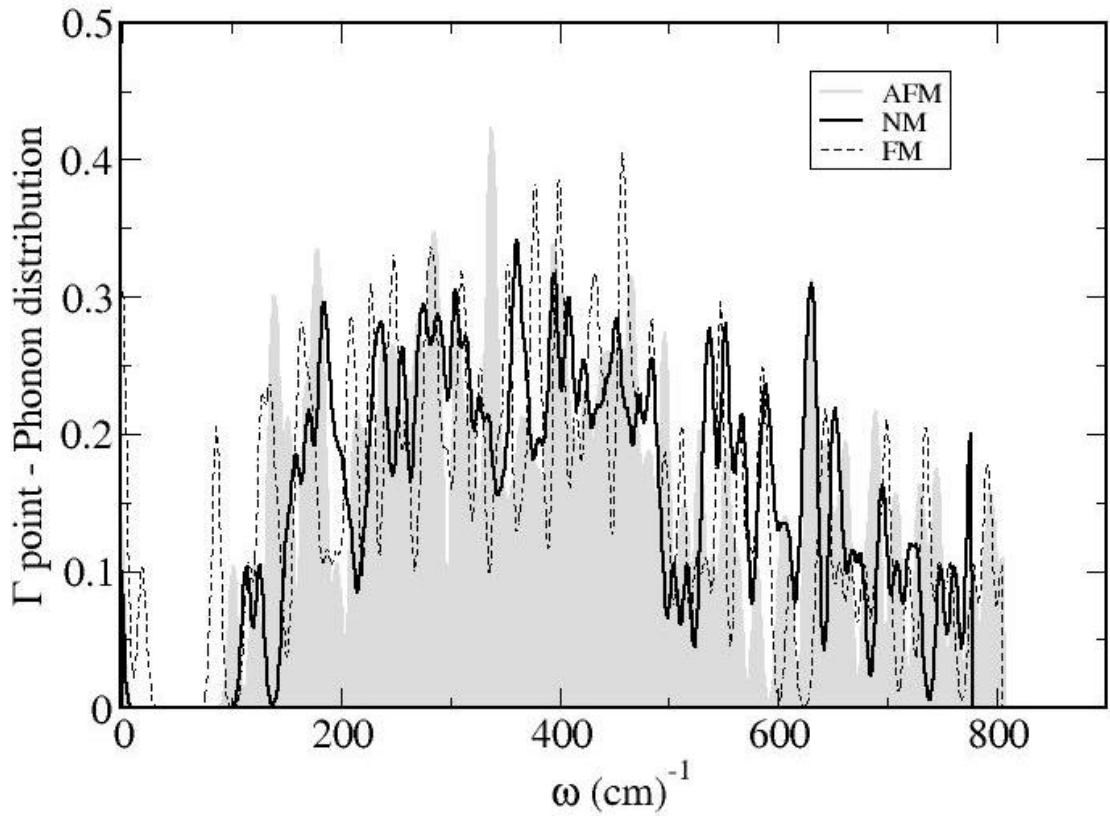

Figure 4.5: Distribution of phonons at $\Gamma$- point for AFM, FM and NM orderings with a Gaussian broadening of ~ 4 cm$^{-1}$.

To understand the interplay between disorder, magnetic ordering and phonons, we determine phonons at $\Gamma$-point for a chemically disordered structure with non-magnetic (NM), FM and AFM ordering (see Fig. 4.5). The spin-phonon coupling is analysed by examining how normal modes depend on the magnetic ordering by examining the correlation matrix between phonon eigen modes of AlFeO$_3$ in two different magnetically ordered states. In the absence of spin-phonon coupling, the phonons would be unaffected by changes in the magnetic order and hence only the diagonal terms would be non-zero in the correlation matrix. Non-zero off-diagonal elements of the correlation matrix clearly uncover the correspondence between eigen modes in different magnetic orders. For example, it determines which phonon modes of the AFM state relate to phonons of the FM state, giving a quantitative idea of mixing between modes due to spin-phonon coupling. The spin-Hamiltonian has the form:

$$H = \frac{1}{2}\sum_{ij} J_{ij} \vec{S}_i \cdot \vec{S}_j \tag{4.1}$$

where, $J_{ij}$ is the exchange interaction between $i^{th}$ and $j^{th}$ ising spins $S_i$ and $S_j$. Only considering the nearest neighbour and isotropic interaction, we reduce $J_{ij}$ to $J$. The change in $J$ due to spin phonon coupling is given by second-order Taylor series expansion of $J$ w.r.t amplitude of atomic displacements ($u_{v\Gamma}$) of the $v^{th}$ $\Gamma$ - phonon mode of the magnetic state [16],

$$J(u_{v\Gamma}) = J_o + \vec{u}_{v\Gamma}(\nabla_u J) + \frac{1}{2}\vec{u}_{v\Gamma}(\nabla^2_u J)\vec{u}_{v\Gamma} \tag{4.2}$$

Summing over all modes gives

$$H = \frac{1}{2}\sum_v \sum_{ij} [J_o + \vec{u}_{v\Gamma}(\nabla_u J) + \frac{1}{2}\vec{u}_{v\Gamma}(\nabla^2_u J)\vec{u}_{v\Gamma}] \cdot \vec{S}_i \cdot \vec{S}_j \tag{4.3}$$

Here, $J_o$ is the bare spin-spin coupling parameter, $\nabla_u J$ corresponds to the force exerted on the system due to change in magnetic ordering from its ground state magnetic configuration and $\nabla^2_u J$ is proportional to the change in phonon frequency ($\Delta$) of $v^{th}$ $\Gamma$ - phonon mode due to change in magnetic ordering. From the spin-Hamiltonian (Eq. 4.1), energies of a single pair of spins in AFM and FM states are given by, $E_{AFM} = -J_o|S|^2$ and $E_{FM} = J_o|S|^2$ respectively. The difference in the energies of AFM and FM states is directly proportional to $J_o$. The unit cell of AlFeO$_3$ used in our simulation contains 8 Fe ions where, the $i^{th}$ Fe ion is connected to $z_i$ number of other Fe ions. This gives, $J_o = (E_{FM} - E_{AFM})/\Sigma_i Z_i * 8 * |S|^2$ here, S = 5/2 and $E_{FM} - E_{AFM} \sim$ 1.5 eV from first principles calculations. Our estimate of the exchange coupling parameter $J_o$ is ~ 6 meV. This value is in good agreement with the one estimated from the two-magnon peak observed in Raman spectrum here. Denoting $\nabla^2_{uv} J$ as $J_2$, the change in phonon frequency ($\Delta$) of the $\lambda^{th}$ $\Gamma$ - point phonon mode is given by [16]

$$\Delta_\lambda = \frac{1}{2\mu_\lambda \omega_\lambda} \sum_v \hat{u}_{v\Gamma} J_2 \hat{u}_{v\Gamma} \qquad (4.4)$$

Here, $\mu_\lambda$ and $\omega_\lambda$ are the reduced mass and frequency of the $\lambda^{th}$ $\Gamma$-phonon mode, respectively. We note large $\Delta$ implies stronger spin-coupling.

The calculations are done for both types of disorder: $Fe_2$ at $Al_2$ site ($Fe_2$-$Al_2$) as well as $Fe_1$ at $Al_2$ site ($Fe_1$-$Al_2$). For $Fe_2$-$Al_2$ type disorder Fig. 4.6 (a) and 4.6 (b) show the changes in $\Gamma$-point phonon frequency ($\Delta$) between FM and AFM, and NM and AFM states, respectively. The corresponding changes for $Fe_1$-$Al_2$ disorder are shown in Fig. 4.6 (c) and (d).

For connecting our results with the experiment, we have listed only those calculated phonon frequencies which are close in frequency to the experimentally observed Raman active phonon modes (refer to Table-4.1). We assume the correlation between the experimentally observed modes which exhibit anomalies at magnetic transition and calculated spin-phonon coupling for modes with frequencies in the vicinity of the observed modes, and carry out the mode assignment. In the case of $Fe_2$-$Al_2$ anti-site disorder, Fig. 4.6 (a) and (b), corresponding to correlation between phonons of FM/NM state with AFM state, $\Delta$ and hence $J_2$ is high for mode frequencies in the neighbourhood of close to modes S1, S4 and S10 for FM-AFM coupling (see Fig. 4.6 (a)) and modes S4, S8, S11 and S12 (see Fig. 4.6 (b)) for NM-AFM coupling. We note that $J_2$ for NM-AFM state coupling is not exactly spin-phonon coupling parameter as in the case of FM-AFM state coupling, but here it gives an estimate of the change in phonon frequencies in going from NM state to AFM ordering. In Fig. 4.2 modes S4, S7, S8 and S10 show sharp changes in frequency at the transition temperature $T_c$ suggesting their strong coupling with spin consistent with our first-principles calculations. Another interesting observation from Fig. 4.6 (b) is that the mode with frequency near S8 shows the largest increase in frequency in going from AFM to the NM state consistent with our experimental observation of most significant hardening of mode S8 with increase in temperature of the AFM state.

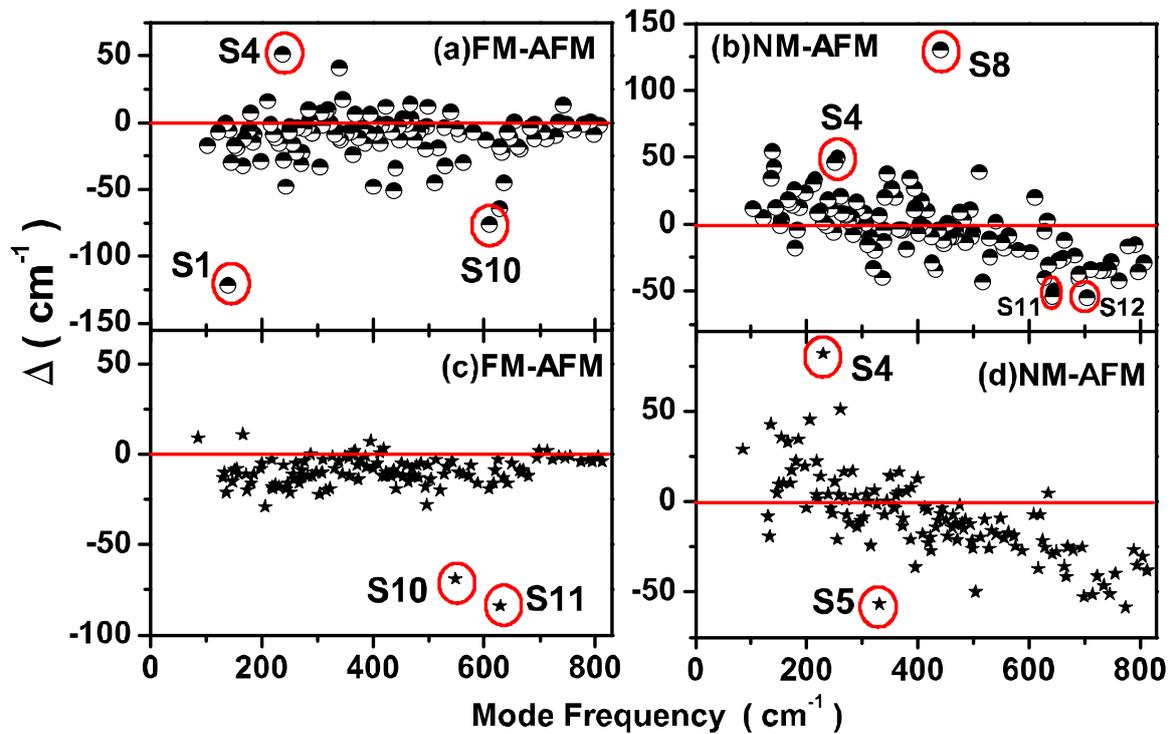

Figure 4.6: Second order spin-phonon coupling in different magnetic states, (a, b) FM/NM-AFM states with $Fe_2$-$Al_2$ anti-site disorder, respectively. And (c, d) FM/NM - AFM states with $Fe_1$-$Al_2$ anti-site disorder, respectively.

We now discuss the effect of $Fe_1$-$Al_2$ anti-site disorder. Figure 4.6 (c) and (d) correspond to changes in frequencies of phonons of FM and NM state correlating with those of the AFM state, respectively. With change in magnetic ordering from FM to AFM state, modes near S10 and S11 (see Fig. 4.6 (c)) exhibit strong second-order coupling with spin. In comparison, modes close to S4 and S5 (see Fig. 4.6 (d)) show large second-order coupling for the transition from NM to AFM state. Mode S4 was observed in our Raman measurements to exhibit a significant hardening across the transition from AFM to the NM state, consistent with our calculated results.

Our first-principles analysis confirms the existence of strong spin-phonon coupling in AlFeO$_3$, and points out that the anomaly in mode S8 is primarily influenced by Fe$_2$-Al$_2$ disorder, while the anomaly in S4 mode is influenced by Fe$_1$-Al$_2$ disorder additionally. Anomalous hardening of the S8 mode is due to strong spin-phonon coupling at the second-order ($J_2$) in AlFeO$_3$.

### 4.1.3 Conclusion

In conclusion, we observed a strong first-order phonon renormalization below the magnetic transition temperature of AlFeO$_3$ due to strong spin-phonon coupling. In addition, high frequency Raman bands between 1100 to 1800 cm$^{-1}$ show pronounced effects of the strong magnetic correlation below $T_c$. In particular, the intensity of mode S15 becomes zero above the transition temperature $T_c$ and hence the mode is attributed to two-magnon Raman scattering. The band position gives an estimate of spin exchange constant $J_0$ to be ~ 5.3 meV, in close agreement with the DFT calculations. With first-principles analysis we have explored the effects of magnetic ordering and (Al, Fe) disorder on phonons. Our results suggest a strong interplay between lattice and magnetic degrees of freedom which are crucial to understand the underlying physics responsible for various exotic physical phenomena in these materials. What is equally noteworthy is that our Raman data shows evidence for a phase transition to a ferroelectric phase below 100 K.

## 4.2 Part-B

# Orbiton-Mediated Multiphonon Raman Scattering in Multiferroic TbMnO$_3$

### 4.2.1 Introduction

The effect of orbital ordering on the Raman spectra of perovskite manganites, RMnO$_3$ (R = Rare Earth), has been investigated both theoretically [28-31] and experimentally [30,32-35]. Three broad bands near 1000 cm$^{-1}$, 1170 cm$^{-1}$ and 1290 cm$^{-1}$ in the Raman spectra of LaMnO$_3$ were attributed to orbiton excitations [30, 36], an assignment still being debated and alternative proposals made [28-29,31,37]. Although in centro-symmetric LaMnO$_3$, Raman modes are not infrared active, infrared absorption [37] shows similar bands as in the Raman spectra, attributing these features to multiphonon scattering instead of orbital excitations. As a result of strong electron-phonon coupling, Allen et al. [28] have proposed that orbitons in LaMnO$_3$ are self-trapped by the local rearrangement of the lattice and hence multiphonon Raman scattering with intensities comparable to the one phonon Raman modes has been predicted. This arises from the Franck-Condon (FC) process via the self-trapped orbitons, suggesting a mixed character of phonons and orbitons to the high frequency modes. This mixed character has also been shown theoretically by other calculations considering the effects of super-exchange and electron-phonon interactions [31,38]. The high intensity ratio (~ 0.1 to 0.4) of the second-order modes to their first-order counterparts has been observed experimentally for LaMnO$_3$ [32] and RMnO$_3$ (R = La, Pr, Ho and Y) [33], supporting the theoretical proposal for the mixed nature of the multiphonon bands. On the other hand, a recent room temperature Raman study of RMnO$_3$ (R = Pr, Eu, Dy, Ho and Y) and O$^{18}$ isotopically substituted EuMnO$_3$ [39] suggests that the high frequency modes are due to second-order scattering involving only Brillouin zone boundary phonons. All these experimental studies have been carried out at room temperature and above. Our present Raman study looks at multiferroic $TbMnO_3$ as a function of temperature from 5 K to 300 K, covering spectral range from 300 $cm^{-1}$ to 1500 $cm^{-1}$ and focuses mainly on the temperature dependence of the two high energy excitations observed at 1168 $cm^{-1}$ and 1328 $cm^{-1}$.

$TbMnO_3$ is orthorhombic (space group *Pbnm*) at room temperature and shows an incommensurate lattice modulation at $T_N$ for sinusoidal antiferromagnetic ordering (with $T_N \sim$ 41 K [40] or $T_N \sim$ 46 K as reported by Bastjan et al. [41]). Ferroelectric order develops at the incommensurate-commensurate transition temperature $T_{FE} \sim$ 27 K [40]. As the temperature is further lowered, rare-earth $Tb^{3+}$ ion-spins also order anti-ferromagnetically at ~ 7 K [40]. As far as first-order phonons are concerned, it has been shown that in $RMnO_3$ (R = La, Nd, Sm, Gd, Dy, Pr and Tb), a few Raman and IR phonons involving oxygen vibrations are anomalous i.e. the phonon frequency decreases as temperature is lowered below $T_N$ [8,16,19,42-43] arising from strong spin-phonon coupling. There have been reports of Raman studies on multiferroic $TbMnO_3$ dealing with only first-order Raman scattering [8, 43-45], but to our knowledge, there is no report of the high frequency excitations in $TbMnO_3$. Here, we present Raman scattering data from an unoriented single crystal of $TbMnO_3$ as a function of temperature in the range 5 K to 300 K. We show that the intensity ratio of the second-order mode to the corresponding first-order is very high and remains constant with temperature as predicted by Allen et al. [28-29] for coupled multiphonons-orbiton modes. In addition, the first-order mode involving oxygen vibrations ($\omega \sim$ 616 $cm^{-1}$) shows anomalous softening below $T_N$ attributed to the strong spin-phonon coupling.

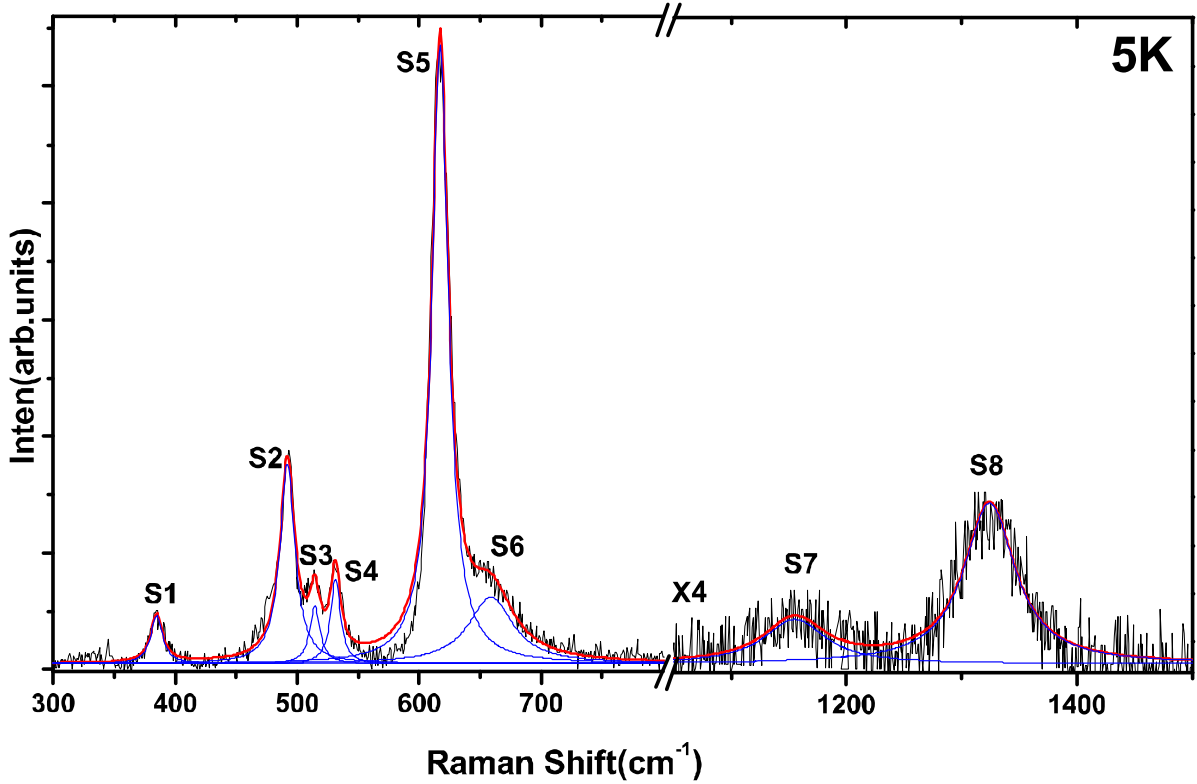

Figure 4.7: Raman spectra of TbMnO$_3$ measured at 5 K. Thick solid line shows the total fit, thin solid lines show the individual Lorentzian fit.

## 4.2.2 Results and Discussion

Figure 4.7 shows Raman spectrum at 5 K displaying 8 modes labeled as S1 to S8. The spectra are fitted to a sum of Lorentzian and the frequencies, linewidths and intensities so obtained are shown in Fig. 4.8 for the first-order Raman modes and in Fig. 4.9 for the multiphonon modes S7 and S8. TbMnO$_3$ has 24 Raman active modes classified as $\Gamma_{Raman} = 7A_g + 5B_{1g} + 7B_{2g} + 5B_{3g}$ [44]. The assignment of low frequency modes S1-S5, given in Table-4.2, has been done following the work of Iliev et al. for TbMnO$_3$ [44]. The origin of mode S6 may be similar to that of the 640 cm$^{-1}$ mode observed in LaMnO$_3$ [46-47], attributed to the disorder-induced phonon density of states [39,48] or second-order Raman scattering [46]. It can also be a disorder induced infrared active phonon mode (transverse optic mode at 641 cm$^{-1}$ and longitudinal optic mode at 657 cm$^{-1}$) observed in infrared studies of TbMnO$_3$ [19], described in detail in Part C-4.3.

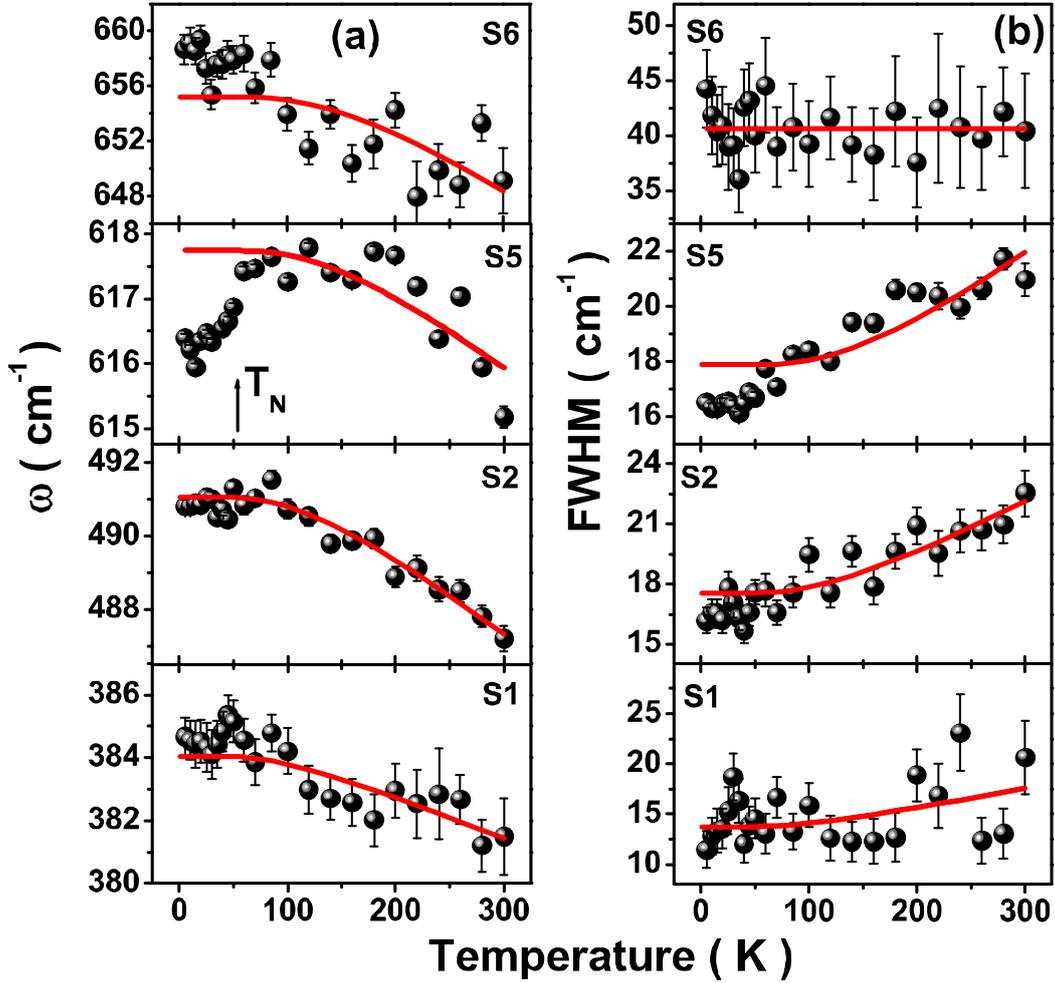

Figure 4.8: Temperature dependence (a) frequencies and (b) linewdiths of the modes S1, S2, S5 and S6. The solid lines are the fitted curve as described in text.

## 4.2.2.1 Temperature Dependence of the First-order Modes

We now discuss the temperature dependence of the first-order phonon modes. In general, the temperature dependent behavior of a phonon mode of frequency '$\omega$' is given as [16]

$$\omega(T) = \omega(0) + (\Delta\omega)_{qh}(T) + (\Delta\omega)_{anh}(T) + (\Delta\omega)_{el\text{-}ph}(T) + (\Delta\omega)_{sp\text{-}ph}(T)$$

(4.5)

$(\Delta\omega)_{qh}(T)$ corresponds to the change in phonon frequency due to a change in the lattice parameters of the unit cell, termed as quasi-harmonic effect. $\Delta\omega_{anh}(T)$ gives the intrinsic anharmonic contributions to the phonon frequency. The effect of renormalization of the phonon frequency (($\Delta\omega)_{el-ph}(T)$) due to electron-phonon coupling is absent in insulating TbMnO$_3$. The last term, $\Delta\omega_{sp-ph}(T)$, represents the change in phonon frequency due to spin-phonon coupling, caused by the modulation of the exchange integral by lattice vibrations [16]. The change in phonon frequency of mode "$i$" due to the change in lattice parameters, i.e. $(\Delta\omega)_{qh}(T)$, can be related to the change in volume if we know Grüneisen parameter $\gamma_i = -(B_0/\omega_i)(\partial\omega_i/\partial P)$, where $B_0$ is the bulk modulus and $\partial\omega_i/\partial P$ is the pressure-derivative of the phonon frequency. For a cubic crystal or isotropically expanded lattice, the change in phonon frequency due to change in volume is given as $(\Delta\omega)_i(T)_{qh}/\omega_i(0) = -\gamma_i(\Delta V(T)/V(0))$. The Grüneisen parameter calculated for RMnO$_3$ [R = Sm, Nd and Pr] is ~ 2 [49-50]. The quasi harmonic contribution in TbMnO$_3$ can be neglected since the fractional change in volume is negligible [51], as has been done in earlier studies of rare earth manganites RMnO$_3$ (R = Gd, Eu, Pr, Nd, Sm, Tb, Dy, Ho, and Y) [8,42].

Solid lines in Fig. 4.8 (a) and (b) are fitted with equations (2.11) and (2.10), respectively. We realize that below the phase transition temperature $T_N$, eqns. (2.10) and (2.11) are not expected to hold good, as is obvious in the temperature dependence of S5 mode (see Fig. 4.8 (a)). Therefore, we fit the data between 50 K to 300 K using eqns. (2.10) and (2.11) and the theoretical curves are extrapolated below 50 K using the fitted parameters given in Table-4.2 (see solid lines in Fig. 4.8). Similar procedure has been adopted in earlier studies of manganites [8,42]. We do not observe any significant signature, within the accuracy of our experiments, of the ferroelectric transition at $T_{FE}$ (~ 27 K) in the temperature dependence of frequencies and linewidths. The fitting parameter C of mode S6 is very high as compared to the other modes, showing that this mode is much more anharmonic.

Table-4.2: List of the experimentally observed phonons frequencies and fitting parameters of a few phonons. Units are in cm$^{-1}$.

| Mode Assignment | Phonon-Frequency | | | | |
| --- | --- | --- | --- | --- | --- |
| | $\omega$(5 K) | $\omega$(0 K) | C | $\Gamma$(0) | D |
| S1 ($A_g$) | 384.9 | 386.1 ± 0.6 | -4.1 ± 0.8 | 10.8 ± 2.7 | 5.9 ± 2.8 |
| S2 ($A_g$) | 491.9 | 495.3 ± 0.3 | -8.5 ± 0.5 | 12.3 ± 0.9 | 10.4 ± 1.4 |
| S3 ($A_g$) | 514.5 | | | | |
| S4 ($B_{2g}$) | 531.3 | | | | |
| S5 ($B_{2g}$) | 616.2 | 620.8 ± 0.7 | -6.2 ± 1.2 | 10.9 ± 1.2 | 14.1 ± 2.1 |
| S6 | 658.4 | 668.8 ± 4.9 | -27.2 ± 8.4 | 40.6 ± 3.2 | 0.09 |
| S7 (Second order) | 1157.1 | | | | |
| S8 (Overtone) | 1327.4 | | | | |

An interesting observation is that the intense mode S5 shows anomalous temperature dependence: the mode shows softening near $T_N \sim 46$ K. Similar anomalous temperature dependence has been observed for a few Raman modes in RMnO$_3$ where R = La [16] and Gd, Pr, Nd, Sm, Dy [8,42], which has been attributed to strong spin-phonon coupling [16]. This is understood as follows: if an ion is displaced from its equilibrium position by "$u$", then the crystal potential is given as U = (0.5)* (k$u^2$) + $\Sigma_{ij}$ J$_{ij}$($u$)S$_i$S$_j$, where k in the first term represents the force constant and the second term arises from spin interactions between the Mn$^{3+}$ spins. The phonon frequency is affected by the additional term ($\Delta\omega$)$_{sp-ph}$(T) = $\lambda$ <S$_i$S$_j$>, where $\lambda = \partial^2 J_{ij}(u)/\partial u^2$ is the spin-phonon coupling coefficient and <S$_i$S$_j$> is the spin-correlation function. The parameter $\lambda$ can be positive or negative and can be different for different phonons. Below $T_N$, the spin correlations build up and hence the spin-phonon coupling becomes important at lower temperatures. The renormalization of the mode S5 frequency starts slightly above $T_N$ (~ 46 K), which can arise from spin fluctuations due to quantum and thermal effects [8].

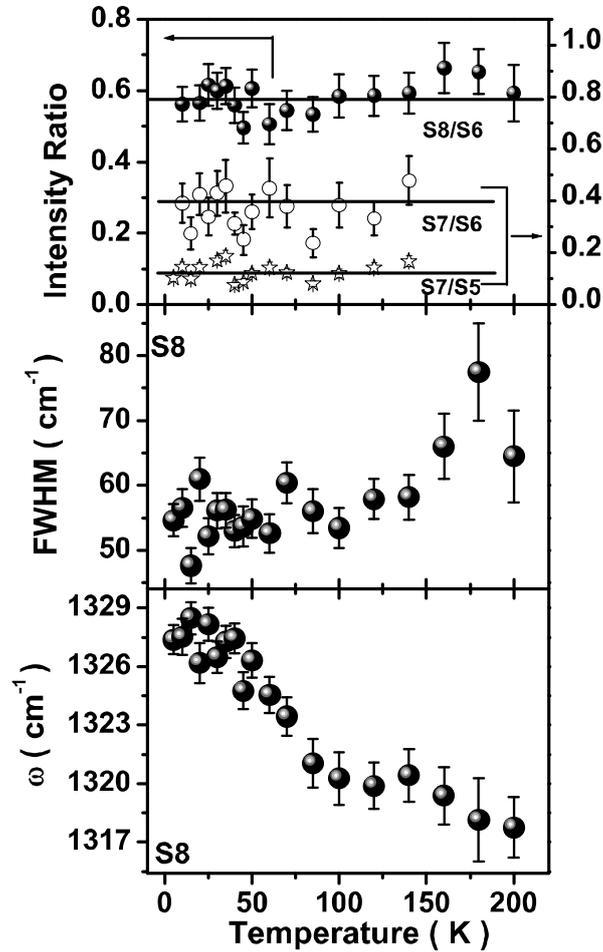

Figure 4.9: Temperature dependence of the intensity ratio of the mode S8 to S6, S7 to S6 and S7 to S5 (Top Panel). Temperature dependence of the Frequency (Lower Panel) and Linewidth (Middle panel) of mode S8. Solid lines in the top panel are guide to the eye.

### 4.2.2.2 Orbiton-Phonon Coupling

We now discuss the two high energy excitations, S7 at 1157 cm$^{-1}$ and S8 at 1328 cm$^{-1}$. Figure 4.9 shows temperature dependence of the frequency and linewidth of S8 mode as well as the intensity ratio of S8 and S6 modes (S8/S6) in the temperature range from 5 K to 200 K.

Above 200 K, the mode S8 is too weak to be analyzed quantitatively as is the case for S7. Mode S7 can be assigned as the second-order Raman mode involving a combination of S2 and S6 or S4 and S5 phonons and S8 as an overtone of S6 (658 cm$^{-1}$) mode. The intensity ratio of S8 to S6 is most interesting, namely, it is ~ 0.6 at all temperatures. The intensity ratio of S7 to S5 is ~ 0.1 and S7 to S6 is ~ 0.4 in the temperature range of 5 K to 140 K. This anomalously large intensity ratio even at low temperatures can only be understood by invoking the mixing of the multiphonon modes with the orbitons [28-29,52]. Figure 4.9 also shows temperature dependence of the frequency and the linewidth of the S8 mode which has yet to be understood quantitatively for the mixed multiphonon-orbiton mode. It will be interesting to explore the role of spin-charge-lattice coupling in understanding multiphonon Raman scattering in multiferroic TbMnO$_3$.

### 4.2.3 Conclusion

In summary, we have carried out a detailed temperature dependence of the first and second-order Raman modes in TbMnO$_3$. The intensity ratio of the second-order phonon (S8) to its first-order counterpart (S6) is unusually high and it remains constant down to 5 K. This anomalous temperature dependence of the intensity ratio is attributed to the mixing of the second-order phonons with the orbitons as theoretically predicted. Four first-order modes (S1, S2, S3 and S4) show normal behavior with temperature, whereas the S5 mode behaves anomalously below $T_N$ attributed to the strong spin-phonon coupling. We submit that the present study brings out yet another example of orbiton mediated multiphonon Raman scattering in the manganite family.

# 4.3 Part-C

# Temperature-Dependent Infrared Reflectivity Studies of Multiferroic TbMnO$_3$: Evidence for Spin-Phonon Coupling

## 4.3.1 Introduction

In this part we present our infrared reflectivity studies of single crystal of TbMnO$_3$ as a function of temperature from 10 K to 300 K. There have not been detailed temperature dependent studies on infrared (IR) phonons in RMnO$_3$, except the work of Paolone et al. [53] wherein they have reported IR mode frequencies only at 300 K and 10 K for undoped and doped LaMnO$_3$. From their tabulated data, it is seen that three modes in undoped and one mode in doped samples show lower frequencies at 10 K as compared to their values at 300 K. Here, we report a detailed temperature dependent study of infrared phonons in TbMnO$_3$ and uncover the role of spin-phonon coupling reflected in the anomalous temperature dependence of the observed phonon modes.

## 4.3.2 Experimental Details

Reflectance spectra of the unoriented single crystal of TbMnO$_3$ were obtained in near normal incidence geometry using a Fourier transform IR spectrometer (Bruker IFS 66v/s) in the frequency range of 50 to 700 cm$^{-1}$. We will focus on the temperature dependence of the infrared active phonons with frequencies above 150 cm$^{-1}$, because of high noise level below 150 cm$^{-1}$. Spectra were collected with an instrumental resolution of 2 cm$^{-1}$. The background reference signal was collected using a gold plated mirror. Temperature dependent reflectance spectra at temperatures ranging from 10 K to 300 K were collected by mounting the sample on the cold finger of a continuous flow liquid helium cryostat (Optistat CF-V, Oxford

Instruments) and the temperature was controlled to an accuracy of ± 0.1 K by a temperature controller (Oxford Instruments).

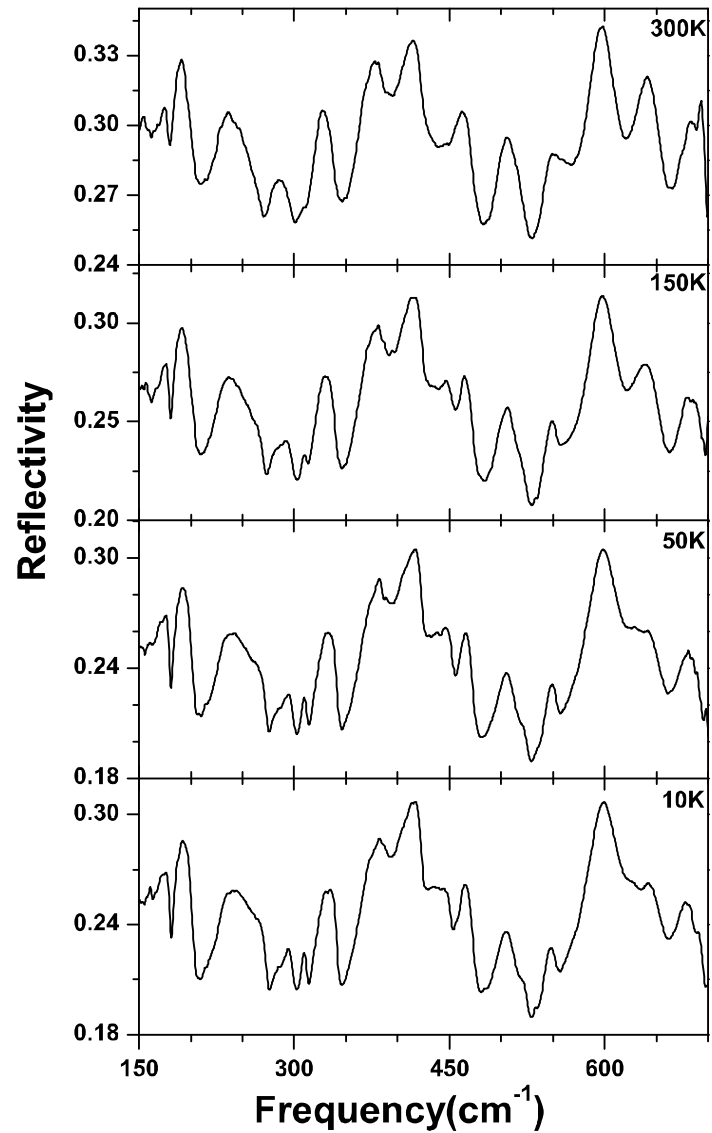

Figure 4.10: Temperature dependent infrared reflectivity of TbMnO$_3$ in the spectral range 150 cm$^{-1}$ to 700 cm$^{-1}$.

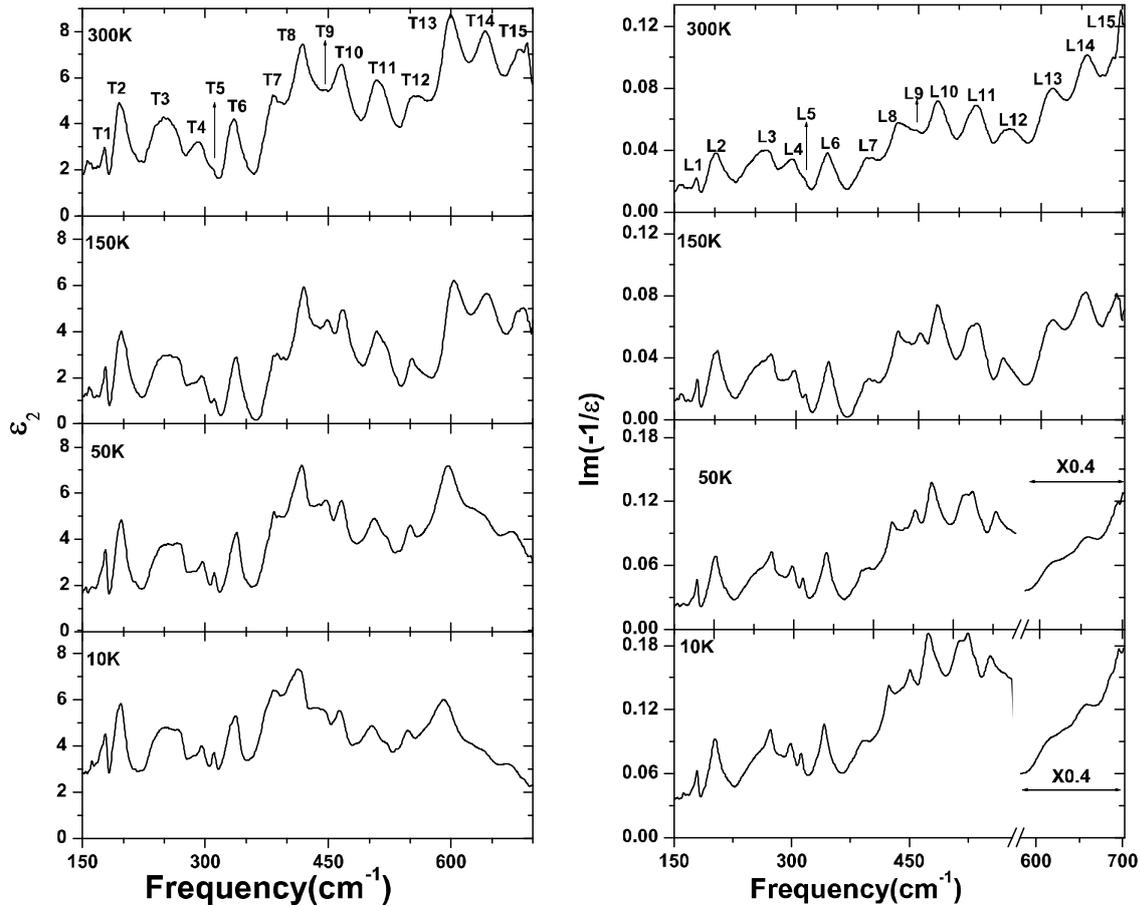

Figure 4.11: (Left Panel) Temperature dependence of the imaginary part of the dielectric function ($\varepsilon_2(\omega)$). (Right Panel) Evolution of the energy loss function (Im(-1/$\varepsilon(\omega)$)) with temperature.

### 4.3.3 Results and Discussion

The temperature dependent reflectivity of TbMnO$_3$ is shown in Fig. 4.10 for a few typical temperatures. The bands in the reflectivity spectra are due to IR active phonon modes. The total numbers of IR active phonon modes of TbMnO$_3$ (space group *Pbnm*), calculated using group theory are 25 and these are classified as $\Gamma_{IR} = 9B_{1u} + 7B_{2u} + 9B_{3u}$ [42]. The dielectric function, $\varepsilon(\omega)$, was obtained by Kramer-Kronig (KK) analysis of the reflectance spectra using the OPUS software (Bruker Optics). The extrapolation procedure used in KK analysis was as follows: reflectivity value R ($\omega$) below 50 cm$^{-1}$ is same as R (50 cm$^{-1}$) and R ($\omega >$ 700 cm$^{-1}$) = R ($\omega$ = 700 cm$^{-1}$). Peak positions in the imaginary part of the dielectric function ($\varepsilon_2(\omega)$) and energy loss function (Im (-1/$\varepsilon(\omega)$)) give the frequencies of the transverse optic (TO) and longitudinal optic (LO) modes, respectively. Temperature dependent $\varepsilon_2(\omega)$ shown in Fig. 4.11 (left panel) for a few typical temperatures clearly reveal fifteen TO modes, labeled as T1 to T15, apart from some weak modes which appear as shoulders (not labeled). It is likely that the band labeled as T3 can be a combination of two modes, as indicated by its lineshape at low temperatures. The corresponding LO modes are identified as L1 to L15 in Im(-1/$\varepsilon(\omega)$), shown in right panel of Fig. 4.11 for a few typical temperatures. We tried to fit $\varepsilon_2(\omega)$ and Im(-1/$\varepsilon(\omega)$) by a sum of fifteen Lorentizian like functions [54-55] to extract the peak positions, linewidth and oscillator strengths of the IR modes. This could not be done due to large number of parameters (15×3 + one for $\varepsilon(\infty)$ + base line parameter = 47). We have, therefore extracted only the peak positions by directly reading from the spectra of $\varepsilon_2(\omega)$ and Im(-1/$\varepsilon(\omega)$). These values are given in table-4.3 for 300 K to give an indication of the LO-TO splitting in TbMnO$_3$. The mode frequencies are close to the observed [53] and calculated [56] frequencies for the infrared phonons of LaMnO$_3$.

We will now discuss the temperature dependence of the LO and TO modes extracted from the reflectivity measurements. It is seen that the temperature dependence of nine modes (modes 1-3, 7, 9, 10, 11, 14 and 15) is very weak and normal i.e. the mode frequency increases by less than 1 or 2 cm$^{-1}$ as temperature is lowered to 10 K. Figure 4.12 and Fig. 4.13 show the temperature dependence of other modes 4, 5, 6, 8, 12 and 13 for both the LO and TO components which reveal significant temperature dependence. We have fitted the modes T4, T5, T6; L4, L5, L6 and L13 using equation (2.11). The solid lines, in Fig. 4.12 and Fig. 4.13, correspond to the fits by equation (2.11), with fitting parameters given in Table-4.3.

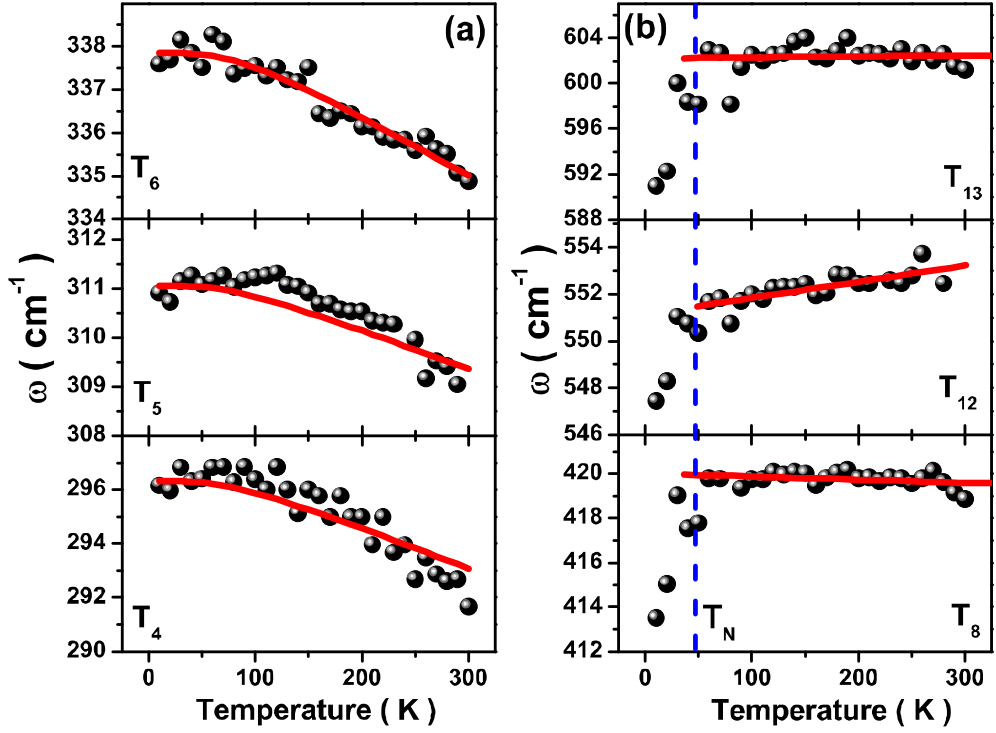

Figure 4.12: Temperature dependence of the TO modes T4, T5, T6, T8, T12 and T13. Solid lines for T4-T6 are the fitted curves as described in the text. The frequencies for the T8, T12 and T13 modes have been fitted by a linear relation above $T_N$.

The most interesting observation from our experiments is that the modes T8, T12, T13 (Fig. 4.12) and L12 (Fig. 4.13) show anomalous temperature dependence: the modes show softening below $T_N$ attributed to the spin-phonon coupling. Similar anomalous temperature dependence has been observed for a few Raman modes in $RMnO_3$ where R = La [57] and Gd, Pr, Nd, Sm, Dy [8, 42], which has been attributed to spin-phonon coupling [16]. We note that the phonon softening for the T13 mode is similar in magnitude to the $B_{2g}$ (~ 604 cm$^{-1}$) mode in $LaMnO_3$ [16] and hence the spin-phonon coupling constant ($\lambda$) is expected to be similar for both the modes.

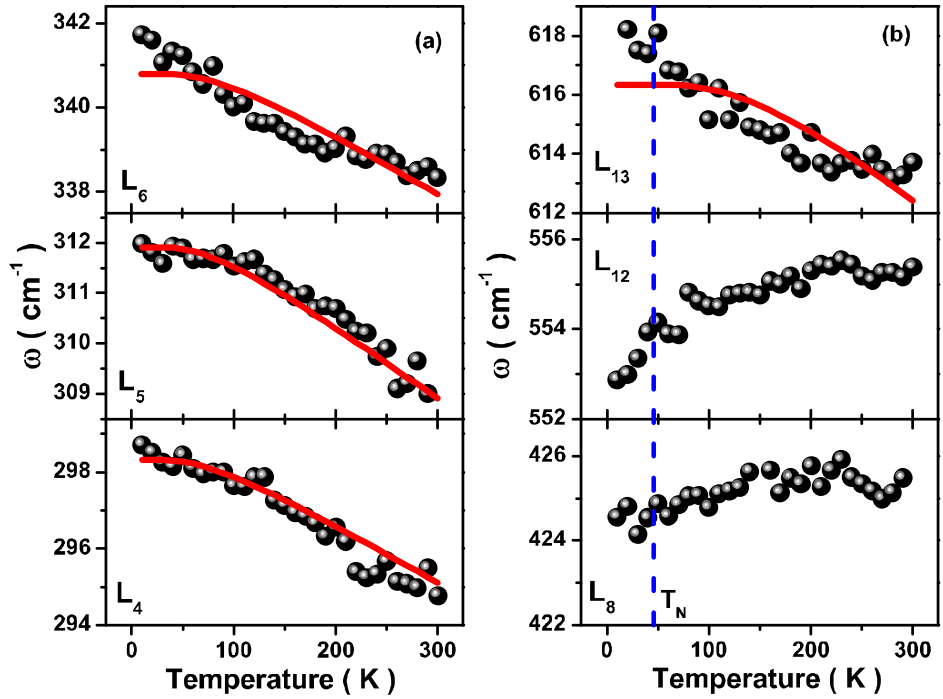

Figure 4.13: Temperature dependence of the LO modes L4, L5, L6, L8, L12 and L13. Solid lines for L4, L5, L6 and L13 are the fitted curves as described in the text.

Table-4.3: List of the experimentally observed phonon frequencies at 300 K and fitting parameters of a few phonons. Units are in cm$^{-1}$.

| Mode label no. | TO mode frequency $\omega(300)$ | $\omega_o$ | C | LO mode frequency $\omega(300)$ | $\omega_o$ | C |
|---|---|---|---|---|---|---|
| 1. | 176 | | | 177 | | |
| 2. | 197 | | | 201 | | |
| 3. | 252 | | | 260 | | |
| 4. | 291 | 298.2 ± 0.9 | -3.4 ± 1.4 | 295 | 300.6 ± 0.4 | -3.1 ± 1.2 |
| 5. | 309 | 312.4 ± 1.1 | -2.7 ± 0.6 | 309 | 313.6 ± 1.2 | -2.1 ± 0.8 |
| 6. | 335 | 339.6 ± 1.3 | -3.6 ± 1.4 | 338 | 342.6 ± 0.4 | -3.5 ± 0.8 |
| 7. | 384 | | | 387 | | |
| 8. | 419 | | | 425 | | |
| 9. | 446 | | | 449 | | |
| 10. | 465 | | | 474 | | |
| 11. | 511 | | | 520 | | |
| 12. | 552 | | | 555 | | |
| 13. | 601 | | | 614 | 623.2 ± 1.1 | -3.6 ± 0.9 |
| 14. | 641 | | | 657 | | |
| 15. | 686 | | | 697 | | |

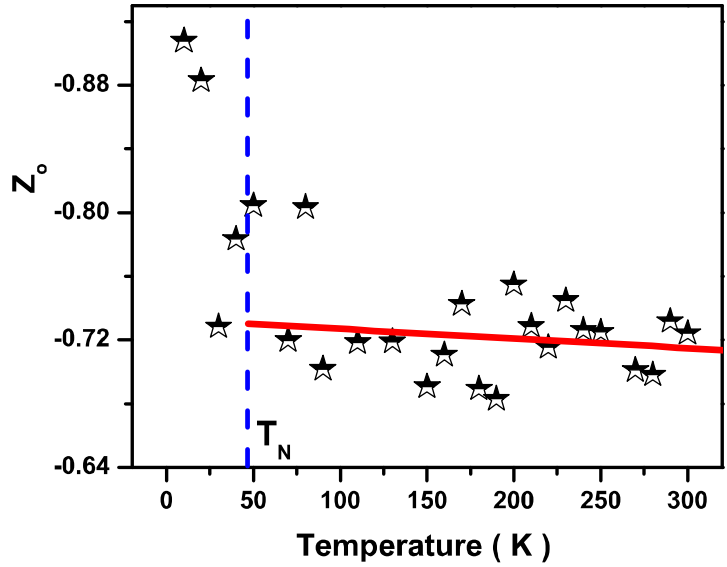

Figure 4.14: Temperature dependence of the effective charge of oxygen ($Z_O$). Solid line shows a linear fit in the temperature range above $T_N$.

We now comment on the temperature dependence of the frequency difference between the LO-TO modes which determines the effective charge of the ions in the lattice. The effective charge of Tb, Mn and O ions in the unit cell with $k$ atoms can be estimated using the relation [58]:

$$\Sigma_k (Z^2_k/M_k) = (\pi V) * \Sigma_j ( \omega^2_{LO,j} - \omega^2_{TO,j} ) \tag{4.6}$$

where V is the unit cell volume, j is an index for the lattice modes and $Z_k$ is the effective charge of the $k^{th}$ ion with mass $M_k$. For the effective charges, there is a sum rule for charge neutrality i.e. $\Sigma_k (Z_k) = 0$. We note that the effective charge calculated using the above equation is not same as the ionic charge, the Szigeti charge or the nominal valence charges [59]. In $TbMnO_3$, where the masses of other constituents are much heavier than the oxygen mass, we can neglect the terms other than the terms for oxygen on the left hand side of equation (4.6). The calculated value of the effective charge of oxygen from equation (4.6) is the average for all the oxygen sites. Using the observed values of the TO and the LO frequencies in the equation (4.6) the temperature dependence of the oxygen effective charge ($Z_O$) is calculated, as shown in Fig. 4.14. The absolute value of $Z_O$ is not crucial, what is more important is its change with temperature. We find that $Z_O$ increases below $T_N$. If $Z_O$ increases, then the induced dipole moment and hence the optical absorption will also increase. The increase in $Z_O$ below $T_N$ suggests a change in the bond length between $Mn^{3+}$ cations and the corresponding bond angles, mediated by the O-ions. This result needs to be understood better.

### 4.3.4 Conclusion

In conclusion, infrared reflectivity measurements of a single crystal of $TbMnO_3$ clearly identify fifteen IR active phonon modes. Out of these, three modes show anomalous behavior below the magnetic phase transition temperature $T_N$ which is attributed to spin-phonon coupling. The effective charge of the oxygen ions shows an increase below $T_N$. We hope that our results on phonon softening and an increase in the effective charge of oxygen ions will motivate theoretical studies of phonons in $TbMnO_3$ and their role in the magneto-electric behavior.

## 4.4 Part-D

# Temperature Dependent Magnetic, Dielectric and Raman Studies of Partially Disordered La$_2$NiMnO$_6$

### 4.4.1 Introduction

There is a recent upsurge of interest in the double-perovskite, La$_2$NiMnO$_6$ (LNMO) due to its rich physics, particularly large dielectric anomaly and the related potential applications [60-65]. Much effort has recently been devoted to a better understanding of the coupling between the magnetic, phononic and electronic degrees of freedom (DOF) because an intricate interplay between these DOF is believed to be responsible for the various novel physical phenomena observed in this system [60]. Magnetization and neutron diffraction studies have established LNMO to be a ferromagnet at near room temperature ($T_C$ ~ 270 K) [66]. It has a monoclinic ($P2_1/n$) structure at ambient temperature and a rhombohedral ($R$-$3$) structure at high temperatures [62,67]. Often, these two phases coexist at room temperature with the majority phase being the monoclinic. However, one can obtain a pure monoclinic phase under stringent synthesis conditions which not only determines the formation of these phases but also controls the formation of lanthanum and oxygen vacancies and anti-site disorder [68-70]. Presence of dielectric anomaly and a large magneto-capacitance (MC) effect has been reported in the temperature interval, 220 K - 280 K near magnetic transition temperature ($T_c$ ~ 280 K) in the highly ordered but biphasic LNMO system [60]. A multiglass behavior and MC effect over a wide range of temperature (150 K - 300 K) have been observed in the partially disordered single phase monoclinic system where the latter has been explained to be due to asymmetric hopping of electrons between Ni$^{2+}$ and Mn$^{4+}$ ions. It is to be noted that dielectric studies on pure rhombohedral phase of LNMO is not known. Meanwhile, an extrinsic magneto-dielectric effect has been reported in nanoparticles of monoclinic LNMO [71].

Further, anti-site disorder between $Ni^{2+}$ and $Mn^{4+}$ ions leads to $Mn^{4+}$-$O^{2-}$-$Mn^{4+}$ and $Ni^{2+}$-$O^{2-}$-$Ni^{2+}$ antiferromagnetic coupling resulting in a spin-glass behavior in the vicinity of 70 K ($T_{sg}$). However, the dominant $Ni^{2+}$-$O^{2-}$-$Mn^{4+}$ super-exchange coupling of the ordered structure remains ferromagnetic below the transition temperature (~ 270 K).

The coupling of soft infrared phonons with Ni and Mn spins via super-exchange interactions has been shown to occur in the rhombohedral phase using first-principles calculations [72]. These soft phonon modes were suggested to be responsible for the observed dielectric anomaly as well as the MD effect near $T_c$ in the multiphase sample [60]. It should be noted that no dielectric anomaly has been reported near $T_c$ in a pure monoclinic phase [65,68-69,71,73]. Since there are no calculations for the monoclinic phase and the fact that the dielectric anomaly is observed only in a multiphase sample, it becomes necessary to understand the origin of dielectric anomaly which may be related to rhombohedral phase or the multiphase nature of the sample.

A significant magneto-dielectric effect in double-perovskite systems imply a strong coupling between the lattice and the magnetic degrees of freedom [64,74]. Therefore, it becomes important to study the role of phonons as a function of temperature using Raman spectroscopy which is a well proven powerful technique to investigate the spin-phonon coupling, charge/orbital ordering, long-range cation ordering and other structural changes in perovskite oxides [10,16,75-77]. Main focus of the already reported studies in the literature has been the first-order Raman modes [61,63,78,79-83] as a function of temperature showing the signatures of spin-phonon coupling; in particular a phonon mode associated with the symmetric stretching vibration of (Ni/Mn)$O_6$ octahedra shows continuous softening down to the lowest temperature in the ferromagnetic phase.

Here, we present our detailed magnetic, dielectric and Raman studies on partial disordered and biphasic LNMO in the temperature range 12 K to 300 K. We observe a broad dielectric anomaly centered around 270 K. The observed first-order symmetric stretching mode (S2) as well its overtone mode (S5) shows anomalous behaviour due to strong spin-phonon coupling. In addition, the anti-symmetric stretching vibration (S1) of the oxygen octahedra shows an anomaly near a temperature associated with spin-glass transition ~ 70 K, as inferred from magnetic susceptibility measurements. The anomalous temperature dependence of the integrated intensity ratio of second to first-order phonon modes evidence the occurring of conventional resonant Raman scattering.

## 4.4.2 Experimental Details

Polycrystalline samples of the composition $La_2NiMnO_6$ were prepared by the standard solid state reaction method by mixing the individual oxides and heating the mixture at high temperatures (900˚C). Final heating was carried at 1400˚C and the sample was cooled slowly (1˚/min) to room temperature. The sample was characterized by *x*-ray diffraction method using the Bruker D8 Discover diffractometer with Cu Kα radiation. Magnetization measurements were carried out using a Superconducting Quantum Interference Device combined with Vibrating Sample (SQUID VSM) and AC susceptibility measurements were performed utilizing the ACMS option in a Physical Property Measuring System (PPMS), Quantum Design, USA. Dielectric measurements were performed using the E4980 Agilent LCR meter using the multifunctional probe utilizing the cryo system of PPMS, with silver paste used as electrodes. Raman spectroscopic measurements were performed at low temperatures in similar ways as described in earlier parts.

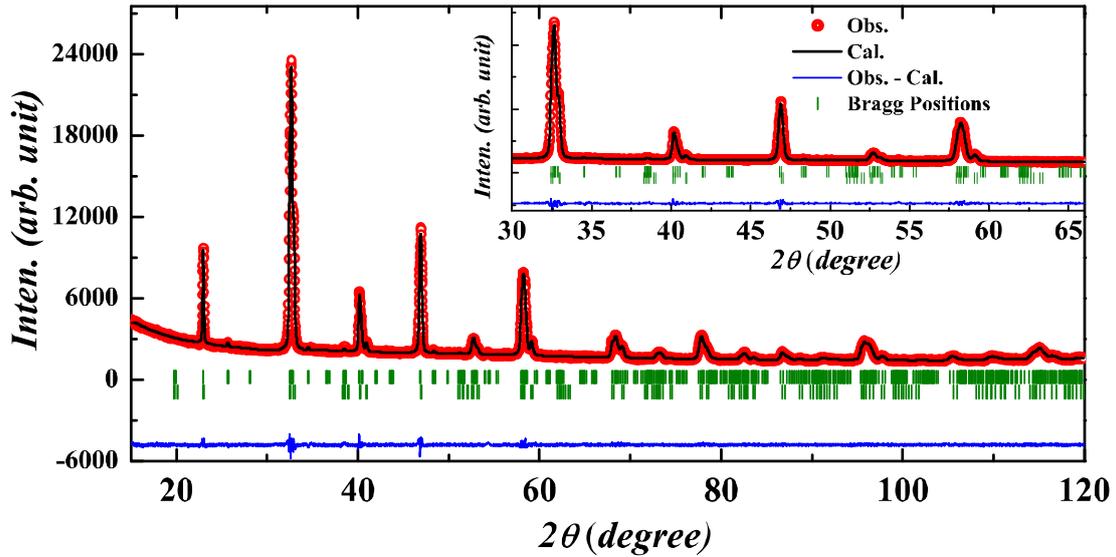

Figure 4.15: Rietveld refinement on *x*-ray diffraction data of partially disordered polycrystalline La$_2$NiMnO$_6$. Inset shows the biphasic nature of the sample.

### 4.4.3 Results and Discussion

Figure 4.15 shows the Rietveld refinement of *x*-ray diffraction data at room temperature. The inset shows the data in a narrow angle range showing the biphasic nature of the sample, fitted to give the coexistence of both the phases (58% (P2$_1$/n) and 42% (R-3)). Figure 4.16 shows the dc magnetization (M) as a function of temperature for zero field cooled (ZFC) and field cooled (FC) at a low field (0.01 T). It could be observed from ZFC as well as FC magnetization that the material shows two ferromagnetic transitions around 270 K and 240 K which is consistent with the presence of two phases as revealed by the *x*-ray diffraction studies. As the ferromagnetic transition temperature in the pure monoclinic phase is found to be near 270 K [66], we attribute the ferromagnetic transition at 240 K to the rhombohedral phase. The saturation magnetic moment obtained from M(H) measurement at 2 K (see inset of Fig. 4.16) is 3.571 μ$_B$/formula unit (f.u.) which signifies anti-site disorder of Ni$^{2+}$ and Mn$^{4+}$ ions.

To probe the nature of magnetism in more detail and to also bring out the relevance of the two phases and their impact on the magnetism, AC susceptibility (χ) measurements were performed at various frequencies ranging from 50 Hz to 500 Hz as shown in Fig. 4.17(a,b). Two significant features can be noticed clearly from the real part ($\chi'$) as well as imaginary part ($\chi''$). There is a frequency-dependent anomaly in the vicinity of 60 K in $\chi''$ (see inset of Fig. 4.17(b)) and a double peak feature in $\chi''$ near the ferromagnetic transition temperatures 240 K and 270 K. There is a frequency dispersion in $\chi''$ below $T_c$ but the peak positions do not shift in temperature with frequency. This may result from the frustration caused by ferromagnetic interactions arising from $Ni^{2+}$ - O - $Mn^{4+}$ coupling and antiferromagnetic interactions due to anti-site disorder. The frequency-dependent cusp in the low temperature regime can be attributed to the spin-glass behavior as reported earlier [69].

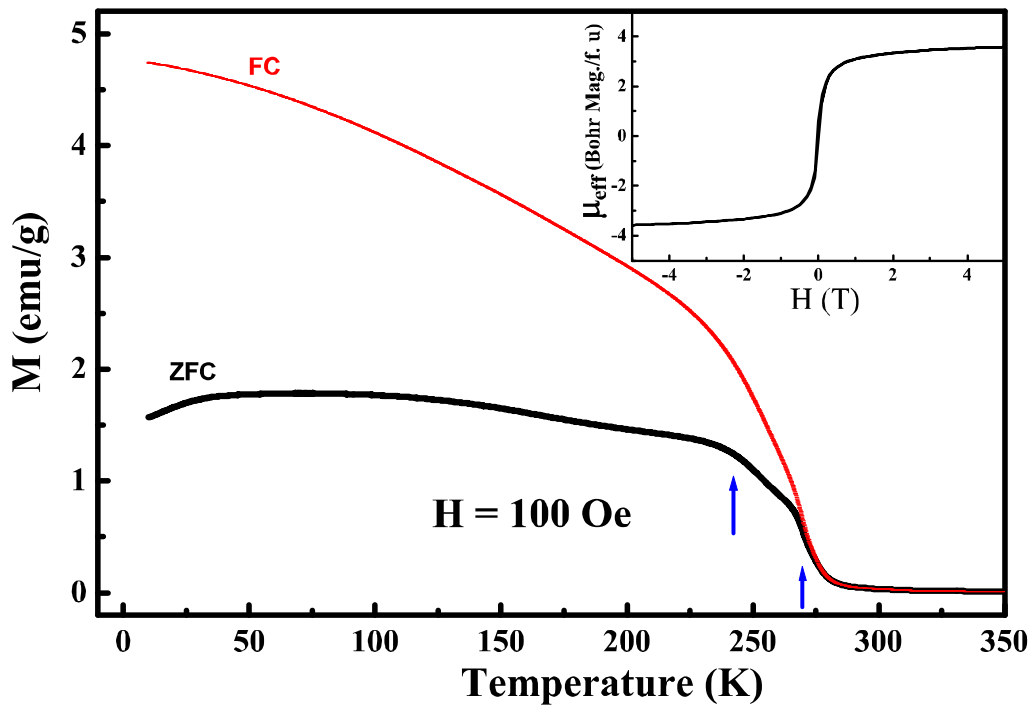

Figure 4.16: Zero-Field-Cooled (ZFC) and Field-Cooled (FC) magnetization as a function of temperature in La$_2$NiMnO$_6$. Inset shows the $\mu_B$/(f.u.); the presence of anti-site disorder can be inferred from the same.

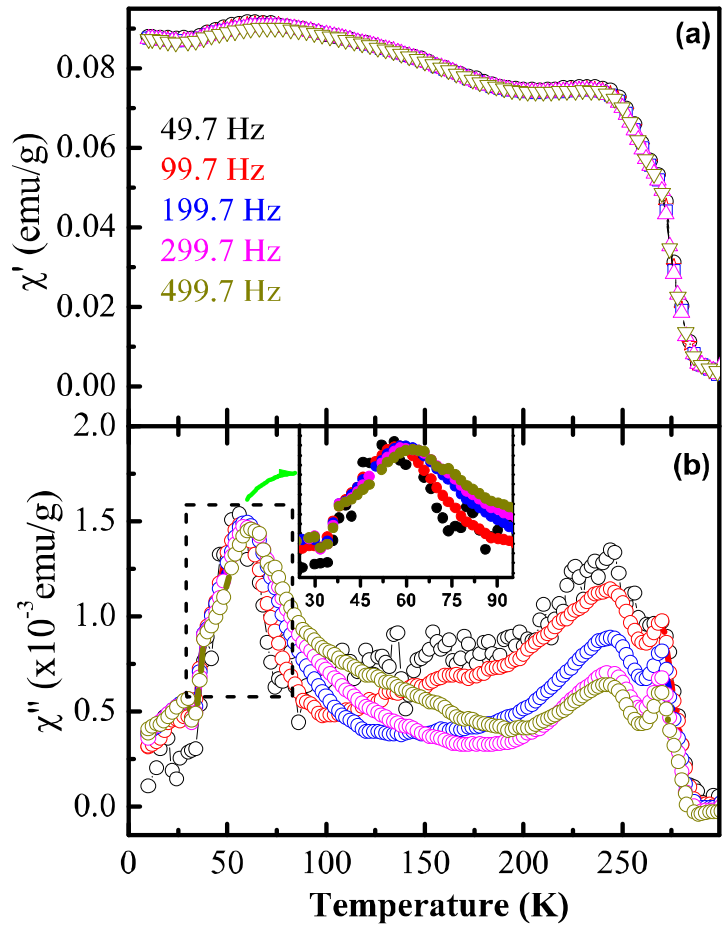

Figure 4.17: (a) Real ($\chi'$) and (b) Imaginary ($\chi''$) part of AC magnetic susceptibility as a function of temperature at various frequencies. Inset in (b) shows the frequency dependenc of the peak around 60 K in a narrow temperature range.

Figure 4.18 shows the frequency-dependence of the real part of dielectric constant ($\varepsilon_r$) and loss ($\tan\delta$) in the temperature range 200 K - 320 K. Dielectric constant is relatively independent of temperature up to 100 K (not shown here), above which it undergoes a Maxwell-Wagner like relaxation [69]. The important feature to note in the present study is the observation of a broad hump in the real part of dielectric constant as well as an anomaly in the loss in the vicinity of $T_{C1}$ which are not seen in the pure monoclinic samples [69,71]. Moreover, it is quite different from the first report of magneto-capacitance in La$_2$NiMnO$_6$ where a step like anomaly was reported at 220 K [60]. As there is no dielectric anomaly in the pure monoclinic sample, we suggest that the dielectric anomaly observed in the present study may be related to the biphasic nature of the sample.

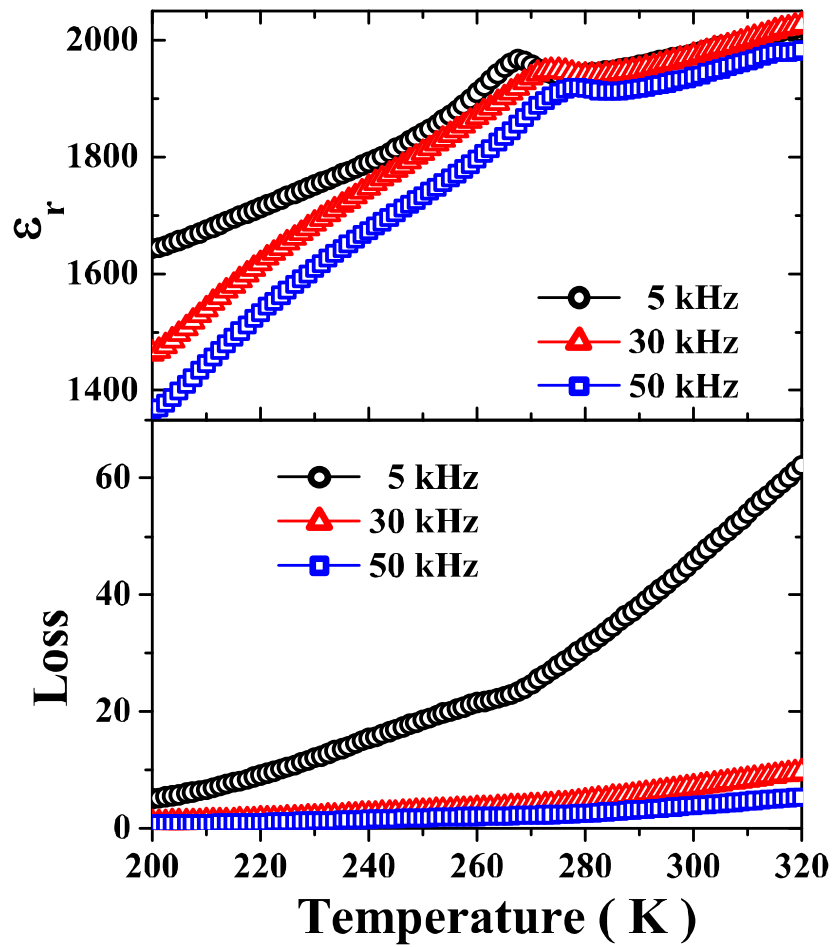

Figure 4.18: Frequency dependence of (a) dielectric constant and (b) loss tangent as a function of temperature.

Figure 4.19 shows the Raman spectrum at 12 K displaying five modes labeled as S1 to S5. The spectra are fitted to a sum of Lorentzians functions. Following the previous studies [61,63,80] two strong modes S1 (~ 530 cm$^{-1}$) and S2 (~ 668 cm$^{-1}$) are assigned to the anti-symmetric and symmetric stretching vibration of the (Ni/Mn)O$_6$ octahedra, respectively. In addition to the first-order Raman bands, we observe three high frequency bands around 1068 (S3), 1206 (S4) and 1324 (S5) cm$^{-1}$. Similar high frequency modes have also been reported for LNMO thin films but only at room temperature [63], where two broad peaks observed near 1210 and 1345 cm$^{-1}$ were attributed to the multiphonon Raman scattering. Mode S3 and S5 can be assigned as the overtone of modes S1 and S2, respectively, and mode S4 as a combination of mode S1 and S2. As second-order phonon scattering involves the phonons over the entire Brillion-zone, therefore the frequencies of the observed second-order phonons are not necessarily double of the first-order phonons at the $\Gamma$ - point. It can be seen from Fig. 4.19 that the two strong first-order modes observed near 530 cm$^{-1}$ and 668 cm$^{-1}$ are slightly asymmetric, which can be explained following the work of Bull et al. [61] as follows. First, the Ni and Mn sites may not be completely ordered, and the different Ni/Mn-O stretching vibrations lie close in frequency, so that the unresolved contributions from different NiO$_6$ and MnO$_6$ environment may be present within the band envelope. Second, the domains with different degrees of Ni and Mn order will result in different contributions to the Ni/Mn-O stretching bonds.

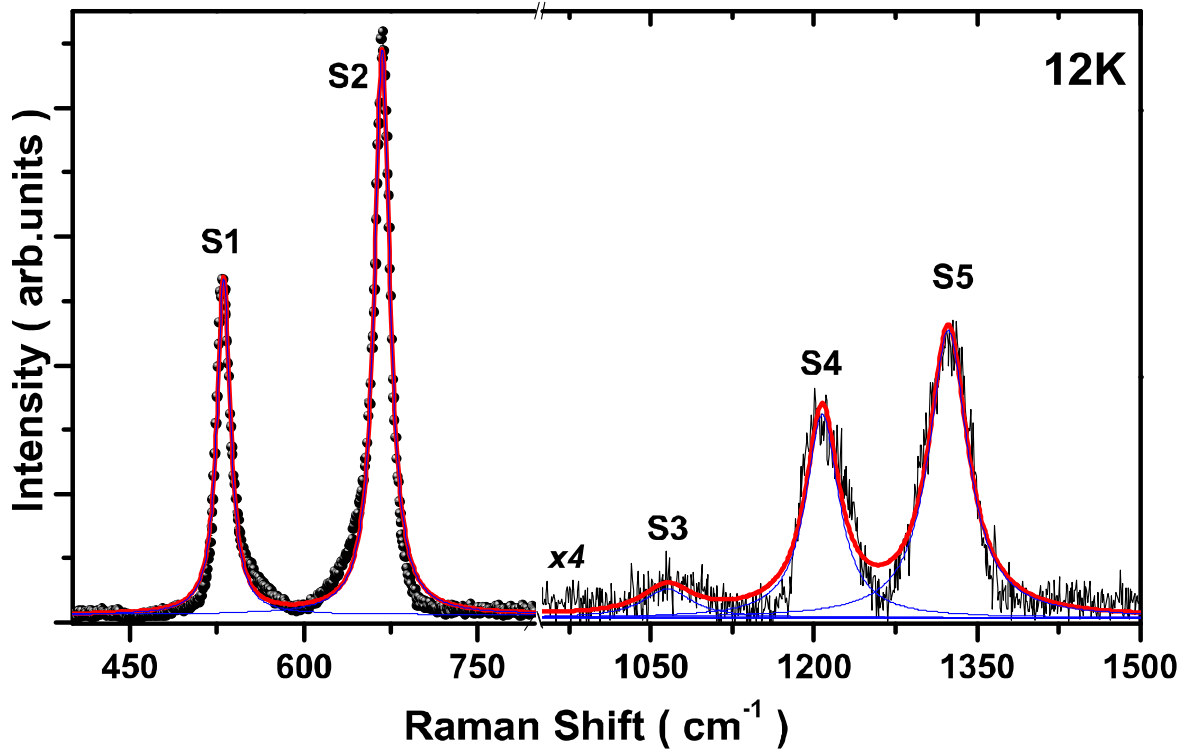

Figure 4.19: Raman spectra of La$_2$NiMnO$_6$ measured at 12 K. Thick solid line shows the total fit, thin solid lines show the individual Lorentzian fit.

In Fig. 4.20 we have plotted the mode frequency and linewidth for the four prominent modes, i.e. S1, S2, S4 and S5, as a function of temperature. The following observations can be made: (i) The symmetrical stretching mode (S2) shows an anomalous softening below paramagnetic-to-ferromagnetic transition temperature ($T_{C1} \sim 270$ K) down to 12 K. The solid line is a linear fit with a slope 0.0156 cm$^{-1}$/K. We note that similar softening below $T_{C1}$ has been reported for the thin films and bulk samples of LNMO and LCMO [78,80,83-84] attributed to the strong spin-phonon coupling in the ferromagnetic phase in line with the earlier experimental as well theoretical studies on similar systems [8,16,19,76-77,80,82]. (ii) In comparison, the temperature dependence of the anti-symmetric stretching mode (S1), being reported for the first time is different: the mode frequency increases with decrease in temperature due to anharmonic effects till 70 K, but shows anomalous softening near $T_{sg}$. The solid lines are linear fits in the two temperature ranges. (iii) The linewidth of the modes S1 and S2 shows normal temperature dependence i.e. linewidth decreases with decrease in temperature, however below $T_{sg}$ the linewidth of S2 mode shows a small anomalous increase, coinciding with the onset of spin-glass state. The solid lines in right panel show a fit to a simple model of cubic anharmonic contribution to the linewidth (see eq. 2.10) described in detail in Chapter 2. The fitted parameters for the modes S1 and S2 are $\Gamma_0^{S1} = 4.5 \pm 0.7$ cm$^{-1}$, $D^{S1} = 9.4 \pm 1.2$ cm$^{-1}$, $\omega_0^{S1} = 534.8 \pm 0.3$ cm$^{-1}$ and $\Gamma_0^{S2} = 6.1 \pm 0.8$ cm$^{-1}$, $D^{S2} = 11.3 \pm 0.9$ cm$^{-1}$, $\omega_0^{S2} = 666.4 \pm 0.5$ cm$^{-1}$, respectively. (iv) The frequencies of the second-order modes S4 (combination of S1 and S2) and S5 (overtone of S2) also show anomalous temperature-dependence, similar to the first-order S2 mode. Solid line for the mode S5 is a linear fit with slope 0.0125 cm$^{-1}$/K. Anomalous temperature dependence of the modes S2, S4 and S5 confirm clear signature of strong spin-phonon coupling. The temperature dependence of the mode S1 shows signature of a change in magnetic ordering below $T_{sg}$.

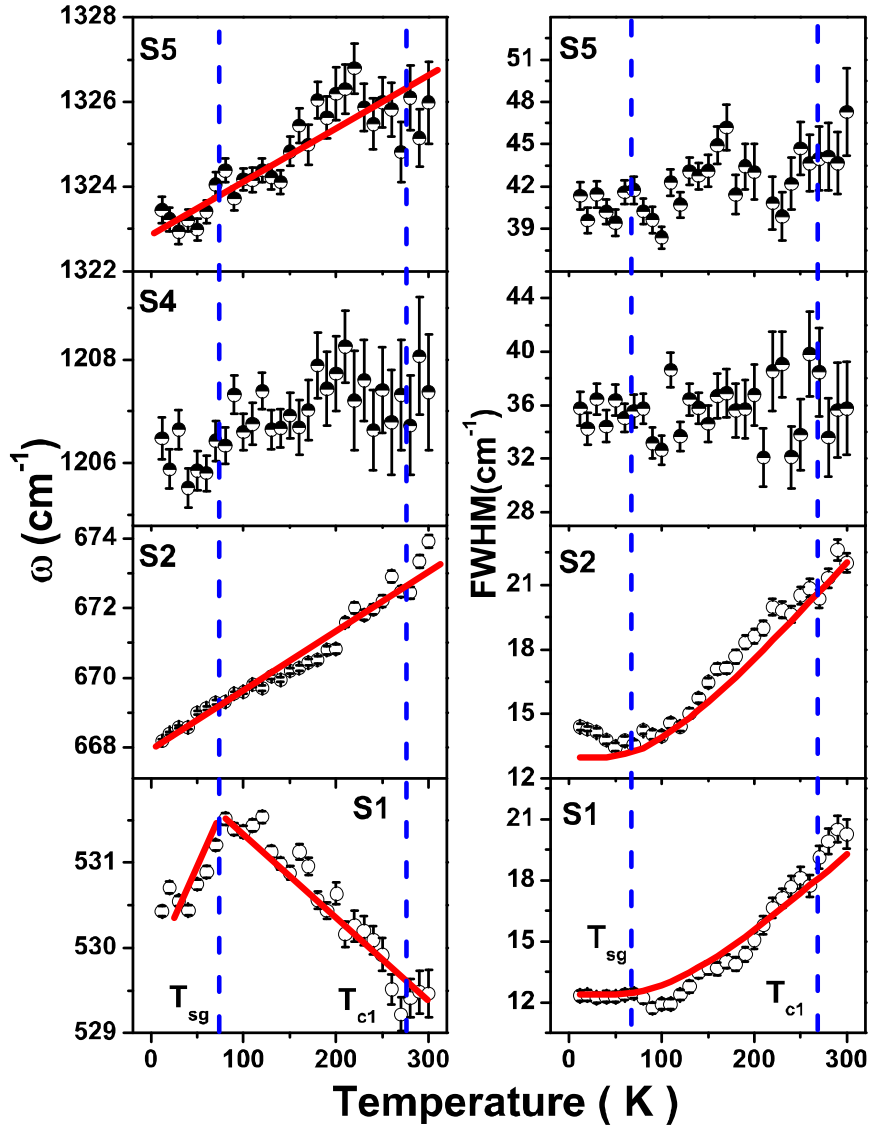

Figure 4.20: Temperature dependence of the modes S1, S2, S4 and S5. The solid lines in left panel for the modes S1, S2 and S5 are linear fits. Solid lines in right panel for the mode S1 and S2 are fitted curves as described in the text.

Figure 4.21 shows the temperature dependence of the integrated-intensity ratio of the second-order modes S5 and S4 with respect to their first-order counter part mode S2. It can be seen that the intensity ratios increase with the decreasing temperature. We note that the anomalous increase in intensity with decreasing temperature does not follow the normal Raman process, where the intensity ratio should have increased with the increasing temperature due to phonon population. The Franck-Condon mechanism should give a constant intensity ratio [28]. The observed intensity ratio can arise due to resonant Raman scattering [33,35,85]. The relevant energy levels involved in resonant Raman scattering using a laser photon energy of 2.41 eV are the Ni and Mn *3d* bands, as shown by the recent DFT calculations [72].

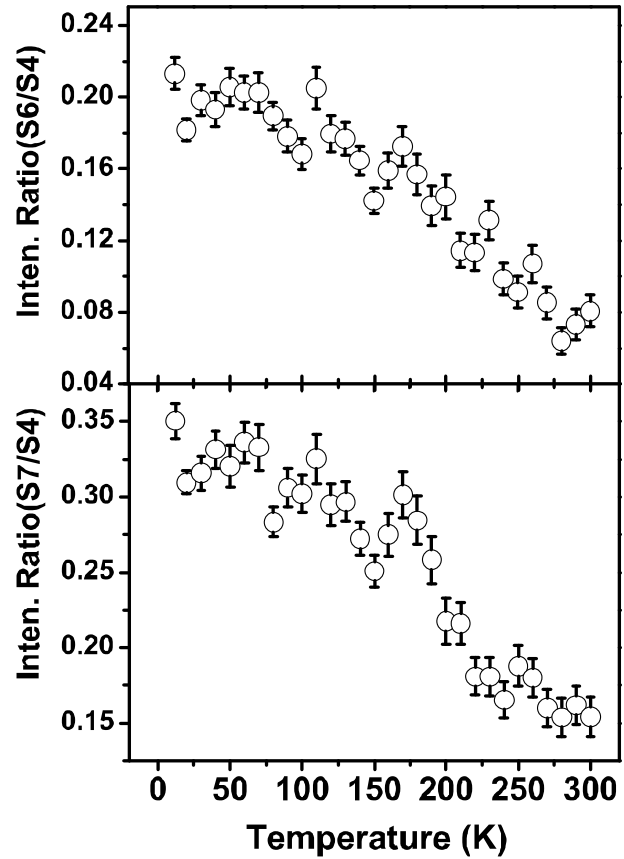

Figure 4.21: Temperature dependence of the integrated-intensity ratio of the mode S5 to S2 and S4 to S2.

## 4.4.4 Conclusion

In conclusion, we observe a dielectric anomaly in partially disordered and biphasic $La_2NiMnO_6$ around 270 K. We also find anomalies in the AC susceptibility near Curie temperature as well as at a lower temperature (~70 K) which has been understood to be due to the emergence of the spin-glass state. The temperature dependent Raman measurements reveal a strong spin-phonon coupling as reflected in the anomalous temperature dependence of the phonon modes. The anomalous increase in the intensity of the second-order phonons with decreasing temperature is attributed to the resonant Raman scattering. Our results obtained here suggest that the interplay between phononic and magnetic degrees of freedom is crucial to understand the underlying physics of double perovskites.

# Bibliography:

[1] W. Eerenstein, N. D. Mathur and J. F. Scott, Nature **442**, 759 (2006).

[2] W. Prellier, M. P. Singh and P. Murugavel, J. Phys. Cond. Matt. **17**, R 803 (2005).

[3] S. W. Cheong and M. Mostovoy, Nature Mater. **6**, 13 (2007).

[4] C. N. R. Rao and C. R. Serrao, J. Mater. Chem. **17**, 4931 (2007).

[5] D. Khomski, Phyics **2**, 20 (2009).

[6] J. H. We, S. J. Kim and C. S. Kim, IEEE Tran. Mag. **42**, 2876 (1991).

[7] F. Bouree, J. L. Baudour, E. Elbadraoui, J. Musso, C. Laurent and A. Rousset, Acta Cryst. B **52**, 217 (1996).

[8] J. Laverdiere, S. Jandl, A. A. Mukhin, V. Yu. Ivanov, V. G. Ivanov and M. N. Iliev, Phys. Rev. B **73**, 214301 (2006).

[9] M. O. Ramirez, M. Krishnamurthi, S. Denev, A. Kumar, S.-Y. Yang, Y.-H. Chu, E. Saiz, J. Seidel, A. P. Pyatakov, A. Bush, D. Viehland, J. Orenstein, R. Ramesh and V. Gopalan, Appl. Phys. Lett. **94**, 161905 (2009).

[10] P. Kumar, S. Saha, D. V. S. Muthu, J. R. Sahu, A. K. Sood and C. N. R. Rao, J. Phys. Cond. Matt. **22**, 115403 (2010).

[11] M. Viswanathan, P. S. Kumar, V. S. Bhadram, C. Narayana, A. K. Bera and S. M. Yusuf, J. Phys. Cond. Matt. **22**, 346006 (2010).

[12] K. Sharma, V. R. Reddy, D. Kothari, A. Gupta, A. Banerjee and V. G. Sathe, J. Phys. Cond. Matt. **22,** 146005 (2010).

[13] S. Mukherjee, R. Gupta and A. Garg, J. Phys. Cond. Matt. **23,** 445403 (2011).

[14] A. Shireen, R. Saha, S. N. Shirodkar, U. V. Waghmare, A. Sundaresan and C. N. R. Rao, arXiv:1112.5848v1.

[15] R. Saha, A. Ajmala, A. K. Bera, S. N. Shirodkar, Y. Sundarayya, N. Kalarikkal, S. M. Yusuf, U. V. Waghmare, A. Sundaresan and C. N. R. Rao, J. Solid State Chem. **184**, 494 (2011).

[16] E. Grando, A. Garcia, J. A. Sanjurjo, C. Rettori, I. Torriani, F. Prado, R. D. Sanchez, A. Caneiro and S. B. Oseroff, Phys. Rev. B **60**, 11879 (1999).

[17] A. P. Litvinchuk, M. N. Iliev, V. N. Popov and M. M. Gospodinov, J. Phys. Cond. Matt. **16,** 809 (2004).

[18] H. Fukumara, N. Hasuike, H. Harima, K. Kisoda, K. Fukae, T. Yoshimura and N. Fujimura, J. Phys. Cond. Matt. **21,** 064218 (2009).

[19] P. Kumar, S. Saha, C. R. Serrao, A. K. Sood and C. N. R. Rao, Pramana J. Phys. **74**, 281

(2010).

[20] M. J. Massey, U. Baier, R. Merlin and W. H. Weber, Phys. Rev. B **41**, 7822 (1990).

[21] S. J. Allen and H. J. Guggenheim, Phy. Rev. Lett. **21**, 1807 (1968).

[22] I. W. Shepherd, Phy. Rev. B **5**, 4524 (1972).

[23] M. O. Ramirez, M. Krishnamurthi, S. Denev, A. Kumar, S.-Y. Yang, Y.-H. Chu, E. Saiz, J. Seidel, A. P. Pyatakov, A. Bush, D. Viehland, J. Orenstein, R. Ramesh and V. Gopalan, Appl. Phys. Lett. **92**, 022511 (2008).

[24] J. P. Perdew, J. A. Chevary, S. H. Vosko, K. A. Jackson, M. R. Pederson, D. J. Singh and C. Fiolhais, Phys. Rev. B **46**, 6671 (1992).

[25] G. Kresse and J. Hafner, Phys. Rev. B **47**, R558 (1993).

[26] G. Kresse and J. Furthmller, Phys. Rev. B **54**, 11 169 (1996).

[27] G. Kresse and D. Joubert, Phys. Rev. B **59**, 1758 (1999).

[28] P. B. Allen and V. Perebeinos, Phys. Rev. Lett. **83**, 4828 (1999).

[29] V. Perebeinos and P. B. Allen, Phys. Rev. B **64**, 085118 (2001).

[30] E. Saito, S. Okamoto, K. T. Takahashi, K. Tobe, K. Yamamoto, T. Kimura, S. Ishihara, S. Maekawa and Y. Tokura, Nature **410**, 180 (2001).

[31] J. Vanden Brink, Phys. Rev. Lett. **87**, 217202 (2001).

[32] R. Kruger, B. Schulz, S. Naler, R. Rauer, D. Budelmann, J. Backstrom, K. H. Kim, S. W. Cheong, V. Perebeinos and M. Rubhausen, Phys. Rev. Lett. **92**, 097203 (2004).

[33] L. Martin-Carron and A. de Andres, Phys. Rev. Lett. **92**, 175501 (2004).

[34] K. Y. Choi, Phys. Rev. Lett. **71**, 174402 (2005).

[35] K. Y. Choi, P. Lemmens, G. Guntherodt, Y. G. Pashkevich, V. P. Gnezdilov, P. Reutler, L. P. -Gaudart, B. Buchner and A. Revcolevschi, Phys. Rev. B **72**, 024301 (2005).

[36] P. B. Allen and V. Perebeinos, Nature **410**, 155 (2001).

[37] M. Gruninger, R. Ruckamp, M. Windt, P. Reutler, C. Zobel, T. Lorenz, A. Freimuth and A. Revcolevschi, Nature **418**, 39 (2002).

[38] J. Bala, A. M. Oles and G. A. Sawatzky, Phys. Rev. B **65**, 184414 (2002).

[39] M. N. Iliev, V. G. Hadjiev, A. P. Litvinchuk, F. Yen, Y.-Q. Wang, Y. Y. Sun, S. Jandl, J. Laverdiere, V. N. Popov and M. M. Gospodinov, Phys. Rev. B **75**, 064303 (2007).

[40] T. Kimura, T. Goto, H. Shintani, K. Ishizaka, T. Arima and Y. Tokura, Nature **426**, 55 (2003).

[41] M. Bastjan, S. G. Singer, G. Neuber, S. Eller, N. Aliouane, D. N. Argyriou, S. L. Cooper and M. Rubhausen, Phys. Rev. B **77,** 193105 (2008).

[42] W. S. Ferreira, J. A. Moreira, A. Almeida, M. R. Chaves, J. P. Araujo, J. B. Oliveira, J.

M. M. Da Silva, M. A. Sa, T. M. Mendonca, P. S. Carvalho, J. Kreisel, J. L. Ribeiro, L. G. Vieira, P. B. Tavares and S. Mendonca, Phys. Rev. B **79**, 054303 (2009).

[43] L. Martin-Carron, A. de Andres, M. J. Martinez-Lope, M. T. Casais and J. A. Alonso, Phys. Rev. B **66**, 174303 (2002).

[44] M. N. Iliev, M. V. Abrashev, J. Laverdiere, S. Jandl, M. M. Gospodinov, Y.-Q. Wang and Y.-Y. Sun, Phys. Rev. B **73**, 064302 (2006).

[45] L. Martin-Carron, J. S. Benitz and A. de Andres, J. Solid State Chem. **171**, 313 (2003).

[46] V. B. Podobedov, A. Weber, D. B. Romero, J. P. Rice and H. D. Drew, Phys. Rev. B **58**, 43 (1998).

[47] M. V. Abrashev, A. P. Litvinchuk, M. N. Iliev, R. L. Meng, V. N. Popov, V. G. Ivanov, R. A. Chakalov and C. Thomson, Phys. Rev. B **59**, 4146 (1999).

[48] M. N. Iliev, M. V. Abrashev, V. N. Popov and V. G. Hadjiev, Phys. Rev. B **67**, 212301 (2003).

[49] R. Choithrani, N. K. Gaur and R. K. Singh, J. Alloys and Compounds **480**, 727 (2009).

[50] R. Choithrani, N. K. Gaur and R. K. Singh, J. Phys. Cond. Matt. **20**, 415201 (2008).

[51] D. Meier, N. Aliouane, D. N. Argyriou, J. A. Mydosh and T. Lorenz, New J. Phys. **9**, 100 (2007).

[52] Light scattering in solids II, Topics in applied physics vol. 50, edited by M. Cardona (Springer-Verleg, Berlin, 1982).

[53] A. Paolone, P. Roy, A. Pimenov, A. Loidl, O. K. Melnikev and A. Y. Shapiro, Phys. Rev. B **61,** 11255 (2000).

[54] C. C. Holmes, T. Vogt, S. M. Shapiro, S. Wakimoto, M. A. Subramanian and A. P. Ramirez, Phys. Rev. B **67** 092106 (2003).

[55] S. Tajima, T. Ido, S. Ishibashi, T. Itoh, H. Eisaki, Y. Mizauo, T. Arima, H. Takagi and S. Uchido, Phys. Rev. B **43,** 10496 (1991).

[56] I. S. Smirnova, Physica B **262,** 247 (1999).

[57] R. V. Aguilar, A. B. Sushkov, C. L. Zhang, Y. J. Choi, S.-W. Cheong and H. D. Drew, Phys. Rev. B **76**, 060404 (2007).

[58] J. F. Scott, Phys. Rev. B **4,** 1360 (1971).

[59] W. Cochran, Nature **191,** 60 (1961).

[60] N. S. Rogado, J. Li, A.W. Sleight and M. A. Subramanian, Adv. Mater. **17**, 2225 (2005).

[61] C. L. Bull and P. C. McMillan, J. Solid State Chem. **177**, 2323 (2004).

[62] C. L. Bull, D. Gleeson and K. S. Knight, J. Phys. Cond. Matt. **15**, 4927 (2003).

[63] H. Guo and J. Burgess, Appl. Phys. Lett. 89, 022509 (2006).

[64] M. P. Singh, K. D. Truong and P. Fournier, Appl. Phys. Lett. **91**, 042504 (2007).

[65] P. Padhan, H. Z. Guo, P. LeClair and A. Gupta, Appl. Phys. Lett. **92**, 022909 (2008).

[66] S.F. Mater, M. A. Subramanian, V. Eyert, M. Whangbo and A. Villesuzanne, J. Magn. Magn.Mater. **308**, 116 (2007).

[67] K. Asai, H. Sekizawa, K. Mizushima and S. Iida, J. Phys. Soc. Jpn. **47**, 1054 (1978).

[68] F. N. Sayed, S.N. Achary and O. D. Jayakumar, J. Mater. Res. **26**, 567 (2011).

[69] D. Choudhury, P. Mandal, R. Mathieu, A. Hazarika, S. Rajan, A. Sundaresan, U. V. Waghmare, R. Knut, O. Karis, P. Nordblad and D. D. Sarma, Phys. Rev. Lett. **108**, 127201 (2012).

[70] R. I. Dass, J. Q. Yan and J.B. Goodenough, Phys. Rev. B **68**, 064415 (2003).

[71] (a) K. D. Chandrasekhar, A. K. Das, C. Mitra and A. Venimadhav, J. Phys. Cond. Matt. **24**, 495901 (2012). (b) K.D. Chandrasekhar, A. K. Das and A. Venimadhav, J. Phys. Cond. Matt. **24**, 376003 (2012).

[72] H. Das, U.V. Waghmare, T. S. Dasgupta and D. D. Sarma, Phys. Rev. Lett. **100**, 186402 (2008).

[73] Z. Zhang, H. Jian, X. Tang, J. Yang, X. Zhu and Y. Sun, Dalton Trans. **41**,11836 (2012).

[74] M. P. Singh, K.D. Truong, S. Jandl and P. Fournier, J. Mag. Mag. Mater. **321**, 1743 (2009) and reference therein.

[75] M. N. Iliev, M. V. Abrashev, H. G. Lee, V. N. Popov, Y. Y. Sun, C. Thomsen, R. L. Meng and C. W. Chu, Phys. Rev. B **57**, 2872 (1998).

[76] R. Gupta, T. V. Pai, A. K. Sood, T. V. R. Ramakrishnan and C. N. R. Rao, Euro Phys. Lett. **58**, 778 (2002).

[77] P. Kumar, D. V. S. Muthu, S. N. Shirodkar, R. Saha, A. Shireen, A. Sundaresan, U. V. Waghmare, A. K. Sood and C. N. R. Rao, Phys. Rev. B **85**, 134449 (2012).

[78] M. P. Singh, K. D. Truong, S. Jandl and P. Fournier, arXiv: 0910.1108.

[79] M. N. Iliev, M. M. Gospodinov, M. P. Singh, J. Meen, K. D. Truong, P. Fournier and S. Jandl, arXiv: 0905.0202.

[80] M. N. Iliev, H. Guo and A. Gupta, Appl. Phys. Lett. **90**, 151914 (2007).

[81] M. P. Singh, K. D. Truong, S. Jandl and P. Fournier, Phys. Rev. B **79**, 224421 (2009).

[82] M. N. Iliev, M. V. Abrashev, M. V. Abrashev, A. P. Litvinchuk, V. G. Hadjiev, H. Guo, and A. Gupta, Phys. Rev. B **75**, 104118 (2007).

[83] M. N. Iliev, M. M. Gospodinov, M. P. Singh, J. Meen, K. D. Truong, P. Fournier and S. Jandl, J. App. Phys. **106**, 023515 (2009).

[84] K. D. Truong, J. Laverdiere, M. P. Singh, S. Jandl and P. Fournier, Phys. Rev. B **76**,

132413 (2007).

[85] A. E. Pantoja, H. J. Trodahl, A. Fainstein, R. G. Pregliasco, R. G. Buckley, G. Balakrishnan, M. R. Lees and D. M. Paul, Phys. Rev. B **63**, 132406 (2001).

# Chapter 5

# Summary and Conclusions

In this thesis, we have worked on two different families of complex materials, namely the iron-based superconductors (FeBS), multiferroic oxides and double perovskite. We studied the nature of quasi-particle excitations, such as phonons, magnons, orbitons etc., in these systems down to liquid 'He' temperature (4 K). Our studies have revealed strong coupling between different degrees of freedom highlighting a prominent role for magnetic and orbital degrees of freedom. Chapter 3 described our detailed studies on FeBS, our studies uncovered the ubiquitous role of spin-phonon coupling in these systems and established the non-degenerate nature of Fe $3d$-orbitals. Chapter 4 described our detailed studies on multiferroic oxides and double perovskite.

In **Part 3.1** we described our detailed temperature dependent Raman and first-principles density functional studies on superconductor $FeSe_{0.82}$. Our studies on this system established the nature of strong coupling between phonons and spin degrees of freedom below the structural and magnetic transition temperature (~100 K). Our studies also established that structural transition in these systems is not the primary order parameter. Rather, magnetic ordering which is a found to be the primary order parameter drives the structural transition. In addition, we also established the non-degenerate nature of Fe $d_{xz}$ and $d_{yz}$ orbitals, which have very prominent role in determining their magnetic, electronic as well as superconducting properties.

**Part 3.2** described our Raman study on superconductor $CeFeAsO_{0.9}F_{0.1}$. We suggested that the anomalous softening of the Raman active phonon mode (S4: 281 cm$^{-1}$, $E_g$, Fe) below $T_c$ is due to an opening of a superconducting gap, yielding $2\Delta/K_BT_c \sim 10$. We also observed signature of strong coupling between crystal field excitations of $Ce^{3+}$ and phonon modes, and their strong coupling also resulted in the observation of high order Raman excitations. In addition, we also observed two high energy excitations similar to those observed for $FeSe_{0.82}$ system clearly indicating the non-degenerate nature of Fe $d_{xz}$ and $d_{yz}$ orbitals.

**Part 3.3**, which is an extension of our studies in part 3.2, described our Raman studies on Y doped $CeFeAsO_{1-x}F_x$. In addition to our observation for $CeFeAsO_{0.9}F_{0.1}$ superconductor, we found the anomalous behavior of one of the Raman active phonon modes (S1: 146 cm$^{-1}$, $A_{1g}$, Ce/Y) below $T_c$ attributed to the strong coupling with the superconducting quasi-particle excitations due to the opening of superconducting gap below $T_c$. The estimated ratio of the superconducting gap to the superconducting transition temperature is ~5, suggesting multiple superconducting gaps in these systems.

In **Part 3.4** we presented our Raman studies on superconductor $Ca_4Al_2O_{5.7}Fe_2As_2$. Our studies revealed strong signature of superconducting fluctuation effect, setting in at almost ~ *2T$_c$*, reflected in anomalous temperature dependence of the observed Raman band. These strong precursor effects observed for the first time in FeBS were attributed to the existence of dynamic superconducting clusters at temperatures as high as *2T$_c$* and to the reduced overlap of the electronic wave function of the Fe-As layers in this compound due to large distance between the pnictides layers. The observed phonon mode also undergoes anomalous blue shift in the recorded temperature range attributed to the strong spin-phonon coupling. It will be exciting to look for these precursor effects in other measurements which may probe the local dynamics.

In **Part 3.5** we described our temperature dependent Raman studies on another family of FeBS, namely $Ca(Fe_{0.95}Co_{0.05})_2As_2$ also termed as "122" systems. From our Raman and first-principles studies we established that all the three observed phonon modes show anomalous temperature dependence due to strong spin-phonon coupling. The anomalous hardening of one of the Raman active mode (E$_g$: Fe and As) below $T_c$ is attributed to the coupling of the mode with the superconducting quasi-particle excitations, yielding an upper limit of *2Δ/k$_B$T$_c$* ~ 15, suggesting that this system belongs to the class of strong coupling superconductors. We note that the estimated ratio of superconducting gap to the transistion temperature in this system is maximum of the so far reported in literature.

**Part 3.6** described our Raman studies on single crystal of Ca(Fe$_{0.97}$Co$_{0.03}$)$_2$As$_2$ in a very wide spectral range from 120-5200 cm$^{-1}$. Our results provided clear evidence of the strong spin-phonon coupling as evidenced from the anomalous temperature dependence of the all the observed first-order phonon modes. The four weak modes observed in the range 400-1200 cm$^{-1}$ established the non-degenerate nature of the Fe *d$_{xz/yz}$* orbitals, similar to our earlier observation for other FeBS. In addition, the high frequency broad Raman band is associated with the coupled spin and orbital degrees of freedom, suggesting very intricate coupling between these two degrees of freedom.

**Part 4.1** described our detailed Raman and first-principles studies on multiferroic AlFeO$_3$. Renormalization of the first-order phonons below the magnetic transition temperature was attributed to the strong spin-phonon coupling. We also observed clear signature of magnetic excitation via observation of two-magnon Raman mode below the magnetic transition temperature, and estimated value of spin exchange constant *J$_0$* is ~ 5.3 meV, in close agreement with our DFT calculations. With first-principles analysis we explored the effects of magnetic ordering and (Al, Fe) disorder on phonons. Our studies revealed that anti-site disorder between Al and Fe results in the spin-state of Fe, and that a strong spin-phonon coupling gives rise to ferroelectric order, which makes it an unusual multiferroic. This potentially opens up a new class of multiferroics, where ionic disorder is central to their magneto-electric properties.

In **Part 4.2** and **4.3** we carried out detailed temperature dependent Raman and infrared reflectivity studies on multiferroic TbMnO$_3$. Our Raman studies of the first and second-order Raman modes revealed the existence of strong spin-phonon coupling and coupled nature of second-order phonons with the orbital degrees of freedom (orbitons) as reflected in the anomalous temperature dependence of the observed Raman modes. Our reflectivity studies also uncovered interesting anomalies in both longitudinal and transverse optic phonon modes below the magnetic transition temperature, $T_N$ (~ 46 K), attributed to the strong spin-phonon coupling.

In **Part 4.4** we presented our detailed temperature dependent magnetic, dielectric and Raman studies on double perovskite La$_2$NiMnO$_6$. Our magnetic and dielectric measurements clearly revealed the signature of two magnetic transitions suggesting existence of both monoclinic and rhombohedral phases. Our Raman studies revealed a strong renormalization of the first as well as second-order Raman modes associated with the (Ni/Mn)O$_6$ octahedra near magnetic transition temperature ($T_{C1}$ ~ 270 K) attributed to the strong spin-phonon coupling. In addition, we observed an anomaly in the temperature dependence of the frequency of the anti-symmetric stretching vibration of the octahedra associated with the spin-glass transition temperature (~70 K).